\documentclass[12pt]{article}
\usepackage[colorlinks, linkcolor=red, citecolor=blue]{hyperref}
\usepackage{amsmath}
\usepackage{amssymb}
\usepackage{float}
\usepackage{mathtools}     
\usepackage{times}
\usepackage{bm}
\usepackage{natbib}
\usepackage{colonequals}
\usepackage{graphicx}
\usepackage{threeparttable}
\usepackage{dcolumn}
\usepackage{multirow}
\usepackage{caption}
\usepackage{color}
\usepackage{authblk}
\usepackage{enumerate}
\usepackage{cmll}
\usepackage{booktabs}
\usepackage{geometry}
\usepackage{setspace}
\usepackage{amsthm}
\usepackage{authblk}

\usepackage{algorithm}
\usepackage{algpseudocode}

\usepackage{mathtools}     
\usepackage{color}

\newtheorem{theorem}{\bf{Theorem}}
\newtheorem{remark}{\bf{Remark}}
\newtheorem{lemma}{\bf{Lemma}}

\newtheorem{proposition}{\bf{Proposition}}

\usepackage{lipsum}
\usepackage{arydshln}

\usepackage{amsmath}
\usepackage{graphicx}
\usepackage{natbib}
\usepackage{url} 
\usepackage{authblk}

\newcommand{\blind}{1}

\usepackage{geometry}
\allowdisplaybreaks[3]
\usepackage{color}
\usepackage{algorithm}
\usepackage{algpseudocode}
\usepackage{epsfig}
\usepackage{dcolumn}
\usepackage{booktabs}
\usepackage{mathtools}
\usepackage{mathrsfs}
\usepackage{array}
\usepackage{scalerel}

\allowdisplaybreaks[4]

\hypersetup{hidelinks}
\graphicspath{{./figures/}}

\newcommand{\cM}{\scaleto{\mathcal{M}}{5pt}}
\newcommand{\cS}{\scaleto{\mathcal{S}}{5pt}}

\newcommand{\bX}{\boldsymbol{X}}

\newcommand{\bR}{\boldsymbol{R}}
\newcommand{\br}{\boldsymbol{r}}
\newcommand{\bs}{\boldsymbol{s}}
\newcommand{\btheta}{\boldsymbol{\theta}}

\newcommand{\bx}{\boldsymbol{x}}

\newcommand{\be}{\boldsymbol{e}}

\newcommand{\bxi}{\boldsymbol{\xi}}

\newcommand{\bzeta}{\boldsymbol{\zeta}}
\newcommand{\bnu}{\boldsymbol{\nu}}

\newcommand{\ba}{\boldsymbol{a}}
\newcommand{\bu}{\boldsymbol{u}}
\newcommand{\bb}{\boldsymbol{b}}
\newcommand{\bg}{\boldsymbol{g}}
\newcommand{\by}{\boldsymbol{y}}
\newcommand{\bY}{\boldsymbol{Y}}
\newcommand{\bS}{\boldsymbol{S}}
\newcommand{\bU}{\boldsymbol{U}}
\newcommand{\bzero}{\boldsymbol{0}}

\newcommand{\bc}{\boldsymbol{c}}
\newcommand{\bl}{\boldsymbol{l}}

\newcommand{\bh}{\boldsymbol{h}}
\newcommand{\bq}{\boldsymbol{q}}

\newcommand{\bZ}{\boldsymbol{Z}}
\newcommand{\bK}{\boldsymbol{K}}

\newcommand{\bM}{\boldsymbol{M}}
\newcommand{\bbm}{\boldsymbol{m}}
\newcommand{\bV}{\boldsymbol{V}}
\newcommand{\bbI}{\mathbf{I}}
\newcommand{\bbF}{\mathbf{F}}

\newcommand{\bbC}{\mathbf{C}}
\newcommand{\bbB}{\mathbf{B}}
\newcommand{\bbP}{\mathbf{P}}
\newcommand{\bbQ}{\mathbf{Q}}
\newcommand{\bbM}{\mathbf{M}}
\newcommand{\bbU}{\mathbf{U}}
\newcommand{\bbD}{\mathbf{D}}
\newcommand{\bbPi}{\mathbf{\Pi}}
\newcommand{\bbGamma}{\mathbf{\Gamma}}
\newcommand{\bbLambda}{\mathbf{\Lambda}}

\newcommand{\bt}{\boldsymbol{t}}
\newcommand{\bv}{\boldsymbol{v}}

\newcommand{\bpsi}{\boldsymbol{\psi}}
\newcommand{\bmu}{\boldsymbol{\mu}}
\newcommand{\bbeta}{\boldsymbol{\beta}}
\newcommand{\bbH}{\mathbf{H}}
\newcommand{\bbA}{\mathbf{A}}
\newcommand{\bbG}{\mathbf{G}}
\newcommand{\bbV}{\mathbf{V}}
\newcommand{\bbOmega}{\mathbf{\Omega}}
\newcommand{\tG}{\widetilde{\mathbf{G}}}
\newcommand{\tV}{\widetilde{\mathbf{V}}}
\newcommand{\tOmega}{\widetilde{\mathbf{\Omega}}}

\newcommand{\tB}{\widetilde{\mathbf{B}}}
\newcommand{\tA}{\widetilde{\mathbf{A}}}

\newcommand{\var}{{\rm var}}
\newcommand{\cov}{{\rm cov}}

\newcommand{\cU}{\mathcal{U}}

\newcommand{\cP}{\mathcal{P}}

\graphicspath{{figures/}} 	

\def\argmax{\mathop{\rm argmax}}
\def\argmin{\mathop{\rm argmin}}

\newcommand{\T}{{\rm T}}

\begin{document}
	\pagenumbering{gobble}

	\def\spacingset#1{\renewcommand{\baselinestretch}%
		{#1}\small\normalsize} \spacingset{1}

	
	\if1\blind
	{
		\title{\bf \fontsize{13}{13}\selectfont A Moment-assisted Approach for Improving Subsampling-based MLE with Large-scale Data
		}
	\author{\normalsize  Miaomiao Su, Qihua Wang, and Ruoyu Wang
	\thanks{Miaomiao Su and Qihua Wang contributed equally to this work, and Ruoyu Wang is the corresponding author. Miaomiao Su is a Lecturer in the School of Mathematical Sciences at Beijing University of Posts and Telecommunications ({\em smm@bupt.edu.cn}). Qihua Wang is a Professor in the School of Statistics and Data Science at Zhejiang Gongshang University, and the Academy of Mathematics and Systems Science, Chinese Academy of Sciences. Ruoyu Wang is an Assistant Professor in the Department of Statistics and Data Science at Tsinghua University ({\em ruoyuwang@mail.tsinghua.edu.cn}). M. Su's research was supported by the National Natural Science Foundation of China (Grant Number 12501405) and the Fundamental Research Fund of Beijing University of Posts and Telecommunications (Program Number 2023RC47). Q. Wang's research was supported by the National Natural Science Foundation of China (General Program 12271510, General Program 11871460, and Program for Innovative Research Group 61621003), and a grant from the Key Laboratory of Random Complex Structures and Data Science, Chinese Academy of Sciences.}}
\date{}
\maketitle
} \fi

\if0\blind
{
\bigskip
\bigskip
\bigskip
\begin{center}
	{\bf \fontsize{13}{13}\selectfont Moment-assisted subsampling method for Cox proportional hazards model with large-scale data}
\end{center}
\medskip
} \fi

\bigskip
\begin{abstract}
The maximum likelihood estimation is computationally demanding for large datasets, particularly when the likelihood function includes integrals. Subsampling can reduce the computational burden, but it often results in efficiency loss.
This paper proposes a moment-assisted subsampling (MAS) method that can improve the estimation efficiency of existing subsampling-based maximum likelihood estimators.
The motivation behind this approach stems from the fact that sample moments can be efficiently computed even if the sample size of the whole data set is huge.
Through the generalized method of moments, the proposed method incorporates informative sample moments of the whole data. 
The MAS estimator can be computed rapidly and is asymptotically normal with a smaller asymptotic variance than the corresponding estimator without incorporating sample moments of the whole data. 
The asymptotic variance of the proposed estimator depends on the specific sample moments incorporated. 
We derive the optimal moment that minimizes the resulting asymptotic variance in terms of Loewner order.
The proposed MAS estimator can achieve the same estimation efficiency as the whole data-based estimator when the optimal moment is incorporated.
Numerical results demonstrate the promising performance of the proposed method in both estimation and computational efficiency compared with existing subsampling methods.
\end{abstract}

\noindent%
{\it Keywords:}  Computational efficiency, Generalized method of moments,  Massive data, One-step approximation, Optimal moment function.
\vfill

\newpage
\pagenumbering{arabic}
\spacingset{1.7} 

\section{Introduction}
The maximum likelihood estimation (MLE) method has evolved into one of the most crucial tools for statisticians.
MLE is asymptotically efficient under very general conditions, providing extensive capabilities for estimation and inference. 
In the current era of big data, there has been a rapid surge in data volume.  Computation of MLE under many commonly used models such as the logistic regression, probit regression, and Poisson regression can be challenging when implementing on large-scale data \citep{Wang2018Subsampling, Yu2022Subsampling}. 
Moreover, the likelihood functions may involve integral operations in many important statistical problems such as the estimation under generalized linear mixed models (GLMMs) \citep{jiang2022usable}, latent variable models \citep{loaiza2022fast,forneron2023sieve}, missing data problems \citep{baker2019maximum}, and the empirical Bayesian framework \citep{atchade2011computational}. MLE faces significant computational hurdles in these cases. 
See Section \ref{subsec: Chicago car crash} for a real data example where the likelihood function for a large-scale dataset involves integral, and see Appendix \ref{app: detail car crash data} for the specific form of the likelihood function in that example. To address computational challenges, researchers might consider leveraging high-performance distributed computing systems. However, the cost and accessibility of such systems can be prohibitive, making them a less viable option for many people in the research community.
Therefore, it is important to develop methods for solving MLE with large-scale data that can be performed on commonly available devices such as laptops or personal computers.

One popular and powerful approach to mitigating the computational burden is subsampling, which draws a subsample from the whole data and performs statistical analysis based on the drawn subsample. Subsampling methods have been applied in various contexts to mitigate the computational difficulty of MLE \citep{Fithian2014Subsampling, Wang2018Subsampling, Wang2020LikeliEfficient}.
It offers a practical solution when analysis is performed on portable devices and is particularly useful in exploratory data analysis where users often make numerous attempts to understand the data and build the model. 
The simplest subsampling method is the uniform subsampling which draws each data point with equal probability.
The uniform subsampling estimator is computationally efficient but suffers from low estimation efficiency \citep{kushilevitz_nisan_1996,Drines2006Subsampling,Wang2020Subsampling}.

The primary objective of subsampling is to overcome computational bottlenecks while preserving, as much as possible, the statistical efficiency of whole data-based estimation.	
Existing subsampling methods mainly focus on designing nonuniform subsampling probabilities to include informative data points with higher probabilities.
A line of the nonuniform subsampling probabilities based methods is uncertainty sampling, which sets inclusion probabilities based on some measure of uncertainty associated with each observation \citep{Fithian2014Subsampling,Han2020Subsampling}.
Another line of the nonuniform subsampling probabilities based subsampling methods designs the nonuniform subsampling probability by minimizing the trace of the asymptotic variance matrix of the resulting estimator \citep{Wang2018Subsampling,Ai2021Subsampling,Yu2022Subsampling}. 
For the sampled data points, the inverse probability weighting method is often employed to estimate the target parameter \citep{Wang2018Subsampling}. 
Most existing subsampling methods primarily concentrate on linear or logistic regression \citep{Drines2006Subsampling,Fithian2014Subsampling,Wang2018Subsampling,Wang2019Subsampling, Wang2019MoreEfficient}. For approaches dealing with general models, the optimal subsampling probability is linked to the score function \citep{ting2018optimal, fan2022nearly}. As a result, when the likelihood and hence the score function involve integrals, computing the optimal subsampling probability for the whole dataset can be time-consuming due to the need for numerical integration.

Recently, some works have shifted focus from designing subsampling probabilities to developing estimation methods that utilize the subsample more effectively than the inverse probability weighting method. 
\cite{Wang2019MoreEfficient} and \cite{Wang2020LikeliEfficient} propose the maximum sampled conditional likelihood method, which achieves higher estimation efficiency than the inverse probability weighting method. However, the maximum sampled conditional likelihood method improves the estimation efficiency only when a nonuniform subsampling probability is adopted. This raises questions about their practicality in scenarios where calculating the optimal subsampling probability is challenging. \cite{fan2022nearly}  improve the inverse probability weighting estimator by incorporating the information in the whole data sample moments to improve the weights via the empirical likelihood method. 

This paper considers maximum likelihood estimation under a parametric conditional density model and proposes a moment-assisted subsampling (MAS) method. The method combines the likelihood score computed from a subsample with inexpensive sample moments computed from the whole dataset through the generalized method of moments. The MAS method is based on the fact that sample moments of the whole data are usually easy to compute, contain the information	of the population moment, and can be informative for the parameter of interest. Hence it can be used to improve the estimation efficiency of existing subsampling-based MLEs. The whole data sample moments consist of the average of some known function vector of observed variables, called the moment function vector, over the whole sample. Clearly, the estimation efficiency of the proposed estimator depends on the moment function vector.	Hence, we also consider the optimal selection problem of the moment function vector in order to achieve the highest estimation efficiency. When the estimating function is nonlinear, the generalized method of moments estimator often lacks a closed form and, hence, may be computationally intensive. To address this, we define the MAS estimator based on a linear approximation to the estimating function. The proposed estimator has an explicit expression, enabling it to be computed with minor additional computational burden compared to the original subsampling estimator. Theoretical analysis reveals the asymptotic normality of the proposed estimator, with a smaller asymptotic variance in Loewner order than the corresponding subsampling estimator without incorporating the whole data sample moment. This justifies the capability of the MAS method to improve the estimation efficiency of subsampling methods. The optimal moment function depends on the data generating process and might be challenging to calculate. We provide several easy-to-compute approximation methods for the optimal moment function. With the estimated optimal moment function and adequately large subsample sizes, the proposed estimator can achieve the same efficiency as the whole data-based MLE. When the estimated optimal moment function is employed and the subsample size is small, the MAS estimator still has a faster convergence rate than	the uniform subsampling estimator, but may not be asymptotically normal. We derive the explicit asymptotic distribution of the MAS estimator and propose an inference procedure tailored to this case. We demonstrate the promising performance of the proposed methods through extensive simulation studies and real data analyses. Our numerical results confirm	the efficiency gains achieved by incorporating whole data sample moments and underscore the substantial efficiency improvement when using the optimal moment function compared to other choices.

The MAS method offers several desirable features. Firstly, it is computationally efficient even when the likelihood function is complicated. In contrast, the optimal subsampling probability is related to the score function \citep{ting2018optimal}, which can be hard to calculate when the likelihood function involves integrals. The MAS method with uniform subsampling probability is particularly advantageous in these situations because it provides a way to improve the efficiency of subsampling estimators without calculating the optimal subsampling probability. 

Secondly, it can improve the estimation efficiency of the subsampling-based MLE under any given subsampling probability. When the optimal subsampling probability is easy to calculate, the proposed method can be used to further improve the estimation efficiency of the optimal subsampling probability-based subsampling estimator.
Moreover, the MAS method can reduce the asymptotic variance of every component of the subsampling estimator. On the contrary, existing nonuniform subsampling probability-based subsampling methods, such as those proposed by \cite{Wang2018Subsampling}, \cite{Yu2022Subsampling}, and \cite{ting2018optimal}, mainly focus on minimizing the trace of the asymptotic variance matrix. While effective in some scenarios, this approach may increase the variance of certain components compared to the uniform subsampling estimator.

The rest of this paper is organized as follows.
In Section \ref{sec: MAS}, we propose the MAS method and establish the asymptotic properties of the MAS estimator. In Section \ref{sec:opth}, we derive the optimal moment function vector, propose a approximation method for the optimal moment function when it is hard to compute, and establish the asymptotic distribution of the MAS estimator with an estimated optimal moment function under different regimes.
Simulation studies were conducted in Section \ref{sec: simulation}, followed by two real data examples in Section \ref{sec: real data}.

\section{Moment Assisted Subsampling Method}\label{sec: MAS}
Denote by $\bX$ a covariate vector and $\bY$ an outcome variable vector.
The data $\{(\bX_{i},\bY_{i})\}_{i=1}^{N}$ are independent and identically distributed observations of $(\bX,\bY)$.
Let $f(\by\mid \bx;\btheta_{0})$ be the conditional density of $\bY$ given $\bX = \bx$, where $\btheta_{0} \in \Theta$ is a $d$-dimensional true parameter and $\Theta$ is the parameter space. 
Then the whole data-based MLE for $\btheta_{0}$ is given by
\begin{equation*}\label{eq: whole}
	\hat{\btheta}=\argmax_{\btheta}\frac{1}{N}\sum_{i=1}^{N}\log f(\bY_{i}\mid \bX_{i};\btheta).
\end{equation*}
The MLE $\hat{\btheta}$ is consistent for $\btheta_{0}$ and asymptotically efficient under very general conditions, but its computation can be time-consuming if the sample size $N$ is extremely large, especially when the likelihood function $f(\by\mid \bx; \btheta)$ involves integrals. 
The likelihood function involves integrals in many important statistical problems. For an illustration, let us consider the mixed effects logistic regression, which is a special case of GLMM and widely adopted in clustered data analysis \citep{CHENG201338,SHIRAZI201610,ijerph192417097}. Suppose $\bY_{i} = (Y_{i}^{(1)}, \dots, Y_{i}^{(t)})^{\T}$ and $\bX_{i} = (\bX_{i}^{(1)\T},\dots, \bX_{i}^{(t)\T})^{\T}$ for some $t \geq 2$ and $i = 1,\dots, N$. Then, the whole data-based log-likelihood function under the mixed effects logistic model with a normal random intercept is
\begin{equation*}\label{eq: glmm likeli}
	\frac{1}{N}\sum_{i=1}^{N}\left[-\frac{1}{2}\log(2\pi\sigma^{2}) +\log\int\prod_{j = 1}^{t}\frac{\exp\{Y_{i}^{(j)}(\bX_{i}^{(j)\T}\bbeta + w)\}}{1+\exp(\bX_{i}^{(j)\T}\bbeta + w)}\exp\left(-\frac{w^{2}}{2\sigma^{2}}\right)dw\right],
\end{equation*} 
where $\btheta = (\sigma, \bbeta^{\T})^{\T}$ is the parameter of interest.
The log-likelihood function involves $N$ integral operations, which can be hard to compute. In certain scenarios,  the log-likelihood function may involve multiple integrals, e.g., under the multilevel mixed models \citep{amatya2018sample}. MLE faces significant computational hurdles in these cases. One popular and powerful approach to mitigating the computational burden is subsampling, which draws a subsample from the whole data and performs statistical analysis based on the drawn subsample.

\subsection{Incorporating Sample Moments}\label{subsec: vanilla MAS}
To be specific, we focus on Poisson subsampling.
For $i=1,\dots,N$, let $\delta_{i}$ be the inclusion indicator for the $i$th observation, where $\delta_{i}=1$ if $(\bX_{i},\bY_{i})$ is included in the subsample and $\delta_{i}=0$ otherwise.
Denote by $n$ the expected subsample size and $\rho_{N} =n/N$ the expected subsampling ratio.
In Poisson subsampling, $\{\delta_{i}\}_{i=1}^{N}$ are  generated from independent Bernoulli trials with $P(\delta_{i}=1\mid \bX_{i},\bY_{i})=\pi(\bX_{i},\bY_{i})$ for $i=1,\dots,N$, where  $\pi(\bx,\by)$ is the specified subsampling probability function satisfying $0< \pi(\bx,\by)\leq 1$ and $E\{\pi(\bX,\bY)\}=\rho_{N}$.  
The uniform subsampling is adopted if one simply set $\pi(\bx,\by) = \rho_{N}$.
Then the actual subample size is $\sum_{i=1}^{N}\delta_{i}$ which satisfies $E(\sum_{i=1}^{N}\delta_{i})=n$.
Throughout this paper, we use $\dot{\bl}$ to denote the gradient of a vector function $\bl$ to the parameter $\btheta$. Additionally, we use $\|\cdot\|$ to denote the Euclid/spectral norm for a vector/matrix. 

A standard way to construct subsampling estimators for $\btheta_{0}$ is to use the subsampling-based MLEs.
For instance, a common subsampling estimator for $\btheta_{0}$ is the inverse probability weighting subsample-based MLE \citep{Wang2018Subsampling} which maximizes the inverse probability weighted log-likelihood function
\begin{equation}\label{eq:ipw likeli}
	\sum_{i\in S}\pi(\bX_{i},\bY_{i})^{-1}\log f(\bY_{i}\mid\bX_{i};\btheta),
\end{equation}
where  $S=\{i\mid \delta_{i}=1,i=1,\dots,N\}$.
In this paper, we propose the MAS method to improve the estimation efficiency of the above subsampling-based MLE. Under regularity conditions, the subsampling-based MLE defined by maximizing \eqref{eq:ipw likeli} can be obtained by solving the estimating equation
\begin{equation}\label{eq: sub}
	\sum_{i\in S}\bu(\bX_{i},\bY_{i};\btheta)=\bzero,
\end{equation}
where $\bu(\bx,\by;\btheta)=\rho_{N}/ \pi(\bx,\by)\bpsi(\bx,\by;\btheta)$, $\bpsi(\bx,\by;\btheta)=\partial\log f(\by\mid \bx;\btheta)/\partial\btheta$ is the score function. We introduce the constant sequence $\rho_{N}$ in the definition of $\bu(\bx,\by;\btheta)$ to ensure that $\var\{\bu(\bX,\bY;\btheta)\}$ does not go to infinity as $N \to \infty$ which facilitates the subsequent demonstrations.  The solution to \eqref{eq: sub} is denoted by $\tilde{\btheta}$ and called the plain subsampling estimator. In effect, the proposed method can be applied to improve any subsampling estimator that can be expressed as the solution of an estimating equation of the form \eqref{eq: sub} with $\bu(\bx,\by;\btheta)$ being an estimating function such that $E\{\delta\bu(\bX, \bY; \btheta_{0})\} = \bzero$.

Next, we improve the efficiency of $\tilde{\btheta}$ defined by \eqref{eq: sub} via incorporating whole data sample moments.  
To incorporate whole data sample moments, we introduce  
\[\hat{\bmu}_{\bh} = N^{-1}\sum_{i=1}^{N}\bh(\bX_{i},\bY_{i}),\]where $\bh$ is some known $q$-dimensional moment function vector that can be specified by the analyst in subsampling for large-scale data problems.
The whole-data moment $\hat{\bmu}_{\bh}$ can be computed efficiently even when $N$ is extremely large provided that the value of $\bh$ is easy to calculate at each observation.
We first develop the proposed method under a general $\bh$. The determination of $\bh$ is investigated in Section \ref{sec:opth}.

Let $\bmu_{\bh} = E\{\bh(\bX,\bY)\}$ be the population counterpart of $\hat{\bmu}_{\bh}$.
We utilize the function $\bs(\bx,\by;\bmu_{\bh}) = \rho_{N}\pi(\bx,\by)^{-1}\{\bh(\bx,\by)-\bmu_{\bh}\}$ to incorporate whole data sample moments. The function $\bs(\bx,\by;\bmu_{\bh})$ is constructed by leveraging the inverse probability weighting technique so that the subsample moment $E\{\delta\bs(\bX,\bY;\bmu_{\bh})\} = \bzero$. Note that the moment $E\{\delta\bs(\bX,\bY;\bmu_{\bh})\}$ is relevant to $\btheta_{0}$ and hence equation $E\{\delta\bs(\bX,\bY;\bmu_{\bh})\} = \bzero$ contains information of $\btheta_{0}$. By plugging in the whole data estimate $\hat{\bmu}_{\bh}$ for $\bmu_{\bh}$, we obtain the estimating function $\bs(\bx,\by;\hat{\bmu}_{\bh})$. We then combine the estimating functions $\bu(\bx, \by; \btheta)$ and $\bs(\bx,\by;\hat{\bmu}_{\bh})$ using the generalized method of moments \citep{hansen1982} to improve efficiency for estimating the parameter $\btheta_{0}$. Specifically, define
\begin{equation*}
	\tilde{\bg}_{\cS}(\btheta)=\frac{1}{n}\sum_{i\in S}\left(\begin{matrix}
		\bu(\bX_{i},\bY_{i};\btheta)\\
		\bs(\bX_{i}, \bY_{i};\hat{\bmu}_{\bh})
	\end{matrix}\right).
\end{equation*}
Then, according the theory of generalized method of moments \citep{newey1994asymptotic}, the parameter of interest $\btheta_{0}$ can be efficiently estimated by solving
\begin{equation}\label{eq:MAS GMM}
	\min_{\btheta} \tilde{\bg}_{\cS}(\btheta)^{\T}\tOmega_{\cS}^{-1}\tilde{\bg}_{\cS}(\btheta),
\end{equation}
where
\begin{equation*}
	\tOmega_{\cS} = 
	\begin{pmatrix}
		\tOmega_{\bu\bu}& \tOmega_{\bu\bs}\\
		\tOmega_{\bs\bu}& \tOmega_{\bs\bs}
	\end{pmatrix}
\end{equation*}
is the estimated asymptotic variance of $\tilde{g}_{\cS}(\btheta_{0})$ with $\tOmega_{\bu\bu} = n^{-1}\sum_{i\in S}\bu(\bX_{i},\bY_{i};\tilde{\btheta})^{\otimes2}$,
\[
\tOmega_{\bu\bs} = \tOmega_{\bs\bu}^{\T} = \frac{1}{n}\sum_{i\in S}\bu(\bX_{i},\bY_{i};\tilde{\btheta})\left[\bs(\bX_{i},\bY_{i};\hat{\bmu}_{\bh})-\rho_{N}\{\bh(\bX_{i},\bY_{i})-\hat{\bmu}_{\bh}\}\right]^{\T},
\]
and
\[
\tOmega_{\bs\bs} = \frac{1}{n}\sum_{i\in S}\bs(\bX_{i},\bY_{i};\hat{\bmu}_{\bh})[\bs(\bX_{i},\bY_{i};\hat{\bmu}_{\bh})-\rho_{N}\{\bh(\bX_{i},\bY_{i})-\hat{\bmu}_{\bh}\}]^{\T}.
\]	
However, when $\tilde{\bg}_{\cS}(\btheta)$ is a nonlinear function of $\btheta$, the solution of \eqref{eq:MAS GMM} usually lacks closed form, and solving for it can be time-consuming. This problem can be mitigated by considering a linear approximation of $\tilde{\bg}_{\cS}(\btheta)$. To be specific, let $\tG_{\cS} = (\tG_{\cS,\bu}^{\T},\bzero^{\T})^{\T}$ where $\tG_{\cS,\bu}=n^{-1}\sum_{i\in S}\dot{\bu}(\bX_{i},\bY_{i};\tilde{\btheta})$.
We consider the following linear approximation of $\tilde{\bg}_{\cS}(\btheta)$
\begin{equation*}
	\tilde{\bg}_{\cS}(\btheta) \approx \tilde{\bg}_{\cS}+\tG_{\cS}^{\T}(\btheta-\tilde{\btheta}),
\end{equation*}
where $\tilde{\bg}_{\cS} = \tilde{\bg}_{\cS}(\tilde{\btheta})$.
Then, the MAS estimator is defined by
\begin{equation*}\label{eq:MAS GMM aproxi}
	\argmin_{\btheta}\left\{\tilde{\bg}_{\cS}+\tG_{\cS}^{\T}(\btheta-\tilde{\btheta})\right\}^{\T}\tOmega_{\cS}^{-1}\left\{\tilde{\bg}_{\cS}+\tG_{\cS}^{\T}(\btheta-\tilde{\btheta})\right\},
\end{equation*}
which has the explicit form
\begin{equation*}\label{est: MAS}
	\tilde{\btheta}_{\cS} = \tilde{\btheta}-\left(\tG_{\cS}^{\T}\tOmega_{\cS}^{-1}\tG_{\cS}\right)^{-1}\tG_{\cS}^{\T}\tOmega_{\cS}^{-1}\tilde{\bg}_{\cS}.
\end{equation*}	   	
Let $\tilde{\bg}_{\bs} =n^{-1}\sum_{i\in S}\bs(\bX_{i},\bY_{i};\hat{\bmu}_{\bh})$. Through some algebras, $\tilde{\btheta}_{\cS}$ has the following simple form
\begin{equation*}\label{est: rMAS}
	\tilde{\btheta}_{\cS} = \tilde{\btheta} + \tG_{\cS,\bu}^{-1}\tOmega_{\bu\bs}\tOmega_{\bs\bs}^{-1}\tilde{\bg}_{\bs}.
\end{equation*}

The proposed MAS estimator $\tilde{\btheta}_{\cS}$ has a calibration or control-variate form. Similar adjustment mechanisms have appeared in empirical likelihood, double sampling, model calibration, and data integration \citep{qin1994empirical,chen2000unified,wu2001model,chatterjee2016constrained,han2019empirical,wang2023survey}.
In survey calibration and data integration, auxiliary information is typically obtained from an data source where only a subset of variables or summary-level information is available. The primary goal in these literature is to improve estimation efficiency by leveraging the auxiliary information, often represented by summary statistics. The purpose of the MAS method, however, is different. In our setting, the complete data vector $(\bX,\bY)$ is observed for every unit, but iteratively evaluating the likelihood or score over the full dataset can be computationally prohibitive. The goal is to construct a easy-to-compute estimator with a high estimation accuracy. 
The distinct objective motivates different estimator constructions. To ensure computational efficiency, our whole data-based auxiliary summary statistic is restricted to the sample mean of a moment function which is computationally cheap. In contrast, existing adjustment methods face no such computational constraints and therefore has no such structural restriction on the summary statistic. Furthermore, the differences in the motivation and problem setting introduce unique theoretical and inferential challenges, which we discuss in the subsequent sections.

Next, we establish the asymptotic properties of $\tilde{\btheta}_{\cS}$. Let 
$\bbG_{\bu} =\rho_{N} ^{-1}E\{\delta\dot{\bu}(\bX,\bY;\btheta_{0})\}$ and $\bbG_{\cS} = (\bbG_{\bu}^{\T},\bzero^{\T})^{\T}$. Define 
$\bbOmega_{\bu\bu} = \rho_{N} ^{-1}\var\{\delta\bu(\bX,\bY;\btheta_{0})\} =\rho_{N} ^{-1}E\{\delta\bu(\bX,\bY;\btheta_{0})^{\otimes2}\},$
\[
\begin{aligned}
	\bbOmega_{\bu\bs} 
	= \bbOmega_{\bs\bu}^{\T}
	& = \rho_{N}^{-1} \cov[\delta\bu(\bX, \bY; \btheta_{0}), \delta \bs(\bX, \bY; \bmu_{\bh}) - \rho_{N}\{\bh(\bX, \bY) - \bmu_{\bh}\}]\\
	& =  \rho_{N}^{-1}E\{\delta\bu(\bX,\bY;\btheta_{0})[\bs(\bX,\bY;\bmu_{\bh})-\rho_{N}\{\bh(\bX,\bY)-\bmu_{\bh}\}]^{\T}\}
\end{aligned}
\]
and
\[
\begin{aligned}
	\bbOmega_{\bs\bs} & =
	\rho_{N}^{-1} \var[\delta \bs(\bX, \bY; \bmu_{\bh}) - \rho_{N}\{\bh(\bX, \bY) - \bmu_{\bh}\}] \\
	& = 
	\rho_{N}^{-1} E\{\delta\bs(\bX,\bY;\bmu_{\bh})[\bs(\bX,\bY;\bmu_{\bh})-\rho_{N}\{\bh(\bX,\bY)-\bmu_{\bh}\}]^{\T}\}.
\end{aligned} 
\]	
Then we have the following asymptotic normality result.

\begin{theorem}\label{theo: sMAS normal}
	Under Conditions (C1)--(C6) in Appendix \ref{app: regularity condtions}, as $n,N\to \infty$, we have
	\begin{equation*}
		\bbV_{\cS,\bh}^{-1/2}(\tilde{\btheta}_{\cS}-\btheta_{0})\to N(\bzero,\bbI)
	\end{equation*}
	in distribution, where $\bbV_{\cS,\bh}=n^{-1}\bbG_{\bu}^{-1}(\bbOmega_{\bu\bu}-\bbOmega_{\bu\bs}\bbOmega_{\bs\bs}^{-1}\bbOmega_{\bs\bu})\bbG_{\bu}^{-1}$ and  $\bbI$ is the  identity matrix.
\end{theorem}

The proposed estimator $\tilde{\btheta}_{\cS}$ is potentially more efficient than the plain subsampling estimator $\tilde{\btheta}$ in \eqref{eq: sub} since $\tilde{\btheta}_{\cS}$ incorporates the whole data-based moment information.
To clarify the efficiency gain, we compare $\bbV_{\cS,\bh}$ with the asymptotic variance $\bbV_{\cP}=n^{-1}\bbG_{\bu}^{-1}\bbOmega_{\bu\bu}\bbG_{\bu}^{-1}$ of the plain subsampling estimator $\tilde{\btheta}$ \citep{Wang2018Subsampling,Wang2020LikeliEfficient}.
It is not hard to see that $\bbV_{\cS,\bh}\leq \bbV_{\cP}$ for any $\bh$ and the inequality holds if $\|\bbG_{\bu}^{-1}\bbOmega_{\bu\bs}\|\neq 0$. 
Thus, the MAS estimator $\tilde{\btheta}_{\cS}$ is more efficient than the plain subsampling estimator $\tilde{\btheta}$ for any $\bh$ such that $\|\bbG_{\bu}^{-1}\bbOmega_{\bu\bs}\|\neq 0$. 
The moment function vector $\bh$ can affect the actual efficiency gain. The determination of $\bh$ is discussed in Section \ref{sec:opth}.

The MAS method can improve the efficiency of the subsampling-based MLE under any given subsampling probability. When the optimal subsampling probability that minimizes the trace of the asymptotic variance of the inverse probability weighted subsampling estimator \citep{Wang2018Subsampling, Yu2022Subsampling,ting2018optimal} is easy to calculate, the MAS method can be applied with the optimal subsampling probability to further improve the optimal subsampling probability-based subsampling estimator. On the other hand, when the optimal subsampling probability is hard to calculate, the MAS method with uniform subsampling probability provides a useful way to improve the efficiency of subsampling estimators without calculating the optimal subsampling probability.

\subsection{An Improved MAS Method}
Notice that $E(\hat{\bmu}_{\bh}) = \bmu_{\bh} = E[E\{\bh(\bX, \bY)\mid \bX\}]$. Let  $\ba(\bx;\btheta) = \int\bh(\bx, \by)f(\by\mid \bx; \btheta)d\by $. If the integration in $\ba(\bx;\btheta)$ can be calculated easily, we may use the moment equation
\begin{equation}\label{eq: population moment}
	E\{\ba(\bX;\btheta_{0})-\hat{\bmu}_{\bh}\} = \bzero
\end{equation}
to construct an estimating function instead of using the estimating function $\bs(\bx, \by; \hat{\bmu}_{\bh})$. 
A modified MAS estimator can be defined similarly to the standard one utilizing the moment condition \eqref{eq: population moment}. 
Based on the subsample, $\hat{\bmu}_{\bh}$ and Equation \eqref{eq: population moment}, an auxiliary estimating equation for $\btheta_{0}$ can be constructed as follows
\begin{equation}\label{eq: aux}
	\sum_{i\in S}\bbm(\bX_{i}, \bY_{i};\btheta,\hat{\bmu}_{\bh})=\bzero,
\end{equation}
where $\bbm(\bX_{i}, \bY_{i};\btheta,\hat{\bmu}_{\bh})=\rho_{N}\pi(\bX_{i},\bY_{i})^{-1}\left\{\ba(\bX_{i};\btheta)-\hat{\bmu}_{\bh}\right\}$. Compared to the moment condition $E\{\delta_i\bs(\bX_i, \bY_i; \hat{\bmu}_{\bh})\} = \bzero$ used in the standard MAS estimator $\tilde{\btheta}_{\cS}$, the moment condition $E\{\delta_i\bbm(\bX_{i}, \bY_{i};\btheta,\hat{\bmu}_{\bh})\} = \bzero$ explicitly incorporates the model $f(\by\mid \bx; \btheta)$ in its construction. This use of model information may potentially lead to efficiency gains.

The modified MAS estimator combines the estimating functions in \eqref{eq: sub} and \eqref{eq: aux} by the generalized method of moments. We first define the sample-level quantities involved in the definition of the modified MAS estimator $\tilde{\btheta}_{\cM}$. Let
\begin{equation*}
	\tilde{\bg}_{\cM}(\btheta)=\frac{1}{n}\sum_{i\in S}\begin{pmatrix}
		\bu(\bX_{i},\bY_{i};\btheta)\\
		\bbm(\bX_{i}, \bY_{i};\btheta,\hat{\bmu}_{\bh})
	\end{pmatrix}, \ \text{and}\
	\tOmega_{\cM}	
	=
	\begin{pmatrix}
		\tOmega_{\bu\bu} & \tOmega_{\bu\bbm}\\
		\tOmega_{\bbm\bu} & \tOmega_{\bbm\bbm}
	\end{pmatrix}
\end{equation*}
be the estimated asymptotic covariance matrix 
of $\tilde{\bg}_{\cM}(\btheta_{0})$ with
\begin{align*}
	\tOmega_{\bu\bu}
	& = \frac{1}{n}\sum_{i\in S}\bu(\bX_{i},\bY_{i};\tilde{\btheta})^{\otimes2},\\
	\tOmega_{\bu\bbm}
	&=\tOmega_{\bbm\bu}^{\T}=\frac{1}{n}\sum_{i\in S}\bu(\bX_{i},\bY_{i};\tilde{\btheta})\left[\bbm(\bX_{i},\bY_{i};\tilde{\btheta},\hat{\bmu}_{\bh})^{\T}-\rho_{N} \left\{\bh(\bX_{i},\bY_{i})-\hat{\bmu}_{\bh}\right\}^{\T}\right],\\
	\tOmega_{\bbm\bbm}
	&=\frac{1}{n}\sum_{i\in S}\Big[\bbm(\bX_{i},\bY_{i};\tilde{\btheta},\hat{\bmu}_{\bh})^{\otimes2}-\rho_{N} \bbm(\bX_{i}, \bY_{i};\tilde{\btheta},\hat{\bmu}_{\bh})\left\{\bh(\bX_{i},\bY_{i})-\hat{\bmu}_{\bh}\right\}^{\T}\\
	&\phantom{\frac{1}{n}\sum_{i\in S}}\qquad-\rho_{N} \left\{\bh(\bX_{i},\bY_{i})-\hat{\bmu}_{\bh}\right\}\bbm(\bX_{i},\bY_{i};\tilde{\btheta},\hat{\bmu}_{\bh})^{\T}\\
	&\phantom{\frac{1}{n}\sum_{i\in S}}\qquad+\rho_{N} ^2\pi(\bX_{i},\bY_{i})^{-1}\left\{\bh(\bX_{i},\bY_{i})-\hat{\bmu}_{\bh}\right\}^{\otimes2}\Big].
\end{align*}
Define $\tG_{\cM} = (\tG_{\cM,\bu}^{\T},\tG_{\bbm}^{\T})^{\T}$, where $\tG_{\cM,\bu} =-n^{-1}\sum_{i\in S}\bu(\bX_{i},\bY_{i};\tilde{\btheta})\bpsi(\bX_{i},\bY_{i};\tilde{\btheta})^{\T}$, and $\tG_{\bbm} = N^{-1}\sum_{i\in S}\pi(\bX_{i},\bY_{i})^{-1}\bh(\bX_{i}, \bY_{i})\bpsi(\bX_{i}, \bY_{i};\tilde{\btheta})^{\T}$. In Appendix \ref{app: detail mMAS}, we provide detailed explanation for the construction of $\tG_{\cM,\bu}$ and $\tG_{\bbm}$.
Let $\tilde{\bg}_{\cM} = \tilde{\bg}_{\cM}(\tilde{\btheta})$. Then, we define the modified MAS estimator as
\begin{equation*}\label{est: mMAS}
	\tilde{\btheta}_{\cM} = \tilde{\btheta}-\left(\tG_{\cM}^{\T}\tOmega_{\cM}^{-1}\tG_{\cM}\right)^{-1}\tG_{\cM}^{\T}\tOmega_{\cM}^{-1}\tilde{\bg}_{\cM}.
\end{equation*}

Next, we establish the asymptotic normality of $\tilde{\btheta}_{\cM}$ and compare its efficiency with $\tilde{\btheta}_{\cS}$. We define the population-level quantities required in the theoretical analysis of $\tilde{\btheta}_{\cM}$. Let $\bbG_{\cM} = (\bbG_{\bu}^{\T},\bbG_{\bbm}^{\T})^{\T}$, where $\bbG_{\bbm} = E[\bh(\bX, \bY)\bpsi(\bX, \bY;\btheta_{0})^{\T}]$. Define $\bbOmega_{\cM}$ be the variance-covariance matrix of 
$\rho_{N}^{-1/2}(\delta\bu(\bX, \bY; \btheta_{0})^{\T}, \delta\bbm(\bX, \bY; \btheta_{0}, \bmu_{\bh})^{\T} - \rho_{N}\{\bh(\bX, \bY) - \bmu_{\bh}\}^{\T})^{\T}.$
Specifically, 
\begin{equation*}\label{eq:Omega}
	\begin{split}
		\bbOmega_{\cM}
		&=
		\begin{pmatrix}
			\bbOmega_{\bu\bu} & \bbOmega_{\bu\bbm}\\
			\bbOmega_{\bbm\bu} & \bbOmega_{\bbm\bbm}
		\end{pmatrix},	
	\end{split}
\end{equation*}
where
$$\bbOmega_{\bu\bbm} =\bbOmega_{\bbm\bu}^{\T} =\rho_{N} ^{-1}E\left(\delta\bu(\bX,\bY;\btheta_{0})\left[\bbm(\bX, \bY;\btheta_{0},\bmu_{\bh})-\rho_{N} \left\{\bh(\bX,\bY)-\bmu_{\bh}\right\}\right]^{\T}\right),$$
and
\begin{align*}
	\bbOmega_{\bbm\bbm}
	&=\rho_{N} ^{-1}E\left\{\delta\bbm(\bX,\bY;\btheta_{0}, \bmu_{\bh})^{\otimes2}\right\}-E\left[\delta\bbm(\bX, \bY;\btheta_{0},\bmu_{\bh})\left\{\bh(\bX,\bY)-\bmu_{\bh}\right\}^{\T}\right]\\
	&\quad-E\left[\delta\left\{\bh(\bX,\bY)-\bmu_{\bh}\right\}\bbm(\bX, Y;\btheta_{0},\bmu_{\bh})^{\T}\right]+\rho_{N} E\left[\left\{\bh(\bX,\bY)-\bmu_{\bh}\right\}^{\otimes2}\right].
\end{align*}

\begin{theorem}\label{theo: mMAS normal}
	Under Conditions (C1)--(C3), (C4)(i), and (C6)--(C9) in Appendix \ref{app: regularity condtions}, as $n,N\to\infty$,  we have
	\begin{equation*}
		\bbV_{\cM,\bh}^{-1/2}(\tilde{\btheta}_{\cM}-\btheta_{0})\to N(\bzero,\bbI)
	\end{equation*}
	in distribution, where $\bbV_{\cM,\bh}=n^{-1}(\bbG_{\cM}^{\T}\bbOmega_{\cM}^{-1}\bbG_{\cM})^{-1}$.
\end{theorem}

Theorem \ref{theo: mMAS normal} establishes the asymptotic normality of $\tilde{\btheta}_{\cM}$. Similar to the discussions behind Theorem \ref{theo: sMAS normal}, we can show that $\bbV_{\cM,\bh}\leq \bbV_{\cP}$ for any $\bh$, where $\bbV_{\cP}=n^{-1}\bbG_{\bu}^{-1}\bbOmega_{\bu\bu}\bbG_{\bu}^{-1}$ is the asymptotic variance of the plain subsampling estimator $\tilde{\btheta}$.
To clarify the efficiency gain of the modified MAS estimator, we compare $\bbV_{\cM,\bh}$ with the asymptotic variance $\bbV_{\cP}$ of the plain subsampling estimator $\tilde{\btheta}$. Recall that $\bbV_{\cM,\bh}=n^{-1}(\bbG_{\cM}^{\T}\bbOmega_{\cM}^{-1}\bbG_{\cM})^{-1}$ and $\bbV_{\cP}=n^{-1}(\bbG_{\bu}^{\T}\bbOmega_{\bu\bu}^{-1}\bbG_{\bu})^{-1}$.
By some algebra, we have
\begin{equation*}
	\bbG_{\cM}^{\T}\bbOmega_{\cM}^{-1}\bbG_{\cM}-\bbG_{\bu}^{\T}\bbOmega_{\bu\bu}^{-1}\bbG_{\bu}=(\bbG_{\bu}^{\T}\bbOmega_{\bu\bu}^{-1}\bbOmega_{\bu\bbm}-\bbG_{\bbm}^{\T})\bbOmega_{\cM}^{22}(\bbG_{\bu}^{\T}\bbOmega_{\bu\bu}^{-1}\bbOmega_{\bu\bbm}-\bbG_{\bbm}^{\T})^{\T}\geq 0,
\end{equation*}
where $\bbOmega_{\cM}^{22}=(\bbOmega_{\bbm\bbm}-\bbOmega_{\bbm\bu}\bbOmega_{\bu\bu}^{-1}\bbOmega_{\bu\bbm})^{-1}$.
This implies that $\bbV_{\cM,\bh}\leq \bbV_{\cP}$ in Loewner order and the inequality holds if $\|\bbG_{\bu}^{\T}\bbOmega_{\bu\bu}^{-1}\bbOmega_{\bu\bbm}-\bbG_{\bbm}^{\T}\| \neq 0$.
In summary, leveraging the whole data sample moment $\hat{\bmu}_{\bh}$ yields a more efficient estimator than the plain subsampling estimator for any specified $\bh(\bx,\by)$ such that $\|\bbG_{\bu}^{\T}\bbOmega_{\bu\bu}^{-1}\bbOmega_{\bu\bbm}-\bbG_{\bbm}^{\T}\| \neq 0$.

The next theorem compares the efficiency of the standard MAS estimator $\tilde{\btheta}_{\cS}$ and the modified MAS estimator $\tilde{\btheta}_{\cM}$. 
Define $\mathcal{H}_{\cS}$ and $\mathcal{H}_{\cM}$  as the sets of functions $\bh$ such that $\bbV_{\cS,\bh}$ and $\bbV_{\cM,\bh}$ exist and are non-singular, respectively.
\begin{theorem}\label{theo: MAS compare}	 	
	Suppose $\pi(\bx,\by)=n/N$. For any moment function vector $\bh\in\mathcal{H}_{\cS}\cap\mathcal{H}_{\cM}$, we have $\bbV_{\cM,\bh}\leq \bbV_{\cS,\bh}$ when $\rho_{N}\leq 1/2$ and $\bbV_{\cM,\bh}\geq \bbV_{\cS,\bh}$ when $\rho_{N}> 1/2$.
\end{theorem}

Theorem \ref{theo: MAS compare} shows that the modified MAS estimator $\tilde{\btheta}_{\cM}$ is more efficient than the standard MAS estimator $\tilde{\btheta}_{\cS}$ when $\rho_{N}\leq 1/2$. The sampling fraction $\rho_{N}$ is typically very small when implementing subsampling methods. In this case, $\tilde{\btheta}_{\cM}$ is more efficient than $\tilde{\btheta}_{\cS}$. Although Theorem \ref{theo: MAS compare} is obtained under the uniform subsampling probability, our simulation results suggest the modified MAS estimator $\tilde{\btheta}_{\cM}$ still exhibits superior efficiency than the standard MAS estimator $\tilde{\btheta}_{\cS}$ when a nonuniform subsampling probability is used and $\rho_{N}$ is small. However, the complexity of the asymptotic variance's expression has precluded us from deriving a formal theoretical result. 

The asymptotic variances $\bbV_{\cS,\bh}$ and $\bbV_{\cM,\bh}$ can be estimated by 
$\tV_{\cS,\bh}=n^{-1}\tG_{\cS,\bu}^{-1}(\tOmega_{\bu\bu}-\tOmega_{\bu\bs}\tOmega_{\bs\bs}^{-1}\tOmega_{\bs\bu})\tG_{\cS,\bu}^{-1}$ and 
$\tV_{\cM,\bh}=n^{-1}(\tG_{\cM}^{\T}\tOmega_{\cM}^{-1}\tG_{\cM})^{-1}$, respectively. 
We prove  $\|\tV_{\cS,\bh}^{-1}\bbV_{\cS,\bh}-\bbI\|=o_{P}(1)$ and  $\|\tV_{\cM,\bh}^{-1}\bbV_{\cM,\bh}-\bbI\|=o_{P}(1)$ in Appendix \ref{app: var est}.

\section{Determination of the Moment Function Vector}\label{sec:opth}

\subsection{Optimal Moment Function Vector}\label{subsec: opt moment}
According to Theorems \ref{theo: sMAS normal} and \ref{theo: mMAS normal}, the asymptotic variances $\bbV_{\cS,\bh}$ and $\bbV_{\cM,\bh}$ are influenced by $\bh(\bx,\by)$.
Recall that $\mathcal{H}_{\cS}$ and $\mathcal{H}_{\cM}$  are the sets of functions $\bh$ such that $\bbV_{\cS,\bh}$ and $\bbV_{\cM,\bh}$ exist and are non-singular.
To guide the choice of $\bh(\bx,\by)$, we theoretically characterize  the optimal moment function vectors that minimize asymptotic variances $\bbV_{\cS,\bh}$ and $\bbV_{\cM,\bh}$ in Loewner order over $\mathcal{H}_{\cS}$ and $\mathcal{H}_{\cM}$, respectively. 

%


\begin{theorem}\label{theo: opt h}
	The moment function $\bh(\bx,\by)$ minimizes $\bbV_{\cS,\bh}$ in Loewner order over $\mathcal{H}_{\cS}$ if and only if there exists some matrix $\bbM$ such that $ \bbM[\bh(\bx,\by)-E\{\bh(\bX,\bY)\}] = \bpsi(\bx,\by;\btheta_{0})$.
	When $\pi(\bx,\by)=n/N$, the moment function $\bh(\bx,\by)$ minimizes $\bbV_{\cM,\bh}$ in Loewner order over $\mathcal{H}_{\cM}$ if and only if there exists some matrix $\bbM$ such that $ \bbM[\bh(\bx,\by)-E\{\bh(\bX,\bY)\}] = \bpsi(\bx,\by;\btheta_{0})$.
\end{theorem}

Theorem \ref{theo: opt h} provides the optimal $\bh$ for the standard MAS estimator under general subsampling probability.
For the modified MAS estimator, Theorem \ref{theo: opt h} provides the optimal $\bh$ when the uniform subsampling is adopted. 
Theorem \ref{theo: opt h} implies that the asymptotic variances $\bbV_{\cS,\bh}$ and $\bbV_{\cM,\bh}$ attain the minimum if we take the score function $\bpsi(\bx,\by;\btheta_{0})$ as the moment function vector.
For the case with a nonuniform subsampling probability, the expression of $\bbV_{\cM,\bh}$ is too complicated to obtain an explicit form of the optimal moment function vector. 
In this case, we still recommend taking the score function as the moment function vector due to its great numerical performance in our simulations.
The score function $\bpsi(\bx,\by;\btheta_{0})$ involves the true model parameter $\btheta_{0}$ which requires estimating.
To address this, we propose to use $\tilde{\bh}^{\rm opt}(\bx,\by) =\bpsi(\bx,\by;\tilde{\btheta})$ instead of $\bpsi(\bx,\by;\btheta_{0})$ in implementation. 

\begin{remark}
	Characterizing optimal estimating or calibration functions is an important topic in the survey sampling literature. For example, \citet{morikawa2025semiparametric} derive optimal estimating functions in survey sampling through an explicit infinite-dimensional Hilbert-space projection, and earlier studies \citep{godambe1976conditional,kim2013weighting} investigate optimal estimating or weighting functions under informative sampling. Closely related to our work, Corollary 1 of \citet{wang2023survey} characterizes the optimal calibration function as the conditional expectation of the full data estimating function given the variables available from the auxiliary source. From this perspective, our Theorem~\ref{theo: opt h} can be viewed as a full-information specialization of this principle: in our setting, the complete data vector is observed, so the relevant information set is the entire observation, and the optimal moment space reduces to the span of the full-data score.	
	Nevertheless, the perspectives of large-scale data subsampling and survey sampling differ substantially. \citet{wang2023survey} focus on how to effectively leverage partial auxiliary information. In contrast, our setting involves fully observed data, and the central challenge is instead to leverage full-data information in a computationally efficient manner. To this end, we take the full-data summaries to be the sample means of the moment functions evaluated over the full dataset, rather than quantities obtained by solving a possibly nonlinear estimating equation based on the full dataset as in \citet{wang2023survey}. The latter can be computationally prohibitive in large-scale settings and hence is not suitable for our problem.
\end{remark}

\begin{remark}
	Define $\bh_{\btheta}(\bX,\bY)=\bpsi(\bX,\bY;\btheta)$ and $\bM(\btheta)=E\{\bh_{\btheta}(\bX,\bY)\}$. 
	For \(\btheta=\btheta_0\), the corresponding population moment is $\bmu_0=\bM(\btheta_0)=0$. At first sight, this may suggest that incorporating the optimal moment function \(\bh_{\btheta_0}\) cannot improve the efficiency of the initial subsampling estimator \(\tilde{\btheta}\). The key point, however, is that the useful information is not directly contained in \(\bM(\btheta_0)\) (which equals 0) itself, but in the moment \(\bM(\tilde{\btheta})\).	To see why \(\bM(\tilde{\btheta})\) is informative,
	a first-order expansion gives
	\[
	\bM(\tilde{\btheta})
	=
	\bbG(\tilde{\btheta}-\btheta_0)
	+
	O_P\!\left(\|\tilde{\btheta}-\btheta_0\|^2\right),
	\]
	where $\bbG$ is the gradient of $\bM$ at $\btheta_0$.
	Therefore, \(\bM(\tilde{\btheta})\) carries first-order information about the estimation error \(\tilde{\btheta}-\btheta_0\). Given $\bM(\tilde{\btheta})$, one can denoise the initial estimator $\tilde{\btheta}$ by defining the estimator
	$\tilde{\btheta} - \widetilde{\bbG}^{-1}\bM(\tilde{\btheta})$	
	with $\widetilde{\bbG}$ being a subsample-based consistent estimator for $\bbG$.
	Then,
	\[
	\tilde{\btheta} - \widetilde{\bbG}^{-1}\bM(\tilde{\btheta}) - \btheta_{0} = (\bbI - \widetilde{\bbG}^{-1}\bbG)(\tilde{\btheta}-\btheta_0) + O_P(\|\tilde{\btheta}-\btheta_0\|^2)
	= o_P(\|\tilde{\btheta}-\btheta_0\|),
	\]
	which demonstrates the effectiveness of the moment information $\bM(\tilde{\btheta})$ in improving the convergence rate of the initial estimator $\tilde{\btheta}$. Next, we show that the MAS estimator is making use of the moment information  $\bM(\tilde{\btheta})$ when the optimal moment function is adopt. 
	Let 
	\begin{align*}
		\bS_{\rm opt} 
		&= -n^{-1}\sum_{i\in S}\bs(\bX_{i}, \bY_{i};\hat{\bmu}_{\bh})\\
		& = -N^{-1}\sum_{i=1}^{N}  \pi(\bX_i,\bY_i)^{-1}\delta_i\bh_{\btheta_0}(\bX_i,\bY_i) + N^{-1}\sum_{i=1}^{N}  \pi(\bX_i,\bY_i)^{-1}\delta_i\times N^{-1}\sum_{i = 1}^{N}\bh_{\btheta_0}(\bX_i,\bY_i)
	\end{align*}
	be the auxiliary moment used in the definition of the MAS estimator when the optimal moment function $\bh_{\btheta_0}$ is adopt. Assuming the subsample size $n$ satisfies $n/N\to 0$, we show in the Appendix \ref{app:illustration of opth} that $\bS_{\rm opt} - \bM(\tilde{\btheta}) = o_P(1/\sqrt{n})$ leveraging the asymptotic expansion of $\tilde{\btheta}$, which indicates that the auxiliary moment equal to the population moment $\bM(\tilde{\btheta})$ up to some higher order terms. Therefore, when the optimal moment function is adopted, the MAS can improve efficiency by (implicitly) leveraging the moment information $\bM(\tilde{\btheta})$. In addition, the above estimation error results for $\bS_{\rm opt} - \bM(\tilde{\btheta})$ remains valid if the moment function $\bh_{\btheta_0}$ in the definition of $\bS_{\rm opt}$ is replaced by $\bh_{\hat{\btheta}_n}$ for any $\sqrt{n}$-consistent estimator $\hat{\btheta}_n$ for $\btheta_{0}$, which provides a justification of the plug-in estimate for the optimal moment function.	
\end{remark}

One may observe that the forms of $\tilde{\btheta}_{\cS}$ and $\tilde{\btheta}_{\cM}$ with the estimated optimal moment function $\tilde{\bh}^{\rm opt}(\bx,\by) =\bpsi(\bx,\by;\tilde{\btheta})$ resemble the one-step update from $\tilde{\btheta}$ using Newton's method. In this sense, the proposed MAS methods with the estimated optimal moment function under uniform subsampling may be viewed as modifications and extensions of Newton’s method, while exhibiting better numerical performance when the subsample size is small.		
More importantly, the MAS methods differ from the Newton one-step update in their greater flexibility. When the score function is computationally expensive to evaluate, as in GLMMs models, the MAS methods can instead incorporate easy-to-compute sample moments instead of the original model score function, while still achieving high estimation efficiency. To improve efficiency while retaining computational advantage, in the next subsection we consider an easy-to-compute parametric approximation to the original score function.		
Please refer to Appendix \ref{app: Newton discuss} and \ref{app: compr with Newton} for more details about the connections and advantages of the MAS methods compared to Newton's method. 
Additionally, please refer to Appendix \ref{app: further comparison} for further comparisons with related works.
\begin{remark}\label{rmk: mMAS opth logit}
	The auxiliary estimating equation \eqref{eq: aux} requires the calculation of the conditional expectation of $\bh(\bX,\bY)$ given $\bX=\bx$, which involves an integral operation. Computing integrals are often numerically time-consuming.
	Fortunately,  for the optimal moment function $\tilde{\bh}^{\rm opt}(\bx,\by)$, the integral $\int\tilde{\bh}^{\rm opt}(\bx,\by)f(\by\mid \bx; \btheta)d\by$ has an explicit expression under many commonly used models, for example, the generalized linear model and the Weibull regression model. 
	Here, we provide the expression under the generalized linear model.
	We consider the generalized linear model with canonical link function
	\begin{equation}\label{eq: glm}
		f(y\mid \bx,\btheta) = w(y)\exp\{y \bx^{\T}\btheta-c(\bx^{\T}\btheta)\}
	\end{equation}
	where $w(\cdot)$ and $c(\cdot)$ are known scalar functions. 
	The model \eqref{eq: glm} reduces to the logistic regression model, when $w(y)=1$ and $c(\bx^{\T}\btheta)=\log\{1+\exp(\bx^{\T}\btheta)\}$.
	Under the generalized linear model, we have $\tilde{\bh}^{\rm opt}(\bx,y) = \{y-\dot{c}(\bx^{\T}\tilde{\btheta})\}\bx$ and $\int\tilde{\bh}^{\rm opt}(\bx,y)f(y\mid \bx; \btheta)dy=\{\dot{c}(\bx^{\T}\btheta)-\dot{c}(\bx^{\T}\tilde{\btheta})\}\bx$, where $\dot{c}$ is the derivative function of $c$. The explicit expression of $\int\tilde{\bh}^{\rm opt}(\bx,y)f(y\mid \bx; \btheta)dy$ is also available under the Weibull regression model. Please refer to Appendix \ref{app: sim Weibull} for more details.
\end{remark}

\subsection{Approximation of the Optimal Moment Function}\label{subsec: approx hopt}

Now, we delve into scenarios where $\bpsi$ is computationally intensive, such as estimating parameters under a GLMM.
When the score function involves integrals, we can calculate some easy-to-compute sample moments instead of the score function of the original model. 
Notably, for any moment function, the MAS methods consistently outperform the plain subsampling estimator in terms of estimation efficiency. 
In order to improve efficiency while keeping the computational efficiency, it is sensible to consider an easy-to-compute parametric approximation of the original score function. 

Building on Theorem~\ref{theo: sMAS normal}, we propose a practical approach that approximates the optimal moment function by minimizing the estimated asymptotic variance over a computationally tractable parametric function class.
Let $W = \pi(\bX,\bY)^{-1} - 1$.
For any functions $\boldsymbol{f}_{1}$ and $\boldsymbol{f}_{2}$, let
\[
\begin{aligned}
	&\bbPi_{\boldsymbol{f}_{2}(\bX, \bY)}\{\boldsymbol{f}_{1}(\bX,\bY)\}\\
	& = E\left\{W\boldsymbol{f}_{1}(\bX,\bY)\boldsymbol{f}_{2}(\bX,\bY)^{\T}\right\}\left[E\left\{W\boldsymbol{f}_{2}(\bX,\bY)\boldsymbol{f}_{2}(\bX,\bY)^{\T}\right\}\right]^{-1}\boldsymbol{f}_{2}(\bX,\bY)
\end{aligned}
\]
be the projection of $\boldsymbol{f}_{1}$ onto the space spanned by $\boldsymbol{f}_{2}$.
The asymptotic variance of the MAS estimator $\tilde{\btheta}_{\cS}$ can be written as 
\begin{equation}\label{eq: variance analysis}
	\begin{split}
		\bbV_{\cS,\bh} 
		& = n^{-1} \bbG_{\bu}^{-1} \left( \bbOmega_{\bu\bu} - \bbOmega_{\bu\bs} \bbOmega_{\bs\bs}^{-1} \bbOmega_{\bs\bu} \right) \bbG_{\bu}^{-1}\\
		&= N^{-1}\bbG_{\bu}^{-1}E\left\{\bpsi(\bX,\bY;\btheta_{0}) \bpsi(\bX,\bY;\btheta_{0})^{\T}\right\}\bbG_{\bu}^{-1} \\
		&\quad + n^{-1} \bbG_{\bu}^{-1} E \left(W\left[\bpsi(\bX,\bY;\btheta_{0}) - \bbPi_{\bh(\bX, \bY)}\{\bpsi(\bX,\bY; \btheta_{0})\} \right]^{\otimes 2} \right) \bbG_{\bu}^{-1}.
	\end{split}
\end{equation}
The only term in $\bbV_{\cS,\bh}$ relevant to $\bh$ is $E \left(W\left[\bpsi(\bX,\bY;\btheta_{0}) - \bbPi_{\bh(\bX, \bY)}\{\bpsi(\bX,\bY; \btheta_{0})\} \right]^{\otimes 2} \right)$ which is positive semi-definite.
When the optimal moment function $\bpsi(\bX,\bY;\btheta_{0})$ is difficult to compute, we consider approximating it using an easy-to-compute parametric working model $\bq(\bx, \by; \bxi)$ where $\bxi$ is the model parameter. 
Motivated by \eqref{eq: variance analysis}, we propose to determine $\bxi$ by minimizing a plug-in estimate of 
\[
\begin{aligned}
	&{\rm tr} \left\{E \left(W\left[\bpsi(\bX,\bY;\btheta_{0}) - \bbPi_{\bq(\bX, \bY; \bxi)}\{\bpsi(\bX,\bY; \btheta_{0})\} \right]^{\otimes 2} \right)\right\}\\
	&= E \left(W\left\|\left[\bpsi(\bX,\bY;\btheta_{0}) - \bbPi_{\bq(\bX, \bY; \bxi)}\{\bpsi(\bX,\bY; \btheta_{0})\} \right]\right\|^{2} \right)
\end{aligned}
\]
based on the subsample $S$ and denote the resulting model parameter by $\tilde{\bxi}$. Although the estimate of the above term may involve integrals, computation of the estimate is not prohibitively hard because only integrals on a small subsample are computed. When the score function itself is easy to compute, of course, one can take $\bq(\bx, \by; \bxi)$ as the score function $\bpsi(\bx, \by; \bxi)$ and take $\tilde{\bxi}$ as the plain subsampling estimator $\tilde{\btheta}$. 

One reasonable parametric class is the score function of a reduced model. For instance, by setting the variance of random effects in a GLMM to be zero, the GLMM reduces to a generalized linear model whose score function is easy to compute. The score of such a generalized linear model serves as a reasonable approximation for the score of the original model. This approximation strategy is adopted in our simulation studies and real data analysis under GLMMs. The MAS estimators are expected to substantially improve estimation efficiency when the reduced model is close to $f(\by\mid\bx;\btheta_{0})$. 
Please refer to Appendix \ref{app: approxi details for glmm} for more details on the working parametric model $\bq(\bx, \by; \bxi)$ to approximate the score function under the GLMM.

The above approach to approximating the optimal moment function is analogous to the use of a working reduced model in model calibration \citep{wu2001model,wang2023survey}. In model calibration and data integration, the working reduced model typically extracts partial information from auxiliary sources. In our setting, however, the working reduced model is introduced solely as a computational device for constructing a moment function that can be evaluated inexpensively over the full dataset, thereby avoiding repeated evaluations of the computationally intensive full-model score. Our numerical studies in Sections~\ref{sec: simulation} and~\ref{sec: real data} confirm that this approximation yields substantial computational benefits when the likelihood involves intractable integrals.

\subsection{Asymptotic Properties with an Estimated Moment Function Vector}

Next, we establish the asymptotic properties of MAS estimators with an estimated moment function which has the form $\tilde{\bh}(\bx,\by) = \bq(\bx,\by;\tilde{\bxi})$, where $\bq(\bx,\by;\bxi)$ is a user-specified working model with unknown parameter $\bxi$ and $\tilde{\bxi}$ is an estimation of $\bxi$ obtained based on the subsample $S$. 
For example, $\tilde{\bh}(\bx,\by)$ can be taken as the estimated optimal moment function $\bpsi(\bx,\by;\tilde{\btheta})$.
We denote the $\tilde{\bh}(\bx,\by)$-based standard and modified MAS estimators by $\tilde{\btheta}_{\cS,\tilde{\bh}}$ and $\tilde{\btheta}_{\cM,\tilde{\bh}}$, respectively.
Let $\bh^{*}(\bx,\by) = \bq(\bx,\by;\bxi^{*})$, where $\bxi^{*}$ is the probability limit of $\tilde{\bxi}$.
For the standard MAS estimator, $\bbOmega_{\bu\bs,\bh^{*}}$, $\bbOmega_{\bs\bs,\bh^{*}}$ and $\bbOmega_{\bs\bu,\bh^{*}}$ are defined in the same way as $\bbOmega_{\bu\bs}$, $\bbOmega_{\bs\bs}$ and $\bbOmega_{\bs\bu}$, respectively, with $\bh$ replaced by $\bh^{*}$.
For the modified MAS estimator, $\bbG_{\cM,\bh^{*}}$ and $\bbOmega_{\cM,\bh^{*}}$ are defined in the same way as $\bbG_{\cM}$ and $\bbOmega_{\cM}$, respectively, with $\bh$ replaced by $\bh^{*}$.

\begin{theorem}\label{theo: asy tilde S}
	Under conditions of Theorem \ref{theo: sMAS normal} and Conditions (C10),(C11) in Appendix \ref{app: regularity condtions}, as $n,N\to\infty$, we have
	\begin{equation*}
		\bbV_{\cS,\bh^{*}}^{-1/2}(\tilde{\btheta}_{\cS,\tilde{\bh}}-\btheta_{0})\to N(\bzero,\bbI)
	\end{equation*}
	in distribution, where $\bbV_{\cS,\bh^{*}}=n^{-1}\bbG_{\bu}^{-1}(\bbOmega_{\bu\bu}-\bbOmega_{\bu\bs,\bh^{*}}\bbOmega_{\bs\bs,\bh^{*}}^{-1}\bbOmega_{\bs\bu,\bh^{*}})\bbG_{\bu}^{-1}$.
\end{theorem}

\begin{theorem}\label{theo: asy tilde M}
	Under conditions of Theorem \ref{theo: mMAS normal} and Conditions (C10)--(C12) in Appendix \ref{app: regularity condtions}, as $n,N\to\infty$,  we have
	\begin{equation*}
		\bbV_{\cM,\bh^{*}}^{-1/2}(\tilde{\btheta}_{\cM,\tilde{\bh}}-\btheta_{0})\to N(\bzero,\bbI)
	\end{equation*}
	in distribution, where $\bbV_{\cM,\bh^{*}}=n^{-1}(\bbG_{\cM,\bh^{*}}^{\T}\bbOmega_{\cM,\bh^{*}}^{-1}\bbG_{\cM,\bh^{*}})^{-1}$.
\end{theorem}	

Theorems \ref{theo: asy tilde S} and \ref{theo: asy tilde M} establish the asymptotic normality of MAS estimators with an estimated moment function vector.
When $\pi(\bx,\by)=n/N$ and the moment function $\tilde{\bh}(\bx,\by) = \bpsi(\bx,\by;\tilde{\btheta})$, we have 
$\bbV_{\cS,\bh^{*}} = N^{-1}[E\{\bpsi(\bX,\bY;\btheta_{0})^{\otimes 2}\}]^{-1}$
under conditions of Theorem \ref{theo: asy tilde S}. This imply that the MAS estimator $\tilde{\btheta}_{\cS,\tilde{\bh}}$ is $\sqrt{N}$-consistent and has the same asymptotic variance as the whole data-based MLE $\hat{\btheta}$. Similar results hold for $\tilde{\btheta}_{\cM}$.

\subsection{Inference with the Optimal Moment Function Vector}

\begin{figure}[!ht]
	\centering
	\includegraphics[width=0.9\linewidth]{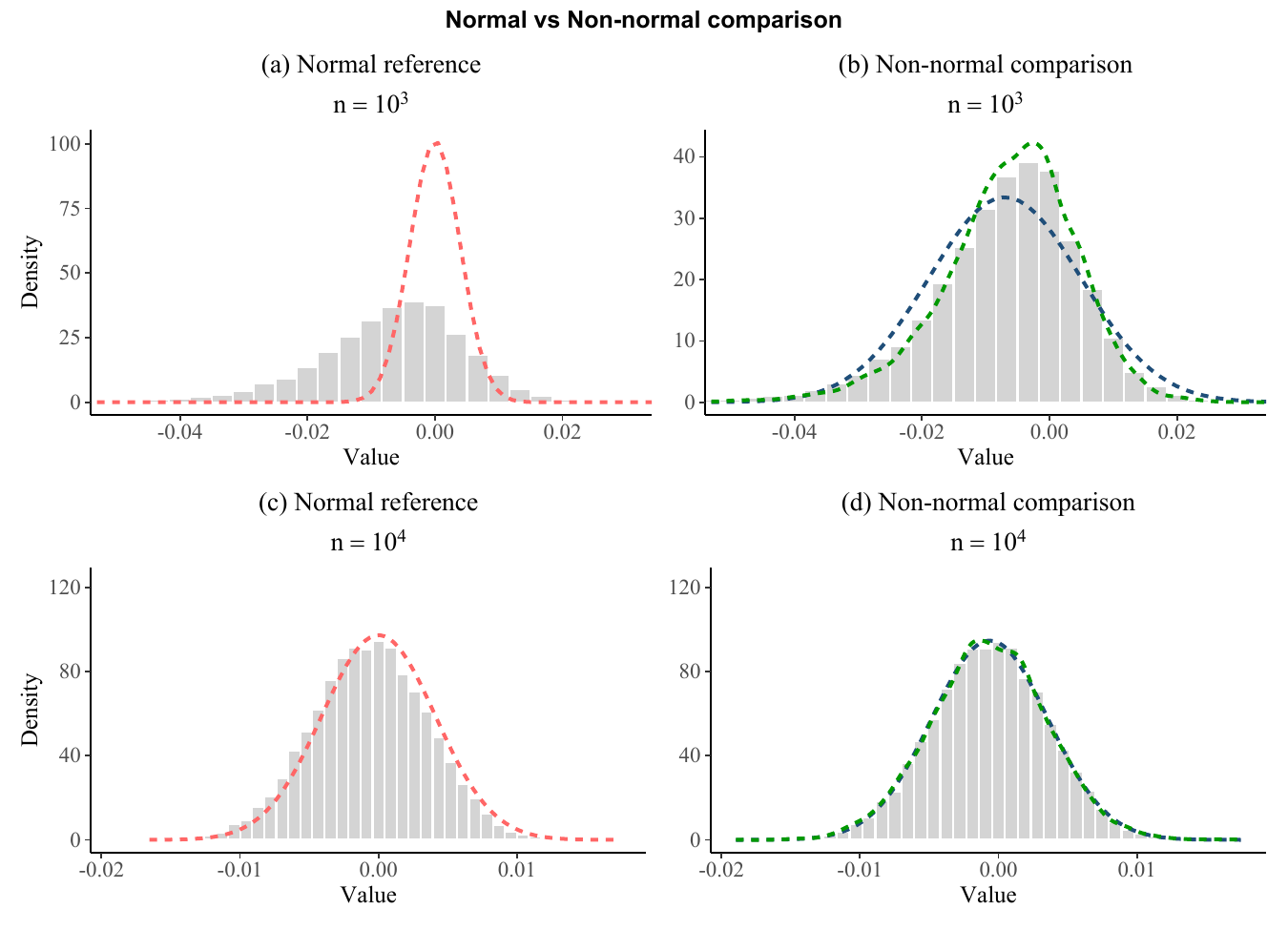}
	\caption{Sampling distribution of the fourth element of the $\tilde{\bh}^{\rm opt}$-based standard MAS estimator (histogram) based on $1,000$ replications of the setting described in Section \ref{subsec: simul logit} ($N=10^6$) with $\bX$ being a 4-dimensional covariates vector, $\alpha_{0}=0.5$ and $\bbeta_{0} = (1, 1, 1, 1)^{\T}$, along with the asymptotic normal approximation from Theorem~\ref{theo: asy tilde S} (left, red dashed line) and the simulated non-normal asymptotic approximation from Theorem~\ref{theo: non-normal S} (right,green dashed line). We also plot the normal distribution with the empirical mean and variance of the estimator for reference (right, dark blue dashed line).}\label{fig: normal vs nonnormal}
\end{figure}

The asymptotic normality results established in Theorems~\ref{theo: sMAS normal}, \ref{theo: mMAS normal}, \ref{theo: asy tilde S} and \ref{theo: asy tilde M} require that  $$n\lambda_{\min}\left(\bbOmega_{\bu\bu} - \bbOmega_{\bu\bs}\bbOmega_{\bs\bs}^{-1}\bbOmega_{\bs\bu}\right)\to\infty$$ or $n\lambda_{\min}\left(\bbOmega_{\bbm\bbm}\right)\to\infty.$ These conditions are necessary to ensure that the linear terms in the asymptotic expansion of the MAS estimators are the leading terms and hence the MAS estimators are asymptotically normal. When the estimated optimal moment function is used, the minimum eigenvalues of $\bbOmega_{\bu\bu} - \bbOmega_{\bu\bs}\bbOmega_{\bs\bs}^{-1}\bbOmega_{\bs\bu}$ and  $\bbOmega_{\bbm\bbm}$ are of order $\rho_N$. In this case, if the subsample size $n$ is small in the sense that $n^{2}/N\not\to\infty$, conditions $n\lambda_{\min}\left(\bbOmega_{\bu\bu} - \bbOmega_{\bu\bs}\bbOmega_{\bs\bs}^{-1}\bbOmega_{\bs\bu}\right)\to\infty$ and $n\lambda_{\min}\left(\bbOmega_{\bbm\bbm}\right)\to\infty$ are violated, and the asymptotic normality may not hold.	
Figure~\ref{fig: normal vs nonnormal} (a) (c) illustrates this phenomenon by presenting the empirical distribution (histogram) of the simulated standard MAS estimator along with the asymptotic normal approximation from Theorem~\ref{theo: asy tilde S}. Figure~\ref{fig: normal vs nonnormal} (a) show that the distribution of the MAS estimator deviates substantially from the normal approximation suggested by Theorem \ref{theo: asy tilde S} and is non-normal when $n$ is not much larger than $\sqrt{N}$.

To address this issue, we derive the asymptotic distribution of the  $\tilde{\bh}^{\rm opt}$-based standard MAS estimators in the following Theorem~\ref{theo: non-normal S}  without imposing any assumptions on the relative magnitude of $n$ and $N$. Figure~\ref{fig: normal vs nonnormal} shows that the asymptotic distribution described in Theorem~\ref{theo: non-normal S} provides a close approximation to the distribution of the MAS estimator. For the $\tilde{\bh}^{\rm opt}$-based modified MAS estimators, we establish the asymptotic distribution in Theorem~\ref{theo: non-normal M} of Appendix~\ref{app: nonnormal asymptotic results}.

Denote by $\operatorname{vec}(\operatorname{upp}(\bbA))$ the vectorization of the upper triangular part of a matrix $\bbA$, arranged row-wise. Let $\dot{\boldsymbol{f}}(\btheta_{0}) = (\partial f_{1}(\btheta)/\partial\btheta,\cdots,\partial f_{d}(\btheta)/\partial\btheta)^{\T}\mid_{\btheta=\btheta_{0}}$, a $d\times d$ matrix, denote the Jacobian matrix of a $d$-dimensional vector-valued function
$\boldsymbol{f}(\btheta)=(f_{1}(\btheta),\ldots,f_{d}(\btheta))^{\T}$ with respect to $\btheta$, evaluated at $\btheta_{0}$. For any $d\times d$ matrix-valued function $\bbF(\btheta)$, define 
\begin{equation*}
	\dot{\bbF}(\btheta_{0}) = 
	\begin{pmatrix}
		\dot{F}_{11}(\btheta_{0})^{\T}& \cdots&\dot{F}_{1d}(\btheta_{0})^{\T} \\
		\vdots & \vdots & \vdots \\
		\dot{F}_{d1}(\btheta_{0})^{\T}& \cdots&\dot{F}_{dd}(\btheta_{0})^{\T} 
	\end{pmatrix}
\end{equation*}
a $d\times d^{2}$ matrix, as its Jacobian with respect to $\btheta$ evaluated at $\btheta_{0}$. Here, $F_{ij}(\btheta)$ denotes the $(i,j)$th entry of $\bbF(\btheta)$, and $\dot{F}_{ij}(\btheta_{0})=\partial F_{ij}(\btheta)/\partial\btheta\mid_{\btheta=\btheta_{0}}$. Let $\ddot{\boldsymbol{f}}(\btheta_{0})$, a $d\times d^{2}$ matrix, denote the derivative of the Jacobian $\dot{\boldsymbol{f}}(\btheta)$ with respect to $\btheta$, evaluated at $\btheta_{0}$. Finally, for any vector $\bv$, denote $\bv^{\otimes 2}=\bv\bv^{\T}$. For any matrices $\bbF_{1}$ and $\bbF_{2}$, let $\bbF_{1}\otimes\bbF_{2}$ be the Kronecker product of $\bbF_{1}$ and $\bbF_{2}$.

\begin{theorem}\label{theo: non-normal S}
	Suppose that the moment function $\tilde{\bh}^{\rm opt}(\bx,\by)$ is used. Under Conditions (C6) and (C1$^{\prime}$)--(C3$^{\prime}$) in Appendix \ref{app: regularity condtions}, as $n,N\to \infty$,
	\begin{equation*}
		\tilde{\btheta}_{\cS}-\btheta_{0}=O_{P}(\max\{n^{-1},N^{-1/2}\})
	\end{equation*}
	and
	\begin{equation*}
		\sup_{t}\Big|P\left(\min\{n,\sqrt{N}\}(\tilde{\btheta}_{\cS}-\btheta_{0})\leq t\right)-P\left(\bl_{\cS,N}(\bbV_{\cS,N}^{1/2}\bZ_{\cS})\leq t\right)\Big|\to 0,
	\end{equation*} 
	where $\bZ_{\cS}$ is a $3d+1+(d^2-d)/2$ dimensional standard normal random vector and $\bbV_{\cS,N}$ is the following symmetric matrix
	\[\bbV_{\cS,N} = 
	\begin{pmatrix}
		V_{\cS,N}^{(11)} & \bV_{\cS,N}^{(12)}& \bV_{\cS,N}^{(13)}& \bV_{\cS,N}^{(14)}\\
		\bV_{\cS,N}^{(12)\T} & \bbV_{\cS,N}^{(22)}& \bbV_{\cS,N}^{(23)}& \bbV_{\cS,N}^{(24)}\\
		\bV_{\cS,N}^{(13)\T} & \bbV_{\cS,N}^{(23)\T}& \bbV_{\cS,N}^{(33)}& \bbV_{\cS,N}^{(34)}\\
		\bV_{\cS,N}^{(14)\T} & \bbV_{\cS,N}^{(24)\T}& \bbV_{\cS,N}^{(34)\T}& \bbV_{\cS,N}^{(44)}
	\end{pmatrix}\]
	with 
	$$V_{\cS,N}^{(11)} = \rho_{N}E(W), \bV_{\cS,N}^{(12)} = \bzero_{d\times 1}^{\T},\bV_{\cS,N}^{(13)} = \rho_{N}E\left[W\bpsi(\bX,\bY;\btheta_{0})^{\T}\right],$$
	$$\bV_{\cS,N}^{(14)} = \rho_{N}E\left[W\operatorname{vec}(\operatorname{upp}(\dot{\bpsi}(\bX,\bY;\btheta_{0})))^{\T}\right],\bbV_{\cS,N}^{(22)} = E\left\{\bpsi(\bX,\bY;\btheta_{0})^{\otimes 2}\right\}, \bbV_{\cS,N}^{(23)} = \sqrt{\rho_{N}}V_{\cS,N}^{(22)},$$ 
	$$\bbV_{\cS,N}^{(24)} = \sqrt{\rho_{N}}E\left\{\bpsi(\bX,\bY;\btheta_{0}) \operatorname{vec}(\operatorname{upp}(\dot{\bpsi}(\bX,\bY;\btheta_{0})))^{\T}\right\},$$
	$$\bbV_{\cS,N}^{(33)} = \rho_{N}E\left\{\pi(\bX,\bY)^{-1}\bpsi(\bX,\bY;\btheta_{0})^{\otimes 2}\right\},$$ $$\bbV_{\cS,N}^{(34)} = \rho_{N}E\left\{\pi(\bX,\bY)^{-1}\bpsi(\bX,\bY;\btheta_{0}) \operatorname{vec}(\operatorname{upp}(\dot{\bpsi}(\bX,\bY;\btheta_{0})))^{\T}\right\},$$
	\begin{equation*}
		\bbV_{\cS,N}^{(44)} 
		= \rho_{N}\left[E\left\{\pi(\bX,\bY)^{-1}\operatorname{vec}(\operatorname{upp}(\dot{\bpsi}(\bX,\bY;\btheta_{0})))^{\otimes 2}\right\}-E\left\{\operatorname{vec}(\operatorname{upp}(\dot{\bpsi}(\bX,\bY;\btheta_{0})))\right\}^{\otimes2}\right],
	\end{equation*}
	and 		
	\begin{equation*}
		\begin{split}
			\bl_{\cS,N}(\bU_{\cS}) 
			&= -c_{1}\bbG_{\bu}^{-1}\bU_{2}+c_{2}\bbG_{\bu}^{-1}U_{1}\bU_{3}+(c_{2}/2)\bbG_{u}^{-1}E\left\{\ddot{\bpsi}(\bX,\bY;\btheta_{0})\right\}\left\{(\bbG_{u}^{-1}\bU_{3})\otimes(\bbG_{u}^{-1}\bU_{3})\right\}\\
			&\quad-c_{2}\bbG_{u}^{-1}\bbU_{C}\bbG_{\bu}^{-1}\bU_{3}-c_{2}\bbG_{\bu}^{-1}\bU_{3}\rho_{N}E\left(\pi^{-1}\bpsi\right)^{\T}\bbOmega_{\bs\bs}^{-1}\bU_{3},
		\end{split}
	\end{equation*}
	with $\bU_{\cS}$ being a $3d+1+(d^2-d)/2$ dimensional vector, $c_{1} = \min\{n,\sqrt{N}\}/\sqrt{N}$, $c_{2}= \min\{n,\sqrt{N}\}/n$, $U_{1}$ being the first component of $\bU_{\cS}$, $\bU_{2}$ consisting of the $2$-th to the $d+1$-th elements of $\bU_{\cS}$, $\bU_{3}$ consisting of the $d+2$-th to the $2d+1$-th elements of $\bU_{\cS}$, $\bU_{4}$ consisting of the $2d+2$-th to $3d+1+(d^2-d)/2$-th elements of $\bU_{\cS}$, $\bbU_{C}$ being a $d\times d$ symmetric matrix with the upper triangular matrix consisted by the elements in $\bU_{4}$ arranged in rows. 
\end{theorem}

Building on the asymptotic distribution established in Theorem~\ref{theo: non-normal S}, a confidence interval for $\btheta_{0}$ based on the standard MAS estimator $\tilde{\btheta}_{\cS}$ can be constructed via a Monte Carlo approximation to the asymptotic distribution as follows:	
\begin{enumerate}[(1)]
	\item Estimate $\bbV_{\cS,N}$ by $\widetilde{\bbV}_{\cS,N}$ which replaces the expectation in $\bbV_{\cS,N}$ by the corresponding sample form based on the subsample indexed by $S$ and replaces the true parameter $\btheta_{0}$ by the plain subsampling estimator $\tilde{\btheta}$;
	\item Generate $n_{s}$ random samples $\{\bZ_{\cS,1},\dots,\bZ_{\cS,n_{s}}\}$ from the $3d+1+(d^{2}-d)/2$ dimensional multivariate  standard normal distribution;
	\item Estimate $\bl_{\cS,N}(\cdot)$ by $\tilde{\bl}_{\cS,N}(\cdot)$ which replaces $\bbG_{\bu}$ by $\tG_{\cS,\bu}$ and the expectation in $\bl_{\cS,N}(\cdot)$ by their sample forms based on the subsample indexed by $S$ and replaces $\btheta_{0}$ by $\tilde{\btheta}$;
	\item For $j = 1,\dots, d$, determine the lower and upper $\alpha/2$ quantiles of $\{\tilde{l}_{\cS,N}^{(j)}(\widetilde{\bbV}_{\cS,N}^{1/2}\bZ_{\cS,i})\}_{i = 1}^{n_{s}}$, denoted by $\tilde{l}_{\cS,N}^{(L,j)}$ and $\tilde{l}_{\cS,N}^{(U,j)}$ respectively, where $\tilde{l}_{\cS,N}^{(j)}$ is the $j$-th component of $\tilde{\bl}_{\cS,N}$;
	\item For $j = 1,\dots, d$, construct the confidence interval for $\btheta_{0,j}$ with confidence level $1-\alpha$ as $[\tilde{\btheta}_{\cS}-\tilde{l}_{\cS,N}^{(U,j)}/\min(\sqrt{N},n),\tilde{\btheta}_{\cS}-\tilde{l}_{\cS,N}^{(L,j)}/\min(\sqrt{N},n)]$.
\end{enumerate}

\subsection{Connections to Survey Sampling and Key Differences}

The subsampling for large scale data is similar to survey sampling in that they both infer some parameter based on a sample from a finite dataset. However, it is worth noting that there are fundamental differences between the set-ups of subsampling in large-scale data and survey sampling. 
In survey sampling, the population is finite but unknown, and inference targets unknown population quantities (e.g., means) under a sampling design. The sampling design is typically chosen to achieve valid design-based inference while controlling data collection cost, whereas computational cost is usually not the primary consideration in constructing the estimator. 
In contrast, in large-scale data subsampling, the whole dataset is fully observed. Subsampling from the whole dataset is introduced primarily for computational efficiency, and the target is to approximate the whole data-based estimator or to infer the parameter of the superpopulation the data come from.

The distinctions between survey sampling and large-scale data subsampling lead to several more specific methodological differences and technical challenges. We next discuss the fundamental aspects in which our framework differs from the survey sampling literature and the new challenges that arise in our setting.

First, the constraint is computational rather than observational. The full data vector is observed for every unit, but  iterative and even direct evaluation of the full likelihood or score may be expensive. In this paper, we show that carefully chosen moment information can make it possible to construct a computationally efficient subsampling estimator whose efficiency is comparable to that of the full-data estimator. To the best of our knowledge, this type of efficiency guarantee is generally not achieved in classical survey sampling and data-integration settings where the whole data-level information is genuinely partial.

Second, the use of estimated optimal moment functions introduces new inferential challenges. In particular, when the subsample size $n$ is not sufficiently large relative to $\sqrt{N}$, the estimator may exhibit non-normal asymptotic behavior. This arises because the optimal moment is (nearly) collinear with the influence function of the initial estimator. As a result, the plug-in estimated optimal moment induces a near-singular covariance matrix, leading to a non-normal limiting distribution when $n$ is not sufficiently large relative to $\sqrt{N}$. Addressing this requires the new asymptotic theory developed in Theorems~\ref{theo: non-normal S} and~\ref{theo: non-normal M}, together with the accompanying Monte Carlo inference procedure. This phenomenon cannot arise when the moment function loses information relative to the full estimating function, as in the calibration and data-integration settings discussed above; it is specific to settings where the estimated optimal moment is (nearly) collinear with the influence function of the initial estimator.

Third, we explicitly address the trade-off between computational cost and statistical efficiency. In particular, when score evaluation is computationally expensive, we develop computationally efficient approximations to the optimal moment function. We also restrict the full-data summaries to sample averages of moment functions evaluated over the entire dataset, rather than quantities obtained by solving possibly nonlinear estimating equations on the full data, as is common in survey sampling \citep{wang2023survey}. The latter approach can itself be computationally prohibitive in large-scale settings and would therefore undermine the computational advantages sought by our method.

\section{Simulation Studies}\label{sec: simulation}

In this section, we conduct simulation studies to evaluate the finite-sample performance of the proposed method. We examine the ability of the MAS framework to improve the estimation efficiency of plain subsampling estimators and assess the advantages of using the optimal moment function.
We consider two subsampling schemes: uniform subsampling, denoted by UNI, and optimal nonuniform subsampling \citep{ting2018optimal}, denoted by NSP. In our simulation setup, we compute the estimated optimal subsampling probabilities following the expression derived by \cite{ting2018optimal}:		
$\tilde{\pi}_i^{\text{opt}} \propto \|\tG_{\cS, \bu}^{-1} \bpsi(\bX_{i}, \bY_{i}; \tilde{\btheta})\|$. For each subsampling scheme, we evaluate the plain estimator without incorporating whole-data sample moments (denoted by Plain), as well as two MAS estimators: the standard MAS estimator and the modified MAS estimator.
We evaluate the MAS method under two models: logistic regression and mixed-effects logistic regression models.
Under the logistic regression model, we consider two moment functions for the MAS estimators:
(i) a moment function for which the resulting estimator $\hat{\bmu}_{\bh}$ corresponds to sufficient statistics of $\btheta_{0}$, denoted by $\bh^{\rm suf}$; and
(ii) the estimated optimal moment function, denoted by $\tilde{\bh}^{\rm opt}$.
Under the mixed-effects logistic regression model, the optimal moment function $\tilde{\bh}^{\rm opt}$ involves intractable integrals and is therefore computationally expensive. Thus, we consider the approximation based on the score function of a reduced model with parameters estimated using the approach described at the end of Subsection~\ref{subsec: approx hopt}. The resulting moment function is denoted by $\tilde{\bh}^{\rm app}$. Consequently, under the mixed-effects logistic regression model, we consider two moment functions for the MAS estimators: $\bh^{\rm suf}$ and $\tilde{\bh}^{\rm app}$.
To distinguish among the resulting MAS estimators, we use the labels sMAS-SUF, sMAS-OPT, and sMAS-APP for the standard MAS estimators with moment functions $\bh^{\rm suf}$, $\tilde{\bh}^{\rm opt}$, and $\tilde{\bh}^{\rm app}$, respectively, and mMAS-SUF, mMAS-OPT, and mMAS-APP for the corresponding modified MAS estimators.

\subsection{Logistic Regression}\label{subsec: simul logit}

Let $\bX = (X_{1},\dots,X_{30})^{\T}$, where $X_{j}$ for $j=1,\dots,30$ are independently and identically distributed from $U(-1,1)$.
Conditional on $\bX$, the outcome $Y$ is generated by the logistic regression with $P(Y=1\mid \bX)=\exp(\alpha_{0}+\bX^{\T}\bbeta_{0})/\{1+\exp(\alpha_{0}+\bX^{\T}\bbeta_{0})\}$ where $\alpha_{0}=0$ and $\bbeta_{0}$ is a $30$-dimensional vector with each component equaling $0.2$. The parameter of interest is $\btheta_{0}=(\alpha_{0},\bbeta_{0}^{\T})^{\T}$.
Under the logistic regression, $\bh^{\rm suf}(\bx,y) = (1,\bx^{\T})^{\T}y$. The specific form of $\tilde{\bh}^{\rm opt}$ can be found in Remark \ref{rmk: mMAS opth logit}.
We fix the whole data size $N= 10^{6}$ and take the subsample size $n$ to be $1000$, $5000$ and $10000$, respectively. 
A pilot subsample of size $1000$ is taken to calculate the optimal nonuniform subsampling probability.

\begin{figure}[!ht]
	\centering
	\includegraphics[width=\linewidth]{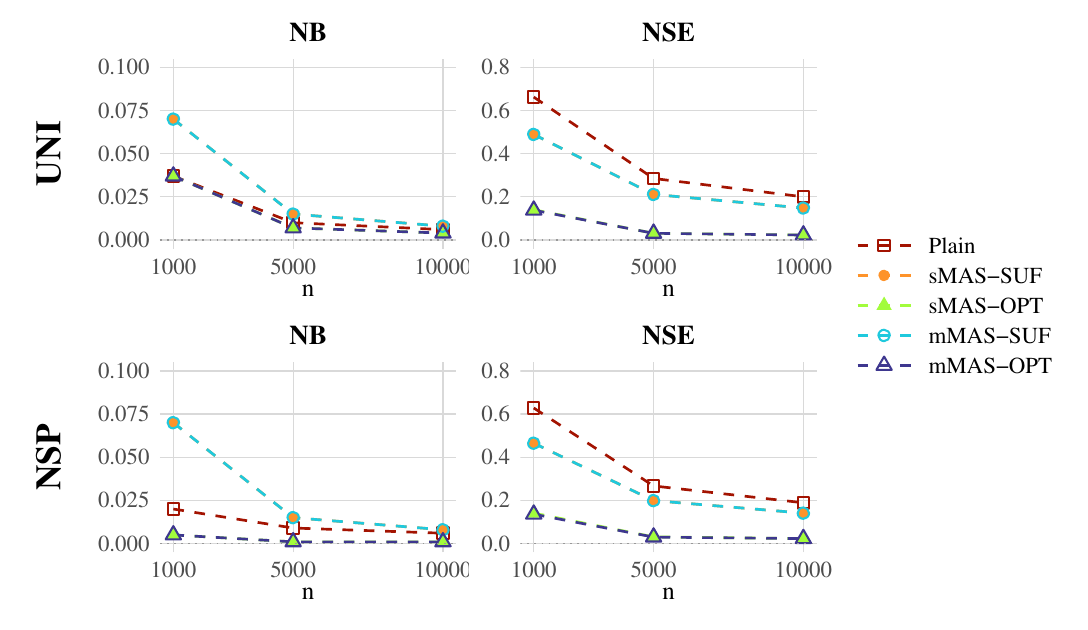}
	\caption{NB and NSE of different estimators under the logistic regression model.}
	\label{fig: logit bias se}
\end{figure}

Figure \ref{fig: logit bias se} presents the norm of bias vector (NB) and standard error (NSE) of different estimators under the logistic regression based on $1000$ repetitions. 
From Fig.~\ref{fig: logit bias se}, we observe that the NBs of the moment-function $\bh^{\rm suf}$–based subsampling estimators, sMAS-SUF and mMAS-SUF, are larger than those of the other estimators, particularly when the subsample size is small. However, the NBs of all subsampling estimators remain substantially smaller than their corresponding NSEs, indicating that the overall estimation error is dominated by the standard error rather than bias.
The NSEs of the standard and modified MAS estimators are significantly smaller than that of the plain estimator under both the uniform subsampling and the optimal nonuniform subsampling no matter which moment function is used.
This phenomenon is consistent with our theory that the estimation efficiency of subsampling methods can be improved by incorporating the whole data sample moments for any moment function under any given subsampling probability.
The moment function $\tilde{\bh}^{\rm opt}$-based subsampling estimators sMAS-OPT and mMAS-OPT perform significantly better than $\bh^{\rm suf}$-based subsampling estimators sMAS-SUF and mMAS-SUF.
This illustrates the superiority of optimal moment function for improving the estimation accuracy compared to other moment functions.

\begin{figure}[!ht]
	\centering
	\includegraphics[width=\linewidth]{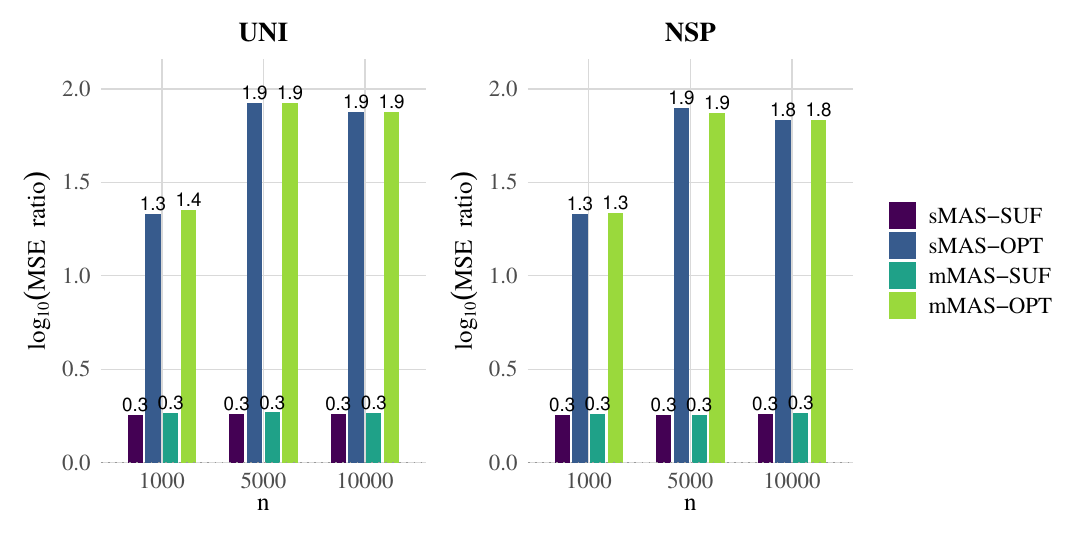}
	\caption{The logarithm of MSE ratio of the plain estimator to the MAS estimators under the logistic regression (larger than zero means the estimator has a smaller mean square error than the plain estimator).}
	\label{fig: logit rela eff}
\end{figure}

We further calculate the mean square error ratio (MSE ratio) of the plain estimator to the MAS estimators. The results are shown in Figure~\ref{fig: logit rela eff}. The MSE ratios are plotted in the $\log$ scale since it can be extremely large in some cases. The MAS estimators outperform the plain estimator in terms of MSE for all subsample sizes, especially when $\tilde{\bh}^{\rm opt}$ is adopted in the MAS estimators. In particular, the log$_{10}$(MSE ratio) can exceed 1.8 when the subsample size is 5000, which corresponds to an MSE ratio greater than 60.

Next, we evaluate the computational performance of the proposed estimators.
We compare the CPU times of the above subsampling estimators and the whole data-based estimator. 
All calculations are implemented in the R programming. 
The computer platform on which calculations are performed is a Windows server with a 52-core processor and 256GB RAM. 
Table \ref{tab: logit time n} presents the CPU times of different estimators under the logistic regression.

\begin{table}[!ht]	
	\caption{\label{tab: logit time n}CPU times (second) of different estimators under the logistic regression.}
	\centering
	\begin{tabular}{ccccccccccccccc}
		\toprule
		&&$n$&Plain& sMAS-SUF&sMAS-OPT& mMAS-SUF&mMAS-OPT\\
		\midrule
		\multirow{3}{15pt}{{UNI}}
		&&1000   &0.080   &     0.157 &         0.290  &      0.166      &    0.283 \\
		&&5000 &0.120     &   0.214      &     0.320  &      0.246    &      0.373 \\
		&&10000 &0.157    &    0.266   &       0.373&         0.310    &      0.433\\ 
		\specialrule{0em}{-4pt}{-4pt}\\
		\multirow{3}{15pt}{{NSP}}
		&&1000   &1.127     &    1.214   &        1.316     &     1.210       &    1.322\\
		&&5000 &1.144   &       1.240    &       1.349  &       1.264     &      1.382\\
		&&10000 &1.172   &      1.284     &      1.394   &      1.332       &    1.447\\		
		\specialrule{0em}{-4pt}{-4pt}\\	
		\multicolumn{5}{c}{Whole data-based estimator: 10.6}\\
		\bottomrule
	\end{tabular}
\end{table}

From Table \ref{tab: logit time n}, it can be seen that the computing times of all subsampling estimators are significantly shorter than that of the whole data-based estimator. 
The computing time of the MAS estimators are close to that of the corresponding plain estimators. This together with the results in Figures \ref{fig: logit bias se} and\ \ref{fig: logit rela eff} shows that the proposed MAS strategy can significantly improve the estimation efficiency of the subsampling estimator with a little extra computational burden.
Additionally, the computing times of the uniform subsampling-based MAS estimators are shorter than that of the nonuniform subsampling-based estimators which need to calculate the optimal subsampling probability over the whole data.

\begin{figure}[!ht]
	\centering
	\includegraphics[width=\linewidth]{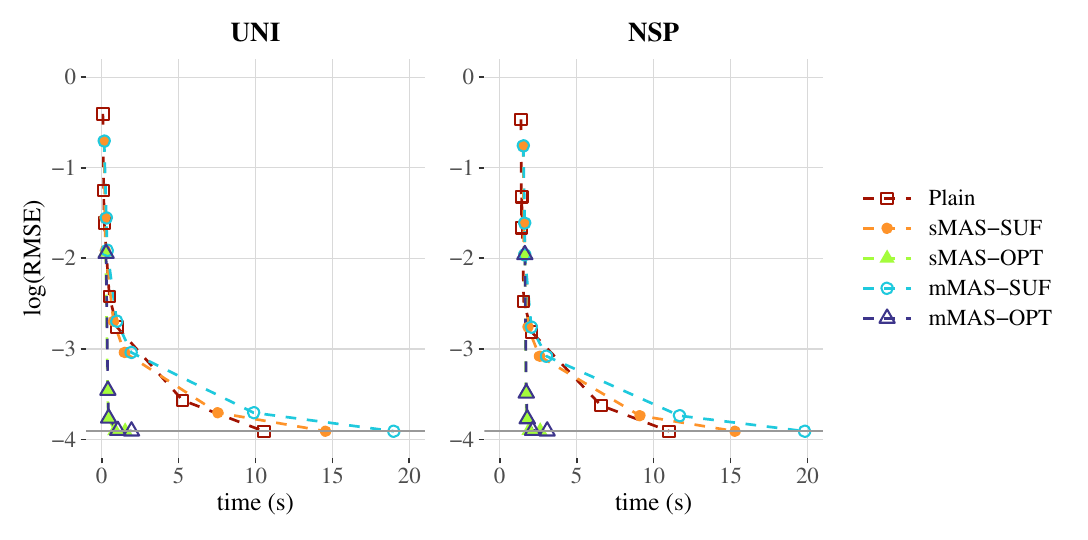}
	\caption{The $\log(\mathrm{RMSE})$ against CPU time for the plain and MAS estimators under the logistic regression.}
	\label{fig: logit accu time uni nsp}
\end{figure}

For a fair comparison, we plot $\log(\mathrm{RMSE})$ against CPU time for the plain and MAS estimators in Figure~\ref{fig: logit accu time uni nsp}, using the accuracy of the whole data–based estimator as the reference. From Fig.~\ref{fig: logit accu time uni nsp}, we observe that, as the subsample size increases, the optimal moment function $\tilde{\bh}^{\rm opt}$–based subsampling estimators, sMAS-OPT and mMAS-OPT, achieve the accuracy of the whole data–based estimator with the least computing time. This indicates that the accuracy gains obtained by incorporating the optimal moment function–based whole data moments are substantial and outweigh the additional computational cost required to compute these moments.
In contrast, the $\bh^{\rm suf}$–based MAS estimators exhibit performance similar to that of the plain subsampling estimator, suggesting that the accuracy improvement from incorporating $\bh^{\rm suf}$–based whole-data moments is largely offset by the associated computational burden. A key reason is that logistic regression is a relatively simple model. When the subsample size is small, the plain subsampling estimator is computationally very fast, making the cost of computing whole data-based sample moments relatively dominant. As the subsample size increases, the marginal accuracy gain from incorporating $\bh^{\rm suf}$–based whole-data moments diminishes.
Consequently, for simple models with easily computable score functions, the MAS estimators based on the optimal moment function are recommended, as they provide the most favorable trade-off between computational efficiency and estimation accuracy.

\begin{figure}[!ht]
	\centering
	\includegraphics[scale=0.85]{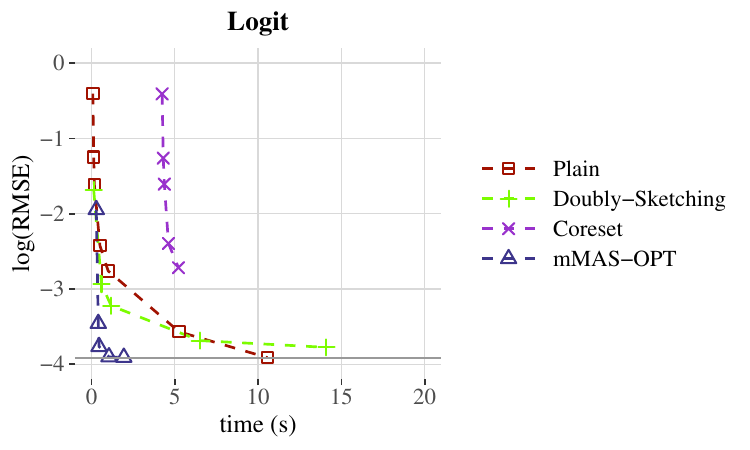}
	\caption{The $\log(\mathrm{RMSE})$ against CPU time for the competitive estimators under the logistic regression.}
	\label{fig: logit better est}
\end{figure}

Two other lines of works that can mitigate the computational burden with large scale data are sketching \citep{pilanci2017newton,ye2021approximate,hou2023generalized} and coreset \citep{huang2021novel,dwivedi2024kernel,munteanu2018coresets}. Unlike the MAS method, matrix sketching operations are usually required to be implemented for many times in sketching methods that accommodate nonlinear models, which may bring about computational disadvantages. On the other hand, even if the final estimation based on the coreset is fast, the construction of the coreset can still be computationally expensive. We next compare the proposed mMAS-OPT estimator with an efficient sketching method \citep{hou2023generalized} and a coreset method with tight theoretical bounds \citep{munteanu2018coresets}.	
To provide a fair comparison of computational and estimation efficiency, we plot $\log(\mathrm{RMSE})$ against CPU time for the uniform-subsampling based plain and mMAS-OPT estimators, the doubly-sketching estimator, and the coreset estimator under the logistic regression model in Figure~\ref{fig: logit better est}. The results show that, for a given computational budget, the MAS estimator achieves substantially smaller RMSEs than the competing subsampling-based methods. Moreover, mMAS-OPT remains the fastest among all methods to attain the estimation accuracy of the whole data-based estimator.

Next, we evaluate the inference performance of the proposed MAS estimators. We compute the average coverage probabilities (CPs) and the average length (ALs) of confidence intervals constructed according to Theorems~\ref{theo: sMAS normal} and \ref{theo: mMAS normal} for the $\bh^{\rm suf}$-based MAS estimators, and according to the procedures derived from Theorems~\ref{theo: non-normal S} and \ref{theo: non-normal M} for the $\tilde{\bh}^{\rm opt}$-based sMAS and mMAS estimators. The results at the 95\% confidence level are summarized in Table~\ref{tab: logit CP}.	

\begin{table}[!ht]	
	\caption{\label{tab: logit CP} CP and AL of uniform subsampling-based palin and MAS estimators under the logistic regression with $N=10^6$ and $n=k\times 10^{3}$.}
	\centering
	\begin{tabular}{ccccccccccccccc}
		\toprule
		& \multicolumn{2}{c}{Plain} & \multicolumn{2}{c}{sMAS-SUF} & \multicolumn{2}{c}{sMAS-OPT}& \multicolumn{2}{c}{mMAS-SUF}&\multicolumn{2}{c}{mMAS-OPT}\\
		\cmidrule{2-3}\cmidrule{4-5}\cmidrule{6-7} \cmidrule{8-9}\cmidrule{10-11}
		$k$& \small{CP} & \small{AL} & \small{CP} & \small{AL}& \small{CP} & \small{AL}& \small{CP} & \small{AL}& \small{CP} & \small{AL}\\
		\midrule
		1 &94.7\%& 0.46   & 91.9\%&  0.34   &93.2\%&   0.08   &91.9\%&  0.34   &93.9\%& 0.08  \\
		5& 94.8\%&  0.20    & 93.6\%&  0.15    &94.6\%&  0.02    & 93.6\%&  0.15  &  95.3\%&  0.02   \\
		10&95.0\%&  0.14    &94.2\%&  0.10   &94.9\%&   0.02   & 94.2\%&  0.10 & 95.6\%&  0.02 \\
		\bottomrule
	\end{tabular}
\end{table}

From Table~\ref{tab: logit CP}, we observe that the proposed MAS estimators achieve coverage probabilities close to the nominal 95\% level, even when the subsample size is small. This suggests that the inference procedures derived from Theorems~\ref{theo: non-normal S} and \ref{theo: non-normal M} provide reliable inference for the parameters of interest.
In addition, the MAS estimators produce shorter average confidence interval lengths—particularly for those based on the optimal moment function—compared with the plain subsampling estimator. This result further demonstrates the advantages of the MAS approach in terms of both reliable inference and improved statistical efficiency.

\subsection{Mixed Effects Logistic Regression}\label{subsec: glmm}

We consider the mixed effects logistic model whose likelihood function involves integrals.
Suppose there are $N$ clusters and $3$ observations in each cluster.
The random effects $W_{i},i=1,\cdots, N$ are independently and identically distributed from $N(0,\sigma_{0}^{2})$ with $\sigma_{0}=1$.
In addition, let $\bX_{i} = (\bX_{i}^{(1)\T},\bX_{i}^{(2)\T},\bX_{i}^{(3)\T})^{\T}$ where $\bX_{i}^{(j)} = (X_{i1}^{(j)},\cdots,X_{i30}^{(j)})^{\T}$ for $j=1,2,3$ and $X_{ik}^{(j)}$ for $k=1,\cdots,30$ are independently and identically distributed from $U(-1,1)$.
Conditional on $W_{i}$ and $\bX_{i}$, the outcome vector $\bY_{i} = (Y_{i}^{(1)},Y_{i}^{(2)},Y_{i}^{(3)})^{\T}$ has binary components generated from the mixed effects logistic model $P(Y_{i}^{(j)}\mid \bX_{i}^{(j)},W_{i}) = \exp(\alpha_{0}+\bX_{i}^{(j)\T}\bbeta_{0}+W_{i})/\{1+\exp(\alpha_{0}+\bX_{i}^{(j)\T}\bbeta_{0}+W_{i})\}$ for $j=1,2,3$ and $i=1,\cdots,N$, where $\alpha_{0}=0$ and $\bbeta_{0}$ is a $30$-dimensional vector with each component equaling $0.2$.
The parameter of interest is $\btheta_{0} = (\sigma_{0},\alpha_{0},\bbeta_{0}^{\T})^{\T}$.
Under the mixed effects logistic regression, $\bh^{\rm suf}(\bx,\by) = \sum_{j = 1}^{3}(1,\bx^{(j)\T})^{\T}y^{(j)}$  is the moment function such that the resulting $\hat{\bmu}_{\bh}$ is a sufficient statistics for $(\alpha_{0},\bbeta_{0}^{\T})^{\T}$.
The estimated optimal moment function $\tilde{\bh}^{\rm opt}$ involves integral operations and is therefore computationally intensive. To address this issue, we adopt fast and tractable approximations. Specifically, we consider the approximated moment function $\tilde{\bh}^{\rm app}$ that combines a Taylor expansion of the first component of the score function for the logistic mixed-effects model at $\sigma_{0}^{2} = 0$ with the score function from the reduced logistic regression model. 
Detailed constructions of $\tilde{\bh}^{\rm app}$ are provided in Appendix~\ref{app: approxi details for glmm}.
We fix the whole data size $N= 10^{6}$ and take the subsample size $n$ to be $1000$, $5000$ and $10000$, respectively. 
In this section, we report only the UNI-based plain and MAS estimators and their inference results. The NSP-based counterparts yield similar results, but incur a substantially higher computational cost. For example, when $n=1000$, they require about 20 times more computation time. This is because evaluating the optimal subsampling probabilities is computationally expensive when the score function involves integral operations.

\begin{figure}[!ht]
	\centering
	\includegraphics[width=\linewidth]{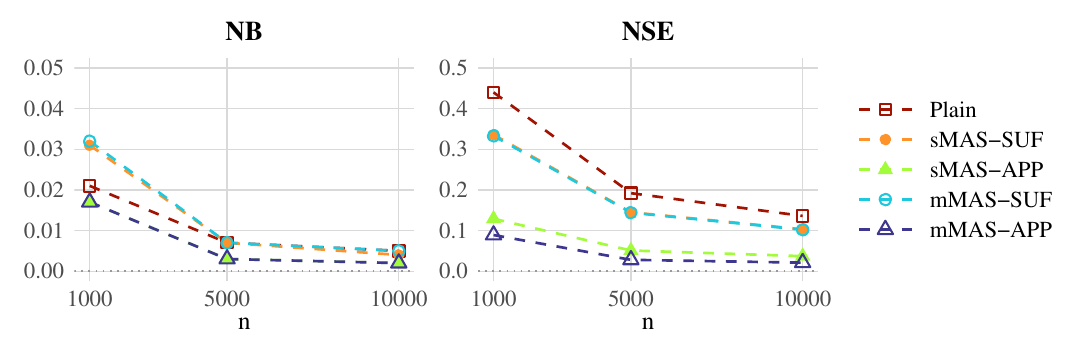}
	\caption{NB and NSE of different estimators under the mixed effects logistic regression.}
	\label{fig: glmm bias se}
\end{figure}

\begin{figure}[!ht]
	\centering
	\includegraphics[scale=1]{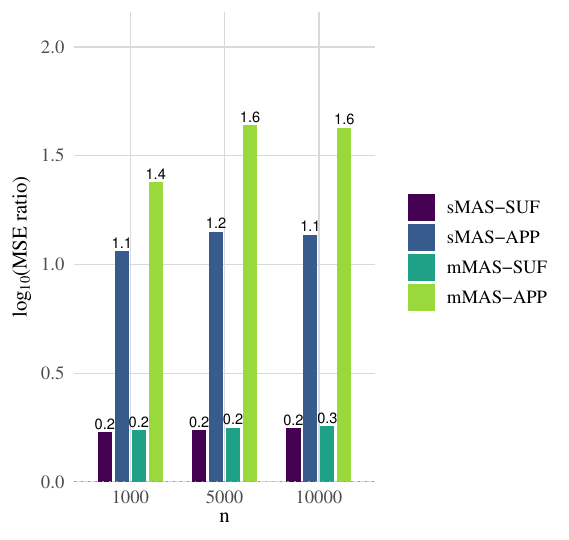}
	\caption{The logarithm of MSE ratio of the plain estimator to the MAS estimators under the mixed effects logistic regression (larger than zero means the estimator has a smaller mean square error than the plain estimator).}
	\label{fig: glmm rela eff}
\end{figure}

We plot the NB and NSE of different estimators under the mixed effects logistic regression based on $1000$ repetitions in Figure~\ref{fig: glmm bias se}, and the mean square error ratio (MSE ratio) of the plain estimator to the MAS estimators in Figure \ref{fig: glmm rela eff}.
As shown in Figure~\ref{fig: glmm bias se}, both the standard and modified MAS estimators generally outperform the plain estimator, with the most pronounced gains achieved by the $\tilde{\bh}^{\rm app}$-based MAS estimators sMAS-APP and mMAS-APP.
In particular, although the plain estimator may exhibit slightly smaller NB than the SUF-based MAS estimators when $n=1000$, it has substantially larger NSE, and hence worse overall accuracy, because the estimation error is driven primarily by variability rather than bias.
Figure~\ref{fig: glmm rela eff} further confirms this advantage: the $\log_{10}$(MSE ratio) is positive for all MAS estimators, indicating that each MAS estimator has a smaller MSE than the plain estimator. Among them, the $\tilde{\bh}^{\rm app}$-based estimator mMAS-APP attains a value of about 1.6 when the subsample size is 5000, corresponding to an MSE ratio of nearly 40. These results demonstrate the substantial efficiency gain achieved by the approximated optimal moment functions in improving estimation accuracy.

\begin{table}[!ht]
	\centering
	\caption{\label{tab: mix logit time n}CPU times (second) of different estimators under the mixed effects logistic model.}
	\begin{tabular}{ccccccccccccccc}
		\toprule
		\small{$n$}&Plain& sMAS-SUF&sMAS-APP& mMAS-SUF&mMAS-APP\\
		\midrule
		1000   &22.542    &    24.459   &     26.426  &    24.598 &      26.628  \\
		5000 & 105.39    &   111.356  &    113.341 &   111.868  &     114.302  \\
		10000 &197.82   &  209.088 &     210.902   &  219.970   &     212.997  \\
		\specialrule{0em}{-4pt}{-4pt}\\
		\multicolumn{4}{c}{Whole data-based estimator: 24657.69}\\
		\bottomrule
	\end{tabular}
\end{table}

\begin{figure}[!ht]
	\centering
	\includegraphics[scale=0.85]{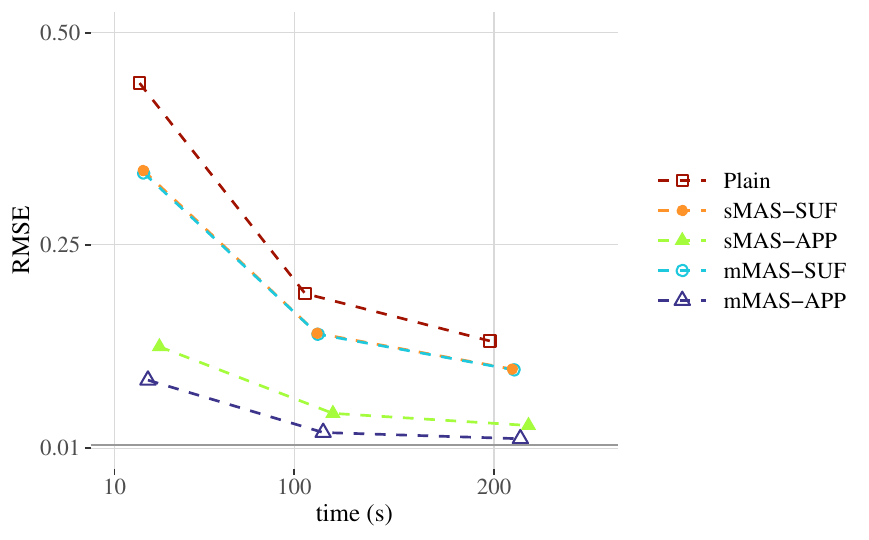}
	\caption{The RMSE against CPU time for the plain and MAS estimators under the mixed effects logistic regression.}
	\label{fig: glmm accu vs time}
\end{figure}

Table~\ref{tab: mix logit time n} reports the CPU times of the subsampling estimators under the mixed-effects logistic regression model.
All subsampling estimators require significantly less computing time than the whole data-based estimator.
Moreover, the MAS estimators incur only a modest additional computational cost relative to their plain counterparts, which demonstrates that the efficiency gains of MAS are achieved without materially sacrificing computational efficiency. To further assess the trade-off between statistical and computational efficiency, Figure~\ref{fig: glmm accu vs time} plots RMSE against CPU time for the subsampling estimators under the mixed-effects logistic regression model. The figure shows that, for a given computational budget, the MAS estimators attain substantially smaller RMSEs than the plain subsampling estimators, with the gains being especially pronounced for the $\tilde{\bh}^{\rm app}$-based MAS estimators.

Next, we evaluate the inference performance of the MAS estimator-based inference procedure under the mixed-effects logistic regression. We report the average coverage probability and average lengths of the confidence intervals across all components of the parameter vector, based on intervals constructed according to Theorems~\ref{theo: asy tilde S} and \ref{theo: asy tilde M}. 
Since $\tilde{\bh}^{\rm app}$ is an approximation of $\tilde{\bh}^{\rm opt}$, the corresponding estimated covariance matrices for the $\tilde{\bh}^{\rm app}$-based estimator may become nearly singular when the subsample size is small. To mitigate this numerical instability, we employ stabilized covariance matrices when constructing confidence intervals for the $\tilde{\bh}^{\rm app}$-based MAS estimators sMAS-APP and mMAS-APP. Specifically, we use $(1 + d/n)\tOmega_{\bu\bu}$ and $(1 + d\log(d)/n)\tOmega_{\bbm\bbm}$ for sMAS-APP and mMAS-APP, respectively.
The inference results at the 95\% confidence level are summarized in Table~\ref{tab: glmm CP}. Both the plain and MAS estimators achieve CPs close to the nominal level, even for relatively small subsample sizes. Meanwhile, the MAS estimators yield substantially shorter ALs than the plain estimator. These findings show that the MAS estimators, especially those based on the approximated optimal moment functions, can maintain the desired confidence level while producing much narrower confidence intervals.

\begin{table}[!ht]	
	\caption{\label{tab: glmm CP}CP and AL of uniform subsampling-based MAS estimators under the mixed effects logistic regression with $N=10^6$ and $n=k\times 10^{3}$.}
	\centering
	\begin{tabular}{ccccccccccccccc}
		\toprule
		& \multicolumn{2}{c}{Plain} & \multicolumn{2}{c}{sMAS-SUF} & \multicolumn{2}{c}{sMAS-APP}&\multicolumn{2}{c}{mMAS-SUF} & \multicolumn{2}{c}{mMAS-APP}\\
		\cmidrule{2-3}\cmidrule{4-5}\cmidrule{6-7} \cmidrule{8-9}\cmidrule{10-11}
		$k$& \small{CP} & \small{AL} & \small{CP} & \small{AL}& \small{CP} & \small{AL}& \small{CP} & \small{AL}& \small{CP} & \small{AL}\\
		\midrule
		1 &95.2 & 0.307 & 93.8\% & 0.229 & 96.8\% &0.099 &92.9\%&0.223 & 97.5\%& 0.071  \\
		5& 95.3 &0.134 &  94.9\%& 0.100 &95.7\%&0.036 & 94.5\%&0.099 & 96.5\%& 0.021   \\
		10& 95.1  & 0.095 & 95.1\%& 0.071  &95.2\%& 0.026  & 94.9\%&0.070 &95.8\%&0.015  \\ 
		\bottomrule
	\end{tabular}
\end{table}

Note that the moment function $\bh^{\mathrm{suf}}$ is constructed based on the sufficient statistics for the fixed-effect parameters $(\alpha_{0}, \bbeta_{0}^{\T})^{\T}$ in the mixed-effects logistic regression model. Consequently, when $\bh^{\mathrm{suf}}$ is used, no information specific to $\sigma_{0}$ is included in the moment conditions. In contrast, the approximated optimal moment function $\tilde{\bh}^{\mathrm{app}}$ leverages the full model information, including contributions from the random-effect variance. We further present separate performance metrics for the estimation of $\sigma_{0}$ in Table \ref{tab: glmm bias se sigma}. 
From Table \ref{tab: glmm bias se sigma}, it can be seen that all estimators have small bias while estimators sMAS-APP and mMAS-APP have significantly smaller standard error than sMAS-SUF and mMAS-SUF, respectively.
The results highlight the advantages of incorporating variance-component information in moment construction.

\begin{table}[!ht]	
	\caption{\label{tab: glmm bias se sigma}Bias and empirical standard error (SE)  of uniform subsampling-based MAS estimators for $\sigma_{0}$ under the mixed effects logistic regression with $N=10^6$ and $n=k\times 10^{3}$.}
	\centering
	\begin{tabular}{ccccccccccccccc}
		\toprule
		& \multicolumn{2}{c}{sMAS-SUF} & \multicolumn{2}{c}{sMAS-APP}&\multicolumn{2}{c}{mMAS-SUF} & \multicolumn{2}{c}{mMAS-APP}\\
		\cmidrule{2-3}\cmidrule{4-5}\cmidrule{6-7} \cmidrule{8-9}
		$k$ & \small{Bias} & \small{SE}& \small{Bias} & \small{SE}& \small{Bias} & \small{SE}& \small{Bias} & \small{SE}\\
		\midrule
		1 &0.090 &0.089 &0.016 &0.031 &0.091 &0.090 &0.019 &0.016\\
		5&0.001 &0.039 &0.003 &0.014 &0.001 &0.039 &0.002 &0.005\\
		10&0.001 &0.027 &0.002 &0.010 &0.001 &0.027 &0.001 &0.004 \\
		\bottomrule
	\end{tabular}
\end{table}

\subsection{Uniform Improvement}

\begin{figure}[!ht]
	\centering
	\includegraphics[width=\linewidth]{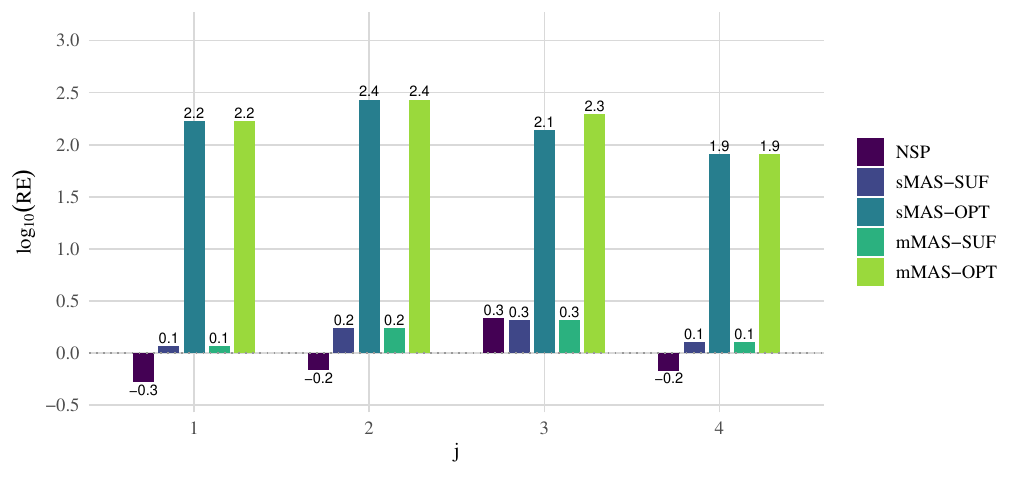}
	\caption{The logarithm of RE of the uniform subsampling-based plain estimator to the MAS estimators and the nonuniform subsampling-based plain estimator for the $j$th dimension under the Weibull regression (larger than zero means the estimator has a smaller variance than the uniform subsampling-based plain estimator).}
	\label{fig: eachdim rela eff}
\end{figure}

The optimal nonuniform subsampling-based method typically focus on minimizing the trace of the asymptotic variance matrix and hence may inadvertently increase the variance of certain components compared to the uniform subsampling estimator.
In contrast, the uniform subsampling-based MAS method defines a more efficient estimator than the uniform subsampling estimator in Lowner order which implies the uniform improvement for each component of the parameter of interest.
In this subsection, we explore the advantage of the MAS method in uniform improvement.
Let $\bX = (X_{1},X_{2})^{\T}$ be a $2$-dimensional covariates vector, where $X_{1} = 0.1T$, $X_{2} = 15\exp(-5X_{1}^{2})T$ and $T$ follows the uniform distribution $U(-1/2,-1/2)$.
Given $\bX$, we generate the outcome $Y$ by Weibull regression $Y = W\exp(-\gamma_{0}-\bX^{\T}\bbeta_{0}/\alpha_{0})$ where $\alpha_{0}=0.5$, $\gamma_{0}=0$, $\bbeta_{0}=(0.2,0.2)^{\T}$ and $W$ follows Weibull distribution with shape parameter $\alpha_{0}$ and scale parameter $1$.
The parameter of interest is $\btheta_{0} = (\alpha_{0},\gamma_{0},\bbeta_{0}^{\T})^{\T}$.
We fix the whole data size $N=10^{6}$ and the subsample size $n=1000$. For the optimal nonuniform subsampling-based subsampling method, we take a pilot subsample of size $100$ to calculate the nonuniform subsampling probability.
Figure \ref{fig: eachdim rela eff} plots the relative efficiency (RE) of the uniform subsampling-based plain estimator to the uniform subsampling-based MAS estimators and the nonuniform subsampling-based plain estimator for each component of the parameter of interest.
In Fig.~\ref{fig: eachdim rela eff}, for simplicity of notation and without causing any confusion, the optimal nonuniform subsampling-based plain estimator we still denote it as NSP, the uniform subsampling-based standard and modified MAS estimators with moment functions $\bh^{\rm suf}$ and $\tilde{\bh}^{\rm opt}$ we still denote them as sMAS-SUF, sMAS-OPT, mMAS-SUF and mMAS-OPT, respectively.

From Fig.~\ref{fig: eachdim rela eff}, it can be seen that the  optimal nonuniform subsampling-based plain estimator has a log value of RE smaller than zero, which means it has a variance larger than that of the uniform subsampling-based plain estimator, for all but the third components. This phenomenon implies the optimal nonuniform subsampling improves the estimation accuracy of the third component at the sacrifice of the estimation accuracy of all the other three components.
In contrast, the log values of RE of the MAS estimators are larger than zero for all components.


	%
	%
	%

\section{Real Data Application}\label{sec: real data}

\subsection{Airline On Time and Delay Dataset}
In this section, we apply the MAS method to analyze the primary factors that affect aircraft delays using the large-scale flight data, which contains nearly 120 million records and takes up 1.6 GB of compressed space and 12 GB when uncompressed. 
The data includes flight arrival and departure details for all commercial flights within the United States from October 1987 through April 2008, and it is available at \texttt{https://www.kaggle.com}.
We delete the records with missing variables. The resulting whole data size is $N=116,212,331$.
The Federal Aviation Administration states that a flight is considered delay if it is no less than fifteen minutes later compared to scheduled.
Therefore we use the delay indicator ($1$: if the arrival delay is fifteen minutes or more; $0$: otherwise) as outcome $Y$ and build a logistic model to describe the relationship between the delay indicator and the following six variables: day/night ($X_1$, $1$: if the departure time is between $7$am and $6$pm; $0$: otherwise), weekend/weekday ($X_{2}$, $1$: if departure occurred in the weekend; $0$: otherwise), the departure delay ($X_{3}$, $1$: if departure delay is 15 minutes or more; $0$: otherwise), and pairwise interaction terms of the three variables ($X_{1}X_{2}$, $X_{1}X_{3}$, $X_{2}X_{3}$).
We denote $\bX = (X_{1},X_{2},X_{3},X_{1}X_{2}, X_{1}X_{3}, X_{2}X_{3})$ as a $6$-dimensional vector of covariates. Then the logistic model has the form $P(Y=1\mid \bX=\bx)=\exp(\alpha_{0}+\bx^{\T}\bbeta_{0})/\{1+\exp(\alpha_{0}+\bx^{\T}\bbeta_{0})\}$,
where $\btheta_{0}=(\alpha_{0},\bbeta_{0}^{\T})^{\T}$ is the parameter of interest.

\begin{figure}[!ht]
	\centering
	\includegraphics[scale=0.8]{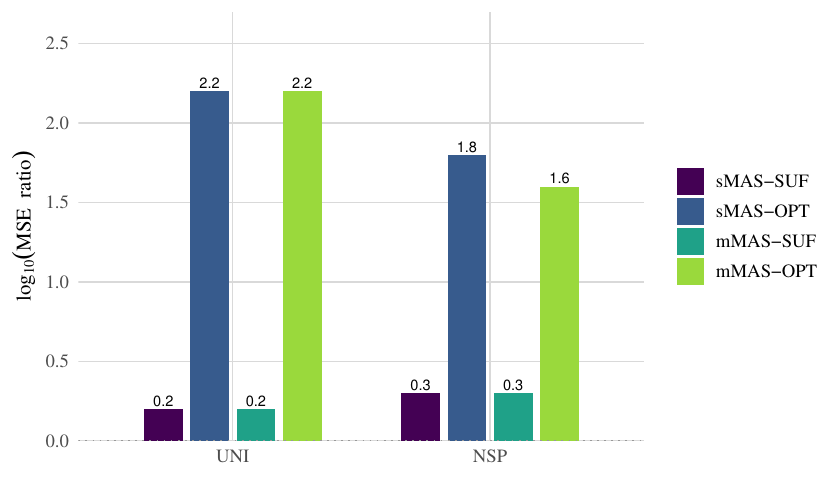}
	\caption{The logarithm of MSE ratio of the plain estimator to the MAS estimators in the airline data (larger than zero means the estimator has a smaller mean square error than the plain estimator).}
	\label{fig: airline rela eff}
\end{figure}

In practice, many researchers do not possess powerful computational resources, e.g., server clusters or distributional systems. Therefore, it is important to provide an estimator that can be carried out on handy computational resources. For the considered dataset, the whole data-based estimator of $\btheta_{0}$ is computationally infeasible for many commonly available devices such as laptops and personal computers.
To mimic the situation with limited computational resources, we conducted the analysis on a laptop with a 4-core processor and 16GB RAM. The program runs out of memory when implementing the whole data-based estimator on the laptop. The subsampling strategy can be adopted to tackle the problem.
We apply the proposed method to estimate $\btheta_{0}$.
For comparison, we calculate the plain subsampling estimators listed at the beginning of Section \ref{sec: simulation}.  
We take the subsample size $n$ to be $50000$.
In addition, a pilot subsample of size $2000$ is drawn to calculate the nonuniform subsampling probability.
Table \ref{tab: real est esd} presents the estimation results (Est) and the estimated standard deviations (ESD).  
In addition, we plot the $\log$ value of MSE ratio of the plain estimator to the MAS estimators in Fig.~\ref{fig: airline rela eff}.

\begin{table}[!ht]
	\centering
	\caption{\label{tab: real est esd}Est and ESD of different subsampling estimators for the $j$th component of $\btheta_{0}$ in the airline data.}
	\begin{tabular}{ccccccccccccccccccccc}
		\toprule
		&&&\multicolumn{2}{c}{Plain}& \multicolumn{2}{c}{sMAS-SUF}& \multicolumn{2}{c}{sMAS-OPT}& \multicolumn{2}{c}{mMAS-SUF}& \multicolumn{2}{c}{mMAS-OPT}\\
		\cmidrule{4-13}
		&&$j$& \small{Est} &\small{ESD}  &\small{Est} &\small{ESD}  &\small{Est} &\small{ESD}  &\small{Est} &\small{ESD} &\small{Est} &\small{ESD}\\
		\midrule
		\multirow{7}{15pt}{{UNI}}
		&&1 &-2.78 &0.068&-2.79 &0.042 &-2.77 &0.004&-2.75 &0.041 &-2.77  &0.004 \\
		&&2  &0.21&0.075   &0.18 &0.054  &0.25 &0.005 &0.19 &0.052  &0.25  &0.005\\
		&&3  &0.33 &0.074  &0.30 &0.052  &0.31 &0.005 &0.27 &0.051  &0.31  &0.005  \\
		&&4  &4.45 &0.087    &4.39 &0.072  &4.42 &0.009 &4.46 &0.070  &4.42  &0.009\\
		&&5 &-0.10 &0.081 &-0.06 &0.067 &-0.10 &0.006&-0.04 &0.065 &-0.10  &0.006 \\
		&&6 &-0.48 &0.078  &-0.46 &0.066 &-0.47 &0.009&-0.53 &0.064 &-0.47  &0.010\\
		&&7  &0.03 &0.081  &0.05 &0.065  &0.03 &0.006 &0.06 &0.065  &0.03  &0.006 \\
		\specialrule{0em}{-3pt}{-3pt}\\
		\multirow{7}{15pt}{{NSP}}
		&&1&-2.74  &0.022 &-2.75  &0.017 &-2.73  &0.002  &-2.77  &0.018 &-2.73  &0.003\\
		&&2  &0.08  &0.029 &0.13  &0.022  &0.12  &0.003 &0.10  &0.023  &0.12  &0.003  \\
		&&3  &0.27  &0.030  &0.25  &0.024  &0.24  &0.004  &0.29  &0.022  &0.24  &0.004\\
		&&4 &4.43  &0.037   &4.44  &0.026  &4.40  &0.005&4.46  &0.026  &4.40  &0.007 \\
		&&5  &-0.01  &0.037 &-0.02  &0.029 &-0.02  &0.004   &-0.04  &0.029 &-0.02&0.004\\
		&&6 &-0.46  &0.039  &-0.48  &0.028 &-0.46  &0.006 &-0.49  &0.027 &-0.46&0.008 \\
		&&7 &0.02  &0.039  &0.00  &0.027  &0.03  &0.004&-0.01  &0.027  &0.03  &0.005 \\
		\bottomrule
	\end{tabular}
\end{table}

\begin{table}[!ht]
	\centering
	\caption{\label{tab: real time n}CPU times (second) of different subsampling estimators in the airline data.}
	\begin{tabular}{ccccccccccccccc}
		\toprule
		&&Plain & sMAS-SUF&sMAS-OPT& mMAS-SUF&mMAS-OPT \\
		\cmidrule{3-7}
		UNI &&32.11   &   33.53  &      76.21 &35.75 &   90.81   \\				
		NSP &&80.91  &      93.02      &       109.12  &104.21      &  120.97\\
		\bottomrule
	\end{tabular}
\end{table}

Table \ref{tab: real est esd} shows that the estimated coefficients given by all subsampling estimators are similar to each other.
However, the plain subsampling estimators have significantly large ESDs than the corresponding MAS estimators for all components of $\btheta_{0}$ no matter which moment function is used. From Fig.~\ref{fig: airline rela eff}, it can be seen that the MSE ratio can be larger than 200.
Table \ref{tab: real time n} presents the CPU times of different subsampling estimators.
Table \ref{tab: real time n} shows that the subsampling methods can be implemented quickly.
The uniform subsampling-based plain estimator takes the shortest computing time, but its efficiency is notably lower than other estimators.
The computing times of the MAS estimators with $\bh^{\rm suf}$ are very close to those of the corresponding plain estimators.
The computing times of the MAS estimators with $\tilde{\bh}^{\rm opt}$ are slightly longer than those of the corresponding plain estimators. Such an increase in computing time is acceptable, given the substantial efficiency improvement of the MAS method.

\subsection{Chicago Car Crash Data}\label{subsec: Chicago car crash}
The Chicago Police Department's E-Crash System recorded about $810,000$ traffic accidents from 2017 to 2023, and it is available at \texttt{https://data.cityofchicago.org}. The car accident data provide valuable information for studying factors that affect the injury levels in a car accident.
In traffic accident records, individuals in the same car form a cluster whose injury levels are typically correlated. In addition, cars in the same accident also form a cluster which may introduce correlations among observations. The GLMM is useful to model the correlation within clusters and is widely employed model for car crash data \citep{CHENG201338,ijerph192417097,SHIRAZI201610}. We utilize a two level mixed effects logistic regression to characterize such correlations and analyze the risk factors that lead individuals to be injured in car accidents. 
We delete the records with missing values and consider a subpopulation containing crashes in which the car does not leave directly. Then, there are about $210$ thousands crashes left.

\begin{table}[!ht]
	\centering
	\caption{\label{tab: crash est esd}Est and ESD of different subsampling estimators for the $j$-th dimension of $\btheta_{0}$ in the car crash data.}
	\begin{tabular}{ccccccccccccccccccccc}
		\toprule
		\multirow{3}{15pt}{{$j$}}
		&\multicolumn{2}{c}{Plain}& \multicolumn{2}{c}{sMAS-SUF}& \multicolumn{2}{c}{sMAS-APP}\\
		\cmidrule{2-7}
		& \small{Est} &\small{ESD} &\small{Est} &\small{ESD}  &\small{Est} &\small{ESD} \\
		\midrule
		1 &2.194    &0.086   &2.191  &0.085  &2.189 &0.085\\
		2 &0.923    &0.091   &0.890  &0.090  &0.897 &0.090\\
		3   &-1.791    &0.123  &-1.857  &0.096    &-1.931 &0.082\\
		4   &-0.496    &0.058  &-0.529  &0.041    &-0.504 &0.030\\
		5 &0.014    &0.063   &0.025  &0.046  &0.000 &0.025\\
		6   &-0.339    &0.066  &-0.378  &0.043    &-0.421 &0.029\\
		7   &-0.625    &0.059  &-0.567  &0.041    &-0.525 &0.026\\
		8   &-0.672    &0.062  &-0.653  &0.045    &-0.538 &0.029\\
		9 &0.139    &0.062   &0.290  &0.041  &0.246 &0.025\\
		10  &-0.533    &0.052  &-0.524  &0.036    &-0.489 &0.025\\
		11   &0.322    &0.064   &0.224  &0.043  &0.226 &0.028\\
		12   &0.183    &0.066   &0.110  &0.047  &0.016 &0.028\\
		13   &1.156    &0.087   &1.143  &0.060  &1.170 &0.050\\
		\bottomrule
	\end{tabular}
\end{table}

\begin{table}[!ht]
	\centering
	\caption{\label{tab: crash time n}CPU times (second) of different estimators in the car crash data.}
	\begin{tabular}{*{4}{c}}
		\toprule
		Plain& sMAS-SUF&sMAS-APP & Whole\\
		\midrule
		4197.78     &     4604.18    &        4586.94 & 108302.88\\
		\bottomrule
	\end{tabular}
\end{table}

The outcome $Y$ is the injury classification ($1$: if injured; $0$: otherwise). We consider ten variables which describe features of vehicles: the first contact point ($X_{1}$, $1$: front; $0$: otherwise), maneuver ($X_{2}$, $1$: straight ahead; $0$: otherwise), make (brand) of the vehicle ($X_{3}$, $1$: electric; $0$: otherwise), traveling direction ($X_{4}$, $1$: north; $0$: otherwise), and use type ($X_{5}$, $1$: personal; $0$: otherwise), and five variables which describe features of people: passenger type ($X_{6}$, $1$: driver; $0$: otherwise), crash time ($X_{7}$, $1$: if the crash time is between $5$am and $10$am or between $5$pm and $10$pm; $0$: otherwise), airbag deployed ($X_{8}$, $1$: not deploy; $0$: otherwise), safety equipment ($X_{9}$, $1$: not use; $0$: otherwise), and sex ($X_{10}$, $1$: male; $0$: otherwise).
More details about the data and model can be found in Appendix \ref{app: detail car crash data}.
The corresponding loss function involves double integral operation which requires excessive computing resources for large-scale data.
We denote the model parameter as $\btheta_{0}$.
The whole data-based MLE for $\btheta_{0}$ is 
$(2.18,  0.90, -1.88, -0.52, -0.02, -0.47, -0.53, -0.55,  0.25,$ $ -0.48,  0.21,  0.04,  1.16).$
The computing time is 108302.88 seconds (30h+) which is too long.
We apply the proposed method to estimate $\btheta_{0}$.
The optimal subsampling probability under the two-level mixed effects logistic regression involves double integral operation over the whole data which is time-consuming. 
In addition, it is difficult to provide an explicit expression for the moment condition \eqref{eq: population moment} in the modified MAS method.
Therefore, we apply the uniform subsampling-based standard MAS method to estimate $\btheta_{0}$. The moment functions used here are the same as those in the simulation. We also calculate the plain subsampling estimator for comparison. 
We take the subsample size $n$ to be $10000$.
Table \ref{tab: crash est esd} presents the estimation results and the estimated standard deviations of subsampling estimators for all components of $\btheta_{0}$.
From Table \ref{tab: crash est esd}, it can be seen that the coefficient estimates produced by the two MAS estimators are generally much closer to those of the whole-data MLE than those obtained from the plain subsampling estimator. Moreover, the standard deviations of plain estimator are significantly larger than those of the two MAS estimators, especially the MAS estimator with $\tilde{\bh}^{\rm opt}$, for most of components.
The recorded CPU times of the three subsampling estimators plain, sMAS-SUF, sMAS-APP are 4197.78, 4604.18, and 4586.94 seconds, respectively, which are similar and shorter than one-twentieth of the whole data-based estimator.

\section{Disscusion}

In this paper, we propose a moment-assisted subsampling framework to improve the statistical efficiency of subsampling-based maximum likelihood estimators under large-scale data. By incorporating easily computable whole-data sample moments within a generalized method of moments formulation, the proposed estimators achieve substantial efficiency gains with only modest additional computational cost. We establish the asymptotic normality of MAS estimator under general moment functions, derive the optimal moment function in the Loewner order, and show that, when the (estimated) optimal moment function is employed and the subsample size is sufficiently large, the MAS estimator attains the same asymptotic efficiency as the whole-data MLE. We further characterize the non-normal asymptotic distribution that arises when the optimal moment function is used and the subsample size is not sufficiently large, and develop a Monte Carlo–based inference procedure tailored to this regime. Extensive simulations and real data analyses demonstrate that MAS achieves a favorable balance between computational efficiency and statistical accuracy, often outperforming uniform subsampling, optimal nonuniform subsampling, and other computationally efficient alternatives.

Despite these advantages, several theoretical and methodological questions remain open. First, a formal efficiency comparison corresponding to Theorem~\ref{theo: MAS compare} in Section~2.1 under general nonuniform sampling probabilities remains to be established. Second, under nonuniform sampling, the asymptotic variance takes a more intricate form, and an explicit 
characterization of the optimal moment function for the modified MAS estimator is still lacking. Although our simulations indicate that using the plug-in version of the score function performs well in practice, a rigorous derivation of the optimal moment under general sampling designs remains an open problem. Addressing this question would deepen the theoretical understanding of the interaction between sampling design and auxiliary moment construction. Third, when the estimated optimal moment function is used and the subsample size is relatively small (e.g., when $n$ is of order $\sqrt{N}$), the MAS estimator may exhibit non-normal limiting behavior. While we derive the corresponding asymptotic distribution and propose a simulation-based inference procedure, the resulting limit result is considerably more complex than a standard normal approximation. Developing simpler yet theoretically valid inference procedures, such as bootstrap or higher-order refinements adapted to the MAS structure, is an interesting direction for future research.

Finally, the MAS framework suggests several broader extensions. One direction is to construct adaptive or machine-learning-based moment functions from rich function classes while controlling overfitting and preserving valid inference. Another is to integrate MAS with modern subsampling, sketching, or coreset techniques to further enhance scalability in nonlinear or high-dimensional models. It would also be of interest to study high-dimensional regimes in which the parameter dimension grows with the sample size, and to combine MAS with regularization techniques. More broadly, the core idea of leveraging whole-data auxiliary moments to enhance subsample-based inference may extend beyond maximum likelihood estimation to other large-scale estimation paradigms, including empirical risk minimization, Bayesian computation, and causal inference with massive observational data. We hope that the MAS framework provides a useful perspective on integrating subsampling and auxiliary information, and stimulates further methodological advances toward scalable and statistically efficient inference in the era of big data.


\newpage

\appendix

\renewcommand{\thesection}{\Alph{section}}

\section{Explanation of the Jacobian Blocks}\label{app: detail mMAS}

The modified MAS (mMAS) estimator leverages model-based information to further improve the efficiency of the subsampling estimator. Accordingly, in constructing the Jacobian blocks of $\widetilde{\bbG}_{\mathcal{M}}$, we estimate them in a model-based way that yields computational convenience (no Hessian is needed to be computed). Specifically, note that for
\begin{equation*}
	\ba(\bx;\btheta)=E\{\bh(\bX,\bY)\mid\bX=\bx\},
\end{equation*}
the derivative of $\ba(\bx;\btheta)$ with respect to $\btheta$ can be expressed as
\begin{equation}
	\begin{split}
		\dot{\ba}(\bx;\btheta) 
		&= \int \bh(\bx,\by)\dot{f}(\by\mid \bx;\btheta)^{\T}d\by \\
		&= \int \bh(\bx,\by)\frac{\dot{f}(\by\mid \bx;\btheta)^{\T}}{f(\by\mid \bx;\btheta)}f(\by\mid \bx;\btheta)d\by\\
		&=E\{\bh(\bX,\bY)\bpsi(\bX,\bY;\btheta)^{\T}\mid \bX=\bx\},
	\end{split}
\end{equation}
where $\bpsi(\bX,\bY;\btheta)=\partial\log f(\bY\mid\bX;\btheta)/\partial\btheta$ is the score function. Taking expectations on both sides yields
\begin{equation}
	E\{\dot{\ba}(\bX;\btheta) \}
	=E\{\bh(\bX,\bY)\bpsi(\bX,\bY;\btheta)^{\T}\}.
\end{equation}
Hence, the Jacobian of the expectation of $\bbm(\bX,\bY;\btheta,\bmu_{\bh})$ with respect to $\btheta$ is naturally approximated by the subsample-based empirical analogue 
$$\tG_{\bbm} = N^{-1}\sum_{i\in S}\pi(\bX_{i},\bY_{i})^{-1}\bh(\bX_{i}, \bY_{i})\bpsi(\bX_{i}, \bY_{i};\tilde{\btheta})^{\T}.$$
Similarly, for the primary estimating function $\bu(\bX,\bY;\btheta)$, we have
\begin{align*}
	&E[\dot{\bpsi}(\bX,\bY;\btheta)\mid \bX=\bx] \\
	&=\int \dot{\bpsi}(\bx,\by;\btheta)f(\by\mid \bx;\btheta)d\by\\
	&=\int\frac{\ddot{f}(\by\mid \bx;\btheta)f(\by\mid \bx;\btheta)-\dot{f}(\by\mid \bx;\btheta)\dot{f}(\by\mid \bx;\btheta)^{\T}}{f(\by\mid \bx;\btheta)^2}f(\by\mid \bx;\btheta)d\by\\
	&=\int\ddot{f}(\by\mid \bx;\btheta)d\by-\int\frac{\dot{f}(\by\mid \bx;\btheta)\dot{f}(\by\mid \bx;\btheta)^{\T}}{f(\by\mid \bx;\btheta)^2}f(\by\mid \bx;\btheta)d\by\\
	&=-E[\bpsi(\bX,\bY;\btheta)\bpsi(\bX,\bY;\btheta)^{\T}],
\end{align*}
which justifies the use of
$$\tG_{\cM,\bu} =-n^{-1}\sum_{i\in S}\bu(\bX_{i},\bY_{i};\tilde{\btheta})\bpsi(\bX_{i},\bY_{i};\tilde{\btheta})^{\T}$$
as a consistent model-based estimator of the corresponding Jacobian block.

For the standard MAS (sMAS) estimator, the auxiliary estimating function does not explicitly depend on the underlying model and is therefore more robust to potential model misspecification (the estimator is consistent for the ``least false parameter" under misspecification). For this reason, the Jacobian $\bbG_{\bu}$ in sMAS is estimated using $\widetilde{\bbG}_{\cS,\bu}$ rather than $\widetilde{\bbG}_{\cM,\bu}$, since the latter may be inconsistent under model misspecification. The above robustness concern does not apply to the mMAS estimator because it is inconsistent no matter with covariance estimator is used so we use $\widetilde{\bbG}_{\mathcal{M},\bu}$ and $\widetilde{\bbG}_{\bbm}$ for computational convenience.

\section{Explanation on the Construction of Variance-Covariance}\label{app: explain var-cov}

In this section, we provide additional details on the derivation of the components of $\bbOmega_{\cS}$ and $\bbOmega_{\cM}$ in the MAS methods.
Note that
\begin{align*}
	\sqrt{n}\tilde{\bg}_{\cS}(\btheta_{0}) 
	&= \sqrt{n}\frac{1}{n}\sum_{i\in S}			
	\begin{pmatrix}
		\bu(\bX_{i},\bY_{i};\btheta_{0})\\
		\bs(\bX_{i},\bY_{i};\hat{\bmu}_{\bh})
	\end{pmatrix}\\
	&= \sqrt{n}\frac{1}{n}\sum_{i=1}^{N}			
	\begin{pmatrix}
		\delta_{i}\bu(\bX_{i},\bY_{i};\btheta_{0})\\
		\delta_{i}\bs(\bX_{i},\bY_{i};\bmu_{\bh}) - \rho_{N}\left\{\bh(\bX_{i},\bY_{i})-\bmu_{\bh}\right\}
	\end{pmatrix} \\
	&\quad-		
	\begin{pmatrix}
		\bzero\\
		\sqrt{n}\left\{\frac{1}{n}\sum_{i=1}^{N}	\frac{\delta_{i}}{\pi(\bX_{i},\bY_{i})}-1\right\}\frac{1}{N}\sum_{i=1}^{N}\left\{\bh(\bX_{i},\bY_{i})-\bmu_{\bh}\right\}
	\end{pmatrix}\\
	&= \sqrt{n}\frac{1}{n}\sum_{i=1}^{N}			
	\begin{pmatrix}
		\delta_{i}\bu(\bX_{i},\bY_{i};\btheta_{0})\\
		\delta_{i}\bs(\bX_{i},\bY_{i};\bmu_{\bh}) - \rho_{N}\left\{\bh(\bX_{i},\bY_{i})-\bmu_{\bh}\right\}
	\end{pmatrix} + O_{P}\left(\frac{1}{\sqrt{N}}\right).
\end{align*}
The matrix $\tOmega_{\bu\bu}$ is the subsampling-based estimation of covariance $$\bbOmega_{\bu\bu}=\rho_{N}^{-1}\var(\delta\bu(\bX,\bY;\btheta_{0})),$$ 
$\tOmega_{\bs\bs}$ is the subsampling-based estimation of covariance $$\bbOmega_{\bs\bs}=\rho_{N}^{-1}\var(\delta\bs(\bX,\bY;\bmu_{\bh}) - \rho_{N}\left\{\bh(\bX,\bY)-\bmu_{\bh}\right\}),$$ and $\tOmega_{\bu\bs}$ is the estimation of covariance $$\bbOmega_{\bu\bs}=\rho_{N}^{-1}\cov(\delta\bu(\bX,\bY;\btheta_{0}),\delta\bs(\bX,\bY;\bmu_{\bh}) - \rho_{N}\left\{\bh(\bX,\bY)-\bmu_{\bh}\right\}).$$

In addition, 
\begin{align*}
	\sqrt{n}\tilde{\bg}_{\cM}(\btheta_{0})
	&=\sqrt{n}\frac{1}{n}\sum_{i=1}^{N}
	\begin{pmatrix}
		\delta_{i}\bu(\bX_{i},\bY_{i};\btheta_{0})\\
		\delta_{i}\bbm(\bX_{i},\bY_{i};\btheta_{0},\bmu_{\bh})-\rho_{N}\left(\bh(\bX_{i},\bY_{i})-\bmu_{\bh}\right)
	\end{pmatrix}\\
	&\quad-		
	\begin{pmatrix}
		\bzero\\
		\sqrt{n}\left\{\frac{1}{n}\sum_{i=1}^{N}	\frac{\delta_{i}}{\pi(\bX_{i},\bY_{i})}-1\right\}\frac{1}{N}\sum_{i=1}^{N}\left\{\bh(\bX_{i},\bY_{i})-\bmu_{\bh}\right\}
	\end{pmatrix}\\
	&=\sqrt{n}\frac{1}{n}\sum_{i=1}^{N}
	\begin{pmatrix}
		\delta_{i}\bu(\bX_{i},\bY_{i};\btheta_{0})\\
		\delta_{i}\bbm(\bX_{i},\bY_{i};\btheta_{0},\bmu_{\bh})-\rho_{N}\left(\bh(\bX_{i},\bY_{i})-\bmu_{\bh}\right)
	\end{pmatrix} + O_{P}\left(\frac{1}{\sqrt{N}}\right).
\end{align*}
The matrix  
$\tOmega_{\bbm\bbm}$ is the subsampling-based estimation of covariance $$\bbOmega_{\bbm\bbm}=\rho_{N}^{-1}\var(\delta\bbm(\bX,\bY;\btheta_{0},\bmu_{\bh})-\rho_{N}\left(\bh(\bX,\bY)-\bmu_{\bh}\right)),$$ and $\tOmega_{\bu\bbm}$ is the estimation of covariance $$\bbOmega_{\bu\bbm}=\rho_{N}^{-1}\cov(\delta\bu(\bX,\bY;\btheta_{0}),\delta\bbm(\bX,\bY;\btheta_{0},\bmu_{\bh})-\rho_{N}\left(\bh(\bX,\bY)-\bmu_{\bh}\right)).$$

\section{Regularity Conditions}\label{app: regularity condtions}

For any matrix $\bbA$, let $\lambda_{\min}(\bbA)$ and $\lambda_{\max}(\bbA)$ be the minimal and maximal singular value of $\bbA$, respectively.
\begin{enumerate}[(C1)]
	\item  The parameter space $\Theta$ is compact and $\btheta_{0}$ is an inner point of $\Theta$.
	\item There exists $L_{1}(\bx,\by)$ such that $\|\bu(\bx,\by;\btheta_{1})-\bu(\bx,\by;\btheta_{2})\|\leq L_{1}(\bx,\by)\|\btheta_{1}-\btheta_{2}\|$ and $\big\|\dot{\bu}(\bx,\by;\btheta_{1})-\dot{\bu}(\bx,\by;\btheta_{2})\big\|\leq L_{1}(\bx,\by)\|\btheta_{1}-\btheta_{2}\|$ with $E\{L_{1}(\bX,\bY)^{2}\}<\infty$.
	\item (i) $E\{\|\bh(\bX,\bY)\|^{2}\}<\infty$; 
	(ii) $\sup_{\btheta}E\{\|\bu(\bX,\bY;\btheta)\|^{2}\}<\infty$, $E\{\|\bu(\bX,\bY;\btheta_{0})\|^{4}\}<\infty$ and $E\{\|\dot{\bu}(\bX,\bY;\btheta_{0})\|^{2}\}<\infty$.
	\item There exist positive constants $c$ and $C$ such that (i) $c<\lambda_{\min}(\bbG_{\bu})\leq\lambda_{\max}(\bbG_{\bu})<C$ and $c<\lambda_{\min}(\bbOmega_{\bu\bu})\leq\lambda_{\max}(\bbOmega_{\bu\bu})<C$; (ii) $\lambda_{\min}(\bbOmega_{\bs\bs})> c$, $\lambda_{\max}(\bbOmega_{\bs\bu})<C$ and $n\lambda_{\min}(\bbOmega_{\bu\bu}-\bbOmega_{\bu\bs}\bbOmega_{\bs\bs}^{-1}\bbOmega_{\bs\bu})\to \infty$.
	\item There exist positive constants $\tau$ and $C$ such that $E\left(\left[|\bb^{\T}\bzeta|/\{\var(\bb^{\T}\bzeta)\}^{1/2}\right]^{2+\tau}\right)\leq C\rho_{N}^{-\tau/2}$ for any nonzero vector $\bb$, where $\bzeta = \delta\bu(\bX,\bY;\btheta_{0})-\rho_{N}\{\delta\pi(\bX,\bY)^{-1}-1\}\bbOmega_{\bu\bs}\bbOmega_{\bs\bs}^{-1}\{\bh(\bX,\bY)-\bmu_{\bh}\}$.
	\item There exist positive constants $c$ and $C$ such that $cn/N\leq \pi(\bx,\by)\leq Cn/N$.	
	\item There exist positive constants $c$ and $C$ such that (i) $c\leq\lambda_{\min}(\bbOmega_{\bu\bu})\leq \lambda_{\max}(\bbOmega_{\bu\bu})\leq C$; (ii) $ \lambda_{\max}(\bbG_{\bbm})\asymp 1$,
	$\lambda_{\max}(\bbOmega_{\bbm\bu})\asymp\alpha_{N}$, $ \lambda_{\min}(\bbOmega_{\bbm\bbm})\asymp\alpha_{N}$, and $ \lambda_{\min}(\bbOmega_{\bbm\bbm}-\bbOmega_{\bbm\bu}\bbOmega_{\bu\bu}^{-1}\bbOmega_{\bbm\bu})\asymp \alpha_{N}$, where $\{\alpha_{N}\}$ is a sequence satisfying $\rho_{N}\lesssim \alpha_{N}\lesssim 1$; (iii)  $E\left[\big\|\left\{\ba(\bX;\btheta_{0})-\bmu_{\bh}\right\}^{\otimes2}\big\|_{F}^{2}\right]\leq C \alpha_{N}^{2}$ and $E\left\{\|\ba(\bX;\btheta_{0})-\bmu_{\bh}\|\|\dot{\ba}(\bX;\btheta_{0})\|\right\}\leq C \alpha_{N}$; (iv) $n\alpha_{N}\to\infty$.
	\item  (i)  $E\{\|\int\bh(\bX,\by)f(\by\mid\bX;\btheta_{0})d\by\|^{4}\}<\infty$ and $E\{\|\int\bh(\bX,\by)\dot{f}(\by\mid\bX;\btheta_{0})^{\T}d\by\|^{2}\}<\infty$. (ii) There exist $L_{2}(\bx)$ such that $\|\int\bh(\bx,\by)\{f(\by\mid\bx;\btheta_{1})-f(\by\mid\bx;\btheta_{2})\}d\by\|\leq L_{2}(\bx)\|\btheta_{1}-\btheta_{2}\|$ and $\|\int\bh(\bx,\by)\{\dot{f}(\by\mid\bx;\btheta_{1})-\dot{f}(\by\mid\bx;\btheta_{2})\}^{\T}d\by\|\leq L_{2}(\bx)\|\btheta_{1}-\btheta_{2}\|$ with $E\{L_{2}(\bX)^{2}\}<\infty$. 
	\item There exist positive constants $\tau$ and $C$ such that $E\left(\left[|\bb^{\T}\bnu|/\{\var(\bb^{\T}\bnu)\}^{1/2}\right]^{2+\tau}\right)\leq C\rho_{N}^{-\tau/2}$ for any nonzero vector $\bb$, where 
	$$\bnu = (\delta\bu(\bX,\bY;\btheta_{0})^{\T},\delta\bbm(\bX, \bY;\btheta_{0},\bmu_{\bh})^{\T}-\rho_{N} \left\{\bh(\bX,\bY)-\bmu_{\bh}\right\}^{\T})^{\T}.$$
\end{enumerate}

Conditions (C2) and (C8)(ii) restrict the smoothness of estimating functions and is required to ensure consistency. Conditions (C3) and (C8)(i) consist of some widely used moment conditions in literature \citep{Vaart2000AS,Yu2022Subsampling}.
Conditions (C5) and (C9) include some moment conditions required by Lindeberg-Feller's central limit theorem. 
Condition (C6) is a regularity condition which excludes extreme subsampling probabilities and is commonly adopted in the subsampling literature \citep{Wang2018Subsampling,Ai2019Subsampling,Yu2022Subsampling,Wang2020LikeliEfficient}.
Conditions (C4) and (C7) are regularity conditions on matrices. More detailed explanations of these conditions are provided in Appendix~\ref{sec: explanation of conditions}.

Let $\bb(\bx;\btheta,\bxi) = \int\bq(\bx,\by;\bxi)f(\by\mid\bx;\btheta)d\by$, $\bt(\bx,\by;\bxi) = \partial\bq(\bx,\by;\bxi)/\partial\bxi$ and $\be(\bx;\btheta,\bxi) = \partial\bb(\bx;\btheta,\bxi)/\partial\bxi$.
To establish the asymptotic properties of $\tilde{\btheta}_{\cM,\tilde{\bh}}$ and $\tilde{\btheta}_{\cS,\tilde{\bh}}$, we impose the following regularity conditions on $\bq$ and $\bxi$. 

\begin{enumerate}
	\item[(C10)] $\|\tilde{\bxi}-\bxi^{*}\|=O_{P}(n^{-1/2})$.
	\item[(C11)] (i) $E\{\|\bt(\bX,\bY;\bxi^{*})\|^{2}\}<\infty$; (ii) There exists $L_{3}(\bx,\by)$ satisfying $E\{L_{3}(\bX,\bY)^{2}\}\leq \infty$ such that $\|\bt(\bx,\by;\bxi_{1})-\bt(\bx,\by;\bxi_{2})\|\leq L_{3}(\bx,\by)\|\bxi_{1}-\bxi_{2}\|$.
	\item[(C12)] (i) $E\{\|\be(\bX;\btheta_{0},\bxi^{*})\|^{2}\}<\infty$; (ii) There exists $L_{4}(\bx,\by)$ satisfying $E\{L_{4}(\bX,\bY)^{2}\}\leq \infty$ such that $\|\bb(\bx,\by;\bxi_{1})-\bb(\bx,\by;\bxi_{2})\|\leq L_{4}(\bx,\by)\|\bxi_{1}-\bxi_{2}\|$; (iii) There exists $L_{5}(\bx)$ satisfying $E\{L_{5}(\bX)^{2}\}\leq \infty$ such that  $\|\be(\bx;\btheta_{1},\bxi_{1})-\be(\bx;\btheta_{1},\bxi_{2})\|\leq L_{5}(\bx)\{\|\btheta_{1}-\btheta_{2}\|+\|\bxi_{1}-\bxi_{2}\|\}$.
\end{enumerate}

To establish Theorems \ref{theo: non-normal S} and \ref{theo: non-normal M}, we introduce the following conditions, some of which are simplified and weaker versions of Conditions~(C1)–(C12) specialized to the case where the moment function is the estimated optimal moment function $\tilde{\bh}^{\rm opt}$.
\begin{enumerate}
	\item[(C1$^{\prime}$)]  There exists $L_{1}(\bx,\by)$ such that $\|\bpsi(\bx,\by;\btheta_{1})-\bpsi(\bx,\by;\btheta_{2})\|\leq L_{1}(\bx,\by)\|\btheta_{1}-\btheta_{2}\|$, $\|\dot{\bpsi}(\bx,\by;\btheta_{1})-\dot{\bpsi}(\bx,\by;\btheta_{2})\|\leq L_{1}(\bx,\by)\|\btheta_{1}-\btheta_{2}\|$, and  $\|\ddot{\bpsi}(\bx,\by;\btheta_{1})-\ddot{\bpsi}(\bx,\by;\btheta_{2})\|\leq L_{1}(\bx,\by)\|\btheta_{1}-\btheta_{2}\|$;
	\item[(C2$^{\prime}$)] $\sup_{\btheta}E\left\{\|\bpsi(\bX,\bY;\btheta)\|^2\right\}<\infty$ and $E\left\{\|\dot{\bpsi}(\bX,\bY;\btheta_{0})\|^2\right\}<\infty$.
	\item[(C3$^{\prime}$)] There exist positive constants $c$ and $C$ such that $c<\lambda_{\min}(\bbOmega_{\bu\bu})\leq\lambda_{\max}(\bbOmega_{\bu\bu})<C$.
\end{enumerate}

\section{Further Explanation of Conditions (C1)--(C12)}\label{sec: explanation of conditions}

In this subsection, we illustrate how the regularity conditions (C1)--(C12) hold under some sufficient conditions and examples of the model $f$, the score $\bpsi$, the moment function $\bh$, and the estimated moment function $\bq$ with parameter $\bxi$. We consider uniform subsampling where $\pi(\bx,\by)=\rho_{N}$. Under this design, Condition (C6) is automatically satisfied. 

Condition (C1) ensures the identifiability of the true parameter $\btheta_{0}$. Condition (C1) holds because $\Theta$ can be chosen as a compact set containing the true parameter $\btheta_0$ in its interior. When $\pi(\bx,\by)=\rho_{N}$, we have $\bu(\bx,\by;\btheta)=\bpsi(\bx,\by;\btheta)$. In this case, Condition (C2) requires the score function to be Lipschitz continuous, and Condition (C3) requires the higher-order moments of the score function and the moment function to be bounded.
As a concrete example, consider a logistic regression model 
\[
P(Y=1\mid \bX=\bx) = \frac{\exp(\bx^\T\btheta)}{1+\exp(\bx^\T\btheta)},
\] 
with covariate vector $\bX\in\mathbb{R}^{p}$ and parameter $\btheta\in\Theta$. 
The score function is
\[
\bpsi(\bx,y;\btheta) = \bx \left\{y - \frac{\exp(\bx^\T\btheta)}{1+\exp(\bx^\T\btheta)}\right\}.
\]
It is easy to verify that $\bpsi(\bx,y;\btheta)$ and its derivatives with respect to $\btheta$ are Lipschitz in $\btheta$.
Condition (C3) holds because the moments $E[\|\bpsi(\bX,Y;\btheta_0)\|^{4}]$ and $E[\|\dot{\bpsi}(\bX,Y;\btheta_0)\|^{2}]$ are finite given $E(\|\bX\|^{4})<\infty$.

Condition (C4) requires
\[
n\lambda_{\min}\!\left(\bbOmega_{\bu\bu}-\bbOmega_{\bu\bs}\bbOmega_{\bs\bs}^{-1}\bbOmega_{\bs\bu}\right)\to\infty.
\]
This condition typically holds because 
$\bbOmega_{\bu\bu}-\bbOmega_{\bu\bs}\bbOmega_{\bs\bs}^{-1}\bbOmega_{\bs\bu}$
is positive definite for most reasonable choices of the moment function. 
A notable exception occurs when the optimal moment function is used. In this case,
$\bbOmega_{\bu\bu}-\bbOmega_{\bu\bs}\bbOmega_{\bs\bs}^{-1}\bbOmega_{\bs\bu} = \rho_{N}\bbOmega_{\bu\bu}$, so the smallest eigenvalue is of order $\rho_{N}$. Consequently, Condition (C4) holds provided that
$n^{2}/N \to \infty$, that is, the subsample size $n$ is not too small relative to the full sample size $N$. 
For the case where the estimated optimal moment function $\tilde{\bh}^{\rm opt}$ is incorporated, we relax this requirement and establish the asymptotic distribution of the standard MAS estimator in Theorem~\ref{theo: non-normal S} without imposing any condition on the relative magnitude of $n$ and $N$.
For the $\tilde{\bh}^{\rm opt}$-based modified MAS estimator, Theorem~\ref{theo: non-normal M} establishes the asymptotic distribution under the weaker condition $n\gtrsim \sqrt{N}$.	

Condition (C5) holds if 
$\bbOmega_{\bu\bu}-\bbOmega_{\bu\bs}\bbOmega_{\bs\bs}^{-1}\bbOmega_{\bs\bu}$ 
is positive definite and the $(2+\tau)$-order moments of $\bpsi$ and $\bh$ are bounded. In this case,
\[
\inf_{\|\bb\|=1}\var(\bb^{\T}\bzeta)
=
\rho_{N}\inf_{\|\bb\|=1}\bb^{\T}
\left(\bbOmega_{\bu\bu}-\bbOmega_{\bu\bs}\bbOmega_{\bs\bs}^{-1}\bbOmega_{\bs\bu}\right)\bb
\ge c\rho_{N}.
\]
Moreover, Condition (C4) implies 
$\lambda_{\max}(\bbOmega_{\bu\bs}\bbOmega_{\bs\bs}^{-1})\le c$.
Using the inequality $|u+v|^{q}\le 2^{q-1}(|u|^{q}+|v|^{q})$, we obtain $E|\bb^{\T}\bzeta|^{2+\tau}\le c\rho_{N}$.
Hence Condition (C5) holds.
For the special case $\bh(\bx,\by)=\bpsi(\bx,\by;\btheta_{0})$, we have 
$\bzeta=\rho_{N}\bpsi$, and it is straightforward to verify that Condition (C5) still holds.

Condition (C7)(ii) holds if
\[
\bbOmega_{\bbm\bbm}-\bbOmega_{\bbm\bu}\bbOmega_{\bu\bu}^{-1}\bbOmega_{\bbm\bu}
\]
is positive definite, in which case $\alpha_{N}\asymp 1$. 
For the case $\bh(\bx,\by)=\bpsi(\bx,\by;\btheta_{0})$ and $\lim_{n,N\to\infty}\rho_{N}=c<1$, we have
\[
\bbOmega_{\bbm\bbm}-\bbOmega_{\bbm\bu}\bbOmega_{\bu\bu}^{-1}\bbOmega_{\bbm\bu}
=
\rho_{N}(1-\rho_{N})\bbOmega_{\bu\bu},
\]
and hence Condition (C7)(ii) holds with $\alpha_{N}\asymp\rho_{N}$. Condition (C7)(iii) holds if the moment function $\bh(\bx,\by)$ is bounded. 
It is also satisfied when $\bh(\bx,\by)=\bpsi(\bx,\by;\btheta_{0})$, since in this case $\ba(\bX;\btheta_{0})=\bzero$.
Conditions (C8) are standard Lipschitz and moment conditions.

For Condition (C9), note that under the uniform design
\[
\bnu =
(\delta\bpsi(\bX,\bY;\btheta_{0}),\,
\delta\left\{\ba(\bX;\btheta_{0})-\bmu_{\bh}\right\}-\rho_{N}\left\{\bh(\bX,\bY)-\bmu_{\bh}\right\}).
\]
If $\bh(\bx,\by)=\bpsi(\bx,\by;\btheta_{0})$, then $\ba(\bX;\btheta_{0})=\bzero$ and
\[
\bnu=(\delta\bpsi(\bX,\bY;\btheta_{0}),-\rho_{N}\bpsi(\bX,\bY;\btheta_{0})).
\]
Let $\bb_{1}$ and $\bb_{2}$ denote the first $p$ and the last $p$ components of a vector $\bb$, respectively. Then
\[
\bb^{\T}\bnu
=
\delta\bb_{1}^{\T}\bpsi(\bX,\bY;\btheta_{0})-\rho_{N}\bb_{2}^{\T}\bpsi(\bX,\bY;\btheta_{0}).
\]
If $\bb_{1}=\bzero$, then
\[
\bb^{\T}\bnu=-\rho_{N}\bb_{2}^{\T}\bpsi(\bX,\bY;\btheta_{0}),
\]
and
\[
\var(\bb^{\T}\bnu)
=
\rho_{N}^{2}\bb_{2}^{\T}\bbOmega_{\bu\bu}\bb_{2}.
\]
If $\bb_{1}\neq\bzero$, we have
\[
\var(\bb^{\T}\bnu)\asymp\rho_{N}, 
\qquad
E|\bb^{\T}\bnu|^{2+\tau}\le c\rho_{N}.
\]
Therefore Condition (C9) holds in this case.
For a general moment function $\bh$ such that $\ba(\bX;\btheta_{0})\neq\bzero$, similar arguments yield
\[
\var(\bb^{\T}\bnu)\asymp\rho_{N}, 
\qquad
E|\bb^{\T}\bnu|^{2+\tau}\le c\rho_{N},
\]
and hence Condition (C9) also holds.

Condition (C10) generally holds because $\tilde{\xi}$ is obtained from a subsample of size $n$, leading to stochastic fluctuations of order $n^{-1/2}$. 
Conditions (C11) and (C12) are standard Lipschitz conditions and are satisfied under mild smoothness assumptions on the score and moment functions.

\section{Illustration of the Optimal Moment Function}\label{app:illustration of opth}

\begin{lemma}
	Under Conditions (C1)--(C3), (C4)(i) and (C6), 
	\begin{equation*}
		\bS_{\rm opt} - \bM(\tilde{\btheta}) = O_P(n^{-1} + N^{-1/2}).
	\end{equation*}
\end{lemma}
\begin{proof}
	By Taylor's expansion and the result in Lemma \ref{lem: AL of initial}, we have
	\begin{align*}
		&\bS_{\rm opt} - \bM(\tilde{\btheta})\\
		& = -\frac{1}{N}\sum_{i=1}^{N}\frac{\delta_{i}}{\pi(\bX_i,\bY_i)}\bh_{\btheta_0}(\bX_i,\bY_i) + \frac{1}{N}\sum_{i=1}^{N}\frac{\delta_{i}}{\pi(\bX_i,\bY_i)}\times \frac{1}{N}\sum_{i = 1}^{N}\bh_{\btheta_0}(\bX_i,\bY_i) - \bM(\tilde{\btheta})\\
		& = \frac{1}{N}\sum_{i=1}^{N}\frac{\delta_{i}}{\pi(\bX_i,\bY_i)}\left\{\bh_{\tilde{\btheta}}(\bX_i,\bY_i) - \bh_{\btheta_0}(\bX_i,\bY_i)\right\}+ \left\{\frac{1}{N}\sum_{i=1}^{N}\frac{\delta_{i}}{\pi(\bX_i,\bY_i)}-1\right\}\frac{1}{N}\sum_{i = 1}^{N}\bh_{\btheta_0}(\bX_i,\bY_i)\\
		& \quad+ \left\{\frac{1}{N}\sum_{i = 1}^{N}\bh_{\btheta_0}(\bX_i,\bY_i) - \bM(\btheta_{0})\right\} - \left\{\bM(\tilde{\btheta})-\bM(\btheta_{0})\right\}\\
		& = \frac{1}{N}\sum_{i=1}^{N}\frac{\delta_{i}}{\pi(\bX_i,\bY_i)}\dot{\bpsi}(\bX_{i},\bY_{i};\btheta_{0})\left(\tilde{\btheta}-\btheta_{0}\right)+ \left\{\frac{1}{N}\sum_{i=1}^{N}\frac{\delta_{i}}{\pi(\bX_i,\bY_i)}-1\right\}\frac{1}{N}\sum_{i = 1}^{N}\bh_{\btheta_0}(\bX_i,\bY_i)\\
		& \quad + \left\{\frac{1}{N}\sum_{i = 1}^{N}\bh_{\btheta_0}(\bX_i,\bY_i) - \bM(\btheta_{0})\right\}- \dot{\bM}(\btheta_{0})(\tilde{\btheta}-\btheta_{0}) + O_{P}(\|\tilde{\btheta}-\btheta_{0}\|^{2})\\
		& = \left\{\frac{1}{N}\sum_{i=1}^{N}\frac{\delta_{i}}{\pi(\bX_i,\bY_i)}\dot{\bpsi}(\bX_{i},\bY_{i};\btheta_{0})-\dot{\bM}(\btheta_{0})\right\}\left(\tilde{\btheta}-\btheta_{0}\right) \\
		&\quad + \left\{\frac{1}{N}\sum_{i=1}^{N}\frac{\delta_{i}}{\pi(\bX_i,\bY_i)}-1\right\}\frac{1}{N}\sum_{i = 1}^{N}\bh_{\btheta_0}(\bX_i,\bY_i)\\
		& \quad + \left\{\frac{1}{N}\sum_{i = 1}^{N}\bh_{\btheta_0}(\bX_i,\bY_i) - \bM(\btheta_{0})\right\}+ O_{P}(\|\tilde{\btheta}-\btheta_{0}\|^{2})\\
		&=O_{P}(n^{-1}+N^{-1/2}).
	\end{align*}
\end{proof}

The above estimation error results for $\bS_{\rm opt} - \bM(\tilde{\btheta})$ remains valid if the moment function $\bh_{\btheta_0}$ in the definition of $\bS_{\rm opt}$ is replaced by $\bh_{\hat{\btheta}_n}$ for any $\sqrt{n}$-consistent estimator $\hat{\btheta}_n$ for $\btheta_{0}$ since
\begin{align*}
	& \frac{1}{N}\sum_{i=1}^{N}\frac{\delta_{i}}{\pi(\bX_i,\bY_i)}\bh_{\hat{\btheta}_{n}}(\bX_i,\bY_i) - \frac{1}{N}\sum_{i=1}^{N}\frac{\delta_{i}}{\pi(\bX_i,\bY_i)}\times \frac{1}{N}\sum_{i = 1}^{N}\bh_{\hat{\btheta}_{n}}(\bX_i,\bY_i)\\
	& \quad-\left\{\frac{1}{N}\sum_{i=1}^{N}\frac{\delta_{i}}{\pi(\bX_i,\bY_i)}\bh_{\btheta_0}(\bX_i,\bY_i) - \frac{1}{N}\sum_{i=1}^{N}\frac{\delta_{i}}{\pi(\bX_i,\bY_i)}\times \frac{1}{N}\sum_{i = 1}^{N}\bh_{\btheta_0}(\bX_i,\bY_i)\right\}\\
	& =\frac{1}{N}\sum_{i=1}^{N}\frac{\delta_{i}}{\pi(\bX_i,\bY_i)}\left\{\bh_{\hat{\btheta}_{n}}(\bX_i,\bY_i) -\bh_{\btheta_0}(\bX_i,\bY_i)\right\} \\
	& \quad-\frac{1}{N}\sum_{i=1}^{N}\frac{\delta_{i}}{\pi(\bX_i,\bY_i)}\times\frac{1}{N}\sum_{i = 1}^{N}\left\{\bh_{\hat{\btheta}_{n}}(\bX_i,\bY_i) -\bh_{\btheta_0}(\bX_i,\bY_i)\right\}\\
	& =\frac{1}{N}\sum_{i=1}^{N}\frac{\delta_{i}}{\pi(\bX_i,\bY_i)}\left\{\bh_{\hat{\btheta}_{n}}(\bX_i,\bY_i) -\bh_{\btheta_0}(\bX_i,\bY_i)\right\}-\frac{1}{N}\sum_{i = 1}^{N}\left\{\bh_{\hat{\btheta}_{n}}(\bX_i,\bY_i) -\bh_{\btheta_0}(\bX_i,\bY_i)\right\} \\
	& \quad
	-\left\{\frac{1}{N}\sum_{i=1}^{N}\frac{\delta_{i}}{\pi(\bX_i,\bY_i)}-1\right\}\times\frac{1}{N}\sum_{i = 1}^{N}\left\{\bh_{\hat{\btheta}_{n}}(\bX_i,\bY_i) -\bh_{\btheta_0}(\bX_i,\bY_i)\right\}\\
	& =\left\{\frac{1}{N}\sum_{i=1}^{N}\frac{\delta_{i}}{\pi(\bX_i,\bY_i)}\dot{\bpsi}(\bX_{i},\bY_{i};\btheta_{0})-\frac{1}{N}\sum_{i = 1}^{N}\dot{\bpsi}(\bX_{i},\bY_{i};\btheta_{0})\right\}\left(\hat{\btheta}_{n}-\btheta_{0}\right) \\
	& \quad
	-\left\{\frac{1}{N}\sum_{i=1}^{N}\frac{\delta_{i}}{\pi(\bX_i,\bY_i)}-1\right\}\times\frac{1}{N}\sum_{i = 1}^{N}\dot{\bpsi}(\bX_{i},\bY_{i};\btheta_{0})\left(\hat{\btheta}_{n}-\btheta_{0}\right) + O_{P}\left(\|\hat{\btheta}_{n}-\btheta_{0}\|^{2}\right)\\
	&=O_{P}(n^{-1}+N^{-1/2}).
\end{align*}

\section{Asymptotic Properties with Optimal Moment Function}\label{app: nonnormal asymptotic results}

In this section, we establish the asymptotic results for the modified MAS estimator $\tilde{\btheta}{\cM}$ with relaxed restrictions on the subsample size. For notational convenience, we define $\bpsi_{2}(\bX,\bY;\btheta) = \bpsi(\bX,\bY;\btheta)^{\otimes 2}$.

\begin{theorem}\label{theo: non-normal M}
	Suppose that the moment function $\tilde{\bh}^{\rm opt}(\bx,\by)$ is used and $n\gtrsim\sqrt{N}$. Under Conditions (C6) and  (C1$^{\prime}$)--(C3$^{\prime}$) in Appendix \ref{app: regularity condtions}, as $n,N\to \infty$,
	\begin{equation*}
		\tilde{\btheta}_{\cM}-\btheta_{0} = O_{P}(\max\{n^{-1},N^{-1/2}\})
	\end{equation*}
	and
	\begin{equation*}
		\sup_{t}\Big|P\left(\min\{n,\sqrt{N}\}(\tilde{\btheta}_{\cM}-\btheta_{0})\leq t\right)-P\left(\bl_{\cM,N}(\bbV_{\cM,N}^{1/2}\bZ_{\cM})\leq t\right)\Big|\to 0,
	\end{equation*}
	where $\bZ_{\cM}$ is a $3d+d^2$ dimensional standard normal random vector and $\bbV_{\cM,N}$ is the following symmetric matrix
	\begin{equation}
		\bbV_{\cM,N} = 
		\begin{pmatrix}
			\bbV_{\cM,N}^{(11)} & \bbV_{\cM,N}^{(12)}& \bbV_{\cM,N}^{(13)}& \bbV_{\cM,N}^{(14)}\\
			\bbV_{\cM,N}^{(12)\T} & \bbV_{\cM,N}^{(22)}& \bbV_{\cM,N}^{(23)}& \bbV_{\cM,N}^{(24)}\\
			\bbV_{\cM,N}^{(13)\T} & \bbV_{\cM,N}^{(23)\T}& \bbV_{\cM,N}^{(33)}& \bbV_{\cM,N}^{(34)}\\
			\bbV_{\cM,N}^{(14)\T} & \bbV_{\cM,N}^{(24)\T}& \bbV_{\cM,N}^{(34)\T}& \bbV_{\cM,N}^{(44)}
		\end{pmatrix},
	\end{equation}
	with 
	$$\bbV_{\cM,N}^{(11)} = E\left\{\bpsi_{2}(\bX,\bY;\btheta_{0})\right\}, \bbV_{\cM,N}^{(12)} = \sqrt{\rho_{N}}V_{\cM,N}^{(11)},$$   
	$$\bbV_{\cM,N}^{(13)} = \sqrt{\rho_{N}}E\left\{\bpsi(\bX,\bY;\btheta_{0}) \operatorname{vec}(\operatorname{upp}(\bpsi_{2}(\bX,\bY;\btheta_{0})))^{\T}\right\},$$ 
	$$\bbV_{\cM,N}^{(14)} = \sqrt{\rho_{N}}E\left\{\rho_{N}\pi(\bX,\bY)^{-1}\bpsi(\bX,\bY;\btheta_{0}) \operatorname{vec}(\operatorname{upp}(\bpsi_{2}(\bX,\bY;\btheta_{0})))^{\T}\right\},$$ 
	$$\bbV_{\cM,N}^{(22)} = \rho_{N}E\left\{\pi(\bX,\bY)^{-1}\bpsi_{2}(\bX,\bY;\btheta_{0})\right\},$$ 
	$$\bbV_{\cM,N}^{(23)} = \rho_{N}E\left\{\pi(\bX,\bY)^{-1}\bpsi(\bX,\bY;\btheta_{0}) \operatorname{vec}(\operatorname{upp}(\bpsi_{2}(\bX,\bY;\btheta_{0})))^{\T}\right\},$$  
	$$\bbV_{\cM,N}^{(24)} = \rho_{N}^{2}E\left\{\pi(\bX,\bY)^{-2}\bpsi(\bX,\bY;\btheta_{0}) \operatorname{vec}(\operatorname{upp}(\bpsi_{2}(\bX,\bY;\btheta_{0})))^{\T}\right\},$$  
	\begin{equation*}
		\begin{split}
			\bbV_{\cM,N}^{(33)} 
			&= \rho_{N}\left[E\left\{\pi(\bX,\bY)^{-1}\operatorname{vec}(\operatorname{upp}(\bpsi_{2}(\bX,\bY;\btheta_{0}))^{\otimes 2}\right\}-E\left\{\operatorname{vec}(\operatorname{upp}(\bpsi_{2}(\bX,\bY;\btheta_{0})))\right\}^{\otimes 2}\right],
		\end{split}
	\end{equation*}
	\begin{equation*}
		\begin{split}
			\bbV_{\cM,N}^{(34)} 
			&= \rho_{N}^{2}\left[E\left\{\pi(\bX,\bY)^{-2}\operatorname{vec}(\operatorname{upp}(\bpsi_{2}(\bX,\bY;\btheta_{0}))^{\otimes 2}\right\}-E\left\{\operatorname{vec}(\operatorname{upp}(\bpsi_{2}(\bX,\bY;\btheta_{0})))\right\}^{\otimes 2}\right],
		\end{split}
	\end{equation*}
	\begin{equation*}
		\begin{split}
			\bbV_{\cM,N}^{(44)} 
			&= \rho_{N}^{3}\left[E\left\{\pi(\bX,\bY)^{-3}\operatorname{vec}(\operatorname{upp}(\bpsi_{2}(\bX,\bY;\btheta_{0}))^{\otimes 2}\right\}-E\left\{\pi(\bX,\bY)^{-1}\operatorname{vec}(\operatorname{upp}(\bpsi_{2}(\bX,\bY;\btheta_{0})))\right\}^{\otimes 2}\right],
		\end{split}
	\end{equation*}
	and
	\begin{equation*}
		\begin{split}
			&\bl_{\cM,N}(\bU_{\cM}) \\
			&= -c_{1}\bbA_{\cM}
			\begin{pmatrix}
				\rho_{N}^{-1/2}\bU_{2}\\
				-\bU_{1}
			\end{pmatrix}
			-c_{2}\bbA_{\cM}
			\begin{pmatrix}
				E\left\{\ddot{\bpsi}(\bX,\bY;\btheta_{0})\right\}/2\\
				-E\left\{\ddot{\bpsi}(\bX,\bY;\btheta_{0})\right\}/2
			\end{pmatrix}
			\{(\bbG_{\bu}^{-1}\bU_{2})\otimes(\bbG_{\bu}^{-1}\bU_{2})\}\\
			&\quad
			-c_{2}\bbA_{\cM}
			\begin{pmatrix}
				-\bU_{C,1}\\
				\bU_{C,1}
			\end{pmatrix}
			\bbG_{\bu}^{-1}\bU_{2}+c_{2}\bbA_{\cM}
			\begin{pmatrix}
				-E\left\{\dot{\bpsi_{2}}(\bX,\bY;\btheta_{0})\right\}\\
				E\left\{\dot{\bpsi_{2}}(\bX,\bY;\btheta_{0})\right\}
			\end{pmatrix}
			\{(\bbG_{\bu}^{-1}\bU_{2})\otimes(\bbG_{\bu}^{-1}\bU_{2})\}\\
			&\quad- c_{1}\br_{N}(\bU_{\cM})
			\begin{pmatrix}
				\rho_{N}^{-1/2}\bU_{2}\\
				-\bU_{1}
			\end{pmatrix}- \br_{N}(\bU_{\cM})
			\begin{pmatrix}
				\frac{1}{2}E(\ddot{\bpsi})\\
				-	\frac{1}{2}E(\ddot{\bpsi})
			\end{pmatrix}\left\{\left(-\bbG_{\bu}^{-1}\bU_{N,2}\right)\otimes\left(-\bbG_{\bu}^{-1}\bU_{2}\right)\right\}\\
			&\quad+\br_{N}(\bU_{\cM})\begin{pmatrix}
				-\bU_{C,1}\\
				\bU_{C,1}
			\end{pmatrix}\left(-\bbG_{\bu}^{-1}\bU_{2}\right)+\br_{N}(\bU_{\cM})
			\begin{pmatrix}
				-E(\dot{\bpsi\bpsi^{\T}})\\
				E(\dot{\bpsi\bpsi^{\T}})
			\end{pmatrix}\left\{\left(-\bbG_{\bu}^{-1}\bU_{2}\right)\otimes\left(-\bbG_{\bu}^{-1}\bU_{2}\right)\right\}
		\end{split}
	\end{equation*}
	with $c_{1} = \min\{n,\sqrt{N}\}/\sqrt{N}$, $c_{2}= \min\{n,\sqrt{N}\}/n$, $\bbA_{\cM} = (\bbG_{\cM}^{\T}\bbOmega_{\cM}^{-1}\bbG_{\cM})^{-1}\bbG_{\cM}^{\T}\bbOmega_{\cM}$, $\bU_{1}$ being the first $d$ components of $\bU_{\cM}$, $\bU_{2}$ consisting of the $d+1$-th to the $2d$-th elements of $\bU_{\cM}$, $\bbU_{3}$ is the $2d+1$-th to the $3d+(d^2-d)/2$-th elements of $\bU_{\cM}$, $\bbU_{4}$ is the $3d+(d^2-d)/2+1$-th to the $3d+d^2$-th elements of $\bU_{\cM}$, $\bbU_{C,1}$ being a $d\times d$ symmetric matrix with the upper triangle matrix consisted by the elements in $\bU_{3}$ arranged in rows, $\bbU_{C,2}$ being a $d\times d$ symmetric matrix with the upper triangle matrix consisted by the elements in $\bU_{4}$ arranged in rows, and $\br_{N}(\cdot)$ is a complex smooth function which can be found in the proof of this theorem.
\end{theorem}

Building on the asymptotic results in Theorem \ref{theo: non-normal M}, the modified MAS estimator $\tilde{\btheta}_{\cM}$-based confidence interval can be constructed in the same manner as those below Theorem \ref{theo: non-normal S}.

\section{Proof of Theorems}\label{app: proof od ths}

In the following proof processes, we simplified notations for convenience. 
Let $\pi_{i} = \pi(\bX_{i},\bY_{i})$, $\bpsi_{i}(\btheta) = \bpsi(\bX_{i},\bY_{i};\btheta)$, $\bpsi_{i} = \bpsi_{i}(\btheta_{0})$, $\bh_{i} = \bh(\bX_{i},\bY_{i})$, $\bu_{i}(\btheta) = \bu(\bX_{i},\bY_{i};\btheta)$, $\bu_{i} = \bu_{i}(\btheta_{0})$, $\bbm_{i} = \bbm(\bX_{i},\bY_{i};\btheta_{0},\bmu_{\bh})$, $\ba_{i}(\btheta) = \ba(\bX_{i};\btheta)$ and $\ba_{i} = \ba_{i}(\btheta_{0})$.  Note that in the proofs, $c$ denotes a generic positive constant that can vary from lines.

\subsection*{Proof of Theorem \ref{theo: sMAS normal}}

By the definition of $\tilde{\btheta}_{\cS}$ and some straightforward algebras, we have
\begin{equation*}
	\begin{split}
		&\bbV_{\cS,\bh}^{-1/2}\left(\tilde{\btheta}_{\cS}-\btheta_{0}\right)\\
		&= \bbV_{\cS,\bh}^{-1/2}\left(\tilde{\btheta}-\btheta_{0}+\bbG_{\bu}^{-1}\bbOmega_{\bu\bs}\bbOmega_{\bs\bs}^{-1}\tilde{\bg}_{\bs}\right) + \bbV_{\cS,\bh}^{-1/2}\left(\tG_{\cS,\bu}^{-1}\tOmega_{\bu\bs}\tOmega_{\bs\bs}^{-1}-\bbG_{\bu}^{-1}\bbOmega_{\bu\bs}\bbOmega_{\bs\bs}^{-1}\right)\tilde{\bg}_{\bs}\\
		& =\bU_{\cS,1}+\bU_{\cS,2}.
	\end{split}
\end{equation*}
We prove the result of Theorem \ref{theo: sMAS normal} by proving $\bU_{\cS,1}\stackrel{d}{\to} N(\bzero,\bbI)$ and $\bU_{\cS,2}=o_{P}(1)$.
We first prove $\bU_{\cS,1}\stackrel{d}{\to} N(\bzero,\bbI)$. Note that
\begin{equation*}
	\begin{split}
		\tilde{\bg}_{\bs} 
		&= \frac{1}{N}\sum_{i=1}^{N}\left(\frac{\delta_{i}}{\pi_{i}}-1\right)\left(\bh_{i}-\bmu_{\bh}\right)+\left(1-\frac{1}{N}\sum_{i=1}^{N}\frac{\delta_{i}}{\pi_{i}}\right)\frac{1}{N}\sum_{i=1}^{N}\left(\bh_{i}-\bmu_{\bh}\right).			
	\end{split}
\end{equation*}
By Chebyshev's inequality and Conditions (C3), (C6), we have $1-N^{-1}\sum_{i=1}^{N}\delta_{i}/\pi_{i}=O_{P}(n^{-1/2})$ and $N^{-1}\sum_{i=1}^{N}\left(\bh_{i}-\bmu_{\bh}\right)=O_{P}(N^{-1/2})$.
Then we have
\begin{equation*}\label{eq: gr}
	\tilde{\bg}_{\bs}=\frac{1}{N}\sum_{i=1}^{N}\left(\frac{\delta_{i}}{\pi_{i}}-1\right)\left(\bh_{i}-\bmu_{\bh}\right)+O_{P}(n^{-1/2}N^{-1/2}).
\end{equation*}
This together with Condition (C4) and Lemma \ref{lem: AL of initial} in Appendix \ref{app: lems} proves
\begin{equation*}
	\begin{split}
		\bU_{\cS,1} 
		&= \bbV_{\cS,\bh}^{-1/2}\frac{1}{N}\sum_{i=1}^{N}\left\{ -\bbG_{\bu}^{-1}\rho_{N}^{-1}\delta_{i}\bu_{i}+\bbG_{\bu}^{-1}\bbOmega_{\bu\bs}\bbOmega_{\bs\bs}^{-1}\left(\frac{\delta_{i}}{\pi_{i}}-1\right)\left(\bh_{i}-\bmu_{\bh}\right)\right\}+ o_{P}(1).
	\end{split}
\end{equation*}
For $i=1,\dots,N$, let
\begin{equation*}
	\begin{split}
		&\bK_{\cS,i} = \bbV_{\cS,\bh}^{-1/2}\frac{1}{N}\left\{-\bbG_{\bu}^{-1}\rho_{N}^{-1}\delta_{i}\bu_{i}+\bbG_{\bu}^{-1}\bbOmega_{\bu\bs}\bbOmega_{\bs\bs}^{-1}\left(\frac{\delta_{i}}{\pi_{i}}-1\right)\left(\bh_{i}-\bmu_{\bh}\right)\right\}.
	\end{split}
\end{equation*}
Then $\bU_{\cS,1} = \sum_{i=1}^{N}\bK_{\cS,i}+o_{P}(1)$. 
We first prove that $\sum_{i=1}^{N}\var(\bK_{\cS,i})\to\bbI$.
Recalling the definition of $\bu_{i}=\rho_{N}\pi_i^{-1}\bpsi_{i}$. Then
\begin{equation*}
	\begin{split}
		\bK_{\cS,i} 
		&= \frac{1}{N}\bbV_{\cS,\bh}^{-1/2}\left\{-\bbG_{\bu}^{-1}\frac{\delta_{i}}{\pi_{i}}\bpsi_{i}+\bbG_{\bu}^{-1}\bbOmega_{\bu\bs}\bbOmega_{\bs\bs}^{-1}\left(\frac{\delta_{i}}{\pi_{i}}-1\right)\left(\bh_{i}-\bmu_{\bh}\right)\right\}.
	\end{split}
\end{equation*}
Then
\begin{equation*}
	\begin{split}
		&\bK_{\mathcal{S},i}\bK_{\mathcal{S},i}^{\T}\\
		& = \frac{1}{N^{2}}\bbV_{\cS,\bh}^{-1/2}\left[\bbG_{\bu}^{-1}\frac{\delta_{i}}{\pi_{i}^{2}}\bpsi_{i}\bpsi_{i}^{\T}\bbG_{u}^{-1}\right.\\
		&\left.\qquad\qquad\qquad\quad+\bbG_{\bu}^{-1}\bbOmega_{\bu\bs}\bbOmega_{\bs\bs}^{-1}\left\{\frac{\delta_{i}}{\pi_{i}^{2}}+1-2\frac{\delta_{i}}{\pi_{i}}\right\}(\bh_{i}-\bmu_{\bh})(\bh_{i}-\bmu_{\bh})^{\T}\bbOmega_{\bs\bs}^{-1}\bbOmega_{\bs\bu}\bbG_{\bu}^{-1}\right.\\
		&\left.\qquad\qquad\qquad\quad-\bbG_{\bu}^{-1}\left\{\frac{\delta_{i}}{\pi_{i}^{2}}-\frac{\delta_{i}}{\pi_{i}}\right\}\bpsi_{i}(\bh_{i}-\bmu_{\bh})^{\T}\bbOmega_{\bs\bs}^{-1}\bbOmega_{\bs\bu}\bbG_{\bu}^{-1}\right.\\
		&\left.\qquad\qquad\qquad\quad-\bbG_{\bu}^{-1}\left\{\frac{\delta_{i}}{\pi_{i}^{2}}-\frac{\delta_{i}}{\pi_{i}}\right\}(\bh_{i}-\bmu_{\bh})\bpsi_{i}^{\T}\bbOmega_{\bs\bs}^{-1}\bbOmega_{\bs\bu}\bbG_{\bu}^{-1}\right]\bbV_{\cS,\bh}^{-1/2}
	\end{split}
\end{equation*}
Then
\begin{align*}
	&\var(\bK_{\mathcal{S},i})\\
	&=E\left(\bK_{\cS,i}\bK_{\cS,i}^{\T}\right)\\
	&=\frac{1}{N^{2}}\bbV_{\cS,\bh}^{-1/2}\left[\bbG_{\bu}^{-1}E\left(\pi^{-1}\bpsi\bpsi^{\T}\right)\bbG_{\bu}^{-1}\right.\\
	&\left.\qquad\qquad\qquad+\bbG_{\bu}^{-1}\bbOmega_{\bu\bs}\bbOmega_{\bs\bs}^{-1}E\left\{(\pi^{-1}-1)(\bh-\bmu_{\bh})(\bh-\bmu_{\bh})^{\T}\right\}\bbOmega_{\bs\bs}^{-1}\bbOmega_{\bs\bu}\bbG_{\bu}^{-1}\right.\\
	&\left.\qquad\qquad\qquad-\bbG_{u}^{-1}E\left\{(\pi^{-1}-1)\bpsi(\bh-\bmu_{\bh})^{\T}\right\}\bbOmega_{\bs\bs}^{-1}\bbOmega_{\bu\bs}\bbG_{\bu}^{-1}\right.\\
	&\left.\qquad\qquad\qquad-\bbG_{u}^{-1}E\left\{(\pi^{-1}-1)(\bh-\bmu_{\bh})\bpsi^{\T}\right\}\bbOmega_{\bs\bs}^{-1}\bbOmega_{\bu\bs}\bbG_{\bu}^{-1}\right]\bbV_{\cS,\bh}^{-1/2}.
\end{align*}
Note that 
$\bbOmega_{\bu\bu}=\rho_{N}E\left(\pi^{-1}\bpsi\bpsi^{\T}\right),\bbOmega_{\bu\bs}=\rho_{N}E\left\{(\pi^{-1}-1)\bpsi(\bh-\bmu_{\bh})^{\T}\right\}=\bbOmega_{\bs\bu}^{\T}$, and $\bbOmega_{\bs\bs}=\rho_{N}E\left\{(\pi^{-1}-1)(\bh-\bmu_{\bh})(\bh-\bmu_{\bh})^{\T}\right\}$, and
\[\bbV_{\cS,\bh} = n^{-1}\bbG_{\bu}^{-1}\left(\bbOmega_{\bu\bu}-\bbOmega_{\bu\bs}\bbOmega_{\bs\bs}^{-1}\bbOmega_{\bs\bu}\right)\bbG_{\bu}^{-1}.\]
Then
\begin{align*}
	&\var(\bK_{\mathcal{S},i})\\
	&=\frac{n}{N^{2}}\left\{\bbG_{\bu}^{-1}\left(\bbOmega_{\bu\bu}-\bbOmega_{\bu\bs}\bbOmega_{\bs\bs}^{-1}\bbOmega_{\bs\bu}\right)\bbG_{\bu}^{-1}\right\}^{-1/2}\\
	&\qquad\qquad\rho_{N}^{-1}\bbG_{\bu}^{-1}\left\{\bbOmega_{\bu\bu}-\bbOmega_{\bu\bs}\bbOmega_{\bs\bs}^{-1}\bbOmega_{\bs\bu}\right\}\bbG_{\bu}^{-1}\left\{\bbG_{\bu}^{-1}\left(\bbOmega_{\bu\bu}-\bbOmega_{\bu\bs}\bbOmega_{\bs\bs}^{-1}\bbOmega_{\bs\bu}\right)\bbG_{\bu}^{-1}\right\}^{-1/2}\\
	&=N^{-1}\bbI.
\end{align*}
Then by Lindeberg-Feller central limit theorem in \cite{Vaart2000AS}, to prove $\sum_{i=1}^{N}\bK_{\cS,i}\to N(\bzero,\bbI)$ in distribution, it suffices to verify that for any given 
$\epsilon>0$, $$\sum_{i=1}^{N}E\{\|\bK_{\cS,i}\|^2 1(\|\bK_{\cS,i}\|>\epsilon)\}\to 0,$$ 
where $1(\cdot)$ is an indicator function. Since for any given $\tau>0$, we have $E\{\|\bK_{\cS,1}\|^2 1(\|\bK_{\cS,1}\|>\epsilon)\}\leq E\{\|\bK_{\cS,1}\|^{2+\tau}/\epsilon^{\tau}\}$.
To prove $\bU_{\cS,1}\stackrel{d}{\to}N(\bzero,\bbI)$, it suffices to verify that $NE(\|\bK_{\cS,1}\|^{2+\tau})=o(1)$ holds for some $\tau>0$.
Let $K_{\cS,1}^{(j)}$ be the $j$th element of $\bK_{\cS,1}$ for $j = 1,\dots,d$.
By the inequality $(t_{1}+t_{2})^{p}\leq 2^{p-1}(|t_{1}|^{p}+|t_{2}|^{p})$ and Condition (C5), we have
\begin{equation*}
	\begin{split}
		NE(\|\bK_{\cS,1}\|^{2+\tau})
		& \leq c NE(|K_{\cS,1}^{(1)}|^{2+\tau}+\dots+|K_{\cS,1}^{(d)}|^{2+\tau})\\
		&\leq c N\rho_{N}^{-\tau/2}\left\{\var(K_{\cS,1}^{(1)})^{1+\tau/2}+\dots+\var(K_{\cS,1}^{(d)})^{1+\tau/2}\right\}.
	\end{split}
\end{equation*}
Since $E(\bK_{\cS,1}\bK_{\cS,1}^{\T})=N^{-1}\bbI$, then it is proved that $NE(\|\bK_{\cS,1}\|^{2+\tau})=O(n^{-\tau/2})=o(1)$.

Next, we prove $\bU_{\cS,2} = o_{P}(1)$. By Chebyshev's inequality and Condition (C2), we can show that $\|\tG_{\cS,\bu}-\bbG_{\bu}\|=O_{P}(n^{-1/2})$, $\|\tOmega_{\bu\bs}-\bbOmega_{\bu\bs}\|=O_{P}(n^{-1/2})$, and  $\|\tOmega_{\bs\bs}-\bbOmega_{\bs\bs}\|=O_{P}(n^{-1/2})$. By the W-H theorem in \cite{WHtheorem},  $\|\tG_{\cS,\bu}^{-1}\tOmega_{\bu\bs}\tOmega_{\bs\bs}^{-1}-\bbG_{\bu}^{-1}\bbOmega_{\bu\bs}\bbOmega_{\bs\bs}^{-1}\|=O_{P}(n^{-1/2})$.
By Chebyshev's inequality and Conditions (C2),(C6), we have $\tilde{\bg}_{\bs} = O_{P}(n^{-1/2})$.
Then by Condition (C4), we have  $\bU_{\cS,2} = o_{P}(1)$.

\subsection*{Proof of Theorem \ref{theo: mMAS normal}}

Let $\bg_{\cM} = \tilde{\bg}_{\cM}(\btheta_{0})$. 
By Taylor's expansion, $\tilde{\bg}_{\cM} = \bg_{\cM}+\bar{\bbG}_{\cM}\left(\tilde{\btheta}-\btheta_{0}\right)$, where $\bar{\bbG}_{\cM} = \partial\tilde{\bg}_{\cM}(\btheta)/\partial\btheta\mid_{\btheta=\bar{\btheta}}$ and $\bar{\btheta}$ is between $\btheta_{0}$ and $\tilde{\btheta}$.
Let $\tA_{\cM} = \left(\tG_{\cM}^{\T}\tOmega_{\cM}^{-1}\tG_{\cM}\right)^{-1}\tG_{\cM}^{\T}\tOmega_{\cM}^{-1}$ and $\bbA_{\cM} = \left(\bbG_{\cM}^{\T}\bbOmega_{\cM}^{-1}\bbG_{\cM}\right)^{-1}\bbG_{\cM}^{\T}\bbOmega_{\cM}^{-1}$.
By the definition of $\tilde{\btheta}_{\cM}$ and straightforward algebras,
\begin{equation*}\label{eq: decompose global}
	\begin{split}
		&\bbV_{\cM,\bh}^{-1/2}(\tilde{\btheta}_{\cM}-\btheta_{0})\\
		& = -\bbV_{\cM,\bh}^{-1/2}\bbA_{\cM}\bg_{\cM}+\bbV_{\cM,\bh}^{-1/2}\bbA_{\cM}\left(\tG_{\cM}-\bar{\bbG}_{\cM}\right)\left(\tilde{\btheta}-\btheta_{0}\right)\\
		&\quad-\bbV_{\cM,\bh}^{-1/2}\left(\tA_{\cM}-\bbA_{\cM}\right)\bg_{\cM}+ \bbV_{\cM,\bh}^{-1/2}\left(\tA_{\cM}-\bbA_{\cM}\right)\left(\tG_{\cM}-\bar{\bbG}_{\cM}\right)\left(\tilde{\btheta}-\btheta_{0}\right)\\
		&=\bU_{\cM,1}+\bU_{\cM,2}+\bU_{\cM,3}+\bU_{\cM,4}.
	\end{split}
\end{equation*}
To prove the asymptotic normality in Theorem \ref{theo: mMAS normal}, it suffices to prove  $\bU_{\cM,1}\stackrel{d}{\to} N(\bzero,\bbI)$ and $\bU_{\cM,i}=o_{P}(1)$ for $i=2,3,4$.
We first prove $\bU_{\cM,1}\stackrel{d}{\to} N(\bzero,\bbI)$. Recalling the definition of $\bg_{\cM}$ and by some algebras,
\begin{equation}\label{eq: M decompose}
	\begin{split}
		&\bbV_{\cM,\bh}^{-1/2}\bbA_{\cM}\bg_{\cM}\\
		&=\bbV_{\cM,\bh}^{-1/2}\bbA_{\cM}\left\{\sum_{i=1}^{N}
		\begin{pmatrix}
			n^{-1}\delta_{i}\bu_{i}\\
			n^{-1}\delta_{i}\bbm_{i}-N^{-1}\left(\bh_{i}-\bmu_{\bh}\right)
		\end{pmatrix}-
		\begin{pmatrix}
			\bzero\\
			\left(N^{-1}\sum_{i=1}^{N}\delta_{i}\pi_{i}^{-1}-1\right)(\hat{\bmu}_{\bh}-\bmu_{\bh})
		\end{pmatrix}\right\}.
	\end{split}
\end{equation}
Rewrite $\bbOmega_{\cM}^{-1}$ as
\begin{equation*}
	\bbOmega_{\cM}^{-1} = 
	\begin{pmatrix}
		\bbB_{\bu\bu}& \bbB_{\bu\bbm}\\
		\bbB_{\bbm\bu}& \bbB_{\bbm\bbm}
	\end{pmatrix},
\end{equation*}
where 
$$\bbB_{\bu\bu} = \bbOmega_{\bu\bu}^{-1}+\bbOmega_{\bu\bu}^{-1}\bbOmega_{\bu\bbm}(\bbOmega_{\bbm\bbm}-\bbOmega_{\bbm\bu}\bbOmega_{\bu\bu}^{-1}\bbOmega_{\bu\bbm})^{-1}\bbOmega_{\bbm\bu}\bbOmega_{\bu\bu}^{-1},$$  $$\bbB_{\bbm\bu}=\bbB_{\bu\bbm}^{\T} = -(\bbOmega_{\bbm\bbm}-\bbOmega_{\bbm\bu}\bbOmega_{\bu\bu}^{-1}\bbOmega_{\bu\bbm})^{-1}\bbOmega_{\bbm\bu}\bbOmega_{\bu\bu}^{-1},$$
and 
$$\bbB_{\bbm\bbm} = (\bbOmega_{\bbm\bbm}-\bbOmega_{\bbm\bu}\bbOmega_{\bu\bu}^{-1}\bbOmega_{\bu\bbm})^{-1}.$$
By Condition (C7), $\bbB_{\bu\bu} \asymp 1$, $\bbB_{\bbm\bu} \asymp 1$, and $\bbB_{\bbm\bbm} \asymp \alpha_{N}^{-1}$. Then 
\begin{equation*}
	\bbG_{\bu}^{\T}\bbB_{\bu\bbm}+\bbG_{\bbm}^{\T}\bbB_{\bbm\bbm} \asymp \alpha_{N}^{-1},
\end{equation*}
\begin{equation*}
	\bbG_{\cM}^{\T}\bbOmega_{\cM}^{-1}\bbG_{\cM} = \bbG_{\bu}^{\T}\bbB_{\bu\bu}\bbG_{\bu}+ \bbG_{\bbm}^{\T}\bbB_{\bbm\bu}\bbG_{\bu} + \bbG_{\bu}^{\T}\bbB_{\bu\bbm}\bbG_{\bbm} + \bbG_{\bbm}^{\T}\bbB_{\bbm\bbm}\bbG_{\bbm}\asymp \alpha_{N}^{-1},
\end{equation*}
and $(\bbG_{\cM}^{\T}\bbOmega_{\cM}^{-1}\bbG_{\cM})^{-1}\asymp\alpha_{N}$.
In addition, by Chebyshev's inequality and Conditions (C3), (C6), (C7), we have $N^{-1}\sum_{i=1}^{N}\delta_{i}\pi_{i}^{-1}-1=O_{P}(n^{-1/2})$ and $\hat{\bmu}_{\bh}-\bmu_{\bh}=O_{P}(N^{-1/2})$. Then
\begin{equation*}
	\begin{split}
		&\bbV_{\cM,\bh}^{-1/2}\bbA_{\cM}
		\begin{pmatrix}
			\bzero\\
			\left(N^{-1}\sum_{i=1}^{N}\delta_{i}\pi_{i}^{-1}-1\right)\left(\hat{\bmu}_{\bh}-\bmu_{\bh}\right)
		\end{pmatrix}\\
		& =  n^{1/2}\left(\bbG_{\cM}^{\T}\bbOmega_{\cM}^{-1}\bbG_{\cM}\right)^{-1/2}\left(\bbG_{\bu}^{\T}\bbB_{\bu\bbm}+\bbG_{\bbm}^{\T}\bbB_{\bbm\bbm}\right)\left(N^{-1}\sum_{i=1}^{N}\delta_{i}\pi_{i}^{-1}-1\right)\left(\hat{\bmu}_{\bh}-\bmu_{\bh}\right)\\
		&=O_{P}(\alpha_{N}^{-1/2}N^{-1/2})=o_{P}(1).
	\end{split}
\end{equation*}
Then 
\begin{equation*}
	\begin{split}
		\bU_{\cM,1}
		&=-\bbV_{\cM,\bh}^{-1/2}\bbA_{\cM}\sum_{i=1}^{N}
		\begin{pmatrix}
			n^{-1}\delta_{i}\bu_{i}\\
			n^{-1}\delta_{i}\bbm_{i}-N^{-1}\left(\bh_{i}-\bmu_{\bh}\right)
		\end{pmatrix}+o_{P}(1).
	\end{split}
\end{equation*}
Let
\begin{equation*}
	\bK_{\cM,i} =-\bbV_{\cM,\bh}^{-1/2}\bbA_{\cM}
	\begin{pmatrix}
		n^{-1}\delta_{i}\bu_{i}\\
		n^{-1}\delta_{i}\bbm_{i}-N^{-1}\left(\bh_{i}-\bmu_{\bh}\right)
	\end{pmatrix}.
\end{equation*}
Then
\begin{equation*}
	\bbV_{\cM,\bh}^{-1/2}\bbA_{\cM}\bg_{\cM}
	=\sum_{i=1}^{N}\bK_{\cM,i}+o_{P}(1).
\end{equation*}
To prove $\bU_{\cM,1}{\to} N(\bzero,\bbI)$ in distribution,  it suffices to prove $\sum_{i=1}^{N}\bK_{\cM,i}{\to}N(\bzero,\bbI)$ in distribution.
We first verify that $\sum_{i=1}^{N}\var(\bK_{\cM,i})\to \bbI$.
Note that
\begin{equation*}
	\begin{split}
		&\bK_{\cM,i}\bK_{\cM,i}^{\T}\\ &=n^{-2}\bbV_{\cM,\bh}^{-1/2}\bbA_{\cM}
		\begin{pmatrix}
			\delta_{i}\bu_{i}\bu_{i}^{\T}&\delta_{i}\bu_{i}\left\{\bbm_{i}-\rho_{N}\left(\bh_{i}-\bmu_{\bh}\right)\right\}\\
			\delta_{i}\left\{\bbm_{i}-\rho_{N}\left(\bh_{i}-\bmu_{\bh}\right)\right\}\bu_{i}^{\T}&\left\{\delta_{i}\bbm_{i}-\rho_{N}\left(\bh_{i}-\bmu_{\bh}\right)\right\}^{\otimes 2}
		\end{pmatrix}\bbA_{\cM}^{\T}\bbV_{\cM,\bh}^{-1/2}.
	\end{split}
\end{equation*}
Recalling the definition of $\bbOmega_{\cM}$, we have
\begin{equation*}
	\begin{split}
		E\left(\bK_{\cM,i}\bK_{\cM,i}^{\T}\right) &=n^{-2}\bbV_{\cM,\bh}^{-1/2}\bbA_{\cM}\rho_{N}\bbOmega_{\cM}\bbA_{\cM}^{\T}\bbV_{\cM,\bh}^{-1/2}.
	\end{split}
\end{equation*}
Further, recalling the definition of 
$\bbA_{\cM}$ and $\bbV_{\cM,\bh}$, we have
\begin{equation*}
	E\left(\bK_{\cM,i}\bK_{\cM,i}^{\T}\right) =N^{-1}\bbI.
\end{equation*}
Then by Lindeberg-Feller central limit theorem in \cite{Vaart2000AS}, to prove $$\sum_{i=1}^{N}\bK_{\cM,i}{\to}N(\bzero,\bbI)$$
in distribution, it suffices to verify that $\sum_{i=1}^{N}E\{\|\bK_{\cM,i}\|^2 1(\|\bK_{\cM,i}\|>\epsilon)\}\to 0$  for any given $\epsilon>0$. Since for any given $\tau>0$, we have $E\{\|\bK_{\cM,1}\|^2 1(\|\bK_{\cM,1}\|>\epsilon)\}\leq E(\|\bK_{\cM,1}\|^{2+\tau}/\epsilon^{\tau})$.
Then it suffices to prove that $NE(\|\bK_{\cM,1}\|^{2+\tau})=o(1)$ for some $\tau>0$.
Let $K_{\cM,1}^{(j)}$ be the $j$th element of $\bK_{\cM,1}$ for $j=1,\dots,d$.
By the inequality $(t_{1}+t_{2})^{p}\leq 2^{p-1}(|t_{1}|^{p}+|t_{2}|^{p})$ and Condition (C9), we have
\begin{equation*}
	\begin{split}
		NE(\|\bK_{\cM,1}\|^{2+\tau})
		& \leq c NE(|K_{\cM,1}^{(1)}|^{2+\tau}+\dots+|K_{\cM,1}^{(d)}|^{2+\tau})\\
		&\leq c N\rho_{N}^{-\tau/2}\{\var(K_{\cM,1}^{(1)})^{1+\tau/2}+\dots+\var(K_{\cM,1}^{(d)})^{1+\tau/2}\}.
	\end{split}
\end{equation*}
Since $E(\bK_{\cM,1}\bK_{\cM,1}^{\T})=N^{-1}\bbI$, then it is proved that $NE(\|\bK_{\cM,1}\|^{2+\tau})=O(n^{-\tau/2})=o(1)$.
Then $\bU_{\cM,1}\stackrel{d}{\to}N(\bzero,\bbI)$.
Now, it remains to prove $\bU_{\cM,i}=o_{P}(1)$ for $i=2,3,4$.

For $\bU_{\cM,2}$, we have
\begin{equation*}
	\begin{split}
		\|\bU_{\cM,2}\|
		&\leq n^{1/2}\Big\|\left(\bbG_{\cM}^{\T}\bbOmega_{\cM}^{-1}\bbG_{\cM}\right)^{-1/2}\Big\|\Big\|\bbG_{\cM}^{\T}\bbOmega_{\cM}^{-1}\left(\tG_{\cM}-\bar{\bbG}_{\cM}\right)\Big\|\|\tilde{\btheta}-\btheta_{0}\|\\
		&\leq cn^{1/2}\alpha_{N}^{1/2}\Big\|\bbG_{\cM}^{\T}\bbOmega_{\cM}^{-1}\left(\tG_{\cM}-\bar{\bbG}_{\cM}\right)\Big\|\|\tilde{\btheta}-\btheta_{0}\|.
	\end{split}
\end{equation*}
Note that
\begin{equation*}
	\begin{split}
		\bbG_{\cM}^{\T}\bbOmega_{\cM}^{-1}\left(\tG_{\cM}-\bar{\bbG}_{\cM}\right)
		& =\bbG_{\bu}^{\T}\bbB_{\bu\bu}\left(\tG_{\cM,\bu}-\bar{\bbG}_{\cM,\bu}\right)+ \bbG_{\bbm}^{\T}\bbB_{\bbm\bu}\left(\tG_{\bbm}-\bar{\bbG}_{\bbm}\right)\\
		&\quad+ \bbG_{\bu}^{\T}\bbB_{\bu\bbm}\left(\tG_{\cM,\bu}-\bar{\bbG}_{\cM,\bu}\right) + \bbG_{\bbm}^{\T}\bbB_{\bbm\bbm}\left(\tG_{\bbm}-\bar{\bbG}_{\bbm}\right).
	\end{split}
\end{equation*}
By the W-H theorem in \cite{WHtheorem} and Lemma \ref{lem: AL of initial}, we can show that $\|\tG_{\cM,\bu}-\bbG_{\bu}\|=O_{P}(n^{-1/2})$ and $\|\tG_{\bbm}-\bbG_{\bbm}\|=O_{P}(n^{-1/2})$.
Then by Conditions (C4) and (C7), we have $$\Big\|\bbG_{\cM}^{\T}\bbOmega_{\cM}^{-1}\left(\tG_{\cM}-\bar{\bbG}_{\cM}\right)\Big\| = O_{P}(n^{-1/2}\alpha_{N}^{-1}).$$
Then $\|\bU_{\cM,2}\|=O_{P}(\alpha_{N}^{-1/2}n^{-1/2})=o_{P}(1)$.
Next, we prove $\bU_{\cM,3}=o_{P}(1)$ and $\bU_{\cM,4}=o_{P}(1)$.
Rewrite $\tOmega_{\cM}^{-1}$ as 
\begin{equation*}
	\tOmega_{\cM}^{-1} = 
	\begin{pmatrix}
		\tB_{\bu\bu}& \tB_{\bu\bbm}\\
		\tB_{\bbm\bu}& \tB_{\bbm\bbm}
	\end{pmatrix},
\end{equation*}
where $$\tB_{\bu\bu} = \tOmega_{\bu\bu}^{-1}+\tOmega_{\bu\bu}^{-1}\tOmega_{\bu\bbm}(\tOmega_{\bbm\bbm}-\tOmega_{\bbm\bu}\tOmega_{\bu\bu}^{-1}\tOmega_{\bu\bbm})^{-1}\tOmega_{\bbm\bu}\tOmega_{\bu\bu}^{-1},$$  
$$\tB_{\bbm\bu}=\tB_{\bu\bbm}^{\T} = -(\tOmega_{\bbm\bbm}-\tOmega_{\bbm\bu}\tOmega_{\bu\bu}^{-1}\tOmega_{\bu\bbm})^{-1}\tOmega_{\bbm\bu}\tOmega_{\bu\bu}^{-1},$$
and
$$\tB_{\bbm\bbm} = (\tOmega_{\bbm\bbm}-\tOmega_{\bbm\bu}\tOmega_{\bu\bu}^{-1}\tOmega_{\bu\bbm})^{-1}.$$
By Lemma \ref{lem: AL of initial}, the Chebyshev's inequality and Conditions (C3), (C7), 
\begin{equation*}
	\begin{split}
		&\tOmega_{\bu\bu}-\bbOmega_{\bu\bu}\\ 
		&= \frac{1}{N}\sum_{i=1}^{N}\frac{\delta_{i}}{\pi_{i}}\rho_{N}\pi_{i}^{-1}\bpsi_{i}(\tilde{\btheta})\bpsi_{i}(\tilde{\btheta})^{\T} - E\left(\rho_{N}\pi^{-1}\bpsi\bpsi^{\T}\right)\\
		& = \frac{1}{N}\sum_{i=1}^{N}\frac{\delta_{i}}{\pi_{i}}\rho_{N}\pi_{i}^{-1}\bpsi_{i}\bpsi_{i}^{\T}- E\left(\rho_{N}\pi^{-1}\bpsi\bpsi^{\T}\right) + \frac{1}{N}\sum_{i=1}^{N}\frac{\delta_{i}}{\pi_{i}}\rho_{N}\pi_{i}^{-1}\dot{\bpsi}_{i}(\tilde{\btheta}-\btheta_{0})\bpsi_{i}^{\T}\\
		&\quad  + \frac{1}{N}\sum_{i=1}^{N}\frac{\delta_{i}}{\pi_{i}}\rho_{N}\pi_{i}^{-1}\bpsi_{i}\left\{\dot{\bpsi}_{i}(\tilde{\btheta}-\btheta_{0})\right\}^{\T} + O_{P}(\|\tilde{\btheta}-\btheta_{0}\|^{2})\\
		&=O_{P}(n^{-1/2}),
	\end{split}
\end{equation*}
and
\begin{equation*}
	\begin{split}
		&\tOmega_{\bu\bbm}-\bbOmega_{\bu\bbm}\\
		&=\frac{1}{N}\sum_{i=1}^{N}\frac{\delta_{i}}{\pi_{i}}\bpsi_{i}(\tilde{\btheta})\left[\rho_{N}\pi_{i}^{-1}\{\ba_{i}(\tilde{\btheta})-\hat{\bmu}_{\bh}\}-\rho_{N}\left\{\bh_{i}-\hat{\bmu}_{\bh}\right\}\right]\\
		&\qquad -\left[E\left\{\rho_{N}\pi^{-1}\bpsi\left(\ba-\bmu_{\bh}\right)^{\T}\right\}-\rho_{N}E\left\{\bpsi(\bh-\bmu_{\bh})^{\T}\right\}\right]\\
		&=\frac{1}{N}\sum_{i=1}^{N}\frac{\delta_{i}}{\pi_{i}}\left\{\bpsi_{i}+\dot{\bpsi}_{i}(\tilde{\btheta}-\btheta_{0})\right\}\left[\rho_{N}\pi_{i}^{-1}(\ba_{i}-\bmu_{\bh})+\rho_{N}\pi_{i}^{-1}\dot{\ba}_{i}(\tilde{\btheta}-\btheta_{0})-\rho_{N}\pi_{i}^{-1}(\hat{\bmu}_{\bh}-\bmu_{\bh})\right.\\
		&\left.\qquad-\rho_{N}(\bh_{i}-\bmu_{\bh})+\rho_{N}(\hat{\bmu}_{\bh}-\bmu_{\bh})\right]^{\T}\\
		&\quad-\left[E\left\{\rho_{N}\pi^{-1}\bpsi\left(\ba-\bmu_{\bh}\right)^{\T}\right\}-\rho_{N}E\left\{\bpsi(\bh-\bmu_{\bh})^{\T}\right\}\right]+o_{P}(n^{-1/2})\\
		&=O_{P}(n^{-1/2}),
	\end{split}
\end{equation*}
where $\dot{\bpsi}_{i}=\partial\bpsi_{i}(\btheta)/\partial \btheta\mid_{\btheta=\btheta_{0}}$ and $\dot{\ba}_{i} = \partial \ba_{i}(\btheta)/\partial \btheta\mid_{\btheta=\btheta_{0}}$, and
\begin{align*}
	&\tOmega_{\bbm\bbm}-\bbOmega_{\bbm\bbm}\\
	&=\frac{1}{N}\sum_{i=1}^{N}\left[\delta_{i}\rho_{N}\pi_{i}^{-2}\left\{\ba_{i}(\tilde{\btheta})-\hat{\bmu}_{\bh}\right\}^{\otimes 2}-\rho_{N}\delta_{i}\pi_{i}^{-1}\left\{\ba_{i}(\tilde{\btheta})-\hat{\bmu}_{\bh}\right\}\left(\bh_{i}-\hat{\bmu}_{\bh}\right)^{\T}\right.\\
	&\left.\quad\qquad\qquad-\rho_{N}\delta_{i}\pi_{i}^{-1}\left(\bh_{i}-\hat{\bmu}_{\bh}\right)\left\{\ba_{i}(\tilde{\btheta})-\hat{\bmu}_{\bh}\right\}^{\T}+\rho_{N}\delta_{i}\pi_{i}^{-1}(\bh_{i}-\hat{\bmu}_{\bh})^{\otimes 2}\right]\\
	&\quad-\left[\rho_{N}E\left\{\pi^{-1}(\ba-\bmu_{\bh})^{\otimes 2}\right\}-\rho_{N}E\left\{(\ba-\bmu_{\bh})(\bh-\bmu_{\bh})^{\T}\right\}-\rho_{N}E\left\{(\bh-\bmu_{\bh})(\ba-\bmu_{\bh})^{\T}\right\}\right.\\
	&\left.\qquad\qquad+\rho_{N}E\left\{\left(\bh-\bmu_{\bh}\right)^{\otimes 2}\right\}\right]\\
	&=\frac{1}{N}\sum_{i=1}^{N}\left[\delta_{i}\rho_{N}\pi_{i}^{-2}\left\{(\ba_{i}-\bmu_{\bh})+\dot{\ba}_{i}(\tilde{\btheta}-\btheta_{0})-(\hat{\bmu}_{\bh}-\bmu_{\bh})\right\}^{\otimes 2}\right.\\
	&\qquad\qquad\left.-\rho_{N}\delta_{i}\pi_{i}^{-1}\left\{(\ba_{i}-\bmu_{\bh})+\dot{\ba}_{i}(\tilde{\btheta}-\btheta_{0})-(\hat{\bmu}_{\bh}-\bmu_{\bh})\right\}\left\{\bh_{i}-\bmu_{\bh}-\left(\hat{\bmu}_{\bh}-\bmu_{\bh}\right)\right\}^{\T}\right.\\
	&\left.\qquad\qquad-\rho_{N}\delta_{i}\pi_{i}^{-1}\left\{\bh_{i}-\bmu_{\bh}-\left(\hat{\bmu}_{\bh}-\bmu_{\bh}\right)\right\}\left\{(\ba_{i}-\bmu_{\bh})+\dot{\ba}_{i}(\tilde{\btheta}-\btheta_{0})-(\hat{\bmu}_{\bh}-\bmu_{\bh})\right\}^{\T}\right.\\
	&\left.\qquad\qquad+\rho_{N}\delta_{i}\pi_{i}^{-1}\left\{\bh_{i}-\bmu_{\bh}-\left(\hat{\bmu}_{\bh}-\bmu_{\bh}\right)\right\}^{\otimes 2}\right]\\
	&\quad-\left[\rho_{N}E\left\{\pi^{-1}(\ba-\bmu_{\bh})^{\otimes 2}\right\}-\rho_{N}E\left\{(\ba-\bmu_{\bh})(\bh-\bmu_{\bh})^{\T}\right\}-\rho_{N}E\left\{(\bh-\bmu_{\bh})(\ba-\bmu_{\bh})^{\T}\right\}\right.\\
	&\left.\qquad\qquad+\rho_{N}E\left\{\left(\bh-\bmu_{\bh}\right)^{\otimes 2}\right\}\right]+O_{P}(n^{-1}).
\end{align*}
Let $\ba_{i}^{(l)}$ and $\bmu_{\bh}^{(l)}$ be the $l$-th elements of $\ba_{i}$ and $\bmu_{\bh}$, respectively.
Then for $l,j=1,\cdots,d$, by Condition (C7)(iii), 
\begin{align*}
	&\var\left\{\frac{1}{N}\sum_{i=1}^{N}\delta_{i}\rho_{N}\pi_{i}^{-2}(\ba_{i}^{(l)}-\bmu_{\bh}^{(l)})(\ba_{i}^{(j)}-\bmu_{\bh}^{(j)})\right\} \\
	&= \frac{1}{N}\rho_{N}^{2}E\left\{\pi^{-3}\left(\ba^{(l)}-\bmu_{\bh}^{(l)}\right)^{2}\left(\ba^{(j)}-\bmu_{\bh}^{(j)}\right)^{2}\right\}\\
	&\leq c\frac{1}{n}E\left(\|(\ba-\bmu_{\bh})^{\otimes2}\|_{F}^{2}\right)\\
	&\leq c\alpha_{N}^{2}/n,
\end{align*}
which implies
\begin{equation*}
	\frac{1}{N}\sum_{i=1}^{N}\delta_{i}\rho_{N}\pi_{i}^{-2}(\ba_{i}-\bmu_{\bh})^{\otimes 2}-\rho_{N}E\left\{\pi^{-1}(\ba-\bmu_{\bh})^{\otimes 2}\right\}=O_{P}(\alpha_{N}n^{-1/2}).
\end{equation*}
In addition,
\begin{align*}
	&\frac{1}{N}\sum_{i=1}^{N}\delta_{i}\rho_{N}\pi_{i}^{-2}\left(\ba_{i}-\bmu_{\bh}\right)\left\{\dot{\ba}_{i}(\tilde{\btheta}-\btheta_{0})\right\}^{\T}\\
	&\leq\frac{1}{N}\sum_{i=1}^{N}\delta_{i}\rho_{N}\pi_{i}^{-2}\|\ba_{i}-\bmu_{\bh}\|\|\dot{\ba}_{i}\|\|\tilde{\btheta}-\btheta_{0}\|\\
	&\leq \Big\|\frac{1}{N}\sum_{i=1}^{N}\delta_{i}\rho_{N}\pi_{i}^{-2}\|\ba_{i}-\bmu_{\bh}\|\|\dot{\ba}_{i}\|-E\left\{\rho_{N}\pi^{-1}\|\ba-\bmu_{\bh}\|\|\dot{\ba}\|\right\}\Big\|\|\tilde{\btheta}-\btheta_{0}\|\\
	&\quad+E\left\{\rho_{N}\pi^{-1}\|\ba-\bmu_{\bh}\|\|\dot{\ba}\|\right\}\|\tilde{\btheta}-\btheta_{0}\|.
\end{align*}
By Condition (C7)(iii), $E\left\{\rho_{N}\pi^{-1}\|\ba-\bmu_{\bh}\|\|\dot{\ba}\|\right\}\leq c\alpha_{N}$ and hence 
\begin{equation*}
	\frac{1}{N}\sum_{i=1}^{N}\delta_{i}\rho_{N}\pi_{i}^{-2}\left(\ba_{i}-\bmu_{\bh}\right)\left\{\dot{\ba}_{i}(\tilde{\btheta}-\btheta_{0})\right\}^{\T}=O_{P}(n^{-1}+\alpha_{N}n^{-1/2}).
\end{equation*}
In addition, 
\begin{align*}
	&\frac{1}{N}\sum_{i=1}^{N}\delta_{i}\rho_{N}\pi_{i}^{-2}(\ba_{i}-\bmu_{\bh})(\hat{\bmu}_{\bh}-\bmu_{\bh})^{\T}\\
	&=\left(\frac{1}{N}\sum_{i=1}^{N}\delta_{i}\rho_{N}\pi_{i}^{-2}(\ba_{i}-\bmu_{\bh})-E\left\{\rho_{N}\pi^{-1}(\ba-\bmu_{\bh})\right\}\right)(\hat{\bmu}_{\bh}-\bmu_{\bh})^{\T}  \\
	&\quad + E\left\{\rho_{N}\pi^{-1}(\ba-\bmu_{\bh})\right\}(\hat{\bmu}_{\bh}-\bmu_{\bh})^{\T}\\
	&=O_{P}(n^{-1/2}N^{-1/2}+\alpha_{N}^{1/2}N^{-1/2})\\
	&=O_{P}(n^{-1}N^{-1/2}+\alpha_{N}n^{-1/2}).
\end{align*}
The last equality is obtained based on the relation $\rho_{N}\lesssim\alpha_{N}$ in Condition (C7)). Then 
\begin{equation*}
	\tOmega_{\bbm\bbm}-\bbOmega_{\bbm\bbm}=O_{P}(\alpha_{N}n^{-1/2}+\rho_{N}n^{-1/2}+n^{-1})=O_{P}(\alpha_{N}n^{-1/2}+n^{-1}).
\end{equation*}
Then, based on the previous results and by matrix differentials,
\begin{align*}
	&\tB_{\bbm\bbm}-\bbB_{\bbm\bbm}\\
	& = -\alpha_{N}^{-1}\left[\left\{\alpha_{N}^{-1}\left(\tOmega_{\bbm\bbm}-\tOmega_{\bbm\bu}\tOmega_{\bu\bu}^{-1}\tOmega_{\bu\bbm}\right)\right\}^{-1}-\left\{\alpha_{N}^{-1}\left(\bbOmega_{\bbm\bbm}-\bbOmega_{\bbm\bu}\bbOmega_{\bu\bu}^{-1}\bbOmega_{\bbm\bu}\right)\right\}^{-1}\right]\\
	&=-\alpha_{N}^{-1}\left[\left\{\alpha_{N}^{-1}\left(\bbOmega_{\bbm\bbm}-\bbOmega_{\bbm\bu}\bbOmega_{\bu\bu}^{-1}\bbOmega_{\bbm\bu}\right)\right\}^{-1}\left\{\alpha_{N}^{-1}\left(\tOmega_{\bbm\bbm}-\tOmega_{\bbm\bu}\tOmega_{\bu\bu}^{-1}\tOmega_{\bu\bbm}\right)\right.\right.\\
	&\left.\left.\qquad\qquad-\alpha_{N}^{-1}\left(\bbOmega_{\bbm\bbm}-\bbOmega_{\bbm\bu}\bbOmega_{\bu\bu}^{-1}\bbOmega_{\bbm\bu}\right)\right\}\left\{\alpha_{N}^{-1}\left(\bbOmega_{\bbm\bbm}-\bbOmega_{\bbm\bu}\bbOmega_{\bu\bu}^{-1}\bbOmega_{\bbm\bu}\right)\right\}^{-1}\right] \\
	&\quad + o_{P}(\alpha_{N}^{-1}\|\alpha_{N}^{-1}\left(\tOmega_{\bbm\bbm}-\tOmega_{\bbm\bu}\tOmega_{\bu\bu}^{-1}\tOmega_{\bu\bbm}\right)-\alpha_{N}^{-1}\left(\bbOmega_{\bbm\bbm}-\bbOmega_{\bbm\bu}\bbOmega_{\bu\bu}^{-1}\bbOmega_{\bbm\bu}\right)\|)\\
	&=O_{P}\left(\alpha_{N}^{-2}\big\|\left(\tOmega_{\bbm\bbm}-\tOmega_{\bbm\bu}\tOmega_{\bu\bu}^{-1}\tOmega_{\bu\bbm}\right)-\left(\bbOmega_{\bbm\bbm}-\bbOmega_{\bbm\bu}\bbOmega_{\bu\bu}^{-1}\bbOmega_{\bbm\bu}\right)\big\|\right),
\end{align*}
and
\begin{align*}
	&\tOmega_{\bbm\bu}\tOmega_{\bu\bu}^{-1}\tOmega_{\bu\bbm}-\bbOmega_{\bbm\bu}\bbOmega_{\bu\bu}^{-1}\bbOmega_{\bu\bbm}\\
	& = \alpha_{N}^{2}\left\{(\alpha_{N}^{-1}\tOmega_{\bbm\bu})\tOmega_{\bu\bu}^{-1}(\alpha_{N}^{-1}\tOmega_{\bu\bbm})-(\alpha_{N}^{-1}\bbOmega_{\bbm\bu})\bbOmega_{\bu\bu}^{-1}(\alpha_{N}^{-1}\bbOmega_{\bu\bbm})\right\}\\
	&=\alpha_{N}^{2}\left\{\left(\alpha_{N}^{-1}\tOmega_{\bbm\bu}-\alpha_{N}^{-1}\bbOmega_{\bbm\bu}\right)\bbOmega_{\bu\bu}^{-1}(\alpha_{N}^{-1}\bbOmega_{\bu\bbm})\right.\\
	&\qquad\qquad\left. + (\alpha_{N}^{-1}\bbOmega_{\bbm\bu})\bbOmega_{\bu\bu}^{-1}\left(\tOmega_{\bu\bu}-\bbOmega_{\bu\bu}\right)\bbOmega_{\bu\bu}^{-1}(\alpha_{N}^{-1}\bbOmega_{\bu\bbm})\right.\\
	&\qquad\qquad\left.+(\alpha_{N}^{-1}\bbOmega_{\bbm\bu})\bbOmega_{\bu\bu}^{-1}\left(\alpha_{N}^{-1}\tOmega_{\bu\bbm}-\alpha_{N}^{-1}\bbOmega_{\bu\bbm}\right)\right.\\
	&\qquad\qquad\left.+o_{P}(\|\alpha_{N}^{-1}\tOmega_{\bu\bbm}-\alpha_{N}^{-1}\bbOmega_{\bu\bbm}\|+\|\tOmega_{\bu\bu}-\bbOmega_{\bu\bu}\|)\right\}\\
	&= O_{P}(\alpha_{N}n^{-1/2}).
\end{align*}
Then $\|\tB_{\bbm\bbm}-\bbB_{\bbm\bbm}\| = O_{P}(\alpha_{N}^{-2}(\alpha_{N}n^{-1/2}+n^{-1})) = O_{P}(\alpha_{N}^{-1}n^{-1/2}+\alpha_{N}^{-2}n^{-1})$.
In addition,
\begin{align*}
	&\tB_{\bbm\bu}-\bbB_{\bbm\bu}\\
	& = -\left\{(\alpha_{N}\tB_{\bbm\bbm})(\alpha_{N}^{-1}\tOmega_{\bbm\bu})\tOmega_{\bu\bu}^{-1}-(\alpha_{N}\bbB_{\bbm\bbm})(\alpha_{N}^{-1}\bbOmega_{\bbm\bu})\bbOmega_{\bu\bu}^{-1}\right\}\\
	& = -\left\{(\alpha_{N}\tB_{\bbm\bbm}-\alpha_{N}\bbB_{\bbm\bbm})(\alpha_{N}^{-1}\bbOmega_{\bbm\bu})\bbOmega_{\bu\bu}^{-1}\right\}\\
	&\quad-\left\{(\alpha_{N}\bbB_{\bbm\bbm})(\alpha_{N}^{-1}\tOmega_{\bbm\bu}-\alpha_{N}^{-1}\bbOmega_{\bbm\bu})\bbOmega_{\bu\bu}^{-1}\right\}\\
	&\quad-\left\{(\alpha_{N}\bbB_{\bbm\bbm})(\alpha_{N}^{-1}\bbOmega_{\bbm\bu})\bbOmega_{\bu\bu}^{-1}(\tOmega_{\bu\bu}-\bbOmega_{\bu\bu})\bbOmega_{\bu\bu}^{-1}\right\}\\
	&\quad+o_{P}(\|\alpha_{N}\tB_{\bbm\bbm}-\alpha_{N}\bbB_{\bbm\bbm}\|+\|\alpha_{N}^{-1}\tOmega_{\bbm\bu}-\alpha_{N}^{-1}\bbOmega_{\bbm\bu}\|+\|\tOmega_{\bu\bu}-\bbOmega_{\bu\bu}\|)\\
	&=O_{P}(\alpha_{N}^{-1}n^{-1/2}),
\end{align*}
and
\begin{align*}
	&\tB_{\bu\bu}-\bbB_{\bu\bu}\\
	& = \tOmega_{\bu\bu}^{-1}-\bbOmega_{\bu\bu}^{-1} + \alpha_{N}\tOmega_{\bu\bu}^{-1}(\alpha_{N}^{-1}\tOmega_{\bu\bbm})(\tB_{\bbm\bu}) - \alpha_{N}\bbOmega_{\bu\bu}^{-1}(\alpha_{N}^{-1}\bbOmega_{\bu\bbm})(\bbB_{\bbm\bu}) \\
	& = \bbOmega_{\bu\bu}^{-1}(\tOmega_{\bu\bu}-\bbOmega_{\bu\bu})\bbOmega_{\bu\bu}^{-1} + \alpha_{N}\bbOmega_{\bu\bu}^{-1}(\tOmega_{\bu\bu}-\bbOmega_{\bu\bu})\bbOmega_{\bu\bu}^{-1}(\alpha_{N}^{-1}\bbOmega_{\bu\bbm})\bbB_{\bbm\bu}\\
	&\quad + \alpha_{N}\bbOmega_{\bu\bu}^{-1}(\alpha_{N}^{-1}\tOmega_{\bu\bbm}-\alpha_{N}^{-1}\bbOmega_{\bu\bbm})\bbB_{\bbm\bu} + \alpha_{N}\bbOmega_{\bu\bu}^{-1}(\alpha_{N}^{-1}\bbOmega_{\bu\bbm})(\tB_{\bbm\bu}-\bbB_{\bbm\bu})\\
	&\quad + o_{P}(\|\tOmega_{\bu\bu}-\bbOmega_{\bu\bu}\|+\alpha_{N}\|\tOmega_{\bu\bu}-\bbOmega_{\bu\bu}\|+\alpha_{N}\|\alpha_{N}^{-1}\tOmega_{\bu\bbm}-\alpha_{N}^{-1}\bbOmega_{\bu\bbm}\| + \alpha_{N}\|\tB_{\bbm\bu}-\bbB_{\bbm\bu}\|)\\
	& = O_{P}(n^{-1/2}).
\end{align*}
Further, 
\begin{align*}
	&(\tG_{\cM}^{\T}\tOmega_{\cM}^{-1}\tG_{\cM})^{-1}-(\bbG_{\cM}^{T}\bbOmega_{\cM}^{-1}\bbG_{\cM})^{-1}\\
	& = \alpha_{N}(\alpha_{N}\bbG_{\cM}^{T}\bbOmega_{\cM}^{-1}\bbG_{\cM})^{-1}(\alpha_{N}\tG_{\cM}^{\T}\tOmega_{\cM}^{-1}\tG_{\cM}-\alpha_{N}\bbG_{\cM}^{T}\bbOmega_{\cM}^{-1}\bbG_{\cM})(\alpha_{N}\bbG_{\cM}^{T}\bbOmega_{\cM}^{-1}\bbG_{\cM})^{-1}\\
	&\quad + o_{P}(\alpha_{N}\|\alpha_{N}\tG_{\cM}^{\T}\tOmega_{\cM}^{-1}\tG_{\cM}-\alpha_{N}\bbG_{\cM}^{T}\bbOmega_{\cM}^{-1}\bbG_{\cM}\|)\\
	&=O_{P}(\alpha_{N}\|\alpha_{N}\tG_{\cM}^{\T}\tOmega_{\cM}^{-1}\tG_{\cM}-\alpha_{N}\bbG_{\cM}^{T}\bbOmega_{\cM}^{-1}\bbG_{\cM}\|)
\end{align*}
and
\begin{align*}
	&\tG_{\cM}^{\T}\tOmega_{\cM}^{-1}\tG_{\cM}-\bbG_{\cM}^{T}\bbOmega_{\cM}^{-1}\bbG_{\cM}\\
	&=\left(\tG_{\cM,\bu}^{\T}\tB_{\bu\bu}\tG_{\cM,\bu}-\bbG_{\bu}^{\T}\bbB_{\bu\bu}\bbG_{\bu}\right)+\left(\tG_{\bbm}^{\T}\tB_{\bbm\bu}\tG_{\cM,\bu}-\bbG_{\bbm}^{\T}\bbB_{\bbm\bu}\bbG_{\bu}\right)\\
	&\quad+\left(\tG_{\cM,\bu}^{\T}\tB_{\bu\bbm}\tG_{\bbm}-\bbG_{\bu}^{\T}\bbB_{\bu\bbm}\bbG_{\bbm}\right)+\left(\tG_{\bbm}^{\T}\tB_{\bbm\bbm}\tG_{\bbm}-\bbG_{\bbm}^{\T}\bbB_{\bbm\bbm}\bbG_{\bbm}\right)\\
	& = \left(\tG_{\cM,\bu}-\bbG_{\bu}\right)^{\T}\bbB_{\bu\bu}\bbG_{\bu}+\bbG_{\bu}^{\T}\left(\tB_{\bu\bu}-\bbB_{\bu\bu}\right)\bbG_{\bu}+\bbG_{\bu}^{\T}\bbB_{\bu\bu}\left(\tG_{\cM,\bu}-\bbG_{\bu}\right)\\		
	&\quad+\left(\tG_{\bbm}-\bbG_{\bbm}\right)^{\T}\bbB_{\bbm\bu}\bbG_{\bu}+\bbG_{\bbm}^{\T}\left(\tB_{\bbm\bu}-\bbB_{\bbm\bu}\right)\bbG_{\bu}+\bbG_{\bbm}^{\T}\bbB_{\bbm\bu}\left(\tG_{\cM,\bu}-\bbG_{\bu}\right)\\
	&\quad+\left(\tG_{\cM,\bu}-\bbG_{\bu}\right)^{\T}\bbB_{\bu\bbm}\bbG_{\bbm}+\bbG_{\bu}^{\T}\left(\tB_{\bu\bbm}-\bbB_{\bu\bbm}\right)\bbG_{\bbm}+\bbG_{\bu}^{\T}\bbB_{\bu\bbm}\left(\tG_{\bbm}-\bbG_{\bbm}\right)\\
	&\quad+\alpha_{N}^{-1}\left(\tG_{\bbm}-\bbG_{\bbm}\right)^{\T}(\alpha_{N}\bbB_{\bbm\bbm})\bbG_{\bbm}+\alpha_{N}^{-1}\bbG_{\bbm}^{\T}\left(\alpha_{N}\tB_{\bbm\bbm}-\alpha_{N}\bbB_{\bbm\bbm}\right)\bbG_{\bbm}\\
	&\quad+\alpha_{N}^{-1}\bbG_{\bbm}^{\T}(\alpha_{N}\bbB_{\bbm\bbm})\left(\tG_{\bbm}-\bbG_{\bbm}\right)+o_{P}(\|\tG_{\cM,\bu}-\bbG_{\bu}\|+\|\tB_{\bu\bu}-\bbB_{\bu\bu}\|\\
	&\quad+\|\tB_{\bu\bbm}-\bbB_{\bu\bbm}\|+\alpha_{N}^{-1}\|\tG_{\bbm}-\bbG_{\bbm}\|+\alpha_{N}^{-1}\|\alpha_{N}\tB_{\bbm\bbm}-\alpha_{N}\bbB_{\bbm\bbm}\|)\\
	&=O_{P}(\alpha_{N}^{-1}n^{-1/2}+\alpha_{N}^{-2}n^{-1}).
\end{align*}
Then
\begin{align*}
	(\tG_{\cM}^{\T}\tOmega_{\cM}^{-1}\tG_{\cM})^{-1}-(\bbG_{\cM}^{T}\bbOmega_{\cM}^{-1}\bbG_{\cM})^{-1}= O_{P}(\alpha_{N}n^{-1/2}+n^{-1}).
\end{align*}
In addition, 
\begin{align*}
	&\left(\tG_{\cM}^{\T}\tOmega_{\cM}^{-1}-\bbG_{\cM}^{\T}\bbOmega_{\cM}^{-1}\right)\\ 
	&= \begin{pmatrix}
		\tG_{\cM,\bu}^{\T}\tB_{\bu\bu}+\tG_{\bbm}^{\T}\tB_{\bbm\bu}-\bbG_{\bu}^{\T}\bbB_{\bu\bu}-\bbG_{\bbm}^{\T}\bbB_{\bbm\bu}\\
		\tG_{\cM,\bu}^{\T}\tB_{\bu\bbm}+\alpha_{N}^{-1}\tG_{\bbm}^{\T}(\alpha_{N}\tB_{\bbm\bbm})-\bbG_{\bu}^{\T}\bbB_{\bu\bbm}-\alpha_{N}^{-1}\bbG_{\bbm}^{\T}(\alpha_{N}\bbB_{\bbm\bbm})
	\end{pmatrix} \\
	&=\begin{pmatrix}
		(\tG_{\cM,\bu}-\bbG_{\bu})^{\T}\bbB_{\bu\bu}+\bbG_{\bu}^{\T}(\tB_{\bu\bu}-\bbB_{\bu\bu})+(\tG_{\bbm}-\bbG_{\bbm})^{\T}\bbB_{\bbm\bu}+\bbG_{\bbm}^{\T}(\tB_{\bbm\bu}-\bbB_{\bbm\bu})\\
		(\tG_{\cM,\bu}-\bbG_{\bu})^{\T}\bbB_{\bu\bbm}+\bbG_{\bu}^{\T}(\tB_{\bu\bbm}-\bbB_{\bu\bbm})+(\tG_{\bbm}-\bbG_{\bbm})^{\T}\bbB_{\bbm\bbm}+\bbG_{\bbm}^{\T}(\tB_{\bbm\bbm}-\bbB_{\bbm\bbm})
	\end{pmatrix} \\
	& \quad + 
	\begin{pmatrix}
		o_{P}(\|\tG_{\cM,\bu}-\bbG_{\bu}\|+\|\tB_{\bu\bu}-\bbB_{\bu\bu}\|+\|\tG_{\bbm}-\bbG_{\bbm}\|+\|\tB_{\bbm\bu}-\bbB_{\bbm\bu}\|)\\
		o_{P}(\|\tG_{\cM,\bu}-\bbG_{\bu}\|+\|\tB_{\bbm\bu}-\bbB_{\bbm\bu}\|+\alpha_{N}^{-1}\|\tG_{\bbm}-\bbG_{\bbm}\|+\|\tB_{\bbm\bbm}-\bbB_{\bbm\bbm}\|)
	\end{pmatrix}\\
	&=\begin{pmatrix}
		O_{P}(\alpha_{N}^{-1}n^{-1/2})\\
		O_{P}(\alpha_{N}^{-1}n^{-1/2}+\alpha_{N}^{-2}n^{-1})
	\end{pmatrix}.
\end{align*}
Since $n\alpha_{N}\to \infty$, then we have 
\begin{align*}
	&\tA_{\cM}-\bbA_{\cM}\\
	& = \left\{(\tG_{\cM}^{\T}\tOmega_{\cM}^{-1}\tG_{\cM})^{-1}-(\bbG_{\cM}^{\T}\bbOmega_{\cM}^{-1}\bbG_{\cM})^{-1}\right\}\bbG_{\cM}^{\T}\bbOmega_{\cM}^{-1}\\
	&\quad+\left\{(\tG_{\cM}^{\T}\tOmega_{\cM}^{-1}\tG_{\cM})^{-1}-(\bbG_{\cM}^{\T}\bbOmega_{\cM}^{-1}\bbG_{\cM})^{-1}\right\}\left(\tG_{\cM}^{\T}\tOmega_{\cM}^{-1}-\bbG_{\cM}^{\T}\bbOmega_{\cM}^{-1}\right)\\
	&\quad+(\bbG_{\cM}^{\T}\bbOmega_{\cM}^{-1}\bbG_{\cM})^{-1}\left(\tG_{\cM}^{\T}\tOmega_{\cM}^{-1}-\bbG_{\cM}^{\T}\bbOmega_{\cM}^{-1}\right)\\
	&=O_{P}\big\{(\alpha_{N}n^{-1/2}+n^{-1})\alpha_{N}^{-1}+(\alpha_{N}n^{-1/2}+n^{-1})(\alpha_{N}^{-1}n^{-1/2}+\alpha_{N}^{-2}n^{-1})+\alpha_{N}(\alpha_{N}^{-1}n^{-1/2}+\alpha_{N}^{-2}n^{-1})\big\}\\
	&=O_{P}(n^{-1/2}+\alpha_{N}^{-1}n^{-1})\\
	&=o_{P}(1),
\end{align*}
and
\begin{align*}
	&(\tA_{\cM}-\bbA_{\cM})\bg_{\cM}\\
	&=(\tA_{\cM}-\bbA_{\cM})
	\begin{pmatrix}
		\frac{1}{n}\sum_{i=1}^{N}\delta_{i}\bu_{i}\\
		\frac{1}{n}\sum_{i=1}^{N}\delta_{i}\bbm_{i}
	\end{pmatrix}-(\tA_{\cM}-\bbA_{\cM})
	\begin{pmatrix}
		\bzero\\
		\frac{1}{N}\sum_{i=1}^{N}\left\{\bh_{i}-\bmu_{\bh}\right\}
	\end{pmatrix}\\
	&\quad-(\tA_{\cM}-\bbA_{\cM})
	\begin{pmatrix}
		\bzero\\
		\left\{\frac{1}{N}\sum_{i=1}^{N}\delta_{i}\pi_{i}^{-1}-1\right\}(\hat{\bmu}_{\bh}-\bmu_{\bh})
	\end{pmatrix}\\
	&=(\tA_{\cM}-\bbA_{\cM})
	\begin{pmatrix}
		\frac{1}{N}\sum_{i=1}^{N}\delta_{i}\pi_{i}^{-1}\bpsi_{i}\\
		\frac{1}{N}\sum_{i=1}^{N}\delta_{i}\pi_{i}^{-1}(\ba_{i}-\bmu_{\bh})
	\end{pmatrix}+O_{P}\{(n^{-1/2}+\alpha_{N}^{-1}n^{-1})(N^{-1/2})\}\\
	& = \left\{(\tG_{\cM}^{\T}\tOmega_{\cM}^{-1}\tG_{\cM})^{-1}-(\bbG_{\cM}^{\T}\bbOmega_{\cM}^{-1}\bbG_{\cM})^{-1}\right\}\bbG_{\cM}^{\T}\bbOmega_{\cM}^{-1}
	\begin{pmatrix}
		\frac{1}{N}\sum_{i=1}^{N}\delta_{i}\pi_{i}^{-1}\bpsi_{i}\\
		\frac{1}{N}\sum_{i=1}^{N}\delta_{i}\pi_{i}^{-1}(\ba_{i}-\bmu_{\bh})
	\end{pmatrix}\\
	&\quad+\left\{(\tG_{\cM}^{\T}\tOmega_{\cM}^{-1}\tG_{\cM})^{-1}-(\bbG_{\cM}^{\T}\bbOmega_{\cM}^{-1}\bbG_{\cM})^{-1}\right\}\left(\tG_{\cM}^{\T}\tOmega_{\cM}^{-1}-\bbG_{\cM}^{\T}\bbOmega_{\cM}^{-1}\right)
	\begin{pmatrix}
		\frac{1}{N}\sum_{i=1}^{N}\delta_{i}\pi_{i}^{-1}\bpsi_{i}\\
		\frac{1}{N}\sum_{i=1}^{N}\delta_{i}\pi_{i}^{-1}(\ba_{i}-\bmu_{\bh})
	\end{pmatrix}\\
	&\quad+(\bbG_{\cM}^{\T}\bbOmega_{\cM}^{-1}\bbG_{\cM})^{-1}\left(\tG_{\cM}^{\T}\tOmega_{\cM}^{-1}-\bbG_{\cM}^{\T}\bbOmega_{\cM}^{-1}\right)
	\begin{pmatrix}
		\frac{1}{N}\sum_{i=1}^{N}\delta_{i}\pi_{i}^{-1}\bpsi_{i}\\
		\frac{1}{N}\sum_{i=1}^{N}\delta_{i}\pi_{i}^{-1}(\ba_{i}-\bmu_{\bh})
	\end{pmatrix}\\
	&=O_{P}\Big(\left(\alpha_{N}n^{-1/2}+n^{-1}\right)\left(n^{-1/2}+\alpha_{N}^{-1}\alpha_{N}^{1/2}n^{-1/2}\right)\\
	&\quad\quad+\left(\alpha_{N}n^{-1/2}+n^{-1}\right)\left\{\alpha_{N}^{-1}n^{-1} +\left(\alpha_{N}^{-1}n^{-1/2}+\alpha_{N}^{-2}n^{-1}\right)n^{-1/2}\alpha_{N}^{1/2}\right\}\\
	&\quad\quad+\left\{n^{-1}+\alpha_{N}(\alpha_{N}^{-1}n^{-1/2}+\alpha_{N}^{-2}n^{-1})n^{-1/2}\alpha_{N}^{1/2}\right\}\Big)\\
	&=o_{P}(n^{-1/2}\alpha_{N}^{1/2}).
\end{align*}
Then 
\begin{align*}
	\|\bU_{\cM,3}\|=n^{1/2}\big\|\left(\bbG_{\cM}^{\T}\bbOmega_{\cM}^{-1}\bbG_{\cM}\right)^{1/2}\big\|\big\|\left(\tA_{\cM}-\bbA_{\cM}\right)\bg_{\cM}\big\| = o_{P}(1).
\end{align*}
In addition,
\begin{equation*}
	\begin{split}
		\|\bU_{\cM,4}\| 
		&= n^{1/2}\|\left(\bbG_{\cM}^{\T}\bbOmega_{\cM}^{-1}\bbG_{\cM}\right)^{1/2}\|\|\tA_{\cM}-\bbA_{\cM}\|\|\tG_{\cM}-\bar{\bbG}_{\cM}\|\|\tilde{\btheta}-\btheta_{0}\|\\
		&=O_{P}(\alpha_{N}^{-1/2}n^{-1/2})=o_{P}(1).
	\end{split}
\end{equation*}

\subsection*{Proof of Theorem \ref{theo: MAS compare}}
By the definitions of $\bbV_{\cS,\bh}$ and $\bbV_{\cM,\bh}$, when $\pi(\bx,\by)=n/N$, we have $\bbOmega_{\bu\bu} = \var(\bpsi)$, $\bbOmega_{\bu\bs} = (1-\rho_{N})\cov(\bpsi,\bh)$, $\bbOmega_{\bs\bs}=(1-\rho_{N})\var(\bh)$, 
$$\bbOmega_{\bu\bbm}= - \rho_{N}\cov(\bpsi,\bh),$$ and $$\bbOmega_{\bbm\bbm} = (1-2\rho_{N})\var(\ba)+\rho_{N}\var(\bh).$$ 
Then
\begin{equation*}
	\begin{split}
		&\bbG_{\cM}^{\T}\bbOmega_{\cM}^{-1}\bbG_{\cM} \\
		& = \bbG_{\bu}^{\T}\bbOmega_{\bu\bu}^{-1}\bbG_{\bu}+(\bbG_{\bu}^{\T}\bbOmega_{\bu\bu}^{-1}\bbOmega_{\bu\bbm}-\bbG_{\bbm}^{\T})(\bbOmega_{\bbm\bbm}-\bbOmega_{\bbm\bu}\bbOmega_{\bu\bu}^{-1}\bbOmega_{\bu\bbm})^{-1}(\bbG_{\bu}^{\T}\bbOmega_{\bu\bu}^{-1}\bbOmega_{\bu\bbm}-\bbG_{\bbm}^{\T})^{\T}\\
		&= \bbOmega_{\bu\bu}+\bbOmega_{\bu\bs}(\bbOmega_{\bbm\bbm}-\bbOmega_{\bbm\bu}\bbOmega_{\bu\bu}^{-1}\bbOmega_{\bu\bbm})^{-1}\bbOmega_{\bu\bs}^{\T},
	\end{split}
\end{equation*}
where the second equality is due to that $\bbG_{\bu} = -\bbOmega_{\bu\bu}$ and $\bbG_{\bbm} =\cov(\bh,\bpsi)$.
In addition, we have
\begin{equation*}
	\begin{split}
		&\bbG_{\bu}^{\T}\left(\bbOmega_{\bu\bu}-\bbOmega_{\bu\bs}\bbOmega_{\bs\bs}^{-1}\bbOmega_{\bs\bu}\right)^{-1}\bbG_{\bu}\\
		& = \bbG_{\bu}^{\T}\bbOmega_{\bu\bu}^{-1}\bbG_{\bu}+\bbG_{\bu}\bbOmega_{\bu\bu}^{-1}\bbOmega_{\bu\bs}\left(\bbOmega_{\bs\bs}-\bbOmega_{\bs\bu}\bbOmega_{\bu\bu}^{-1}\bbOmega_{\bu\bs}\right)^{-1}\bbOmega_{\bs\bu}\bbOmega_{\bu\bu}^{-1}\bbG_{\bu}\\
		&=\bbOmega_{\bu\bu}+\bbOmega_{\bu\bs}\left(\bbOmega_{\bs\bs}-\bbOmega_{\bs\bu}\bbOmega_{\bu\bu}^{-1}\bbOmega_{\bu\bs}\right)^{-1}\bbOmega_{\bu\bs}^{\T}.
	\end{split}
\end{equation*}
Let $\bbP = \cov(\bh,\bpsi)\var(\bpsi)^{-1}$. 
Note that $$\var(\bh)=\var(\ba)+\var(\bh-\ba)$$
and
\begin{equation*}
	\var(\bh-\ba)=\var(\bh-\ba-\bbP \bpsi)+\bbP \cov(\bpsi,\bh).
\end{equation*}
Then we have 
\begin{equation*}
	\begin{split}
		&\bbOmega_{\bbm\bbm}-\bbOmega_{\bbm\bu}\bbOmega_{\bu\bu}^{-1}\bbOmega_{\bu\bbm}\\
		& = (1-2\rho_{N})\var(\ba)+\rho_{N}\var(\bh)-\rho_{N}^{2}\bbP \cov(\bpsi,\bh)\\
		&=(1-\rho_{N})\var(\ba)+\rho_{N}(1-\rho_{N})\var(\bh-\ba)+\rho_{N}^{2}\var(\bh-\ba-\bbP\bpsi),
	\end{split}
\end{equation*}
and
\begin{equation*}
	\begin{split}
		&\bbOmega_{\bs\bs}-\bbOmega_{\bs\bu}\bbOmega_{\bu\bu}^{-1}\bbOmega_{\bu\bs}\\
		&= (1-\rho_{N})\var(\bh)-(1-\rho_{N})^{2}\bbP \cov(\bpsi,\bh)\\
		&=(1-\rho_{N})\left[\var(\bh)-(1-\rho_{N})\left\{\var(\bh)-\var(\ba)-\var(\bh-\ba-\bbP\bpsi)\right\}\right]\\
		&=(1-\rho_{N})\var(\ba)+\rho_{N}(1-\rho_{N})\var(\bh-\ba)\\
		&\quad+(1-\rho_{N})^{2}\var(\bh-\ba-\bbP\bpsi).
	\end{split}
\end{equation*}

When $0\leq\rho_{N}\leq 1/2$, we have $$\bbOmega_{\bbm\bbm}-\bbOmega_{\bbm\bu}\bbOmega_{\bu\bu}^{-1}\bbOmega_{\bu\bbm}\leq\bbOmega_{\bs\bs}-\bbOmega_{\bs\bu}\bbOmega_{\bu\bu}^{-1}\bbOmega_{\bu\bs}.$$
Then $$\bbG_{\bu}^{\T}\left(\bbOmega_{\bu\bu}-\bbOmega_{\bu\bs}\bbOmega_{\bs\bs}^{-1}\bbOmega_{\bs\bu}\right)^{-1}\bbG_{\bu}\geq \bbG_{\cM}^{\T}\bbOmega_{\cM}^{-1}\bbG_{\cM}$$
and hence $\bbV_{\cM,\bh}\leq\bbV_{\cS,\bh}$.

When $1/2\leq\rho_{N}\leq 1$, we have $$\bbOmega_{\bbm\bbm}-\bbOmega_{\bbm\bu}\bbOmega_{\bu\bu}^{-1}\bbOmega_{\bu\bbm}\geq\bbOmega_{\bs\bs}-\bbOmega_{\bs\bu}\bbOmega_{\bu\bu}^{-1}\bbOmega_{\bu\bs}.$$
Then $$\bbG_{\bu}^{\T}\left(\bbOmega_{\bu\bu}-\bbOmega_{\bu\bs}\bbOmega_{\bs\bs}^{-1}\bbOmega_{\bs\bu}\right)^{-1}\bbG_{\bu}\leq \bbG_{\cM}^{\T}\bbOmega_{\cM}^{-1}\bbG_{\cM}$$
and hence $\bbV_{\cM,\bh}\geq\bbV_{\cS,\bh}$.

\subsection*{Proof of Theorem \ref{theo: opt h}}
We first prove Theorem \ref{theo: opt h} for $\bbV_{\cS,\bh}$. Recalling the definition of $\bV_{\cS,\bh}$, we have
\begin{equation*}
	\begin{split}
		\bbV_{\cS,\bh}
		&= n^{-1}\bbG_{\bu}^{-1}(\bbOmega_{\bu\bu}-\bbOmega_{\bu\bs}\bbOmega_{\bs\bs}^{-1}\bbOmega_{\bs\bu})\bbG_{\bu}^{-1}\\
		& = \frac{1}{N}\bbG_{\bu}^{-1}E\left[\left\{\rho_{N}^{-1}\delta\bu-\bbOmega_{\bu\bs}\bbOmega_{\bs\bs}^{-1}(\delta\pi^{-1}-1)(\bh-\bmu_{\bh})\right\}^{\otimes2}\right]\bbG_{\bu}^{-1}.
	\end{split}
\end{equation*}
Without loss generality,  we assume $\bmu_{\bh}=0$ since $\bbV_{\cS,\bh}$ is invariant if $\bh(\bx,\by)$ is replaced by $\bh(\bx,\by)-\bc$ for any $q$ dimensional constant vector $\bc$.
Then we have
\begin{align*}	
	&E\left[\left\{\rho_{N}^{-1}\delta\bu-\bbOmega_{\bu\bs}\bbOmega_{\bs\bs}^{-1}\left(\delta\pi^{-1}-1\right)\bh\right\}^{\otimes2}\right]\\
	& =E\left(\pi^{-1}\bpsi^{\otimes2}\right)+E\left\{\left(\pi^{-1}-1\right)\left(\bbOmega_{\bu\bs}\bbOmega_{\bs\bs}^{-1}\bh\right)^{\otimes 2}\right\}-2E\left\{\left(\pi^{-1}-1\right)\bpsi\left(\bbOmega_{\bu\bs}\bbOmega_{\bs\bs}^{-1}\bh\right)^{\T}\right\}\\
	&=E\left\{\bpsi^{\otimes2}\right\}+E\left\{\left(\pi^{-1}-1\right)\left(\bpsi-\bbOmega_{\bu\bs}\bbOmega_{\bs\bs}^{-1}\bh\right)^{\otimes2}\right\}\\
	& \geq E\left(\bpsi^{\otimes2}\right).
\end{align*}
Then $\bbV_{\cS,\bh}$ attains the minimum if and only if $\bpsi=\bbOmega_{\bu\bs}\bbOmega_{\bs\bs}^{-1}\bh$. 
Recalling the definition of $\bbOmega_{\bu\bs}$ and $\bbOmega_{\bs\bs}$, it is easy to verify that $\bpsi=\bbOmega_{\bu\bs}\bbOmega_{\bs\bs}^{-1}\bh$ if and only if  $\bpsi(\bx,\by;\btheta_{0})=\bbM\bh(\bx,\by)$ for some matrix $\bbM$.

Next, we prove Theorem \ref{theo: opt h} for $\bbV_{\cM,\bh}$.
We first prove that $\bbV_{\cM,\bh}\geq N^{-1}\bbOmega_{\bu\bu}^{-1}$ for any $\bh(\bx,\by)$, and then prove that the equality holds if and only if $\bpsi(\bx,\by;\btheta_{0})=\bbM\bh(\bx,\by)$ for some matrix $\bbM$.
Recalling the definition of $\bbV_{\cM,\bh}$, to prove $\bbV_{\cM,\bh}\geq N^{-1}\bbOmega_{\bu\bu}^{-1}$ for any $\bh(\bx,\by)$, it suffices to prove $\bbG_{\cM}^{\T}\bbOmega_{\cM}^{-1}\bbG_{\cM}\leq \rho_{N}^{-1}\bbOmega_{\bu\bu}$ for any $\bh(\bx,\by)$. 
When $\pi(\bx,\by)=n/N$, we have $\bu(\bx,\by;\btheta)= \bpsi(\bx,\by;\btheta)$,  $\bbG_{\bu}=-E\left(\bpsi^{\otimes2}\right)$, $\bbG_{\bbm}=E(\bh\bpsi^{\T})$,  $\bbOmega_{\bu\bu}=E(\bpsi^{\otimes2})$, $\bbOmega_{\bu\bbm}=\bbOmega_{\bbm\bu}^{\T} = -\rho_{N} E(\bpsi\bh^{\T})$ and
$$\bbOmega_{\bbm\bbm}=(1-2\rho_{N} )E\left[\left\{\ba(\bX;\btheta)-\bmu_{\bh}\right\}^{\otimes2}\right]+\rho_{N} \left\{E(\bh-\bmu_{\bh})\right\}^{\otimes2}.$$
Then we have 
\begin{equation*}
	\begin{split}
		\bbG_{\cM}^{\T}\bbOmega_{\cM}^{-1}\bbG_{\cM} 
		= \bbOmega_{\bu\bu}+(1-\rho_{N} )^{2}E(\bpsi\bh^{\T})\left(\bbOmega_{\bbm\bbm}-\bbOmega_{\bbm\bu}\bbOmega_{\bu\bu}^{-1}\bbOmega_{\bu\bbm}\right)^{-1}E\left(\bh\bpsi^{\T}\right).
	\end{split}
\end{equation*}
Without loss generality, we assume $\bmu_{\bh}=\bzero$ since $\bbG_{\cM}^{\T}\bbOmega_{\cM}^{-1}\bbG_{\cM}$ is invariant if $\bh(\bx,\by)$ is replaced by $\bh(\bx,\by)-\bc$ for any $q$-dimensional constant vector $\bc$. For any $q$-dimensional function vector $\bh$, there exists $\bbM^{\prime}\in\mathbb{R}^{q\times d}$ and $q$ dimensional function vector $\bh^{\prime}$ satisfying $E\left(\bh^{\prime}\bpsi^{\T}\right)=0$ such that $\bh = \bbM^{\prime}\bpsi+\bh^{\prime}$.
Then 
\begin{equation*}
	\begin{split}
		\bbG_{\cM}^{\T}\bbOmega_{\cM}^{-1}\bbG_{\cM}
		&= \bbOmega_{\bu\bu}+\frac{1-\rho_{N}}{\rho_{N}}\bbOmega_{\bu\bu} \bbM^{\prime\T}\left\{\bbM^{\prime}\bbOmega_{\bu\bu} \bbM^{\prime\T}+\bbF(\bh^{\prime})\right\}^{-1}\bbM^{\prime}\bbOmega_{\bu\bu},
	\end{split}
\end{equation*}
where $$\bbF(\bh^{\prime}) = \frac{1-2\rho_{N}}{\rho_{N}-\rho_{N}^2}E\left[\left\{E\left(\bh^{\prime}\mid \bX\right)\right\}^{\otimes2}\right]+\frac{1}{1-\rho_{N}}E\left(\bh^{\prime\otimes2}\right).$$
Denote $\Upsilon_{\bh} = \bbM^{\prime\T}\left\{\bbM^{\prime}\bbOmega_{\bu\bu} \bbM^{\prime\T}+\bbF(\bh^{\prime})\right\}^{-1}\bbM^{\prime}$. 
Then to prove $\bbG_{\cM}^{\T}\bbOmega_{\cM}^{-1}\bbG_{\cM}\leq \rho_{N}^{-1}\bbOmega_{\bu\bu}$ for any $\bh$, it suffices to prove $\Upsilon_{\bh}\leq \bbOmega_{\bu\bu}^{-1}$ for any $\bh$.

Denote $\bbQ_{0} = \bbM^{\prime}\bbOmega_{\bu\bu} \bbM^{\prime\T}$ and $\bbQ_{1} = \bbF(\bh^{\prime})$. 
Then we have 
$\Upsilon_{\bh} = \bbM^{\prime\T}(\bbQ_{0}+\bbQ_{1})^{-1}\bbM^{\prime}$. According to the result of Theorem 5 in \cite{Wu1980Inverse}, we have $(\bbQ_{0}+\bbQ_{1})^{-1}\leq \bbQ_{0}^{-}$ where $\bbQ_{0}^{-}$ means the generalized inverse of $\bbQ_{0}$. Then we have $\bbM^{\prime\T}(\bbQ_{0}+\bbQ_{1})^{-1}\bbM^{\prime}\leq \bbM^{\prime\T}\bbQ_{0}^{-}\bbM^{\prime}$. Note that $\bbM^{\prime}$ can be represents as $\bbQ_{0}\bbC$ where $\bbC$ is a $q\times d$ matrix since the column space of $\bbM^{\prime}$ belongs to column space of $\bbQ_{0}$. Then we have $\bbM^{\prime\T}\bbQ_{0}^{-}\bbM^{\prime}=\bbC^{\T}\bbQ_{0}\bbQ_{0}^{-}\bbQ_{0}\bbC=\bbC^{\T}\bbQ_{0}\bbQ_{0}^{+}\bbQ_{0}\bbC=\bbM^{\prime\T}\bbQ_{0}^{+}\bbM^{\prime}$ where $\bbQ_{0}^{+}$ represents the Moore-Penrose inverse of $\bbQ_{0}$. Combing these results, we have
\begin{equation}\label{eq: 1}
	\Upsilon_{\bh}\leq \bbM^{\prime\T}(\bbM^{\prime}\bbOmega_{\bu\bu} \bbM^{\prime\T})^{+}\bbM^{\prime}.
\end{equation}

In addition, we can show that there exists  a full row rank matrix $\bbP_{1}\in \mathbb{R}^{q_{1}\times d}$ such that
\begin{equation}\label{eq: 2}
	\bbM^{\prime\T}(\bbM^{\prime}\bbOmega_{\bu\bu} \bbM^{\prime\T})^{+}\bbM^{\prime}=\bbP_{1}^{\T}(\bbP_{1}\bbOmega_{\bu\bu} \bbP_{1}^{\T})^{-1}\bbP_{1},
\end{equation}
If $\bbM^{\prime}$ is full row rank, then $\bbP_{1}=\bbM^{\prime}$. If $\bbM^{\prime}$ is not full row rank, there exists an invertible matrix $\bbU\in \mathbb{R}^{q\times q}$ such that $\bbU\bbM^{\prime}=(\bbP_{1}^{\T},\bzero^{\T})^{\T}$ and $\bbP_{1}$ is full row rank, and hence
\[
\bbM^{\prime\T}(\bbM^{\prime}\bbOmega_{\bu\bu} \bbM^{\prime\T})^{+}\bbM^{\prime}= (\bbU\bbM^{\prime})^{\T}\{\bbU\bbM^{\prime}\bbOmega_{\bu\bu} (\bbU\bbM^{\prime})^{\T}\}^{+}\bbU\bbM^{\prime}=\bbP_{1}^{\T}(\bbP_{1}\bbOmega_{\bu\bu} \bbP_{1}^{\T})^{-1}\bbP_{1}.
\]

If $\bbP_{1}$ is invertible, we have $\bbP_{1}^{\T}(\bbP_{1}\bbOmega_{\bu\bu} \bbP_{1}^{\T})^{-1}\bbP_{1}=\bbOmega_{\bu\bu}^{-1}$ which together \eqref{eq: 1} and \eqref{eq: 2} proves $\Upsilon_{\bh}\leq \bbOmega_{\bu\bu}^{-1}$ for any $\bh(\bx,\by)$. 
If $\bbP_{1}$ is not invertible, since $\bbP_{1}$ is full row rank, then we have $q_{1}\leq d$. In addition, since the row vectors of  $\bbP_{1}$ are linearly independent, then we can complement row vectors of  $\bbP_{1}$ as a group of bases of $\mathbb{R}^{d}$. Specifically, we can choose $\bbP_{2}$ such that $\bbP_{3} = (\bbP_{1}^{\T},\bbP_{2}^{\T})^{\T}$ is an invertible matrix where $\bbP_{2}\in \mathbb{R}^{(d-q_{1})\times d}$ satisfying $\bbP_{2}\bbOmega_{\bu\bu} \bbP_{1}^{\T}=0$. Then we have 
\begin{equation}\label{eq: 3}
	\bbP_{1}^{\T}(\bbP_{1}\bbOmega_{\bu\bu} \bbP_{1}^{\T})^{-1}\bbP_{1} + \bbP_{2}^{\T}(\bbP_{2}\bbOmega_{\bu\bu} \bbP_{2}^{\T})^{-1}\bbP_{2}=\bbP_{3}^{\T}(\bbP_{3}\bbOmega_{\bu\bu} \bbP_{3}^{\T})^{-1}\bbP_{3}=\bbOmega_{\bu\bu}^{-1}.
\end{equation}
Combining \eqref{eq: 1}--\eqref{eq: 3}, we have $\Upsilon_{\bh}\leq \bbOmega_{\bu\bu}^{-1}$ for any $\bh(\bx,\by)$.

Now, to prove Theorem \ref{theo: opt h}, it remains to show that the $\Upsilon_{\bh}= \bbOmega_{\bu\bu}^{-1}$ holds if and only if $\bpsi(\bx,\by;\btheta_{0})=\bbM\bh(\bx,\by)$ for some matrix $\bbM$.
By the definition of $\Upsilon_{\bh}$, this is equivalent to proving $\bbM^{\prime\T}(\bbQ_{0}+\bbQ_{1})^{-1}\bbM^{\prime} = \bbOmega_{\bu\bu}^{-1}$  holds if and only if $\bpsi(\bx,\by;\btheta_{0})=\bbM\bh(\bx,\by)$ for some matrix $\bbM$.
We first prove that if there are some matrix $\bbM$ such that $\bpsi(\bx,\by;\btheta_{0})=\bbM\bh(\bx,\by)$, we have $\bbM^{\prime\T}(\bbQ_{0}+\bbQ_{1})^{-1}\bbM^{\prime} = \bbOmega_{\bu\bu}^{-1}$.
Since $\bpsi(\bx,\by;\btheta_{0})=\bbM\bh(\bx,\by)$ and $\bh(\bx,\by) = \bbM^{\prime}\bpsi(\bx,\by;\btheta_{0})+\bh^{\prime}(\bx,\by)$, then we have $\bpsi(\bx,\by;\btheta_{0})=\bbM\bbM^{\prime}\bpsi(\bx,\by;\btheta_{0})+\bbM\bh^{\prime}(\bx,\by)$ and hence $(\bbM\bbM^{\prime}-\bbI)\bbOmega_{\bu\bu}=\bzero$.
Then we have $\bbM\bbM^{\prime}=\bbI$ and $\bbM\bh^{\prime}(\bx,\by)=\bzero$. 
Since $\bbM\bbM^{\prime}=\bbI$ implies $\bbM$ is full row rank, then we can choose $\bbM_{1}$ satisfying $\bbM_{1}\bbM^{\prime}=\bzero$ such that $\bbM_{2} = (\bbM^{\T},\bbM_{1}^{\T})^{\T}$ is an invertible matrix.
In addition, $\bbM\bh^{\prime}(\bx,\by)=\bzero$ implies $\bbM\bbQ_{1}=\bzero$.
Based on these results, by the definition of $\Upsilon_{\bh}$ and some simple algebras, we have 
$\bbM^{\prime\T}(\bbQ_{0}+\bbQ_{1})^{-1}\bbM^{\prime}  =\bbM^{\prime\T}\bbM_{2}^{\T}(\bbM_{2}\bbQ_{0}\bbM_{2}^{\T}+\bbM_{2}\bbQ_{1}\bbM_{2}^{\T})^{-1}\bbM_{2}\bbM^{\prime} = \bbOmega_{\bu\bu}^{-1}$.
Next, we prove that if $\bbM^{\prime\T}(\bbQ_{0}+\bbQ_{1})^{-1}\bbM^{\prime} = \bbOmega_{\bu\bu}^{-1}$, there are some matrix $\bbM$ such that $\bpsi(\bx,\by;\btheta_{0})=\bbM\bh(\bx,\by)$.
Since $\bbM^{\prime\T}(\bbQ_{0}+\bbQ_{1})^{-1}\bbM^{\prime} = \bbOmega_{\bu\bu}^{-1}$, by multiplying both sides by $\bbM^{\prime}\bbOmega_{\bu\bu}$ and $\bbOmega_{\bu\bu}\bbM^{\prime}$ respectively, we have $\bbQ_{0}(\bbQ_{0}+\bbQ_{0})^{-1}\bbQ_{0}$.
By conducting a singular value decomposition on $\bbQ_{0}$, we have $\bbQ_{0}=\bbU_{0}^{\T}\bbLambda\bbU_{0}$ where
\begin{equation*}
	\bbLambda = \begin{pmatrix}
		\bbLambda_{0} &\bzero^{\T}\\
		\bzero & \bzero
	\end{pmatrix}.
\end{equation*}
Then we have
$\bbLambda\left(\lambda+\bbU_{0}^{\T}\bbQ_{1}\bbU_{0}\right)^{-1}\bbLambda=\bbLambda$.
Let 
$$\bbD = \begin{pmatrix}
	\bbD_{11} & \bbD_{12}\\
	\bbD_{21} & \bbD_{22}
\end{pmatrix}.$$
Then we have $\bbLambda_{1}(\bbLambda_{1}+\bbD_{11}-\bbD_{12}\bbD_{22}^{-1}\bbD_{21})^{-1}\bbLambda_{1}=\bbLambda_{1}$.
Since $\bbLambda_{1}$ is invertible, then we have $\bbD_{11}-\bbD_{12}\bbD_{22}^{-1}\bbD_{21} = \bzero$ which is equivalent to 
$$
(\bbI,-\bbD_{12}\bbD_{22}^{-1})\bbU_{0}^{\T}\bbQ_{1}\bbU_{0}\begin{pmatrix}
	\bbI\\
	-\bbD_{22}^{-1}\bbD_{12}
\end{pmatrix} = \bzero.
$$
For convenience, we denote $\bbGamma^{\T} = (\bbI,-\bbD_{12}\bbD_{22}^{-1})\bbU_{0}^{\T}$.
Then we have $\bbGamma^{\T}\bbQ_{1}\bbGamma = \bzero$.
Since $\bbQ_{1} = \rho_{N}^{-1}\var\left\{E\left(\bh^{\prime}\mid\bX\right)\right\}+(1-\rho_{N})^{-1}E\left\{\var\left(\bh^{\prime}\mid \bX\right)\right\}$, then we have $\bbGamma^{\T}\bh^{\prime} = \bzero$, and hence $\bbGamma^{\T}\bh = \bbGamma^{\T}\bbM^{\prime}\bpsi$.
Note that $\bbGamma^{\T}\bbQ_{0}\bbGamma = \bbLambda_{1}$ and $\bbGamma^{\T}\bbQ_{0}\bbGamma = \bbGamma^{\T}\bbM^{\prime}\bbOmega_{\bu\bu}\bbM^{\prime\T}\bbGamma$.
Then we have $\bbGamma^{\T}\bbM^{\prime}\bbOmega_{\bu\bu}\bbM^{\prime\T}\bbGamma = \bbLambda_{1}$. Note that $\bbLambda_{1}$ is invertible. We have $\bbGamma^{\T}\bbM^{\prime}$ is invertible. Thus, $(\bbGamma^{\T}\bbM^{\prime})^{-1}\bbGamma^{\T}\bh =\bpsi$.
This shows that there exists matrix $\bbM=(\bbGamma^{\T}\bbM^{\prime})^{-1}\bbGamma^{\T}$ such that $\bpsi=\bbM\bh$.

\subsection*{Proof of Theorem \ref{theo: asy tilde S}}

Let  $\tilde{\bh}_{i} = \tilde{\bh}(\bX_{i},\bY_{i})$, $\bh_{i}^{*} = \bh^{*}(\bX_{i},\bY_{i})$, $\bt_{i}(\bxi) = \bt(\bX_{i},\bY_{i};\bxi)$, $\tilde{\bs}_{i}(\bmu) = \rho_{N}\pi_{i}^{-1}\left(\tilde{\bh}_{i}-\bmu\right)$, and $\bq_{i}(\bxi) = \bq(\bX_{i},\bY_{i};\bxi)$.
Let $\tilde{\bg}_{\bs,\tilde{\bh}} =n^{-1}\sum_{i=1}^{N}\delta_{i}\tilde{\bs}_{i}(\hat{\bmu}_{\tilde{\bh}})$, 
$$\tOmega_{\bu\bs,\tilde{\bh}} = \frac{1}{n}\sum_{i=1}^{N}\delta_{i}\bu_{i}(\tilde{\btheta})\left\{\tilde{\bs}_{i}(\hat{\bmu})-\rho_{N}\left(\tilde{\bh}_{i}-\hat{\bmu}\right)\right\}^{\T}$$
and
$$\tOmega_{\bs\bs,\tilde{\bh}} = \frac{1}{n}\sum_{i=1}^{N}\delta_{i}\tilde{\bs}_{i}(\hat{\bmu})\left\{\tilde{\bs}_{i}(\hat{\bmu})-\rho_{N}\left(\tilde{\bh}_{i}-\hat{\bmu}_{\tilde{\bh}}\right)\right\}^{\T}.$$	
Then
\begin{equation*}
	\tilde{\btheta}_{\cS,\tilde{\bh}} = \tilde{\btheta} - \tG_{\cS,\bu}^{-1}\tOmega_{\bu\bs,\tilde{\bh}}\tOmega_{\bs\bs,\tilde{\bh}}^{-1}\tilde{\bg}_{\bs,\tilde{\bh}}.
\end{equation*}
Under Conditions (C10), (C11), $\tOmega_{\bu\bs,\tilde{\bh}} - \tOmega_{\bu\bs,\bh^{*}} = O_{P}(n^{-1/2})$, $\tOmega_{\bs\bs,\tilde{\bh}}-\tOmega_{\bs\bs,\bh^{*}} = O_{P}(n^{-1/2})$, and
\begin{align*}
	&\tilde{\bg}_{\bs,\tilde{\bh}} - \tilde{\bg}_{\bs,\bh^{*}}\\
	& = \frac{1}{N}\sum_{i=1}^{N}\left(\delta_{i}\pi_{i}^{-1}-1\right)\bt_{i}(\bxi^{*})\left(\tilde{\bxi}-\bxi^{*}\right)+ \frac{1}{N}\sum_{i=1}^{N}\left(\delta_{i}\pi_{i}^{-1}-1\right)\left\{\bt_{i}(\bar{\bxi})-\bt(\bxi^{*})\right\}\left(\tilde{\bxi}-\bxi^{*}\right)\\
	&\quad + \left(1-\frac{1}{N}\sum_{i=1}^{N}\delta_{i}\pi_{i}^{-1}\right)\frac{1}{N}\sum_{i=1}^{N}\left\{\bq_{i}(\tilde{\bxi})-\bq_{i}(\bxi^{*})\right\}\\
	&=O_{P}(n^{-1}).
\end{align*}
In addition, according to the proof of Theorem \ref{theo: sMAS normal}, $\tilde{\bg}_{\bs,\bh^{*}}=O_{P}(n^{-1/2})$. Then $\tilde{\btheta}_{\cS,\tilde{\bh}} = \tilde{\btheta} - \tG_{\cS,\bu}^{-1}\tOmega_{\bu\bs,\bh^{*}}\tOmega_{\bs\bs,\bh^{*}}^{-1}\tilde{\bg}_{\bs,\bh^{*}}+O_{P}(n^{-1})$.
The result in Theorem \ref{theo: asy tilde S} can be proved by applying the result in Theorem \ref{theo: sMAS normal}.

\subsection*{Proof of Theorem \ref{theo: asy tilde M}}

Let $\tilde{\bh}_{i} = \tilde{\bh}(\bX_{i},\bY_{i})$, $\bh_{i}^{*} = \bh^{*}(\bX_{i},\bY_{i})$, $\bb_{i}(\btheta,\bxi) = \bb(\bX_{i};\btheta,\bxi)$, $\be_{i}(\btheta,\bxi) = \be(\bX_{i};\btheta,\bxi)$, $\bt_{i}(\bxi) = \bt(\bX_{i},\bY_{i};\bxi)$.
Let $\tilde{\bg}_{\cM,\tilde{\bh}}$, $\tG_{\cM,\tilde{\bh}}$ and  $\tOmega_{\cM,\tilde{\bh}}$ be defined in the same way as $\tilde{\bg}_{\cM}$, $\tG_{\cM}$ and $\tOmega_{\cM}$, respectively, with $\bh$ replaced by $\tilde{\bh}$.       
Let $\tA_{\cM,\tilde{\bh}} = \left(\tG_{\cM,\tilde{\bh}}^{\T}\tOmega_{\cM,\tilde{\bh}}^{-1}\tG_{\cM,\tilde{\bh}}\right)^{-1}\tG_{\cM,\tilde{\bh}}^{\T}\tOmega_{\cM,\tilde{\bh}}^{-1}$ and $\bbA_{\cM,\bh^{*}} = \left(\bbG_{\cM,\bh^{*}}^{\T}\bbOmega_{\cM,\bh^{*}}^{-1}\bbG_{\cM,\bh^{*}}\right)^{-1}\bbG_{\cM,\bh^{*}}^{\T}\bbOmega_{\cM,\bh^{*}}^{-1}$.
Then
\begin{align*}
	&\tilde{\btheta}_{\cM,\tilde{\bh}}-\btheta_{0} \\
	& = \tilde{\btheta}-\tA_{\cM,\tilde{\bh}}\tilde{\bg}_{\cM,\tilde{\bh}}-\btheta_{0}\\
	& = \tA_{\cM,\tilde{\bh}}\tG_{\cM,\tilde{\bh}}\left(\tilde{\btheta}-\btheta_{0}\right)-\tA_{\cM,\tilde{\bh}}\tilde{\bg}_{\cM,\tilde{\bh}}\\
	& = -\bbA_{\cM,\bh^{*}}\tilde{\bg}_{\cM,\tilde{\bh}}-\left(\tA_{\cM,\tilde{\bh}}-\bbA_{\cM,\bh^{*}}\right)\tilde{\bg}_{\cM,\tilde{\bh}}\\
	&\quad+\bbA_{\cM,\bh^{*}}\tG_{\cM,\tilde{\bh}}\left(\tilde{\btheta}-\btheta_{0}\right)+\left(\tA_{\cM,\tilde{\bh}}-\bbA_{\cM,\bh^{*}}\right)\tG_{\cM,\tilde{\bh}}\left(\tilde{\btheta}-\btheta_{0}\right)\\
	& = -\bbA_{\cM,\bh^{*}}\bg_{\cM,\bh^{*}}-\bbA_{\cM,\bh^{*}}\left(\bg_{\cM,\tilde{\bh}}-\bg_{\cM,\bh^{*}}\right)\\
	&\quad-\bbA_{\cM,\bh^{*}}\left(\tilde{\bg}_{\cM,\tilde{\bh}}-\bg_{\cM,\tilde{\bh}}\right)+\bbA_{\cM,\bh^{*}}\tG_{\cM,\tilde{\bh}}\left(\tilde{\btheta}-\btheta_{0}\right)\\
	&\quad-\left(\tA_{\cM,\tilde{\bh}}-\bbA_{\cM,\bh^{*}}\right)\bg_{\cM,\bh^{*}}-\left(\tA_{\cM,\tilde{\bh}}-\bbA_{\cM,\bh^{*}}\right)\left(\bg_{\cM,\tilde{\bh}}-\bg_{\cM,\bh^{*}}\right)\\
	&\quad-\left(\tA_{\cM,\tilde{\bh}}-\bbA_{\cM,\bh^{*}}\right)\left(\tilde{\bg}_{\cM,\tilde{\bh}}-\bg_{\cM,\tilde{\bh}}\right)+\left(\tA_{\cM,\tilde{\bh}}-\bbA_{\cM,\bh^{*}}\right)\tG_{\cM,\tilde{\bh}}\left(\tilde{\btheta}-\btheta_{0}\right)\\
	& = -\bbA_{\cM,\bh^{*}}\bg_{\cM,\bh^{*}}-\bbA_{\cM,\bh^{*}}\left(\bg_{\cM,\tilde{\bh}}-\bg_{\cM,\bh^{*}}\right)\\
	&\quad+\bbA_{\cM,\bh^{*}}\left(\tG_{\cM,\tilde{\bh}}-\bar{G}_{\cM,\tilde{\bh}}\right)\left(\tilde{\btheta}-\btheta_{0}\right)\\
	&\quad-\left(\tA_{\cM,\tilde{\bh}}-\bbA_{\cM,\bh^{*}}\right)\bg_{\cM,\bh^{*}}-\left(\tA_{\cM,\tilde{\bh}}-\bbA_{\cM,\bh^{*}}\right)\left(\bg_{\cM,\tilde{\bh}}-\bg_{\cM,\bh^{*}}\right)\\
	&\quad+\left(\tA_{\cM,\tilde{\bh}}-\bbA_{\cM,\bh^{*}}\right)\left(\tG_{\cM,\tilde{\bh}}-\bar{G}_{\cM,\tilde{\bh}}\right)\left(\tilde{\btheta}-\btheta_{0}\right)\\
	&=J_{\cM,1}+J_{\cM,2}+J_{\cM,3}+J_{\cM,4}+J_{\cM,5}+J_{\cM,6}.
\end{align*}	   
According to Theorem \ref{theo: mMAS normal},  $\bbV_{\cM,\bh^{*}}^{-1/2}J_{\cM,1}\stackrel{d}{\to}N(\bzero,\bbI)$.
Then to prove the result in Theorem \ref{theo: asy tilde M}, it suffices to prove that $\bbV_{\cM,\bh^{*}}^{-1/2}J_{\cM,i}=o_{P}(1), i=2,\cdots,6$. Through similar arguments as those used in the proof of Theorem \ref{theo: asy tilde M}, we can show that $\bbV_{\cM,\bh^{*}}^{-1/2}J_{\cM,3}, \bbV_{\cM,\bh^{*}}^{-1/2}J_{\cM,4}, \bbV_{\cM,\bh^{*}}^{-1/2}J_{\cM,6}$ are $o_{P}(1)$s. 
To prove that $\bbV_{\cM,\bh^{*}}^{-1/2}J_{\cM,2}$ and $\bbV_{\cM,\bh^{*}}^{-1/2}J_{\cM,5}$ are $o_{P}(1)$s, it is sufficient to prove that $\bg_{\cM,\tilde{\bh}}-\bg_{\cM,\bh^{*}}=O_{P}(n^{-1})$. 
Note that $\|\bg_{\cM,\tilde{\bh}}- \bg_{\cM,\bh^{*}}\|=\|\bg_{\bbm,\tilde{\bh}}-\bg_{\bbm,\bh^{*}}\|$. 
It suffices to prove $\bg_{\bbm,\tilde{\bh}}-\bg_{\bbm,\bh^{*}}=O_{P}(n^{-1})$.
Through some algebras, 
\begin{align*}
	&\bg_{\bbm,\tilde{\bh}}- \bg_{\bbm,\bh^{*}}\\ 
	&=\frac{1}{N}\sum_{i=1}^{N}\left(\delta_{i}\pi_{i}^{-1}-1\right)\left\{\bb_{i}(\btheta_{0},\tilde{\bxi})-\bb_{i}(\btheta_{0},\bxi^{*})\right\}+\frac{1}{N}\sum_{i=1}^{N}\left(\delta_{i}\pi_{i}^{-1}-1\right)\left(\hat{\bmu}_{\bh^{*}}-\hat{\bmu}_{\tilde{\bh}}\right)\\
	&\quad+\frac{1}{N}\sum_{i=1}^{N}\left\{\bb_{i}(\btheta_{0},\tilde{\bxi})-\bb_{i}(\btheta_{0},\bxi^{*})-\left(\tilde{\bh}_{i}-\bh_{i}^{*}\right)\right\}\\
	&= Q_{\cM,1}+Q_{\cM,2}+Q_{\cM,3}.
\end{align*}
For $Q_{\cM,1}$, we have
\begin{align*}
	Q_{\cM,1}
	& = \frac{1}{N}\sum_{i=1}^{N}\left(\delta_{i}\pi_{i}^{-1}-1\right)\be_{i}(\btheta_{0},\bxi^{*})(\tilde{\bxi}-\bxi^{*})\\
	&\quad+\frac{1}{N}\sum_{i=1}^{N}\left(\delta_{i}\pi_{i}^{-1}-1\right)\left\{\be_{i}(\btheta_{0},\bar{\bxi})-\be_{i}(\btheta_{0},\bxi^{*})\right\}(\tilde{\bxi}-\bxi^{*})\\
	&=Q_{\cM,11}+Q_{\cM,12}.
\end{align*}
By Chebyshev's inequality and Conditions (C6) and (C12), we have $$\frac{1}{N}\sum_{i=1}^{N}\left(\delta_{i}\pi_{i}^{-1}-1\right)\be_{i}(\btheta_{0},\bxi^{*})=O_{P}(n^{-1/2})$$
Then by Condition (C10), we have $Q_{\cM,11} = O_{P}(n^{-1})$.
By Conditions (C10), (C12), we have $Q_{\cM,12}=O_{P}(n^{-1})$. 
Then $Q_{\cM,1} = O_{P}(n^{-1})$.
By Conditions (C6), (C10), we have $Q_{\cM,2} = O_{P}(n^{-1})$.
In addition,
\begin{align*}
	Q_{\cM,3} 
	&= \frac{1}{N}\sum_{i=1}^{N}\left\{\be_{i}\left(\btheta_{0},\bxi^{*}\right)-\bt_{i}(\bxi^{*})\right\}(\tilde{\bxi}-\bxi^{*})\\
	&\quad+\frac{1}{N}\sum_{i=1}^{N}\left\{\be_{i}\left(\btheta_{0},\bar{\bxi}\right)-\be_{i}\left(\btheta_{0},\bxi^{*}\right)\right\}(\tilde{\bxi}-\bxi^{*})\\
	&\quad-\frac{1}{N}\sum_{i=1}^{N}\left\{\bt_{i}\left(\bar{\bxi}\right)-\bt_{i}\left(\bxi^{*}\right)\right\}(\tilde{\bxi}-\bxi^{*})\\
	&=Q_{\cM,31}+Q_{\cM,32}+Q_{\cM,33}.
\end{align*}
Note that $E\{\bt(\bxi^{*})\} =E\{\be(\btheta_{0},\bxi^{*})\}$.
Then by Chebyshev's inequality, we have $Q_{\cM,31}=O_{P}(n^{-1})$.
By Conditions (C11), (C12), we have $Q_{\cM,32}=O_{P}(n^{-1})$, $Q_{\cM,33}=O_{P}(n^{-1})$ and $Q_{\cM,34}=O_{P}(n^{-1})$. Then $Q_{\cM,3} = O_{P}(n^{-1})$.


\subsection*{Proof of Theorem \ref{theo: non-normal S}}

Denote $\hat{\bmu} = N^{-1}\sum_{i=1}^{N}\bpsi_{i}(\tilde{\btheta})$.
By the definition of $\tilde{\btheta}$,  $N^{-1}\sum_{i=1}^{N}\delta_{i}\pi_{i}^{-1}\bpsi_{i}(\tilde{\btheta})=0$. With the estimated optimal moment function $\tilde{\bh}^{\rm opt} = \bpsi(\tilde{\btheta})$, it follows that 
$$\tilde{\bg}_{\bs} = N^{-1}\sum_{i=1}^{N}\frac{\delta_{i}}{\pi_{i}}\left\{\bpsi_{i}(\tilde{\btheta})-\hat{\bmu}\right\}=-\frac{1}{N}\sum_{i=1}^{N}\frac{\delta_{i}}{\pi_{i}}\hat{\bmu},$$
$$\tG_{\cS,\bu} = N^{-1}\sum_{i=1}^{N}\frac{\delta_{i}}{\pi_{i}}\dot{\bpsi}_{i}(\tilde{\btheta}),$$ 
$$\tOmega_{\bs\bs}= \rho_{N}\frac{1}{N}\sum_{i=1}^{N}\frac{\delta_{i}}{\pi_{i}}(\pi_{i}^{-1}-1)\left[\left\{\bpsi_{i}(\tilde{\btheta})-\hat{\bmu}\right\}\left\{\bpsi_{i}(\tilde{\btheta})-\hat{\bmu}\right\}^{\T}\right],$$
and
$$\tOmega_{\bu\bs}= \rho_{N}\frac{1}{N}\sum_{i=1}^{N}\frac{\delta_{i}}{\pi_{i}}(\pi_{i}^{-1}-1)\bpsi_{i}(\tilde{\btheta})\left\{\bpsi_{i}(\tilde{\btheta})-\hat{\bmu}\right\}^{\T}.$$
It can be observed that
$$\tOmega_{\bu\bs} = \tOmega_{\bs\bs} + \left\{\hat{\bmu}\frac{1}{N}\sum_{i=1}^{N}\rho_{N}\frac{\delta_{i}}{\pi_{i}}\pi_{i}^{-1}\bpsi_{i}(\tilde{\btheta})^{\T}-\frac{1}{N}\sum_{i=1}^{N}\rho_{N}\frac{\delta_{i}}{\pi_{i}}(\pi_{i}^{-1}-1)\hat{\bmu}\hat{\bmu}^{\T}\right\}.$$ 
Recalling the definition of standard MAS estimator, $\tilde{\btheta}_{\cS} = \tilde{\btheta}+\tG_{\cS,\bu}^{-1}\tOmega_{\bu\bs}\tOmega_{\bs\bs}^{-1}\tilde{\bg}_{\bs}$, and applying the Taylor's expansion of $\bpsi_{i}(\tilde{\btheta})$ around $\btheta_{0}$,
\begin{align}\label{rMAS: opth expansion}
	&\tilde{\btheta}_{\cS} - \btheta_{0} \notag\\
	& = \tilde{\btheta} - \btheta_{0} + \bbG_{u}^{-1}\tilde{\bg}_{\bs} + \left(\tG_{\cS,\bu}^{-1}\tOmega_{\bu\bs}\tOmega_{\bs\bs}^{-1}-\bbG_{\bu}^{-1}\right)\tilde{\bg}_{\bs}\notag\\
	& = \bbG_{\bu}^{-1}\bbG_{\bu}\left(\tilde{\btheta}-\btheta_{0}\right)- \bbG_{\bu}^{-1}\frac{1}{N}\sum_{i=1}^{N}\frac{\delta_{i}}{\pi_{i}}\frac{1}{N}\sum_{i=1}^{N}\bpsi_{i} - \bbG_{\bu}^{-1}\frac{1}{N}\sum_{i=1}^{N}\frac{\delta_{i}}{\pi_{i}}\frac{1}{N}\sum_{i=1}^{N}\dot{\bpsi}_{i}\left(\tilde{\btheta}-\btheta_{0}\right)\notag\\
	&\quad- \bbG_{\bu}^{-1}\frac{1}{N}\sum_{i=1}^{N}\frac{\delta_{i}}{\pi_{i}}\frac{1}{2N}\sum_{i=1}^{N}\ddot{\bpsi}_{i}\left\{\left(\tilde{\btheta}-\btheta_{0}\right)\otimes\left(\tilde{\btheta}-\btheta_{0}\right)\right\} +\tG_{\cS,\bu}^{-1}\left(\tOmega_{\bu\bs}\tOmega_{\bs\bs}^{-1}-\bbI\right)\tilde{\bg}_{\bs}\notag\\
	&\quad - \tG_{\cS,\bu}^{-1}\left(\tG_{\cS,\bu}-\bbG_{\bu}\right)\bbG_{\bu}^{-1}\tilde{\bg}_{\bs}+ o_{P}(n^{-1})\notag\\
	& = -\bbG_{\bu}^{-1}\frac{1}{N}\sum_{i=1}^{N}\frac{\delta_{i}}{\pi_{i}}\frac{1}{N}\sum_{i=1}^{N}\bpsi_{i}-\bbG_{\bu}^{-1}\left(\frac{1}{N}\sum_{i=1}^{N}\frac{\delta_{i}}{\pi_{i}}\frac{1}{N}\sum_{i=1}^{N}\dot{\bpsi}_{i}-\bbG_{\bu}\right)\left(\tilde{\btheta}-\btheta_{0}\right)\notag\\
	&\quad-\bbG_{\bu}^{-1}\frac{1}{N}\sum_{i=1}^{N}\frac{\delta_{i}}{\pi_{i}}\frac{1}{2N}\sum_{i=1}^{N}\ddot{\bpsi}_{i}\left\{\left(\tilde{\btheta}-\btheta_{0}\right)\otimes\left(\tilde{\btheta}-\btheta_{0}\right)\right\}-\tG_{\cS,\bu}^{-1}\left(\tG_{\cS,\bu}-\bbG_{\bu}\right)\bbG_{\bu}^{-1}\tilde{\bg}_{\bs}\notag\\
	&\quad -\tG_{\cS,\bu}^{-1}\hat{\bmu}\frac{1}{N}\sum_{i=1}^{N}\rho_{N}\frac{\delta_{i}}{\pi_{i}}\pi_{i}^{-1}\bpsi_{i}^{\T}\tOmega_{\bs\bs}^{-1}\hat{\bmu}+ o_{P}(n^{-1}).
\end{align}
It is straightforward to verify
\begin{equation*}
	\frac{1}{N}\sum_{i=1}^{N}\frac{\delta_{i}}{\pi_{i}}\frac{1}{N}\sum_{i=1}^{N}\bpsi_{i} = \frac{1}{N}\sum_{i=1}^{N}\bpsi_{i} + O_{P}(n^{-1/2}N^{-1/2}),
\end{equation*}
\begin{equation*}
	\frac{1}{N}\sum_{i=1}^{N}\frac{\delta_{i}}{\pi_{i}}\frac{1}{N}\sum_{i=1}^{N}\dot{\bpsi}_{i}-\bbG_{\bu} = \left(\frac{1}{N}\sum_{i=1}^{N}\frac{\delta_{i}}{\pi_{i}}-1\right)E(\dot{\bpsi})+O_{P}(N^{-1/2}),
\end{equation*}
\begin{equation*}
	\frac{1}{N}\sum_{i=1}^{N}\frac{\delta_{i}}{\pi_{i}}\frac{1}{N}\sum_{i=1}^{N}\ddot{\bpsi} = E(\ddot{\bpsi}) + O_{P}(n^{-1/2}),
\end{equation*}
\begin{equation*}
	\tG_{\cS,\bu}-\bbG_{\bu} = \left(\frac{1}{N}\sum_{i=1}^{N}\frac{\delta_{i}}{\pi_{i}}\dot{\bpsi}_{i}-\bbG_{\bu}\right) + \frac{1}{N}\sum_{i=1}^{N}\frac{\delta_{i}}{\pi_{i}}\ddot{\bpsi}_{i}\left(\tilde{\btheta}-\btheta_{0}\right)+O_{P}(n^{-1}),
\end{equation*}
and
\begin{equation*}
	\tilde{\bg}_{\bs} = -\frac{1}{N}\sum_{i=1}^{N}\bpsi_{i} - E(\dot{\bpsi})\left(\tilde{\btheta}-\btheta_{0}\right)+O_{P}(n^{-1}).
\end{equation*}
Note that $\bbG_{\bu} = E(\dot{\bpsi})$. Then by Lemma \ref{lem: AL of initial}, $\tilde{\btheta} - \btheta_{0} = -\bbG_{\bu}^{-1}N^{-1}\sum_{i=1}^{N}\delta_{i}\pi_{i}^{-1}\bpsi_{i} + O_{P}(n^{-1})$, substituting into \eqref{rMAS: opth expansion}, we have
\begin{align*}
	&\tilde{\btheta}_{\cS} - \btheta_{0}\\
	& = -\bbG_{\bu}^{-1}\frac{1}{N}\sum_{i=1}^{N}\bpsi_{i}-\left(\frac{1}{N}\sum_{i=1}^{N}\frac{\delta_{i}}{\pi_{i}}-1\right)\left(\tilde{\btheta}-\btheta_{0}\right)-\bbG_{\bu}^{-1}\frac{1}{2}E(\ddot{\bpsi})\left\{\left(\tilde{\btheta}-\btheta_{0}\right)\otimes\left(\tilde{\btheta}-\btheta_{0}\right)\right\}\\
	&\quad-\bbG_{\bu}^{-1}\left\{\frac{1}{N}\sum_{i=1}^{N}\frac{\delta_{i}}{\pi_{i}}\dot{\bpsi}_{i}-\bbG_{\bu}\right\}\bbG_{\bu}^{-1}\left\{-E(\dot{\bpsi})\left(\tilde{\btheta}-\btheta_{0}\right)\right\}\\
	&\quad+\bbG_{\bu}^{-1} E(\ddot{\bpsi})\left\{\left(\tilde{\btheta}-\btheta_{0}\right)\otimes\left(\tilde{\btheta}-\btheta_{0}\right)\right\}\\
	&\quad -\left(\tilde{\btheta}-\btheta_{0}\right)\rho_{N}E\left(\pi^{-1}\bpsi\right)^{\T}\bbOmega_{\bs\bs}^{-1}\bbG_{\bu}\left(\tilde{\btheta}-\btheta_{0}\right) + o_{P}(\max\{n^{-1},N^{-1/2}\})\\
	&= -\bbG_{\bu}^{-1}\frac{1}{N}\sum_{i=1}^{N}\bpsi_{i}+\left(\frac{1}{N}\sum_{i=1}^{N}\frac{\delta_{i}}{\pi_{i}}-1\right)\bbG_{\bu}^{-1}\frac{1}{N}\sum_{i=1}^{N}\frac{\delta_{i}}{\pi_{i}}\bpsi_{i}\\
	&\quad+\frac{1}{2}\bbG_{\bu}^{-1}E(\ddot{\bpsi})\left\{\left(\bbG_{\bu}^{-1}\frac{1}{N}\sum_{i=1}^{N}\frac{\delta_{i}}{\pi_{i}}\bpsi_{i}\right)\otimes\left(\bbG_{\bu}^{-1}\frac{1}{N}\sum_{i=1}^{N}\frac{\delta_{i}}{\pi_{i}}\bpsi_{i}\right)\right\}\\
	&\quad-\bbG_{\bu}^{-1}\left\{\frac{1}{N}\sum_{i=1}^{N}\frac{\delta_{i}}{\pi_{i}}\dot{\bpsi}_{i}-\bbG_{\bu}\right\}\bbG_{\bu}^{-1}\frac{1}{N}\sum_{i=1}^{N}\frac{\delta_{i}}{\pi_{i}}\bpsi_{i}\\
	&\quad -\bbG_{\bu}^{-1}\left(\frac{1}{N}\sum_{i=1}^{N}\frac{\delta_{i}}{\pi_{i}}\bpsi_{i}\right)\rho_{N}E\left(\pi^{-1}\bpsi\right)^{\T}\bbOmega_{\bs\bs}^{-1}\left(\frac{1}{N}\sum_{i=1}^{N}\frac{\delta_{i}}{\pi_{i}}\bpsi_{i}\right)+ o_{P}(\max\{n^{-1},N^{-1/2}\}).
\end{align*}
Define
\begin{equation*}
	\bU_{\cS,N} = \frac{1}{N}\sum_{i=1}^{N}
	\begin{pmatrix}
		\sqrt{n}(\delta_{i}\pi_{i}^{-1}-1)\\
		\sqrt{N}\bpsi_{i}\\
		\sqrt{n}\delta_{i}\pi_{i}^{-1}\bpsi_{i}\\
		\sqrt{n}\left\{\delta_{i}\pi_{i}^{-1}\operatorname{vec}(\operatorname{upp}(\dot{\bpsi}_{i}))-\operatorname{vec}(\operatorname{upp}(E(\dot{\bpsi})))\right\}\\
	\end{pmatrix}.
\end{equation*}
Let $U_{\cS,N,1}$ be the first component of $\bU_{\cS,N}$, $\bU_{\cS,N,2}$ the $2$-th to the $d+1$-th elements of $\bU_{\cS,N}$, $\bU_{\cS,N,3}$ the $d+2$-th to the $2d+1$-th elements of $\bU_{\cS,N}$, $\bU_{\cS,N,4}$ the $2d+2$-th to $3d+1+(d^2-d)/2$-th elements of $\bU_{\cS,N}$.
Let $\bbU_{\cS,N,C}$ be the $d\times d$ symmetric matrix whose upper triangle matrix consisted by the elements in $\bU_{\cS,N,4}$ arranged in rows.
Define 
\begin{equation*}
	\begin{split}
		\bl_{\cS,N}(\bU_{\cS,N}) 
		&= -c_{1}\bbG_{\bu}^{-1}\bU_{\cS,N,2}+c_{2}\bbG_{\bu}^{-1}U_{\cS,N,1}\bU_{\cS,N,3}+\frac{c_{2}}{2}\bbG_{\bu}^{-1}E(\ddot{\bpsi})\left\{(\bbG_{\bu}^{-1}\bU_{\cS,N,3})\otimes(\bbG_{\bu}^{-1}\bU_{\cS,N,3})\right\}\\
		&\quad-c_{2}\bbG_{\bu}^{-1}\bbU_{\cS,N,C}\bbG_{\bu}^{-1}\bU_{\cS,N,3}-c_{2}\bbG_{\bu}^{-1}\bU_{\cS,N,3}\rho_{N}E\left(\pi^{-1}\bpsi\right)^{\T}\bbOmega_{\bs\bs}^{-1}\bU_{\cS,N,3},
	\end{split}
\end{equation*}
where $c_{1} = \min\{n,\sqrt{N}\}/\sqrt{N}$ and $c_{2} = \min\{n,\sqrt{N}\}/n$.  
Then
\begin{equation*}
	\min\{n,\sqrt{N}\}(\tilde{\btheta}_{\cS} - \btheta_{0}) = l_{\cS,N}(\bU_{\cS,N})+o_{P}(1).
\end{equation*}

Next, we prove 
\begin{equation}\label{eq: continous mapping dist S}
	\sup_{t}\big|P(\bl_{\cS,N}(\bU_{\cS,N})<t)-P(\bl_{\cS,N}(\bbV_{\cS,N}^{1/2}\bZ_{\cS})<t)\big|\to 0,
\end{equation}
where $\bZ_{\cS}$ is a $3d+1+(d^2-d)/2$-dimensional standard normal random vector.
Consider the eigenvalue decomposition $\bbV_{\cS,N} = \bbQ_{\cS,N}\bbLambda_{\cS,N}\bbQ_{\cS,N}^{\T}$ where $\bbQ_{\cS,N}$ is a $(3d+1+(d^2-d)/2)\times r$ matrix of eigenvectors and $\bbLambda_{\cS,N}$ is a $r\times r$ diagonal matrix containing the positive eigenvalues of $\bbV_{\cS,N}$, with $r = \operatorname{rank}(\bbV_{\cS,N})$.
Note that $\bbV_{\cS,N}^{1/2}\bZ_{\cS}$ and $\bbQ_{\cS,N}\bbLambda_{\cS,N}^{1/2}\bZ_{\cS}$ share the same distribution. Therefore, establishing \eqref{eq: continous mapping dist S} is equivalent to proving
\begin{equation}\label{eq: continous mapping dist S 1}
	\sup_{t}\big|P({\bl_{\cS,N}(\bU_{\cS,N})<t})-P({\bl_{\cS,N}(\bbQ_{\cS,N}\bbLambda_{\cS,N}^{1/2}\bZ_{\cS})<t})\big|\to 0.
\end{equation}
Before proceeding to prove \eqref{eq: continous mapping dist S 1}, we first establish that $\bbLambda_{\cS,N}^{-1/2}\bbQ_{\cS,N}^{\T}\bU_{\cS,N}\stackrel{d}{\to}N(\bzero,\bbI)$.
Note that $\bbLambda_{\cS,N}^{-1/2}\bbQ_{\cS,N}^{\T}\bU_{\cS,N} = \sum_{i=1}^{N}\bK_{\cU,i}$, where
\begin{equation*}
	\bK_{\cU,i} = \bbLambda_{\cS,N}^{-1/2}\bbQ_{\cS,N}^{\T}N^{-1}
	\begin{pmatrix}
		\sqrt{n}(\delta_{i}\pi_{i}^{-1}-1)\\
		\sqrt{N}\bpsi_{i}\\
		\sqrt{n}\delta_{i}\pi_{i}^{-1}\bpsi_{i}\\
		\sqrt{n}\left\{\delta_{i}\pi_{i}^{-1}\operatorname{vec}(\operatorname{upp}(\dot{\bpsi}_{i}))-\operatorname{vec}(\operatorname{upp}(E(\dot{\bpsi})))\right\}\\
	\end{pmatrix}.
\end{equation*}
Recalling the definition of $\bbV_{\cS,N}$, it follows directly that $\sum_{i=1}^{N}\var(\bK_{\cU,i})=\bbI$.
By the Lindeberg–Feller central limit theorem \citep{Vaart2000AS}, to establish $\sum_{i=1}^{N}\bK_{\cU,i}\stackrel{d}{\to}N(\bzero,\bbI)$, it suffices to verify that for any $\epsilon>0$, $\sum_{i=1}^{N}E\{\|\bK_{\cU,i}\|^2 1(\|\bK_{\cU,i}\|>\epsilon)\}\to 0$, 
where $1(\cdot)$ denotes the indicator function.
Since for any $\tau>0$, $$E\{\|\bK_{\cU,1}\|^2 1(\|\bK_{\cU,1}\|>\epsilon)\}\leq E\{\|\bK_{\cU,1}\|^{2+\tau}/\epsilon^{\tau}\},$$
it suffices to show that $NE(\|\bK_{\cU,1}\|^{2+\tau})=o(1)$ for some $\tau>0$.
By the inequality $(b_{1}+b_{2})^{p}\le2^{p-1}(|b_{1}|^{p}+|b_{2}|^{p})$,
\begin{equation*}
	\begin{split}
		&NE(\|\bK_{\cU,1}\|^{2+\tau})\\
		&\leq \frac{\|\bbLambda_{\cS,N}^{-1/2}\bbQ_{\cS,N}^{\T}\|^{2+\tau}}{N^{1+\tau}}\left\{n^{1+\tau/2}E(\pi^{-1-\tau})+N^{1+\tau/2}E(\|\bpsi\|^{2+\tau})+n^{1+\tau/2}E(\pi^{-1-\tau}\|\bpsi\|^{2+\tau})\right.\\
		&\left.\qquad\qquad\qquad\qquad\qquad+n^{1+\tau/2}E(\pi^{-1-\tau}\|\operatorname{vec}(\operatorname{upp}(\dot{\bpsi}))\|^{2+\tau})\right\}\\
		&=O(N^{-\tau/2}+n^{-\tau/2}) =o(1),
	\end{split}
\end{equation*}
which verifies the Lindeberg condition.
Now, we prove \eqref{eq: continous mapping dist S 1} which equals to prove 
\begin{equation}\label{eq: continous mapping dist}
	\sup_{t}\big|P(\bl_{\cS,N}^{\dagger}(\bU_{\cS,N}^{\dagger})<t)-P(\bl_{\cS,N}^{\dagger}(\bZ_{\cS})<t)\big|\to 0,
\end{equation}
where $\bl_{\cS,N}^{\dagger}(\cdot) = \bl_{\cS,N}(\bbQ_{\cS,N}\bbLambda_{\cS,N}^{1/2}\cdot)$ and $\bU_{\cS,N}^{\dagger} = \bbLambda_{\cS,N}^{-1/2}\bbQ_{\cS,N}^{\T}\bU_{\cS,N}$.
Note that the mapping $\bl_{\cS,N}^{\dagger}$ is uniformly continuous and bounded on a compact set.
This implies that the function class $\{\bl_{\cS,N}^{\dagger}\}$ is uniformly bounded and equicontinuous.
Then according to Arzelà–Ascoli theorem, every subsequence of $\{\bl_{\cS,N}^{\dagger}\}$ itself has a uniformly convergent subsequence.

For any $\epsilon>0$, consider a compact set $T = [-C,C]^{3d+1+(d^2-d)/2}$ for some constant $C$ such that $P(\bZ_{\cS}\notin T)\leq \epsilon$.
Let $\bZ_{\cS}^{\dagger} = 1(\bZ_{\cS}\in T)\bZ_{\cS}$ and $\bU_{\cS,N}^{\ddagger} =  1(\bU_{\cS,N}^{\dagger}\in T)\bU_{\cS,N}^{\dagger}$. 
Note that $\bU_{\cS,N}^{\dagger}\stackrel{d}{\to}\bZ_{\cS}$. Then for any $\epsilon>0$, with sufficiently large $N$, $|P(\bU_{\cS,N}^{\dagger}\in \mathcal{C})-P(\bZ_{\cS}\in\mathcal{C})|\leq \epsilon$ holds for any set $\mathcal{C}\in \mathbb{R}^{3d+1+(d^2-d)/2}$.
Then 
\begin{equation*}
	\begin{split}
		P(\bU_{\cS,N}^{\ddagger}\in \mathcal{C})\leq P(\bU_{\cS,N}^{\dagger}\in T\cap \mathcal{C}) + P(\bU_{\cS,N}^{\dagger}\notin T)\leq P(\bZ_{\cS}\in T\cap \mathcal{C})+3\epsilon,
	\end{split}
\end{equation*}
and 
\begin{equation*}
	\begin{split}
		P(\bU_{\cS,N}^{\ddagger}\in \mathcal{C})\geq P(\bU_{\cS,N}^{\ddagger}\in T\cap \mathcal{C})\geq P(\bU_{\cS,N}^{\dagger}\in T\cap \mathcal{C})\geq P(\bZ_{\cS}\in T\cap \mathcal{C})-\epsilon.
	\end{split}
\end{equation*}
On the other hand,
\begin{equation*}
	\begin{split}
		P(\bZ_{\cS}^{\dagger}\in \mathcal{C})\leq P(\bZ_{\cS}\in T\cap \mathcal{C}) + P(\bZ_{\cS}\notin T)\leq P(\bZ_{\cS}\in T\cap \mathcal{C})+\epsilon,
	\end{split}
\end{equation*}
and 
\begin{equation*}
	\begin{split}
		P(\bZ_{\cS}^{\dagger}\in \mathcal{C})\geq P(\bZ_{\cS}^{\dagger}\in T\cap \mathcal{C})\geq P(\bZ_{\cS}\in T\cap \mathcal{C}).
	\end{split}
\end{equation*}
Then
\begin{equation*}
	|P(\bU_{\cS,N}^{\ddagger}\in \mathcal{C})-P(\bZ_{\cS}^{\dagger}\in \mathcal{C})|\leq 3\epsilon
\end{equation*}
and $\bU_{\cS,N}^{\ddagger}\stackrel{d}{\to}\bZ_{\cS}^{\dagger}$ is proved.
Suppose $l_{\cS,N^{\prime}}^{\dagger}$ is a uniformly convergent subsequence of $l_{\cS,N}^{\dagger}$.
Then 
\begin{equation}\label{eq: truncated convergence}
	\begin{split}
		&\sup_{t}\big|P(\bl_{\cS,N^{\prime}}^{\dagger}(\bU_{\cS,N^{\prime}}^{\dagger})\leq t)-P(\bl_{\cS,N^{\prime}}^{\dagger}(\bZ_{\cS})\leq t)\big|\\
		&\leq \sup_{t}\big|P(\bl_{\cS,N^{\prime}}^{\dagger}(\bU_{\cS,N^{\prime}}^{\ddagger})\leq t)-P(\bl_{\cS,N^{\prime}}^{\dagger}(\bZ_{\cS}^{\dagger})\leq t)\big|\\
		&+\sup_{t}\big|P(\bl_{\cS,N^{\prime}}^{\dagger}(\bU_{\cS,N^{\prime}}^{\dagger})\leq t)-P(\bl_{\cS,N^{\prime}}^{\dagger}(\bU_{\cS,N^{\prime}}^{\ddagger})\leq t)\big|\\
		&+\sup_{t}\big|P(\bl_{\cS,N^{\prime}}^{\dagger}(\bZ_{\cS}^{\dagger})\leq t)-P(\bl_{\cS,N^{\prime}}^{\dagger}(\bZ_{\cS})\leq t)\big|
	\end{split}
\end{equation}
By the definition of $\bZ_{\cS}^{\dagger}$ and $\bU_{\cS,N}^{\ddagger}$,
\begin{equation*}
	P(\bl_{\cS,N}^{\dagger}(\bU_{\cS,N}^{\ddagger})\neq \bl_{\cS,N}^{\dagger}(\bU_{\cS,N}^{\dagger}))\leq P(\bU_{\cS,N}^{\dagger}\notin T)\leq 2\epsilon,
\end{equation*}
and
\begin{equation*}
	P(\bl_{\cS,N}^{\dagger}(\bZ_{\cS}^{\dagger})\neq \bl_{\cS,N}^{\dagger}(\bZ_{\cS}))\leq P(\bZ_{\cS}\notin T)\leq \epsilon.
\end{equation*}
In addition, the first term at the right side of \eqref{eq: truncated convergence} converges to zero which is guaranteed by the uniform convergence of $l_{\cS,N^{\prime}}^{\dagger}$, $\bU_{\cS,N}^{\ddagger}\stackrel{d}{\to}\bZ_{\cS}^{\dagger}$ and the continuous mapping theorem in page 259 of \cite{Vaart2000AS}. Combining these, we have $\sup_{t}\big|P(l_{\cS,N^{\prime}}^{\dagger}(\bU_{\cS,N^{\prime}}^{\dagger})\leq t)-P(\bl_{\cS,N^{\prime}}^{\dagger}(\bZ_{\cS})\leq t)\big|\leq 5\epsilon$ for sufficiently large $N^{\prime}$. Then \eqref{eq: continous mapping dist} holds; otherwise if for any $N_{k}$, there are $N_{k}^{\prime}$ such that
\begin{equation}\label{eq: subseq div}
	\sup_{t}\big|P(\bl_{\cS,N_{k}^{\prime}}^{\dagger}(\bU_{\cS,N_{k}^{\prime}}^{\dagger})\leq t)-P(\bl_{\cS,N_{k}^{\prime}}^{\dagger}(\bZ_{\cS})\leq t)\big|>\epsilon,
\end{equation}
then there exists a subsequence of $\{\bl_{\cS,N}^{\dagger}\}$ such that \eqref{eq: subseq div} holds.
This leads to a contradict to that every subsequence of $\{\bl_{\cS,N}^{\dagger}\}$ itself has a uniformly convergent subsequence.

%

\subsection*{Proof of Theorem \ref{theo: non-normal M}}

With the estimated optimal moment function $\bpsi(\tilde{\btheta})$, we have
\begin{equation*}
	\tilde{\bg}_{\cM} = 
	\begin{pmatrix}
		\frac{1}{N}\sum_{i=1}^{N}\delta_{i}\pi_{i}^{-1}\bpsi_{i}(\tilde{\btheta})\\
		-\frac{1}{N}\sum_{i=1}^{N}\delta_{i}\pi_{i}^{-1}\hat{\bmu}
	\end{pmatrix},
\end{equation*}
\begin{equation*}
	\tG_{\cM} = 
	\begin{pmatrix}
		-\frac{1}{N}\sum_{i=1}^{N}\delta_{i}\pi_{i}^{-1}\bpsi_{i}(\tilde{\btheta})\bpsi_{i}(\tilde{\btheta})^{\T}\\
		\frac{1}{N}\sum_{i=1}^{N}\delta_{i}\pi_{i}^{-1}\bpsi_{i}(\tilde{\btheta})\bpsi_{i}(\tilde{\btheta})^{\T}
	\end{pmatrix},
\end{equation*}
and
\begin{equation*}
	\tOmega_{\cM} = 
	\begin{pmatrix}
		\tOmega_{\bu\bu} & \tOmega_{\bu\bbm}\\
		\tOmega_{\bbm\bu} &\tOmega_{\bbm\bbm}
	\end{pmatrix},
\end{equation*}
where $\hat{\bmu} = N^{-1}\sum_{i=1}^{N}\bpsi_{i}(\tilde{\btheta})$, $\tOmega_{\bu\bu} = \rho_{N}N^{-1}\sum_{i=1}^{N}\delta_{i}\pi_{i}^{-2}\bpsi_{i}(\tilde{\btheta})\bpsi_{i}(\tilde{\btheta})^{\T}$, 
$$\tOmega_{\bu\bbm} = \tOmega_{\bbm\bu}^{\T} = -\rho_{N}\frac{1}{N}\sum_{i=1}^{N}\frac{\delta_{i}}{\pi_{i}}\bpsi_{i}(\tilde{\btheta})\bpsi_{i}(\tilde{\btheta})^{\T}-\rho_{N}\frac{1}{N}\sum_{i=1}^{N}\frac{\delta_{i}}{\pi_{i}}\pi_{i}^{-1}\bpsi_{i}(\tilde{\btheta})\hat{\bmu}^{\T},$$ and
$$\tOmega_{\bbm\bbm} = \rho_{N}\frac{1}{N}\sum_{i=1}^{N}\frac{\delta_{i}}{\pi_{i}}\bpsi_{i}(\tilde{\btheta})\bpsi_{i}(\tilde{\btheta})^{\T}+\rho_{N}\frac{1}{N}\sum_{i=1}^{N}\frac{\delta_{i}}{\pi_{i}}\left(\pi_{i}^{-1}-1\right)\hat{\bmu}\hat{\bmu}^{\T}.$$
The corresponding population forms are
\begin{equation*}
	\bbG_{\cM} = \begin{pmatrix}
		-E\left(\bpsi\bpsi^{\T}\right)\\
		E\left(\bpsi\bpsi^{\T}\right)
	\end{pmatrix}
\end{equation*}
and
\begin{equation*}
	\bbOmega_{\cM} = \begin{pmatrix}
		\bbOmega_{\bu\bu}& \bbOmega_{\bu\bbm}\\
		\bbOmega_{\bbm\bu}& \bbOmega_{\bbm\bbm}\\
	\end{pmatrix},
\end{equation*}
where $\bbOmega_{\bu\bu} = \rho_{N}E\left(\pi^{-1}\bpsi\bpsi^{\T}\right)$, $\bbOmega_{\bu\bbm} = \bbOmega_{\bbm\bu}^{\T} = -\rho_{N}E\left(\bpsi\bpsi^{\T}\right)$, and $\bbOmega_{\bbm\bbm} = \rho_{N}E\left(\bpsi\bpsi^{\T}\right)$.
By Lemma \ref{lem: AL of initial}, $\|\tilde{\btheta}-\btheta_{0}\|=O_{P}(n^{-1/2})$. It is easy to show that $\|\tOmega_{\bu\bu}-\bbOmega_{\bu\bu}\| = O_{P}(n^{-1/2})$, $\|\tOmega_{\bbm\bu}-\bbOmega_{\bbm\bu}\| = O_{P}(n^{-1/2})$, and $\|\tOmega_{\bbm\bbm}-\bbOmega_{\bbm\bbm}\| = O_{P}(\rho_{N}n^{-1/2}+n^{-1})$.
Recalling the definition of $\tilde{\btheta}_{\cM}$, by some simple algebras,
\begin{equation}\label{eq: nonnormal mMAS decomp}
	\tilde{\btheta}_{\cM} - \btheta_{0} 
	= -\bbA_{\cM}\tilde{\bg}_{\cM} - (\tA_{\cM}-\bbA_{\cM})\tilde{\bg}_{\cM} + \bbA_{\cM}\tG_{\cM}\left(\tilde{\btheta}-\btheta_{0}\right) + (\tA_{\cM}-\bbA_{\cM})\tG_{\cM}\left(\tilde{\btheta}-\btheta_{0}\right),
\end{equation}
where  $\tA_{\cM}=(\tG_{\cM}^{\T}\tOmega_{\cM}^{-1}\tG_{\cM})^{-1}\tG_{\cM}^{\T}\tOmega_{\cM}^{-1}$ and $\bbA_{\cM}=(\bbG_{\cM}^{\T}\bbOmega_{\cM}^{-1}\bbG_{\cM})^{-1}\bbG_{\cM}^{\T}\bbOmega_{\cM}^{-1}$. We first analyze the sum of the first and third terms at the right side of \eqref{eq: nonnormal mMAS decomp}. 
By Conditions (C3$^{\prime}$) and (C6), $\|\bbA_{\cM}\|=O(1)$. By Taylor's expansion of $\bpsi_{i}(\tilde{\btheta})$ at $\btheta_{0}$ and $E\left(\dot{\bpsi}\right)=-E\left(\bpsi\bpsi^{\T}\right)$,
\begin{align*}
	&-\bbA_{\cM}\tilde{\bg}_{\cM} + \bbA_{\cM}\tG_{\cM}\left(\tilde{\btheta}-\btheta_{0}\right)\\
	& = - \bbA_{\cM}
	\begin{pmatrix}
		\frac{1}{N}\sum_{i=1}^{N}\delta_{i}\pi_{i}^{-1}\bpsi_{i}\\
		-\frac{1}{N}\sum_{i=1}^{N}\delta_{i}\pi_{i}^{-1}\frac{1}{N}\sum_{i=1}^{N}\bpsi_{i} 
	\end{pmatrix}- \bbA_{\cM}
	\begin{pmatrix}
		\frac{1}{N}\sum_{i=1}^{N}\delta_{i}\pi_{i}^{-1}\dot{\bpsi}_{i}\left(\tilde{\btheta}-\btheta_{0}\right) \\
		-\frac{1}{N}\sum_{i=1}^{N}\delta_{i}\pi_{i}^{-1} \frac{1}{N}\sum_{i=1}^{N}\dot{\bpsi}_{i}\left(\tilde{\btheta}-\btheta_{0}\right)
	\end{pmatrix}\\
	&\quad- \bbA_{\cM}
	\begin{pmatrix}
		\frac{1}{2N}\sum_{i=1}^{N}\delta_{i}\pi_{i}^{-1}\ddot{\bpsi}_{i}\left\{\left(\tilde{\btheta}-\btheta_{0}\right)\otimes\left(\tilde{\btheta}-\btheta_{0}\right)\right\}\\
		-\frac{1}{N}\sum_{i=1}^{N}\delta_{i}\pi_{i}^{-1}\frac{1}{2N}\sum_{i=1}^{N}\ddot{\bpsi}_{i}\left\{\left(\tilde{\btheta}-\btheta_{0}\right)\otimes\left(\tilde{\btheta}-\btheta_{0}\right)\right\}
	\end{pmatrix}\\
	&\quad+\bbA_{\cM}
	\begin{pmatrix}
		-\frac{1}{N}\sum_{i=1}^{N}\delta_{i}\pi_{i}^{-1}\bpsi_{i}(\tilde{\btheta})\bpsi_{i}(\tilde{\btheta})^{\T}\\
		\frac{1}{N}\sum_{i=1}^{N}\delta_{i}\pi_{i}^{-1}\bpsi_{i}(\tilde{\btheta})\bpsi_{i}(\tilde{\btheta})^{\T}
	\end{pmatrix}\left(\tilde{\btheta}-\btheta_{0}\right) + o_{P}(n^{-1})\\
	&= - \bbA_{\cM}
	\begin{pmatrix}
		\frac{1}{N}\sum_{i=1}^{N}\delta_{i}\pi_{i}^{-1}\bpsi_{i}\\
		-\frac{1}{N}\sum_{i=1}^{N}\bpsi_{i}
	\end{pmatrix}- \bbA_{\cM}
	\begin{pmatrix}
		\frac{1}{N}\sum_{i=1}^{N}\delta_{i}\pi_{i}^{-1}\dot{\bpsi}_{i}-E(\dot{\bpsi})\\
		-\frac{1}{N}\sum_{i=1}^{N}\dot{\bpsi}_{i} + E(\dot{\bpsi})
	\end{pmatrix}\left(\tilde{\btheta}-\btheta_{0}\right)\\
	&\quad  - \bbA_{\cM}
	\begin{pmatrix}
		\frac{1}{2N}\sum_{i=1}^{N}\delta_{i}\pi_{i}^{-1}\ddot{\bpsi}_{i}\\
		-\frac{1}{2N}\sum_{i=1}^{N}\ddot{\bpsi}_{i}
	\end{pmatrix}\left\{\left(\tilde{\btheta}-\btheta_{0}\right)\otimes\left(\tilde{\btheta}-\btheta_{0}\right)\right\} \\
	&\quad  + \bbA_{\cM}
	\begin{pmatrix}
		-	\frac{1}{N}\sum_{i=1}^{N}\delta_{i}\pi_{i}^{-1}\bpsi_{i}\bpsi_{i}^{\T}+E(\bpsi\bpsi^{\T})\\
		\frac{1}{N}\sum_{i=1}^{N}\delta_{i}\pi_{i}^{-1}\bpsi_{i}\bpsi_{i}^{\T} - E(\bpsi\bpsi^{\T})
	\end{pmatrix}\left(\tilde{\btheta}-\btheta_{0}\right)  \\
	&\quad + \bbA_{\cM}
	\begin{pmatrix}
		-	\frac{1}{N}\sum_{i=1}^{N}\delta_{i}\pi_{i}^{-1}\dot{(\bpsi_{i}\bpsi_{i}^{\T})}\\
		\frac{1}{N}\sum_{i=1}^{N}\delta_{i}\pi_{i}^{-1}\dot{(\bpsi_{i}\bpsi_{i}^{\T})}
	\end{pmatrix}\left\{\left(\tilde{\btheta}-\btheta_{0}\right)\otimes\left(\tilde{\btheta}-\btheta_{0}\right)\right\} +o_{P}(\max\{n^{-1},N^{-1/2}\}).
\end{align*}
Rewrite
\begin{equation*}
	\bbOmega_{\cM}^{-1} = 
	\begin{pmatrix}
		\bbB_{\bu\bu}& \bbB_{\bu\bbm}\\
		\bbB_{\bbm\bu}& \bbB_{\bbm\bbm}
	\end{pmatrix},
\end{equation*}
where 
$$\bbB_{\bu\bu} = \bbOmega_{\bu\bu}^{-1}+\bbOmega_{\bu\bu}^{-1}\bbOmega_{\bu\bbm}(\bbOmega_{\bbm\bbm}-\bbOmega_{\bbm\bu}\bbOmega_{\bu\bu}^{-1}\bbOmega_{\bu\bbm})^{-1}\bbOmega_{\bbm\bu}\bbOmega_{\bu\bu}^{-1},$$  $$\bbB_{\bbm\bu}=\bbB_{\bu\bbm}^{\T} = -(\bbOmega_{\bbm\bbm}-\bbOmega_{\bbm\bu}\bbOmega_{\bu\bu}^{-1}\bbOmega_{\bu\bbm})^{-1}\bbOmega_{\bbm\bu}\bbOmega_{\bu\bu}^{-1},$$
$$\bbB_{\bbm\bbm} = (\bbOmega_{\bbm\bbm}-\bbOmega_{\bbm\bu}\bbOmega_{\bu\bu}^{-1}\bbOmega_{\bu\bbm})^{-1}.$$
Then we have
\begin{equation*}
	\begin{split}
		&\bbA_{\cM}
		\begin{pmatrix}
			\frac{1}{N}\sum_{i=1}^{N}\delta_{i}\pi_{i}^{-1}\dot{\bpsi}_{i}-E(\dot{\bpsi})\\
			-\frac{1}{N}\sum_{i=1}^{N}\dot{\bpsi}_{i} + E(\dot{\bpsi})
		\end{pmatrix}\left(\tilde{\btheta}-\btheta_{0}\right)\\
		& = (\bbG_{\cM}^{\T}\bbOmega_{\cM}^{-1}\bbG_{\cM})^{-1}\left(\bbG_{\bu}^{\T}\bbB_{\bu\bu}+\bbG_{\bbm}\bbB_{\bbm\bu}\right) \left\{\frac{1}{N}\sum_{i=1}^{N}\frac{\delta_{i}}{\pi_{i}}\dot{\bpsi}_{i}-E(\dot{\bpsi})\right\}\left(\tilde{\btheta}-\btheta_{0}\right)\\
		&\quad- (\bbG_{\cM}^{\T}\bbOmega_{\cM}^{-1}\bbG_{\cM})^{-1}\left(\bbG_{\bu}^{\T}\bbB_{\bu\bbm}+\bbG_{\bbm}\bbB_{\bbm\bbm}\right)\left\{\frac{1}{N}\sum_{i=1}^{N}\dot{\bpsi}_{i}-E(\dot{\bpsi})\right\}\left(\tilde{\btheta}-\btheta_{0}\right),
	\end{split}
\end{equation*}
It is easy to show that $\|(\bbG_{\cM}^{\T}\bbOmega_{\cM}^{-1}\bbG_{\cM})^{-1}\|=O(\rho_{N})$, $\|\bbG_{\bu}^{\T}\bbB_{\bu\bu}+\bbG_{\bbm}\bbB_{\bbm\bu}\| = O(1)$, and $\|\bbG_{\bu}^{\T}\bbB_{\bu\bbm}+\bbG_{\bbm}\bbB_{\bbm\bbm}\| = O(\rho_{N}^{-1})$.
In addition,  $N^{-1}\sum_{i=1}^{N}\delta_{i}\pi_{i}^{-1}\dot{\bpsi}_{i}-E(\dot{\bpsi}) = O_{P}(n^{-1/2})$ and $N^{-1}\sum_{i=1}^{N}\dot{\bpsi}_{i}-E(\dot{\bpsi}) = O_{P}(N^{-1/2})$.
Then 
\begin{equation*}
	\bbA_{\cM}
	\begin{pmatrix}
		\frac{1}{N}\sum_{i=1}^{N}\delta_{i}\pi_{i}^{-1}\dot{\bpsi}_{i}-E(\dot{\bpsi})\\
		-\frac{1}{N}\sum_{i=1}^{N}\dot{\bpsi}_{i} + E(\dot{\bpsi})
	\end{pmatrix}\left(\tilde{\btheta}-\btheta_{0}\right)=O_{P}(\rho_{N}n^{-1}+n^{-1/2}N^{-1/2}) = o_{P}(\max\{n^{-1},N^{-1/2}\}),
\end{equation*}
and hence
\begin{equation*}
	\begin{split}
		&-\bbA_{\cM}\tilde{\bg}_{\cM} + \bbA_{\cM}\tG_{\cM}\left(\tilde{\btheta}-\btheta_{0}\right)\\
		&= - \bbA_{\cM}
		\begin{pmatrix}
			\frac{1}{N}\sum_{i=1}^{N}\delta_{i}\pi_{i}^{-1}\bpsi_{i}\\
			-\frac{1}{N}\sum_{i=1}^{N}\bpsi_{i}
		\end{pmatrix} - \bbA_{\cM} 
		\begin{pmatrix}
			\frac{1}{2N}\sum_{i=1}^{N}\delta_{i}\pi_{i}^{-1}\ddot{\bpsi}_{i}\\
			-\frac{1}{2N}\sum_{i=1}^{N}\ddot{\bpsi}_{i}
		\end{pmatrix}\left\{\left(\tilde{\btheta}-\btheta_{0}\right)\otimes\left(\tilde{\btheta}-\btheta_{0}\right)\right\}\\
		&\quad   + \bbA_{\cM}
		\begin{pmatrix}
			-	\frac{1}{N}\sum_{i=1}^{N}\delta_{i}\pi_{i}^{-1}\bpsi_{i}\bpsi_{i}^{\T}+E[\bpsi\bpsi^{\T}]\\
			\frac{1}{N}\sum_{i=1}^{N}\delta_{i}\pi_{i}^{-1}\bpsi_{i}\bpsi_{i}^{\T} - E[\bpsi\bpsi^{\T}]
		\end{pmatrix}\left(\tilde{\btheta}-\btheta_{0}\right)  \\
		&\quad + \bbA_{\cM}
		\begin{pmatrix}
			-	\frac{1}{N}\sum_{i=1}^{N}\delta_{i}\pi_{i}^{-1}\dot{(\bpsi_{i}\bpsi_{i}^{\T})}\\
			\frac{1}{N}\sum_{i=1}^{N}\delta_{i}\pi_{i}^{-1}\dot{(\bpsi_{i}\bpsi_{i}^{\T})}
		\end{pmatrix}\left\{\left(\tilde{\btheta}-\btheta_{0}\right)\otimes\left(\tilde{\btheta}-\btheta_{0}\right)\right\} +o_{P}(\max\{n^{-1},N^{-1/2}\}).
	\end{split}
\end{equation*}

Next, we analyze the sum of the second and fourth terms at the right side of \eqref{eq: nonnormal mMAS decomp}. 
Note that
\begin{equation*}
	\begin{split}
		(\tA_{\cM}-\bbA_{\cM})
		& = \left\{(\tG_{\cM}^{\T}\tOmega_{\cM}^{-1}\tG_{\cM})^{-1}-(\bbG_{\cM}^{\T}\bbOmega_{\cM}^{-1}\bbG_{\cM})^{-1}\right\}\bbG_{\cM}^{\T}\bbOmega_{\cM}^{-1}\\
		&\quad+\left\{(\tG_{\cM}^{\T}\tOmega_{\cM}^{-1}\tG_{\cM})^{-1}-(\bbG_{\cM}^{\T}\bbOmega_{\cM}^{-1}\bbG_{\cM})^{-1}\right\}\left(\tG_{\cM}^{\T}\tOmega_{\cM}^{-1}-\bbG_{\cM}^{\T}\bbOmega_{\cM}^{-1}\right)\\
		&\quad+(\bbG_{\cM}^{\T}\bbOmega_{\cM}^{-1}\bbG_{\cM})^{-1}\left(\tG_{\cM}^{\T}\tOmega_{\cM}^{-1}-\bbG_{\cM}^{\T}\bbOmega_{\cM}^{-1}\right).
	\end{split}
\end{equation*}
Recalling the expression of $\tOmega_{\cM}^{-1}$, we have
\begin{equation*}
	\tOmega_{\cM}^{-1} = 
	\begin{pmatrix}
		\tB_{\bu\bu}& \tB_{\bu\bbm}\\
		\tB_{\bbm\bu}& \tB_{\bbm\bbm}
	\end{pmatrix},
\end{equation*}
where 
$$\tB_{\bbm\bbm} = (\tOmega_{\bbm\bbm}-\tOmega_{\bbm\bu}\tOmega_{\bu\bu}^{-1}\tOmega_{\bu\bbm})^{-1},$$
$$\tB_{\bbm\bu}=\tB_{\bu\bbm}^{\T} = -\tB_{\bbm\bbm}\tOmega_{\bu\bbm})^{-1}\tOmega_{\bbm\bu}\tOmega_{\bu\bu}^{-1},$$
$$\tB_{\bu\bu} = \tOmega_{\bu\bu}^{-1}-\tOmega_{\bu\bu}^{-1}\tOmega_{\bu\bbm}\tB_{\bbm\bu}.$$  
By matrix differentials,
\begin{align*}
	&\tB_{\bbm\bbm}-\bbB_{\bbm\bbm}\\
	& = -\rho_{N}^{-1}\left[\left\{\rho_{N}^{-1}\left(\tOmega_{\bbm\bbm}-\tOmega_{\bbm\bu}\tOmega_{\bu\bu}^{-1}\tOmega_{\bu\bbm}\right)\right\}^{-1}-\left\{\rho_{N}^{-1}\left(\bbOmega_{\bbm\bbm}-\bbOmega_{\bbm\bu}\bbOmega_{\bu\bu}^{-1}\bbOmega_{\bbm\bu}\right)\right\}^{-1}\right]\\
	&=-\rho_{N}^{-1}\left[\left\{\rho_{N}^{-1}\left(\bbOmega_{\bbm\bbm}-\bbOmega_{\bbm\bu}\bbOmega_{\bu\bu}^{-1}\bbOmega_{\bbm\bu}\right)\right\}^{-1}\left\{\rho_{N}^{-1}\left(\tOmega_{\bbm\bbm}-\tOmega_{\bbm\bu}\tOmega_{\bu\bu}^{-1}\tOmega_{\bu\bbm}\right)\right.\right.\\
	&\left.\left.\qquad\qquad-\rho_{N}^{-1}\left(\bbOmega_{\bbm\bbm}-\bbOmega_{\bbm\bu}\bbOmega_{\bu\bu}^{-1}\bbOmega_{\bbm\bu}\right)\right\}\left\{\rho_{N}^{-1}\left(\bbOmega_{\bbm\bbm}-\bbOmega_{\bbm\bu}\bbOmega_{\bu\bu}^{-1}\bbOmega_{\bbm\bu}\right)\right\}^{-1}\right] \\
	&\quad + o_{P}(\rho_{N}^{-1}\|\rho_{N}^{-1}\left(\tOmega_{\bbm\bbm}-\tOmega_{\bbm\bu}\tOmega_{\bu\bu}^{-1}\tOmega_{\bu\bbm}\right)-\rho_{N}^{-1}\left(\bbOmega_{\bbm\bbm}-\bbOmega_{\bbm\bu}\bbOmega_{\bu\bu}^{-1}\bbOmega_{\bbm\bu}\right)\|)\\
	&=O_{P}\left(\rho_{N}^{-2}\|\left(\tOmega_{\bbm\bbm}-\tOmega_{\bbm\bu}\tOmega_{\bu\bu}^{-1}\tOmega_{\bu\bbm}\right)-\left(\bbOmega_{\bbm\bbm}-\bbOmega_{\bbm\bu}\bbOmega_{\bu\bu}^{-1}\bbOmega_{\bbm\bu}\right)\|\right),
\end{align*}
and
\begin{equation*}
	\begin{split}
		&\tOmega_{\bbm\bu}\tOmega_{\bu\bu}^{-1}\tOmega_{\bu\bbm}-\bbOmega_{\bbm\bu}\bbOmega_{\bu\bu}^{-1}\bbOmega_{\bu\bbm}\\
		& = \rho_{N}^{2}\left\{(\rho_{N}^{-1}\tOmega_{\bbm\bu})\tOmega_{\bu\bu}^{-1}(\rho_{N}^{-1}\tOmega_{\bu\bbm})-(\rho_{N}^{-1}\bbOmega_{\bbm\bu})\bbOmega_{\bu\bu}^{-1}(\rho_{N}^{-1}\bbOmega_{\bu\bbm})\right\}\\
		&=\rho_{N}^{2}\left\{\left(\rho_{N}^{-1}\tOmega_{\bbm\bu}-\rho_{N}^{-1}\bbOmega_{\bbm\bu}\right)\bbOmega_{\bu\bu}^{-1}(\rho_{N}^{-1}\bbOmega_{\bu\bbm}) + (\rho_{N}^{-1}\bbOmega_{\bbm\bu})\bbOmega_{\bu\bu}^{-1}\left(\tOmega_{\bu\bu}-\bbOmega_{\bu\bu}\right)\bbOmega_{\bu\bu}^{-1}(\rho_{N}^{-1}\bbOmega_{\bu\bbm})\right.\\
		&\qquad\qquad\left.+(\rho_{N}^{-1}\bbOmega_{\bbm\bu})\bbOmega_{\bu\bu}^{-1}\left(\rho_{N}^{-1}\tOmega_{\bu\bbm}-\rho_{N}^{-1}\bbOmega_{\bu\bbm}\right)+o_{P}(\|\rho_{N}^{-1}\tOmega_{\bu\bbm}-\rho_{N}^{-1}\bbOmega_{\bu\bbm}\|+\|\tOmega_{\bu\bu}-\bbOmega_{\bu\bu}\|)\right\}\\
		&= O_{P}(\rho_{N}n^{-1/2}).
	\end{split}
\end{equation*}
Recalling that $\|\tOmega_{\bbm\bbm}-\bbOmega_{\bbm\bbm}\| = O_{P}(\rho_{N}n^{-1/2}+n^{-1})$, then we have $\|\tB_{\bbm\bbm}-\bbB_{\bbm\bbm}\| = O_{P}(\rho_{N}^{-1}n^{-1/2}+\rho_{N}^{-2}n^{-1})$.
In addition,
\begin{equation*}
	\begin{split}
		&\tB_{\bu\bbm}-\bbB_{\bu\bbm}\\
		& = -\left\{(\rho_{N}\tB_{\bbm\bbm})(\rho_{N}^{-1}\tOmega_{\bbm\bu})\tOmega_{\bu\bu}^{-1}-(\rho_{N}\bbB_{\bbm\bbm})(\rho_{N}^{-1}\bbOmega_{\bbm\bu})\bbOmega_{\bu\bu}^{-1}\right\}\\
		& = -\left\{\left(\rho_{N}\tB_{\bbm\bbm}-\rho_{N}\bbB_{\bbm\bbm}\right)(\rho_{N}^{-1}\bbOmega_{\bbm\bu})\bbOmega_{\bu\bu}^{-1}\right\}-\left\{(\rho_{N}\bbB_{\bbm\bbm})\left(\rho_{N}^{-1}\tOmega_{\bbm\bu}-\rho_{N}^{-1}\bbOmega_{\bbm\bu}\right)\bbOmega_{\bu\bu}^{-1}\right\}\\
		&\qquad\quad-\left\{(\rho_{N}\bbB_{\bbm\bbm})(\rho_{N}^{-1}\bbOmega_{\bbm\bu})\bbOmega_{\bu\bu}^{-1}\left(\tOmega_{\bu\bu}-\bbOmega_{\bu\bu}\right)\bbOmega_{\bu\bu}^{-1}\right\}\\
		&\quad+o_{P}(\|\rho_{N}\tB_{\bbm\bbm}-\rho_{N}\bbB_{\bbm\bbm}\|+\|\rho_{N}^{-1}\tOmega_{\bbm\bu}-\rho_{N}^{-1}\bbOmega_{\bbm\bu}\|+\|\tOmega_{\bu\bu}-\bbOmega_{\bu\bu}\|)\\
		&=O_{P}(\rho_{N}^{-1}n^{-1/2}),
	\end{split}
\end{equation*}
and
\begin{equation*}
	\begin{split}
		&\tB_{\bu\bu}-\bbB_{\bu\bu}\\
		& = \tOmega_{\bu\bu}^{-1}-\bbOmega_{\bu\bu}^{-1} + \rho_{N}\tOmega_{\bu\bu}^{-1}(\rho_{N}^{-1}\tOmega_{\bu\bbm})\tB_{\bbm\bu} - \rho_{N}\bbOmega_{\bu\bu}^{-1}(\rho_{N}^{-1}\bbOmega_{\bu\bbm})\bbB_{\bbm\bu} \\
		& = \bbOmega_{\bu\bu}^{-1}(\tOmega_{\bu\bu}-\bbOmega_{\bu\bu})\bbOmega_{\bu\bu}^{-1} + \rho_{N}\bbOmega_{\bu\bu}^{-1}(\tOmega_{\bu\bu}-\bbOmega_{\bu\bu})\Omega_{\bu\bu}^{-1}(\rho_{N}^{-1}\bbOmega_{\bu\bbm})\bbB_{\bbm\bu}\\
		&\quad + \rho_{N}\bbOmega_{\bu\bu}^{-1}(\rho_{N}^{-1}\tOmega_{\bu\bbm}-\rho_{N}^{-1}\bbOmega_{\bu\bbm})\bbB_{\bbm\bu} + \rho_{N}\Omega_{\bu\bu}^{-1}(\rho_{N}^{-1}\bbOmega_{\bu\bbm})(\tB_{\bbm\bu}-\bbB_{\bbm\bu})\\
		&\quad + o_{P}(\rho_{N}\|\tOmega_{\bu\bu}-\bbOmega_{\bu\bu}\|+\rho_{N}\|\rho_{N}^{-1}\tOmega_{\bu\bbm}-\rho_{N}^{-1}\bbOmega_{\bu\bbm}\| + \rho_{N}\|\tB_{\bbm\bu}-\bbB_{\bbm\bu}\|)\\
		& = O_{P}(n^{-1/2}).
	\end{split}
\end{equation*}
Further, 
\begin{equation*}
	\begin{split}
		&(\tG_{\cM}^{\T}\tOmega_{\cM}^{-1}\tG_{\cM})^{-1}-(\bbG_{\cM}^{T}\bbOmega_{\cM}^{-1}\bbG_{\cM})^{-1}\\
		& = \rho_{N}(\rho_{N}\bbG_{\cM}^{T}\bbOmega_{\cM}^{-1}\bbG_{\cM})^{-1}(\rho_{N}\tG_{\cM}^{\T}\tOmega_{\cM}^{-1}\tG_{\cM}-\rho_{N}\bbG_{\cM}^{T}\bbOmega_{\cM}^{-1}\bbG_{\cM})(\rho_{N}\bbG_{\cM}^{T}\bbOmega_{\cM}^{-1}\bbG_{\cM})^{-1}\\
		&\quad + o_{P}(\rho_{N}\|\rho_{N}\tG_{\cM}^{\T}\tOmega_{\cM}^{-1}\tG_{\cM}-\rho_{N}\bbG_{\cM}^{T}\bbOmega_{\cM}^{-1}\bbG_{\cM}\|)\\
		&=O_{P}(\rho_{N}^{2}\|\tG_{\cM}^{\T}\tOmega_{\cM}^{-1}\tG_{\cM}-\bbG_{\cM}^{T}\bbOmega_{\cM}^{-1}\bbG_{\cM}\|),
	\end{split}
\end{equation*}
and
\begin{align*}
	&\tG_{\cM}^{\T}\tOmega_{\cM}^{-1}\tG_{\cM}-\bbG_{\cM}^{T}\bbOmega_{\cM}^{-1}\bbG_{\cM}\\
	&=\left(\tG_{\cM,\bu}^{\T}\tB_{\bu\bu}\tG_{\cM,\bu}-\bbG_{\bu}^{\T}\bbB_{\bu\bu}\bbG_{\bu}\right)+\left(\tG_{\bbm}^{\T}\tB_{\bbm\bu}\tG_{\cM,\bu}-\bbG_{\bbm}^{\T}\bbB_{\bbm\bu}\bbG_{\bu}\right)\\
	&\quad+\left(\tG_{\cM,\bu}^{\T}\tB_{\bu\bbm}\tG_{\bbm}-\bbG_{\bu}^{\T}\bbB_{\bu\bbm}\bbG_{\bbm}\right)+\left(\tG_{\bbm}^{\T}\tB_{\bbm\bbm}\tG_{\bbm}-\bbG_{\bbm}^{\T}\bbB_{\bbm\bbm}\bbG_{\bbm}\right)\\
	& = \left(\tG_{\cM,\bu}-\bbG_{\bu}\right)^{\T}\bbB_{\bu\bu}\bbG_{\bu}+\bbG_{\bu}^{\T}\left(\tB_{\bu\bu}-\bbB_{\bu\bu}\right)\bbG_{\bu}+\bbG_{\bu}^{\T}\bbB_{\bu\bu}\left(\tG_{\cM,\bu}-\bbG_{\bu}\right)\\		
	&\quad+\left(\tG_{\bbm}-\bbG_{\bbm}\right)^{\T}\bbB_{\bbm\bu}\bbG_{\bu}+\bbG_{\bbm}^{\T}\left(\tB_{\bbm\bu}-\bbB_{\bbm\bu}\right)\bbG_{\bu}+\bbG_{\bbm}^{\T}\bbB_{\bbm\bu}\left(\tG_{\cM,\bu}-\bbG_{\bu}\right)\\
	&\quad+\left(\tG_{\cM,\bu}-\bbG_{\bu}\right)^{\T}\bbB_{\bu\bbm}\bbG_{\bbm}+\bbG_{\bu}^{\T}\left(\tB_{\bu\bbm}-\bbB_{\bu\bbm}\right)\bbG_{\bbm}+\bbG_{\bu}^{\T}\bbB_{\bu\bbm}\left(\tG_{\bbm}-\bbG_{\bbm}\right)\\
	&\quad+\rho_{N}^{-1}\left(\tG_{\bbm}-\bbG_{\bbm}\right)^{\T}(\rho_{N}\bbB_{\bbm\bbm})\bbG_{\bbm}+\rho_{N}^{-1}\bbG_{\bbm}^{\T}\left(\rho_{N}\tB_{\bbm\bbm}-\rho_{N}\bbB_{\bbm\bbm}\right)\bbG_{\bbm}\\
	&\quad+\rho_{N}^{-1}\bbG_{\bbm}^{\T}(\rho_{N}\bbB_{\bbm\bbm})\left(\tG_{\bbm}-\bbG_{\bbm}\right)+o_{P}(\|\tG_{\cM,\bu}-\bbG_{\bu}\|+\|\tB_{\bu\bu}-\bbB_{\bu\bu}\|\\
	&\quad+\|\tB_{\bu\bbm}-\bbB_{\bu\bbm}\|+\rho_{N}^{-1}\|\tG_{\bbm}-\bbG_{\bbm}\|+\rho_{N}^{-1}\|\rho_{N}\tB_{\bbm\bbm}-\rho_{N}\bbB_{\bbm\bbm}\|)\\
	&=O_{P}(\rho_{N}^{-1}n^{-1/2}+\rho_{N}^{-2}n^{-1}).
\end{align*}
Then
\begin{equation*}
	(\tG_{\cM}^{\T}\tOmega_{\cM}^{-1}\tG_{\cM})^{-1}-(\bbG_{\cM}^{T}\bbOmega_{\cM}^{-1}\bbG_{\cM})^{-1}= O_{P}(\rho_{N}n^{-1/2}+n^{-1}).
\end{equation*}
In addition, 
\begin{align*}
	&\left(\tG_{\cM}^{\T}\tOmega_{\cM}^{-1}-\bbG_{\cM}^{\T}\bbOmega_{\cM}^{-1}\right)\\ 
	&= \begin{pmatrix}
		\tG_{\cM,\bu}^{\T}\tB_{\bu\bu}+\tG_{\bbm}^{\T}\tB_{\bbm\bu}-\bbG_{\bu}^{\T}\bbB_{\bu\bu}-\bbG_{\bbm}^{\T}\bbB_{\bbm\bu}\\
		\tG_{\cM,\bu}^{\T}\tB_{\bu\bbm}+\tG_{\bbm}^{\T}\tB_{\bbm\bbm}-\bbG_{\bu}^{\T}\bbB_{\bu\bbm}-\bbG_{\bbm}^{\T}\bbB_{\bbm\bbm}
	\end{pmatrix} \\
	&=\begin{pmatrix}
		(\tG_{\cM,\bu}-\bbG_{\bu})^{\T}\bbB_{\bu\bu}+\bbG_{\bu}^{\T}(\tB_{\bu\bu}-\bbB_{\bu\bu})+(\tG_{\bbm}-\bbG_{\bbm})^{\T}\bbB_{\bbm\bu}+\bbG_{\bbm}^{\T}(\tB_{\bbm\bu}-\bbB_{\bbm\bu})\\
		(\tG_{\cM,\bu}-\bbG_{\bu})^{\T}\bbB_{\bu\bbm}+\bbG_{\bu}^{\T}(\tB_{\bu\bbm}-\bbB_{\bu\bbm})+(\tG_{\bbm}-\bbG_{\bbm})^{\T}\bbB_{\bbm\bbm}+\bbG_{\bbm}^{\T}(\tB_{\bbm\bbm}-\bbB_{\bbm\bbm})
	\end{pmatrix} \\
	& \quad + 
	\begin{pmatrix}
		o_{P}(\|\tG_{\cM,\bu}-\bbG_{\bu}\|+\|\tB_{\bu\bu}-\bbB_{\bu\bu}\|+\|\tG_{\bbm}-\bbG_{\bbm}\|+\|\tB_{\bbm\bu}-\bbB_{\bbm\bu}\|)\\
		o_{P}(\|\tG_{\cM,\bu}-\bbG_{\bu}\|+\|\tB_{\bbm\bu}-\bbB_{\bbm\bu}\|+\|\tG_{\bbm}-\bbG_{\bbm}\|+\|\tB_{\bbm\bbm}-\bbB_{\bbm\bbm}\|)
	\end{pmatrix}\\
	&=\begin{pmatrix}
		O_{P}(\rho_{N}^{-1}n^{-1/2})\\
		O_{P}(\rho_{N}^{-1}n^{-1/2}+\rho_{N}^{-2}n^{-1})
	\end{pmatrix}.
\end{align*}
Since $n^{2}/N>C$, then
\begin{equation*}
	\begin{split}
		&\tA_{\cM}-\bbA_{\cM}\\
		& = \left\{(\tG_{\cM}^{\T}\tOmega_{\cM}^{-1}\tG_{\cM})^{-1}-(\bbG_{\cM}^{\T}\bbOmega_{\cM}^{-1}\bbG_{\cM})^{-1}\right\}\bbG_{\cM}^{\T}\bbOmega_{\cM}^{-1}\\
		&\quad+\left\{(\tG_{\cM}^{\T}\tOmega_{\cM}^{-1}\tG_{\cM})^{-1}-(\bbG_{\cM}^{\T}\bbOmega_{\cM}^{-1}\bbG_{\cM})^{-1}\right\}\left(\tG_{\cM}^{\T}\tOmega_{\cM}^{-1}-\bbG_{\cM}^{\T}\bbOmega_{\cM}^{-1}\right)\\
		&\quad+(\bbG_{\cM}^{\T}\bbOmega_{\cM}^{-1}\bbG_{\cM})^{-1}\left(\tG_{\cM}^{\T}\tOmega_{\cM}^{-1}-\bbG_{\cM}^{\T}\bbOmega_{\cM}^{-1}\right)\\
		&=O_{P}\left\{(\rho_{N}n^{-1/2}+n^{-1})\rho_{N}^{-1}+(\rho_{N}n^{-1/2}+n^{-1})(\rho_{N}^{-1}n^{-1/2}+\rho_{N}^{-2}n^{-1})+(n^{-1/2}+\rho_{N}^{-1}n^{-1})\right\}\\
		&=O_{P}(1).
	\end{split}
\end{equation*}
By Taylor's expansion, 
\begin{align*}
	&- (\tA_{\cM}-\bbA_{\cM})\tilde{\bg}_{\cM} + (\tA_{\cM}-\bbA_{\cM})\tG_{\cM}\left(\tilde{\btheta}-\btheta_{0}\right)\\
	&= - (\tA_{\cM}-\bbA_{\cM})
	\begin{pmatrix}
		\frac{1}{N}\sum_{i=1}^{N}\delta_{i}\pi_{i}^{-1}\bpsi_{i}\\
		-\frac{1}{N}\sum_{i=1}^{N}\bpsi_{i}
	\end{pmatrix}- (\tA_{\cM}-\bbA_{\cM})
	\begin{pmatrix}
		\frac{1}{N}\sum_{i=1}^{N}\delta_{i}\pi_{i}^{-1}\dot{\bpsi}_{i}-E(\dot{\bpsi})\\
		-\frac{1}{N}\sum_{i=1}^{N}\dot{\bpsi}_{i} + E(\dot{\bpsi})
	\end{pmatrix}\left(\tilde{\btheta}-\btheta_{0}\right)\\
	&\quad  - (\tA_{\cM}-\bbA_{\cM})
	\begin{pmatrix}
		\frac{1}{2N}\sum_{i=1}^{N}\delta_{i}\pi_{i}^{-1}\ddot{\bpsi}_{i}\\
		-\frac{1}{2N}\sum_{i=1}^{N}\ddot{\bpsi}_{i}
	\end{pmatrix}\left\{\left(\tilde{\btheta}-\btheta_{0}\right)\otimes\left(\tilde{\btheta}-\btheta_{0}\right)\right\} \\
	&\quad + (\tA_{\cM}-\bbA_{\cM})
	\begin{pmatrix}
		-	\frac{1}{N}\sum_{i=1}^{N}\delta_{i}\pi_{i}^{-1}\bpsi_{i}\bpsi_{i}^{\T}+E[\bpsi\bpsi^{\T}]\\
		\frac{1}{N}\sum_{i=1}^{N}\delta_{i}\pi_{i}^{-1}\bpsi_{i}\bpsi_{i}^{\T} - E[\bpsi\bpsi^{\T}]
	\end{pmatrix}\left(\tilde{\btheta}-\btheta_{0}\right)  \\
	&\quad +  (\tA_{\cM}-\bbA_{\cM})
	\begin{pmatrix}
		-	\frac{1}{N}\sum_{i=1}^{N}\delta_{i}\pi_{i}^{-1}\dot{(\bpsi_{i}\bpsi_{i}^{\T})}\\
		\frac{1}{N}\sum_{i=1}^{N}\delta_{i}\pi_{i}^{-1}\dot{(\bpsi_{i}\bpsi_{i}^{\T})}
	\end{pmatrix}\left\{\left(\tilde{\btheta}-\btheta_{0}\right)\otimes\left(\tilde{\btheta}-\btheta_{0}\right)\right\} +o_{P}(n^{-1}).
\end{align*}
Note that
\begin{align*}
	&(\tA_{\cM}-\bbA_{\cM})
	\begin{pmatrix}
		\frac{1}{N}\sum_{i=1}^{N}\delta_{i}\pi_{i}^{-1}\bpsi_{i}\\
		-\frac{1}{N}\sum_{i=1}^{N}\bpsi_{i}
	\end{pmatrix}\\
	& = \left\{(\tG_{\cM}^{\T}\tOmega_{\cM}^{-1}\tG_{\cM})^{-1}-(\bbG_{\cM}^{\T}\bbOmega_{\cM}^{-1}\bbG_{\cM})^{-1}\right\}\bbG_{\cM}^{\T}\bbOmega_{\cM}^{-1}
	\begin{pmatrix}
		\frac{1}{N}\sum_{i=1}^{N}\delta_{i}\pi_{i}^{-1}\bpsi_{i}\\
		-\frac{1}{N}\sum_{i=1}^{N}\bpsi_{i}
	\end{pmatrix}\\
	&\quad+\left\{(\tG_{\cM}^{\T}\tOmega_{\cM}^{-1}\tG_{\cM})^{-1}-(\bbG_{\cM}^{\T}\bbOmega_{\cM}^{-1}\bbG_{\cM})^{-1}\right\}\left(\tG_{\cM}^{\T}\tOmega_{\cM}^{-1}-\bbG_{\cM}^{\T}\bbOmega_{\cM}^{-1}\right)
	\begin{pmatrix}
		\frac{1}{N}\sum_{i=1}^{N}\delta_{i}\pi_{i}^{-1}\bpsi_{i}\\
		-\frac{1}{N}\sum_{i=1}^{N}\bpsi_{i}
	\end{pmatrix}\\
	&\quad+(\bbG_{\cM}^{\T}\bbOmega_{\cM}^{-1}\bbG_{\cM})^{-1}\left(\tG_{\cM}^{\T}\tOmega_{\cM}^{-1}-\bbG_{\cM}^{\T}\bbOmega_{\cM}^{-1}\right)
	\begin{pmatrix}
		\frac{1}{N}\sum_{i=1}^{N}\delta_{i}\pi_{i}^{-1}\bpsi_{i}\\
		-\frac{1}{N}\sum_{i=1}^{N}\bpsi_{i}
	\end{pmatrix}\\
	&=O_{P}\Big\{(\rho_{N}n^{-1/2}+n^{-1})(n^{-1/2}+\rho_{N}^{-1}N^{-1/2})+(\rho_{N}n^{-1/2}+n^{-1})(\rho_{N}^{-1}n^{-1}\\
	&\quad\quad +\rho_{N}^{-2}n^{-1}N^{-1/2})+(n^{-1}+\rho_{N}^{-1}n^{-1}N^{-1/2})\Big\}\\
	&=O_{P}(\max\{n^{-1},N^{-1/2}\}).
\end{align*}
Similarly, 
\begin{equation*}
	\begin{split}
		&(\tA_{\cM}-\bbA_{\cM})
		\begin{pmatrix}
			\frac{1}{N}\sum_{i=1}^{N}\delta_{i}\pi_{i}^{-1}\dot{\bpsi}_{i}-E(\dot{\bpsi})\\
			-\frac{1}{N}\sum_{i=1}^{N}\dot{\bpsi}_{i} + E(\dot{\bpsi})
		\end{pmatrix}\left(\tilde{\btheta}-\btheta_{0}\right) \\
		&= O_{P}\Big\{(\rho_{N}n^{-1/2}+n^{-1})(n^{-1/2}+\rho_{N}^{-1}N^{-1/2})n^{-1/2}+(\rho_{N}n^{-1/2}+n^{-1})(\rho_{N}^{-1}n^{-1}\\
		&\quad\quad\quad +\rho_{N}^{-2}n^{-1}N^{-1/2})n^{-1/2}+(n^{-1}+\rho_{N}^{-1}n^{-1}N^{-1/2})n^{-1/2}\Big\}\\
		&=o_{P}(\max\{n^{-1},N^{-1/2}\}),
	\end{split}
\end{equation*}
\begin{equation*}
	\begin{split}
		&(\tA_{\cM}-\bbA_{\cM})
		\begin{pmatrix}
			\frac{1}{2N}\sum_{i=1}^{N}\delta_{i}\pi_{i}^{-1}\ddot{\bpsi}_{i}\\
			-\frac{1}{2N}\sum_{i=1}^{N}\ddot{\bpsi}_{i}
		\end{pmatrix}\left\{\left(\tilde{\btheta}-\btheta_{0}\right)\otimes\left(\tilde{\btheta}-\btheta_{0}\right)\right\}\\
		&= O_{P}\Big\{(\rho_{N}n^{-1/2}+n^{-1})n^{-1}+(\rho_{N}n^{-1/2}+n^{-1})(\rho_{N}^{-1}n^{-3/2}+\rho_{N}^{-2}n^{-2})+(n^{-3/2}+\rho_{N}^{-1}n^{-2})\Big\}\\
		&=O_{P}(\max\{n^{-1},N^{-1/2}\}),
	\end{split}
\end{equation*}
and
\begin{equation*}
	\begin{split}
		&(\tA_{\cM}-\bbA_{\cM})
		\begin{pmatrix}
			-\frac{1}{N}\sum_{i=1}^{N}\delta_{i}\pi_{i}^{-1}\dot{(\bpsi_{i}\bpsi_{i}^{\T})}\\
			\frac{1}{N}\sum_{i=1}^{N}\delta_{i}\pi_{i}^{-1}\dot{(\bpsi_{i}\bpsi_{i}^{\T})}
		\end{pmatrix}\left\{\left(\tilde{\btheta}-\btheta_{0}\right)\otimes\left(\tilde{\btheta}-\btheta_{0}\right)\right\}\\
		&=O_{P}\Big\{(\rho_{N}n^{-1/2}+n^{-1})n^{-1}+(\rho_{N}n^{-1/2}+n^{-1})(\rho_{N}^{-1}n^{-3/2}+\rho_{N}^{-2}n^{-2})+(n^{-3/2}+\rho_{N}^{-1}n^{-2})\Big\}\\
		&=O_{P}(\max\{n^{-1},N^{-1/2}\}).
	\end{split}
\end{equation*}
Then 
\begin{equation*}
	\begin{split}
		- (\tA_{\cM}-\bbA_{\cM})\tilde{\bg}_{\cM} + (\tA_{\cM}-\bbA_{\cM})\tG_{\cM}\left(\tilde{\btheta}-\btheta_{0}\right)=O_{P}(\max\{n^{-1},N^{-1/2}\}).
	\end{split}
\end{equation*}
Based on the above expansions, we have
\begin{equation*}
	\begin{split}
		&\min\{n,\sqrt{N}\}(\tA_{\cM}-\bbA_{\cM})
		\begin{pmatrix}
			\frac{1}{N}\sum_{i=1}^{N}\delta_{i}\pi_{i}^{-1}\bpsi_{i}\\
			-\frac{1}{N}\sum_{i=1}^{N}\bpsi_{i}
		\end{pmatrix}= c_{1}\bR_{N}(\bbA_{\cM})
		\begin{pmatrix}
			\rho_{N}^{-1/2}\sqrt{n}\frac{1}{N}\sum_{i=1}^{N}\delta_{i}\pi_{i}^{-1}\bpsi_{i}\\
			-\sqrt{N}\frac{1}{N}\sum_{i=1}^{N}\bpsi_{i}
		\end{pmatrix}+o_{P}(1),
	\end{split}
\end{equation*}
\begin{equation*}
	\begin{split}
		&\min\{n,\sqrt{N}\}(\tA_{\cM}-\bbA_{\cM})
		\begin{pmatrix}
			\frac{1}{2N}\sum_{i=1}^{N}\delta_{i}\pi_{i}^{-1}\ddot{\bpsi}_{i}\\
			-\frac{1}{2N}\sum_{i=1}^{N}\ddot{\bpsi}_{i}
		\end{pmatrix}\left\{\left(\tilde{\btheta}-\btheta_{0}\right)\otimes\left(\tilde{\btheta}-\btheta_{0}\right)\right\}\\
		& = c_{2}\bR_{N}(\bbA_{\cM})
		\begin{pmatrix}
			\frac{1}{2}E(\ddot{\bpsi})\\
			-	\frac{1}{2}E(\ddot{\bpsi})
		\end{pmatrix}\left\{\left(-\bbG_{\bu}^{-1}\sqrt{n}\frac{1}{N}\sum_{i=1}^{N}\delta_{i}\pi_{i}^{-1}\bpsi_{i}\right)\otimes\left(-\bbG_{\bu}^{-1}\sqrt{n}\frac{1}{N}\sum_{i=1}^{N}\delta_{i}\pi_{i}^{-1}\bpsi_{i}\right)\right\}+o_{P}(1),
	\end{split}
\end{equation*}
\begin{equation*}
	\begin{split}
		&\min\{n,\sqrt{N}\}(\tA_{\cM}-\bbA_{\cM})
		\begin{pmatrix}
			-	\frac{1}{N}\sum_{i=1}^{N}\delta_{i}\pi_{i}^{-1}\bpsi_{i}\bpsi_{i}^{\T}+E[\bpsi\bpsi^{\T}]\\
			\frac{1}{N}\sum_{i=1}^{N}\delta_{i}\pi_{i}^{-1}\bpsi_{i}\bpsi_{i}^{\T} - E[\bpsi\bpsi^{\T}]
		\end{pmatrix}\left(\tilde{\btheta}-\btheta_{0}\right)\\
		& = c_{2}\bR_{N}(\bbA_{\cM})\begin{pmatrix}
			-	\sqrt{n}\left\{\frac{1}{N}\sum_{i=1}^{N}\delta_{i}\pi_{i}^{-1}\bpsi_{i}\bpsi_{i}^{\T}-E\left(\bpsi\bpsi^{\T}\right)\right\}\\
			\sqrt{n}\left\{\frac{1}{N}\sum_{i=1}^{N}\delta_{i}\pi_{i}^{-1}\bpsi_{i}\bpsi_{i}^{\T} - E\left(\bpsi\bpsi^{\T}\right)\right\}
		\end{pmatrix}\left(-\bbG_{\bu}^{-1}\sqrt{n}\frac{1}{N}\sum_{i=1}^{N}\delta_{i}\pi_{i}^{-1}\bpsi_{i}\right)+o_{P}(1),
	\end{split}
\end{equation*}
\begin{equation*}
	\begin{split}
		&\min\{n,\sqrt{N}\}(\tA_{\cM}-\bbA_{\cM})
		\begin{pmatrix}
			\frac{1}{2N}\sum_{i=1}^{N}\delta_{i}\pi_{i}^{-1}\ddot{\bpsi}_{i}\\
			-\frac{1}{2N}\sum_{i=1}^{N}\ddot{\bpsi}_{i}
		\end{pmatrix}\left\{\left(\tilde{\btheta}-\btheta_{0}\right)\otimes\left(\tilde{\btheta}-\btheta_{0}\right)\right\}\\
		&=c_{2}\bR_{N}(\bbA_{\cM})
		\begin{pmatrix}
			-E(\dot{\bpsi\bpsi^{\T}})\\
			E(\dot{\bpsi\bpsi^{\T}})
		\end{pmatrix}\left\{\left(-\bbG_{\bu}^{-1}\sqrt{n}\frac{1}{N}\sum_{i=1}^{N}\delta_{i}\pi_{i}^{-1}\bpsi_{i}\right)\otimes\left(-\bbG_{\bu}^{-1}\sqrt{n}\frac{1}{N}\sum_{i=1}^{N}\delta_{i}\pi_{i}^{-1}\bpsi_{i}\right)\right\}\\
		&\quad+o_{P}(1),
	\end{split}	
\end{equation*}
where 
$$\bR_{N}(\bbA_{\cM}) = \bR_{N}((\bbG_{\cM}^{\T}\bbOmega_{\cM}^{-1}\bbG_{\cM})^{-1})\{\bbG_{\cM}^{\T}\bbOmega_{\cM}^{-1}+\bR_{N}(\bbG_{\cM}^{\T}\bbOmega_{\cM}^{-1})\}+(\bbG_{\cM}^{\T}\bbOmega_{\cM}^{-1}\bbG_{\cM})^{-1}\bR_{N}(\bbG_{\cM}^{\T}\bbOmega_{\cM}^{-1}),$$
$$\bR_{N}((\bbG_{\cM}^{\T}\bbOmega_{\cM}^{-1}\bbG_{\cM})^{-1}) =(\bbG_{\cM}^{\T}\bbOmega_{\cM}^{-1}\bbG_{\cM})^{-1} \bR_{N}(\bbG_{\cM}^{\T}\bbOmega_{\cM}^{-1}\bbG_{\cM})(\bbG_{\cM}^{\T}\bbOmega_{\cM}^{-1}\bbG_{\cM})^{-1},$$
\begin{equation*}
	\begin{split}
		\bR_{N}(\bbG_{\cM}^{\T}\bbOmega_{\cM}^{-1}\bbG_{\cM})
		& =  \bR_{N}\left(\bbG_{\bu}\right)^{\T}\bbB_{\bu\bu}\bbG_{\bu}+\bbG_{\bu}^{\T}\bR_{N}\left(\bbB_{\bu\bu}\right)\bbG_{\bu}+\bbG_{\bu}^{\T}\bbB_{\bu\bu}\bR\left(\bbG_{\bu}\right)\\		
		&\quad+\bR_{N}\left(\bbG_{\bbm}\right)^{\T}\bbB_{\bbm\bu}\bbG_{\bu}+\bbG_{\bbm}^{\T}\bR_{N}\left(\bbB_{\bbm\bu}\right)\bbG_{\bu}+\bbG_{\bbm}^{\T}\bbB_{\bbm\bu}\bR_{N}\left(\bbG_{\bu}\right)\\
		&\quad+\bR_{N}\left(\bbG_{\bu}\right)^{\T}\bbB_{\bu\bbm}\bbG_{\bbm}+\bbG_{\bu}^{\T}\bR_{N}\left(\bbB_{\bu\bbm}\right)\bbG_{\bbm}+\bbG_{\bu}^{\T}\bbB_{\bu\bbm}\bR_{N}\left(\bbG_{\bbm}\right)\\
		&\quad+\bR_{N}\left(\bbG_{\bbm}\right)^{\T}\bbB_{\bbm\bbm}\bbG_{\bbm}+\bbG_{\bbm}^{\T}\bR_{N}\left(\bbB_{\bbm\bbm}\right)\bbG_{\bbm}+\bbG_{\bbm}^{\T}\bbB_{\bbm\bbm}\bR_{N}(\bbG_{\bbm}),
	\end{split}
\end{equation*}
\begin{equation*}
	\begin{split}
		&\bR_{N}(\bbB_{\bbm\bbm})\\ &=-\left[\left(\bbOmega_{\bbm\bbm}-\bbOmega_{\bbm\bu}\bbOmega_{\bu\bu}^{-1}\bbOmega_{\bbm\bu}\right)^{-1}\left\{\bR_{N}(\bbOmega_{\bbm\bbm})-\bR_{N}(\bbOmega_{\bbm\bu})\bbOmega_{\bu\bu}^{-1}\bbOmega_{\bu\bbm}\right.\right.\\
		&\left.\left.\qquad-\bbOmega_{\bbm\bu}\bbOmega_{\bu\bu}^{-1}\bR_{N}(\bbOmega_{\bu\bu})\bbOmega_{\bu\bu}^{-1}\bbOmega_{\bu\bbm}-\bbOmega_{\bbm\bu}\bbOmega_{\bu\bu}^{-1}\bR_{N}(\bbOmega_{\bu\bbm})\right\}\left(\bbOmega_{\bbm\bbm}-\bbOmega_{\bbm\bu}\bbOmega_{\bu\bu}^{-1}\bbOmega_{\bbm\bu}\right)^{-1}\right], 
	\end{split}
\end{equation*}
\begin{equation*}
	\bR_{N}(\bbB_{\bbm\bu}) =\bR_{N}(\bbB_{\bbm\bbm})\bbOmega_{\bbm\bu}\bbOmega_{\bu\bu}^{-1}+\bbB_{\bbm\bbm}\bR_{N}(\bbOmega_{\bbm\bu})\bbOmega_{\bu\bu}^{-1}+\bbB_{\bbm\bbm}\bbOmega_{\bbm\bu}\bbOmega_{\bu\bu}^{-1}\bR_{N}(\bbOmega_{\bu\bu})\bbOmega_{\bu\bu}^{-1},
\end{equation*}
\begin{equation*}
	\begin{split}
		\bR_{N}(\bbB_{\bu\bu}) &=\bbOmega_{\bu\bu}^{-1}\bR_{N}(\bbOmega_{\bu\bu})\bbOmega_{\bu\bu}^{-1}+\bbOmega_{\bu\bu}^{-1}\bR_{N}(\bbOmega_{\bu\bu})\bbOmega_{\bu\bu}^{-1}\bbOmega_{\bu\bbm}\bbB_{\bbm\bu}+\bbOmega_{\bu\bu}^{-1}\bR_{N}(\bbOmega_{\bu\bbm})\bbB_{\bbm\bu}\\
		&\quad+\bbOmega_{\bu\bu}^{-1}\bbOmega_{\bu\bbm}\bR_{N}(\bbB_{\bbm\bu}),
	\end{split}
\end{equation*}
\begin{equation*}
	\bR_{N}(\bbG_{\bu}) = n^{-1/2}\left[-\sqrt{n}\left(\frac{1}{N}\sum_{i=1}^{N}\frac{\delta_{i}}{\pi_{i}}\bpsi_{i}\bpsi_{i}^{\T}-E\left(\bpsi\bpsi^{\T}\right)\right)-E\left(\dot{(\bpsi\bpsi^{\T})}\right)\left(-\bbG_{\bu}^{-1}\sqrt{n}\frac{1}{N}\sum_{i=1}^{N}\delta_{i}\pi_{i}^{-1}\bpsi_{i}\right)\right],
\end{equation*}
\begin{equation*}
	\bR_{N}(\bbG_{\bbm}) =-\bR_{N}(\bbG_{\bu}),
\end{equation*}
\begin{equation*}
	\begin{split}
		\bR_{N}(\bbOmega_{\bu\bu}) 
		& =n^{-1/2}\left[\sqrt{n}\left\{\frac{1}{N}\sum_{i=1}^{N}\frac{\delta_{i}}{\pi_{i}}\rho_{N}\pi_{i}^{-1}
		\bpsi_{i}\bpsi_{i}^{\T}-E\left\{\rho_{N}
		\pi^{-1}
		\bpsi\bpsi^{\T}\right\}\right\}\right.\\
		&\left.\quad+E\left\{\rho_{N}\pi^{-1}\dot{(\bpsi\bpsi^{\T})}\right\}\left(-\bbG_{\bu}^{-1}\sqrt{n}\frac{1}{N}\sum_{i=1}^{N}\delta_{i}\pi_{i}^{-1}\bpsi_{i}\right)\right],
	\end{split}
\end{equation*}
\begin{align*}
	&\bR_{N}(\bbOmega_{\bu\bbm}) \\
	& =-\rho_{N}\left(n^{-1/2}\left\{\sqrt{n}\left(\frac{1}{N}\sum_{i=1}^{N}\frac{\delta_{i}}{\pi_{i}}\bpsi_{i}\bpsi_{i}^{\T}-E\left(\bpsi\bpsi^{\T}\right)\right)+E\left(\dot{(\bpsi\bpsi^{\T})}\right)\left(-\bbG_{\bu}^{-1}\sqrt{n}\frac{1}{N}\sum_{i=1}^{N}\delta_{i}\pi_{i}^{-1}\bpsi_{i}\right)\right\}\right)\\
	&\quad-\rho_{N}E\left\{\pi^{-1}\bpsi\right\}\left\{\rho_{N}^{1/2}\sqrt{N}\frac{1}{N}\sum_{i=1}^{N}\bpsi_{i}+E\left(\dot{\bpsi}\right)\left(-\bbG_{\bu}^{-1}\sqrt{n}\frac{1}{N}\sum_{i=1}^{N}\delta_{i}\pi_{i}^{-1}\bpsi_{i}\right)\right\}^{\T},
\end{align*}
and
\begin{align*}
	\bR_{N}(\bbOmega_{\bbm\bbm}) 
	& =\rho_{N}n^{-1/2}\left[\sqrt{n}\left\{\frac{1}{N}\sum_{i=1}^{N}\frac{\delta_{i}}{\pi_{i}}\bpsi_{i}\bpsi_{i}^{\T}-E\left(\bpsi\bpsi^{\T}\right)\right\}+E\left(\dot{(\bpsi\bpsi^{\T})}\right)\left(-\bbG_{\bu}^{-1}\sqrt{n}\frac{1}{N}\sum_{i=1}^{N}\delta_{i}\pi_{i}^{-1}\bpsi_{i}\right)\right]\\
	&\quad+\rho_{N}n^{-1}E\left(\pi^{-1}-1\right)\left\{\rho_{N}^{1/2}\sqrt{N}\frac{1}{N}\sum_{i=1}^{N}\bpsi_{i}+E\left(\dot{\bpsi}\right)\left(-\bbG_{\bu}^{-1}\sqrt{n}\frac{1}{N}\sum_{i=1}^{N}\delta_{i}\pi_{i}^{-1}\bpsi_{i}\right)\right\}^{\otimes 2}.
\end{align*}
Then we have 
\begin{align*}
	&\min\{n,\sqrt{N}\}\left(\tilde{\btheta}-\btheta_{0}\right)\\
	&=- c_{1}\bbA_{\cM}
	\begin{pmatrix}
		\rho_{N}^{-1/2}\sqrt{n}\frac{1}{N}\sum_{i=1}^{N}\delta_{i}\pi_{i}^{-1}\bpsi_{i}\\
		-\sqrt{N}\frac{1}{N}\sum_{i=1}^{N}\bpsi_{i}
	\end{pmatrix} \\
	&- c_{2}\bbA_{\cM}
	\begin{pmatrix}
		\frac{1}{2N}\sum_{i=1}^{N}\delta_{i}\pi_{i}^{-1}\ddot{\bpsi}_{i}\\
		-\frac{1}{2N}\sum_{i=1}^{N}\ddot{\bpsi}_{i}
	\end{pmatrix}\left\{\left(-\bbG_{\bu}^{-1}\sqrt{n}\frac{1}{N}\sum_{i=1}^{N}\delta_{i}\pi_{i}^{-1}\bpsi_{i}\right)\otimes\left(-\bbG_{\bu}^{-1}\sqrt{n}\frac{1}{N}\sum_{i=1}^{N}\delta_{i}\pi_{i}^{-1}\bpsi_{i}\right)\right\} \\
	&\quad + c_{2}\bbA_{\cM}
	\begin{pmatrix}
		-	\sqrt{n}\left\{\frac{1}{N}\sum_{i=1}^{N}\delta_{i}\pi_{i}^{-1}\bpsi_{i}\bpsi_{i}^{\T}-E[\bpsi\bpsi^{\T}]\right\}\\
		\sqrt{n}\left\{\frac{1}{N}\sum_{i=1}^{N}\delta_{i}\pi_{i}^{-1}\bpsi_{i}\bpsi_{i}^{\T} - E[\bpsi\bpsi^{\T}]\right\}
	\end{pmatrix}\left(-\bbG_{\bu}^{-1}\sqrt{n}\frac{1}{N}\sum_{i=1}^{N}\delta_{i}\pi_{i}^{-1}\bpsi_{i}\right)  \\
	&\quad + c_{2}\bbA_{\cM}
	\begin{pmatrix}
		-	E[\dot{(\bpsi\bpsi^{\T})}]\\
		E[\dot{(\bpsi\bpsi^{\T})}]
	\end{pmatrix}\left\{\left(-\bbG_{\bu}^{-1}\sqrt{n}\frac{1}{N}\sum_{i=1}^{N}\delta_{i}\pi_{i}^{-1}\bpsi_{i}\right)\otimes\left(-\bbG_{\bu}^{-1}\sqrt{n}\frac{1}{N}\sum_{i=1}^{N}\delta_{i}\pi_{i}^{-1}\bpsi_{i}\right)\right\}\\
	&\quad- c_{1}\bR_{N}(\bbA_{\cM})
	\begin{pmatrix}
		\rho_{N}^{-1/2}\sqrt{n}\frac{1}{N}\sum_{i=1}^{N}\delta_{i}\pi_{i}^{-1}\bpsi_{i}\\
		-\sqrt{N}\frac{1}{N}\sum_{i=1}^{N}\bpsi_{i}
	\end{pmatrix}\\
	&\quad- c_{2}\bR_{N}(\bbA_{\cM})
	\begin{pmatrix}
		\frac{1}{2}E(\ddot{\bpsi})\\
		-	\frac{1}{2}E(\ddot{\bpsi})
	\end{pmatrix}\left\{\left(-\bbG_{\bu}^{-1}\sqrt{n}\frac{1}{N}\sum_{i=1}^{N}\delta_{i}\pi_{i}^{-1}\bpsi_{i}\right)\otimes\left(-\bbG_{\bu}^{-1}\sqrt{n}\frac{1}{N}\sum_{i=1}^{N}\delta_{i}\pi_{i}^{-1}\bpsi_{i}\right)\right\}\\
	&\quad+c_{2}\bR_{N}(\bbA_{\cM})\begin{pmatrix}
		-	\sqrt{n}\left\{\frac{1}{N}\sum_{i=1}^{N}\delta_{i}\pi_{i}^{-1}\bpsi_{i}\bpsi_{i}^{\T}-E[\bpsi\bpsi^{\T}]\right\}\\
		\sqrt{n}\left\{\frac{1}{N}\sum_{i=1}^{N}\delta_{i}\pi_{i}^{-1}\bpsi_{i}\bpsi_{i}^{\T} - E[\bpsi\bpsi^{\T}]\right\}
	\end{pmatrix}\left(-\bbG_{\bu}^{-1}\sqrt{n}\frac{1}{N}\sum_{i=1}^{N}\delta_{i}\pi_{i}^{-1}\bpsi_{i}\right)\\
	&\quad+c_{2}\bR_{N}(\bbA_{\cM})
	\begin{pmatrix}
		-E(\dot{\bpsi\bpsi^{\T}})\\
		E(\dot{\bpsi\bpsi^{\T}})
	\end{pmatrix}\left\{\left(-\bbG_{\bu}^{-1}\sqrt{n}\frac{1}{N}\sum_{i=1}^{N}\delta_{i}\pi_{i}^{-1}\bpsi_{i}\right)\otimes\left(-\bbG_{\bu}^{-1}\sqrt{n}\frac{1}{N}\sum_{i=1}^{N}\delta_{i}\pi_{i}^{-1}\bpsi_{i}\right)\right\}\\
	&\quad +o_{P}(1).
\end{align*}
Let
\begin{align*}
	\bU_{\cM,N} = \frac{1}{N}\sum_{i=1}^{N}
	\begin{pmatrix}
		\sqrt{N}\bpsi_{i}\\
		\sqrt{n}\delta_{i}\pi_{i}^{-1}\bpsi_{i}\\
		\sqrt{n}\left(\delta_{i}\pi_{i}^{-1}\operatorname{vec}(\operatorname{upp}(\bpsi_{i}\bpsi_{i}^{\T}))-\operatorname{vec}(\operatorname{upp}(E[\bpsi\bpsi^{\T}]))\right)\\
		\sqrt{n}\left(\rho_{N}\delta_{i}\pi_{i}^{-2}\operatorname{vec}(\operatorname{upp}(\bpsi_{i}\bpsi_{i}^{\T}))-\operatorname{vec}(\operatorname{upp}(E[\rho_{N}\pi^{-1}\bpsi\bpsi^{\T}]))\right)
	\end{pmatrix}.
\end{align*}
Let $\bU_{\cM,N,1}$ be the first $d$ components of $\bU_{\cM,N}$, $\bU_{\cM,N,2}$ be the $d+1$-th to the $2d$-th elements of $\bU_{\cM,N}$, $\bbU_{\cM,N,3}$ be the $2d+1$-th to the $3d+(d^2-d)/2$-th elements of $\bU_{\cM,N}$, $\bbU_{\cM,N,4}$ be the $3d+(d^2-d)/2+1$-th to the $3d+d^2$-th elements of $\bU_{\cM,N}$, $\bbU_{\cM,N,C,1}$ be a $d\times d$ symmetric matrix with the upper triangle matrix consisted by the elements in $\bU_{\cM,N,3}$ arranged in rows, $\bbU_{\cM,N,C,2}$ be a $d\times d$ symmetric matrix with the upper triangle matrix consisted by the elements in $\bU_{\cM,4}$ arranged in rows.
Let $\br_{N}(\cdot)$ be a function of $\bU_{\cM,N}$ defined by replacing 
$$\sqrt{n}\frac{1}{N}\sum_{i=1}^{N}\delta_{i}\pi_{i}^{-1}\bpsi_{i},\sqrt{N}\frac{1}{N}\sum_{i=1}^{N}\bpsi_{i},$$
$$\sqrt{n}\left(\frac{1}{N}\sum_{i=1}^{N}\frac{\delta_{i}}{\pi_{i}}\bpsi_{i}\bpsi_{i}^{\T}-E\left(\bpsi\bpsi^{\T}\right)\right),$$
$$\sqrt{n}\left\{\frac{1}{N}\sum_{i=1}^{N}\frac{\delta_{i}}{\pi_{i}}\rho_{N}\pi_{i}^{-1}
\bpsi_{i}\bpsi_{i}^{\T}-E\left\{\rho_{N}
\pi^{-1}
\bpsi\bpsi^{\T}\right\}\right\},$$
in
$\bR_{N}(\bbG_{\bu})$, $\bR_{N}(\bbOmega_{\bu\bu}) $, $\bR_{N}(\bbOmega_{\bu\bbm}) $ and $\bR_{N}(\bbOmega_{\bbm\bbm})$
with $\bU_{\cM,N,2}$, $\bU_{\cM,N,1}$, $\bU_{\cM,N,C,1}$ and $\bU_{\cM,N,C,2}$, respectively. Let
\begin{align*}
	&\bl_{\cM,N}(\bU_{\cM,N}) \\
	&= -c_{1}\bbA_{\cM}
	\begin{pmatrix}
		\rho_{N}^{-1/2}\bU_{\cM,N,2}\\
		-\bU_{\cM,N,1}
	\end{pmatrix}
	-c_{2}\bbA_{\cM}
	\begin{pmatrix}
		E(\ddot{\bpsi})/2\\
		-E(\ddot{\bpsi})/2
	\end{pmatrix}\left\{(E(\dot{\bpsi})^{-1}\bU_{\cM,N,2})\otimes(E(\dot{\bpsi})^{-1}\bU_{\cM,N,2})\right\}\\
	&\quad
	-c_{2}\bbA_{\cM}
	\begin{pmatrix}
		-\bU_{\cM,N,C,1}\\
		\bU_{\cM,N,C,1}
	\end{pmatrix}
	E(\dot{\bpsi})^{-1}\bU_{\cM,N,2}\\
	&\quad+c_{2}\bbA_{\cM}
	\begin{pmatrix}
		-E[\dot{\bpsi\bpsi^{\T}}]\\
		E[\dot{\bpsi\bpsi^{\T}}]
	\end{pmatrix}\left\{(E(\dot{\bpsi})^{-1}\bU_{N,2})\otimes(E(\dot{\bpsi})^{-1}\bU_{N,2})\right\}\\
	&\quad- c_{1}\br_{N}(\bU_{\cM,N})
	\begin{pmatrix}
		\frac{1}{N}\sum_{i=1}^{N}\delta_{i}\pi_{i}^{-1}\bpsi_{i}\\
		-\frac{1}{N}\sum_{i=1}^{N}\bpsi_{i}
	\end{pmatrix}\\
	&\quad- c_{2}\br_{N}(\bU_{\cM,N})
	\begin{pmatrix}
		\frac{1}{2}E(\ddot{\bpsi})\\
		-	\frac{1}{2}E(\ddot{\bpsi})
	\end{pmatrix}\left\{\left(-\bbG_{\bu}^{-1}\bU_{\cM,N,2}\right)\otimes\left(-\bbG_{\bu}^{-1}\bU_{\cM,N,2}\right)\right\}\\
	&\quad+c_{2}\br_{N}(\bU_{\cM,N})\begin{pmatrix}
		-\bU_{\cM,N,C,1}\\
		\bU_{\cM,N,C,1}
	\end{pmatrix}\left(-\bbG_{\bu}^{-1}\bU_{\cM,N,2}\right)\\
	&\quad+c_{2}\br_{N}(\bU_{\cM,N})
	\begin{pmatrix}
		-E(\dot{\bpsi\bpsi^{\T}})\\
		E(\dot{\bpsi\bpsi^{\T}})
	\end{pmatrix}\left\{\left(-\bbG_{\bu}^{-1}\bU_{\cM,N,2}\right)\otimes\left(-\bbG_{\bu}^{-1}\bU_{\cM,N,2}\right)\right\}.
\end{align*}
Then  $\min\{n,\sqrt{N}\}\left(\tilde{\btheta}-\btheta_{0}\right)=\bl_{\cM,N}(\bU_{N})+o_{P}(1)$.
Next, we prove 
\begin{equation}\label{eq: continous mapping dist M}
	\sup_{t}\big|P(\bl_{\cM,N}(\bU_{\cM,N})<t)-P(\bl_{\cM,N}(\bbV_{\cM,N}^{1/2}\bZ_{\cM})<t)\big|\to 0,
\end{equation} 
where $\bZ_{\cM}$ is a $3d+d^2$-dimensional standard normal random vector.
Consider the eigenvalue decomposition $\bbV_{\cM,N} = \bbQ_{\cM,N}\bbLambda_{\cM,N}\bbQ_{\cM,N}^{\T}$, where $\bbQ_{\cM,N}$ is a $(3d+(d^2-d)/2)\times r$  matrix of eigenvectors and $\bbLambda_{\cM,N}$ is a $r\times r$ diagonal matrix containing the positive eigenvalues of $\bbV_{\cM,N}$, with $r=\operatorname{rank}(\bbV_{\cM,N})$. 
Since $\bbQ_{\cM,N}\bbLambda_{\cM,N}^{1/2}\bZ_{\cM}$ shares the same distribution as $\bbV_{\cM,N}^{1/2}\bZ_{\cM}$, therefore, proving \eqref{eq: continous mapping dist M} is equivalent to proving 
\begin{equation}\label{eq: continous mapping dist M 1}
	\sup_{t}\big|P(\bl_{\cM,N}(\bU_{\cM,N})<t)-P(\bl_{\cM,N}(\bbQ_{\cM,N}\bbLambda_{\cM,N}^{1/2}\bZ_{\cM})<t)\big|\to 0.
\end{equation} 
Using the same arguments as those used in the proof of Theorem \ref{theo: non-normal S}, we can show that $\bbLambda_{\cM,N}^{-1/2}\bbQ_{\cM,N}^{\T}\bU_{\cM,N}\stackrel{d}{\to}N(\bzero,\bbI)$.
Next, we prove \eqref{eq: continous mapping dist M 1} which equals to prove 
\begin{equation*}
	\sup_{t}\big|P(\bl_{\cM,N}^{\dagger}(\bU_{\cM,N}^{\dagger})<t)-P(\bl_{\cM,N}^{\dagger}(\bZ)<t)\big|\to 0,
\end{equation*}
where $\bl_{\cM,N}^{\dagger}(\cdot) = \bl_{\cM,N}(\bbQ_{\cM,N}\bbLambda_{\cM,N}^{1/2}\cdot)$, $\bU_{\cM,N}^{\dagger} = \bbLambda_{\cM,N}^{-1/2}\bbQ_{\cM,N}^{\T}\bU_{\cM,N}$ and $\bZ_{\cM}$ is a $3d+(d^2-d)/2$-dimensional standard normal random vector.
The proof processes are similar to those of Theorem \ref{theo: non-normal S}. Here we only need to verify that $\bl_{\cM,N}^{\dagger}(\cdot)$ is bounded on any tight set. That $\bl_{\cM,N}^{\dagger}(\cdot)$ is bounded on any tight set can be concluded from that $\|(\bbG_{\cM}^{\T}\bbOmega_{\cM}^{-1}\bbG_{\cM})^{-1}\| = O(\rho_{N})$, $\|\bbB_{\bu\bu}\|=O(1)$, $\|\bbB_{\bu\bbm}\|=O(1)$, $\|\bbB_{\bbm\bbm}\|=O(\rho_{N}^{-1})$, and $\|\bbA_{\cM}\|=O(1)$.


\section{Lemmas}\label{app: lems}


\begin{lemma}\label{lem: AL of initial}
	Under Conditions (C1)--(C3), (C4)(i) and (C6), we have
	\begin{equation}\label{eq: AL}
		\tilde{\btheta}-\btheta_{0} = -\bbG_{\bu}^{-1}\frac{1}{n}\sum_{i=1}^{N}\delta_{i}\bu_{i} + O_{P}(n^{-1}).
	\end{equation}
\end{lemma}
\begin{proof}
	We first prove $\tilde{\btheta}-\btheta_{0}=o_{P}(1)$, and then prove the result \eqref{eq: AL}.	
	Note that $\tilde{\btheta}$ and $\btheta_{0}$ are solutions to $$n^{-1}\sum_{i=1}^{N}\delta_{i}\bu_{i}(\btheta)=\bzero$$ and $$E\left\{\rho_{N}^{-1}\pi\bu(\bX,\bY;\btheta)\right\}=\bzero,$$ respectively.
	By Theorem 5.7 in \cite{Vaart2000AS} and Condition (C1), it is sufficient to prove that 
	\begin{equation}\label{eq: uniform 1}
		\begin{aligned}
			&\sup_{\btheta}\Bigg|\left\|\frac{1}{n}\sum_{i=1}^{N}\delta_{i}\bu_{i}(\btheta)\right\|^2-\left\|\rho_{N}^{-1}E[\pi\bu(\bX,\bY;\btheta)]\right\|^2\Bigg|=o_{P}(1).
		\end{aligned}
	\end{equation}
	Note that
	\begin{equation*}
		\begin{aligned}
			&\left\{\frac{1}{n}\sum_{i=1}^{N}\delta_{i}\bu_{i}(\btheta)^{\T}\right\}\left\{\frac{1}{n}\sum_{i=1}^{N}\delta_{i}\bu_{i}(\btheta)\right\}\\
			=&\frac{1}{n^2}\sum_{i=1}^{N}\delta_{i}\bu_{i}(\btheta)^{\T}\bu_{i}(\btheta)+\frac{1}{n^2}\sum_{i\neq j}\delta_{i}\delta_{j}\bu_{i}(\btheta)^{\T}\bu(\bX_{j},\bY_{j};\btheta).
		\end{aligned}
	\end{equation*}
	Therefore, to prove \eqref{eq: uniform 1}, it suffices to prove
	\begin{equation}\label{eq: uniform 1-1}
		\begin{aligned}
			&\sup_{\btheta}\Big|\frac{1}{n^2}\sum_{i=1}^{N}\delta_{i}\bu_{i}(\btheta)^{\T}\bu_{i}(\btheta)\Big|=o_{P}(1)
		\end{aligned}
	\end{equation}
	and
	\begin{equation}\label{eq: uniform 1-2}
		\begin{aligned}
			&\sup_{\btheta}\Big|\frac{1}{n^2}\sum_{i\neq j}\delta_{i}\delta_{j}\bu_{i}(\btheta)^{\T}\bu(\bX_{j},\bY_{j};\btheta)\\
			&\phantom{\sup_{\btheta}}-\rho_{N}^{-2}E\{\pi\bu(\bX,\bY;\btheta)^{\T}\}E\{\pi\bu(\bX,\bY;\btheta)\}\Big|=o_{P}(1).
		\end{aligned}
	\end{equation}
	First, \eqref{eq: uniform 1-1} can be proved by Markov inequality since 
	\begin{equation*}\label{eq: Pn1}
		\begin{aligned}
			&E\left\{\sup_{\btheta}\Big|\frac{1}{n^2}\sum_{i=1}^{N}\delta_{i}\bu_{i}(\btheta)^{\T}\bu_{i}(\btheta)\Big|\right\}\\
			=&E\left\{\frac{1}{n^2}\sum_{i=1}^{N}\delta_{i}\sup_{\btheta}\|\bu_{i}(\btheta)\|^2\right\}\\
			= &E\left\{\frac{N}{n^{2}}\pi\sup_{\btheta} \|\bu(\bX,\bY;\btheta)\|^2\right\}\\
			=&O(n^{-1})
		\end{aligned}
	\end{equation*}
	by Conditions (C3) and (C6). Next, we prove \eqref{eq: uniform 1-2}.
	For any $\btheta^{\prime}$ in $\Theta$, let
	\begin{equation}
		\begin{split}
			\Delta_{r}(\bX_{1},\bY_{1},\bX_{2},\bY_{2};\btheta^{\prime}) 
			&= \sup_{\btheta\in B(\btheta^{\prime},r)}\bu(\bX_{1},\bY_{1};\btheta)^{\T}\bu(\bX_{2},\bY_{2};\btheta)\\
			&\phantom{\sup_{\btheta\in B(\btheta^{\prime},r)}}-\inf_{\btheta\in B(\btheta^{\prime},r)}\bu(\bX_{1},\bY_{1};\btheta)^{\T}\bu(\bX_{2},\bY_{2};\btheta).
		\end{split}
	\end{equation}
	By Condition (C2)(i), we have $\Delta_{r}(\bX_{1},\bY_{1},\bX_{2},\bY_{2};\btheta^{\prime})\to 0$ a.s. as $r\to 0$. In addition, note that $\Delta_{r}(\bX_{1},\bY_{1},\bX_{2},\bY_{2};\btheta^{\prime})\leq 2\sup_{\btheta}|\bu(\bX_{1},\bY_{1};\btheta)^{\T}\bu(\bX_{2},\bY_{2};\btheta)|$. By Condition (C3) and dominated convergence theorem, we have $E\{\Delta_{r}(\bX_{1},\bY_{1},\bX_{2},\bY_{2};\btheta^{\prime})\}\to 0$ as $r\to 0$.
	Thus for any $\btheta$ and $\epsilon>0$, there exists a $r_{\epsilon(\btheta)}$ such that $E\{\Delta_{r_{\epsilon(\btheta)}}(\bX_{1},\bY_{1},\bX_{2},\bY_{2};\btheta)\}<\epsilon$. 
	Obviously, $\Theta$ can be covered by $\{B(\btheta,r_{\epsilon(\btheta)}):\btheta\in \Theta\}$.
	Since $\Theta$ is compact, this implies that $\Theta$ can be covered by a finite sub-cover $\cup_{k=1}^{K} B(\btheta_{k},r_{\epsilon(\btheta_{k})})$. 
	Note that
	\begin{align*}
		&\sup_{\btheta\in \Theta}\Big[\frac{1}{n^2}\sum_{i\neq j}\delta_{i}\delta_{j}\bu_{i}(\btheta)^{\T}\bu(\bX_{j},\bY_{j};\btheta)\\
		&\phantom{\sup_{\btheta\in \Theta}}-\rho_{N}^{-2}E\{\pi\bu(\bX,\bY;\btheta)^{\T}\}E\{\pi\bu(\bX,\bY;\btheta)\}\Big]\\
		=&\max_{k}\sup_{\btheta\in B(\btheta_{k},r_{\epsilon(\btheta_{k})})}\left[\frac{1}{n^2}\sum_{i\neq j}\delta_{i}\delta_{j}\bu_{i}(\btheta)^{\T}\bu(\bX_{j},\bY_{j};\btheta)\right.\\
		&\left.\phantom{\sup_{\btheta\in \Theta}}-\rho_{N}^{-2}E\{\pi\bu(\bX,\bY;\btheta)^{\T}\}E\{\pi\bu(\bX,\bY;\btheta)\}\right]\\
		\leq & \max_{k}\left[\frac{1}{n^2}\sum_{i\neq j}\delta_{i}\delta_{j}\sup_{\btheta\in B(\btheta_{k},r_{\epsilon(\btheta_{k})})}\bu_{i}(\btheta)^{\T}\bu(\bX_{j},\bY_{j};\btheta)\right.\\
		&\left.\phantom{\max_{k}}-E\left\{\pi(\bX_{1},\bY_{1})\pi(\bX_{2},\bY_{2})\sup_{\btheta\in B(\btheta_{k},r_{\epsilon(\btheta_{k})})}\bu(\bX_{1},\bY_{1};\btheta)^{\T}\bu(\bX_{2},\bY_{2};\btheta)\right\}\right]\\
		&+\max_{k}\rho_{N}^{-2}E\left[\pi(\bX_{1},\bY_{1})\pi(\bX_{2},\bY_{2})\Big\{\sup_{\btheta\in B(\btheta_{k},r_{\epsilon(\btheta_{k})})}\bu(\bX_{1},\bY_{1};\btheta)^{\T}\bu(\bX_{2},\bY_{2};\btheta)\right.\\
		&\left.\phantom{\max_{k}\rho_{N}^{-2}\pi(\bX_{1},\bY_{1})\pi(\bX_{2},\bY_{2})\bu(\bX_{1},\bY_{1};\btheta)}-\inf_{\btheta\in B(\btheta_{k},r_{\epsilon(\btheta_{k})})}\bu(\bX_{1},\bY_{1};\btheta)^{\T}\bu(\bX_{2},\bY_{2};\btheta)\Big\}\right]\\
		= & U_{1}+U_{2}.
	\end{align*}
	For any $k$,  by standard calculations for the variance of U-statistics, we have
	\begin{equation*}
		\begin{aligned}
			&\var\left\{\frac{1}{n^{2}}\sum_{i\neq j}\delta_{i}\delta_{j}\sup_{\btheta\in B(\btheta_{k},r_{\epsilon(\btheta_{k})})}\bu_{i}(\btheta)^{\T}\bu(\bX_{j},\bY_{j};\btheta)\right\}=O\left(n^{-1}\right)
		\end{aligned}
	\end{equation*}
	by Conditions (C3) and (C6).
	By Chebyshev inequality, we have $U_{1}=o_{P}(1)$.
	In addition, according to the previous discussion, we have $U_{3}<\epsilon$.
	This proves 
	\begin{equation}\label{eq: consis sup}
		\begin{aligned}
			&\sup_{\btheta\in \Theta}\left[\frac{1}{n^2}\sum_{i\neq j}\delta_{i}\delta_{j}\bu_{i}(\btheta)^{\T}\bu(\bX_{j},\bY_{j};\btheta)\right.\\
			&\left.\phantom{\sup_{\btheta\in \Theta}}-\rho_{N}^{-2}E\{\pi\bu(\bX,\bY;\btheta)^{\T}\}E\{\pi\bu(\bX,\bY;\btheta)\}\right]\\
			\leq & o_{P}(1)+\epsilon.
		\end{aligned}
	\end{equation}
	Similarly, it can be proved 
	\begin{equation}\label{eq: consis inf}
		\begin{aligned}
			&\inf_{\btheta\in \Theta}\left[\frac{1}{n^2}\sum_{i\neq j}\delta_{i}\delta_{j}\bu_{i}(\btheta)^{\T}\bu(\bX_{j},\bY_{j};\btheta)\right.\\
			&\left.\phantom{\sup_{\btheta\in \Theta}}-\rho_{N}^{-2}E\{\pi\bu(\bX,\bY;\btheta)^{\T}\}E\{\pi\bu(\bX,\bY;\btheta)\}\right] \\
			\geq & o_{P}(1)-\epsilon.
		\end{aligned}
	\end{equation}
	Then \eqref{eq: consis sup} and \eqref{eq: consis inf} together prove \eqref{eq: uniform 1-2}. 
	
	Next, we prove 
	\begin{equation}\label{eq: AL 2}
		\tilde{\btheta}-\btheta_{0} = -\bbG_{\bu}^{-1}n^{-1}\sum_{i=1}^{N}\delta_{i}\bu_{i} + O_{P}(n^{-1}).
	\end{equation}
	Based on the result in step 1 and the following Taylor's expansion, we have
	\begin{equation}\label{eq: decomp}
		\begin{split}
			0
			&=\frac{1}{n}\sum_{i=1}^{N}\delta_{i}\bu(\bX_{i},\bY_{i};\tilde{\btheta})\\
			&=\frac{1}{n}\sum_{i=1}^{N}\delta_{i}\bu_{i}+\frac{1}{n}\sum_{i=1}^{N}\delta_{i}\dot{\bu}(\bX_{i},\bY_{i};\bar{\btheta})\left(\tilde{\btheta}-\btheta_{0}\right).
		\end{split}
	\end{equation}
	By Condition (C2), we have
	\begin{equation}\label{eq: inv}
		\begin{split}
			&\big\|\frac{1}{n}\sum_{i=1}^{N}\delta_{i}\dot{\bu}(\bX_{i},\bY_{i};\bar{\btheta})-\frac{1}{n}\sum_{i=1}^{N}\delta_{i}\dot{\bu}(\bX_{i},\bY_{i};\btheta_{0})\big\|\\
			&\leq\frac{1}{n}\sum_{i=1}^{N}\delta_{i}L_{1}(\bX_{i},\bY_{i})\|\tilde{\btheta}-\btheta_{0}\|.
		\end{split}
	\end{equation}
	Note that $E\{\frac{1}{n}\sum_{i=1}^{N}\delta_{i}L_{1}(\bX_{i},\bY_{i})\}=E\{\rho_{N} \pi L_{1}(\bX,\bY)\}\leq cE\{L_{1}(\bX\bY)^{2}\}<\infty$ by Conditions (C2) and (C6).
	Then by Markov's inequality, we have $\frac{1}{n}\sum_{i=1}^{N}\delta_{i}L_{1}(\bX_{i},\bY_{i})=O_{P}(1)$ which together with \eqref{eq: inv} proves
	\begin{equation}\label{eq: inv 1}
		\begin{aligned}
			&\big\|\frac{1}{n}\sum_{i=1}^{N}\delta_{i}\dot{\bu}(\bX_{i},\bY_{i};\bar{\btheta})-\frac{1}{n}\sum_{i=1}^{N}\delta_{i}\dot{\bu}(\bX_{i},\bY_{i};\btheta_{0})\big\|=O_{P}(\|\tilde{\btheta}-\btheta_{0}\|).
		\end{aligned}
	\end{equation}
	Further, note that for the $j$th row and $l$th column component of $\dot{\bu}(\bX_{i},\bY_{i};\btheta_{0})$, by Conditions (C3) and (C6), we have
	\begin{equation*}
		\var\left\{\frac{1}{n}\sum_{i=1}^{N}\delta_{i}\dot{\bu}^{(jl)}(\bX_{i},\bY_{i};\btheta_{0})\right\}
		\leq \frac{2N}{n^{2}}E\left\{\pi\dot{\bu}^{(jl)}(\bX,\bY;\btheta_{0})\right\}
		=O\left(n^{-1}\right).
	\end{equation*}
	Then we have 
	\begin{equation}\label{eq: inv 2}
		\begin{aligned}
			\frac{1}{n}\sum_{i=1}^{N}\delta_{i}\dot{\bu}(\bX_{i},\bY_{i};\bar{\btheta}) 
			&= \rho_{N}^{-1}E\left\{\pi\dot{\bu}(\bX,\bY;\btheta_{0})\right\}+O_{P}(n^{-1/2})\\
			&=\bbG_{\bu}+O_{P}(n^{-1/2}).
		\end{aligned}
	\end{equation}
	Equations \eqref{eq: inv 1} and \eqref{eq: inv 2} together prove
	\begin{equation}\label{eq: inv 3}
		\begin{aligned}
			\frac{1}{n}\sum_{i=1}^{N}\delta_{i}\dot{\bu}(\bX_{i},\bY_{i};\bar{\btheta}) 
			&= \rho_{N}^{-1}E\left\{\pi\dot{\bu}(\bX,\bY;\btheta_{0})\right\}+O_{P}(\|\tilde{\btheta}-\btheta_{0}\|)+O_{P}(n^{-1/2})\\
			&= \bbG_{\bu}+O_{P}(\|\tilde{\btheta}-\btheta_{0}\|)+O_{P}(n^{-1/2}).
		\end{aligned}
	\end{equation}
	This together with \eqref{eq: decomp} proves 
	\begin{equation}\label{eq: decomp 2}
		\begin{aligned}
			0&= 
			\frac{1}{n}\sum_{i=1}^{N}\delta_{i}\bu_{i}
			+\bbG_{\bu}\left(\tilde{\btheta}-\btheta_{0}\right)\\
			&\phantom{=}+O_{P}(\|\tilde{\btheta}-\btheta_{0}\|^2)+O_{P}(n^{-1/2}\|\tilde{\btheta}-\btheta_{0}\|).
		\end{aligned}
	\end{equation}
	In addition, by Chebyshev's inequality and Conditions (C3) and (C6), we can prove that $$n^{-1}\sum_{i=1}^{N}\delta_{i}\bu_{i}=O_{P}(n^{-1/2})$$ and this together with \eqref{eq: decomp 2} proves $\|\tilde{\btheta}-\btheta_{0}\|=O_{P}(n^{-1/2})$.
	By Condition (C4), then \eqref{eq: AL 2} is proved.
\end{proof}

\section{Variance Estimation}\label{app: var est}
\begin{proposition}
	Under Conditions (C1)--(C4) and (C6)--(C8), we have $\|\tV_{\cS,\bh}^{-1}\bbV_{\cS,\bh}-\bbI\|=o_{P}(1)$ and $\|\tV_{\cM,\bh}^{-1}\bbV_{\cM,\bh}-\bbI\|=o_{P}(1)$.
\end{proposition}
\begin{proof}
	The result can be directly obtained from Lemma \ref{lem: AL of initial} and Slustky theorem.
\end{proof}

\section{Comparisons with Newton's Method}\label{app: Newton discuss}
When the subsampling probability $\pi(\bx,\by) = n/N$ and the moment function $\tilde{\bh}(\bx,\by) = \bpsi(\bx,\by;\tilde{\btheta})$, the standard MAS estimator 
\[
\tilde{\btheta}_{\cS,\tilde{\bh}} = \tilde{\btheta}-(1-s)\tG_{\cS,\bu}^{-1}\hat{\bmu},
\] 
and the modified MAS estimator 
\[
\tilde{\btheta}_{\cM,\tilde{\bh}} = \tilde{\btheta}-\{1-s/(1+s)\}\tG_{\cM,\bu}^{-1}\hat{\bmu},
\]
where $s=\hat{\bmu}^{\T}(\tOmega_{\bu\bu}+\hat{\bmu}\hat{\bmu}^{\T})^{-1}\hat{\bmu}$ and $\hat{\bmu} =  N^{-1}\sum_{i=1}^{N}\nabla_{\btheta}\log f(\bY_{i}\mid \bX_{i};\btheta)\big|_{\btheta = \tilde{\btheta}}$ is the gradient of the whole data-based log-likelihood function. Notice that $\tG_{\cS,\bu}$, $\tG_{\cM,\bu}$, and the whole data-based Hessian matrix $\widehat{\bbH} = N^{-1}\sum_{i=1}^{N}\nabla_{\btheta}^{2}\log f(\bY_{i}\mid \bX_{i};\btheta)\big|_{\btheta = \tilde{\btheta}}$ all converge to the population hessian matrix $E\{\nabla_{\btheta}^{2}\log f(\bY\mid\bX;\btheta)\}\big|_{\btheta = \btheta_{0}}$. Hence, they are expected to be closed to each other. Then, the forms of $\tilde{\btheta}_{\cS,\tilde{\bh}}$ and $\tilde{\btheta}_{\cM,\tilde{\bh}}$ resemble the one-step update from $\tilde{\btheta}$ using Newton's method, which is given by $\tilde{\btheta} - \widehat{\bbH}^{-1}\hat{\bmu}$. Next, we explain the connections and differences between the proposed estimators and the Newton estimator $\tilde{\btheta} - \widehat{\bbH}^{-1}\hat{\bmu}$.		
The intuition behind $\tilde{\btheta}_{\cS,\tilde{\bh}}$ and $\tilde{\btheta}_{\cM,\tilde{\bh}}$ is similar, and we take $\tilde{\btheta}_{\cS,\tilde{\bh}}$ as an example to illustrate.
Newton's method approximates the whole data log-likelihood function by the second order Taylor expansion
\begin{equation}\label{eq: quad approx}
	\frac{1}{2}(\btheta-\tilde{\btheta})^{\T}\widehat{\bbH}(\btheta-\tilde{\btheta})+\hat{\bmu}^{\T}(\btheta-\tilde{\btheta}),
\end{equation}
and the Newton estimator $\tilde{\btheta} - \widehat{\bbH}^{-1}\hat{\bmu}$ is the minimum point of the quadratic form \eqref{eq: quad approx}.
Compared to the Newton estimator, the standard MAS estimator $\tilde{\btheta}_{\cS,\tilde{\bh}}$ uses $\tG_{\cS,\bu}$ in place of the whole data-based $\widehat{\bbH}$ and adopts a data-adaptive step size $1-s$. In this sense, $\tilde{\btheta}_{\cS,\tilde{\bh}}$ can be viewed as an approximate one-step Newton update with an adaptive step size. Notice that the quadratic approximation \eqref{eq: quad approx} and the one-step Newton update $\tilde{\btheta} - \widehat{\bbH}^{-1}\hat{\bmu}$ work well only when $\tilde{\btheta}$ is close to the whole data MLE $\hat{\btheta}$  \citep{nocedal1999numerical}; otherwise, the Newton update is not guaranteed to improve over $\tilde{\btheta}$. 
When $\tilde{\btheta}$ is close to $\hat{\btheta}$, $s$ is small according to its definition, and a large step is employed for the update in $\tilde{\btheta}_{\cS,\tilde{\bh}}$. When $\tilde{\btheta}$ is far from $\hat{\btheta}$, the Newton update may not work well. Then, $\tilde{\btheta}_{\cS,\tilde{\bh}}$ automatically adopts a small step size because $s$ can be large if $\tilde{\btheta}$ is far from $\hat{\btheta}$.
Thanks to the adaptive step size,
$\tilde{\btheta}_{\cS, \tilde{\bh}}$ can exhibit a better  performance compared to the Newton estimator especially when $n$ is small and $\tilde{\btheta}$ is not guaranteed to be close to $\hat{\btheta}$. This phenomenon is demonstrated through numerical results in Appendix \ref{app: compr with Newton}. Similar results hold for $\tilde{\btheta}_{\cM,\tilde{\bh}}$.

Moreover, $\tilde{\btheta}_{\cS,\tilde{\bh}}$ and $\tilde{\btheta}_{\cM,\tilde{\bh}}$ distinguish themselves from the Newton estimator by their better flexibility. The Newton estimator possesses good properties only when $\hat{\bmu}$ is the gradient of the whole data log-likelihood function, that is $\hat{\bmu} =  N^{-1}\sum_{i=1}^{N}\bpsi(\bX_{i},\bY_{i};\tilde{\btheta})$, which can be hard-to-compute under certain models.
In contrast, the standard and modified MAS estimators can accommodate general whole data sample moments with guaranteed efficiency gain. See Appendix \ref{app: compr with Newton} for a numerical demonstration of this advantage.

\section{Further Comparisons with Related Work}\label{app: further comparison}

Recently, \cite{fan2022nearly} proposes an empirical likelihood-based subsampling method that can incorporate the whole data sample moments to improve efficiency. Our work differs from \cite{fan2022nearly} in four key ways.

First, in order to optimize the use of moment information, this work focuses on the choice of moment function, while \cite{fan2022nearly} delve to the design of nonuniform subsampling probability and hence faces the same difficulties as other nonuniform subsampling probability-based methods. As shown in our numerical results, the choice of whole data sample moments is essential for the efficiency of the resulting estimator. We derive the optimal sample moments which is unstudied in \cite{fan2022nearly}. Our estimator can achieve the same efficiency as the whole data-based estimator when the optimal moment function is adopted as discussed at the end of Section \ref{sec:opth} in the main text. This desirable property is not achieved by \cite{fan2022nearly}.  Additionally, we propose several reasonable approaches to balance computational complexity with the statistical benefits of using the whole data sample moments. 

Second, our method takes the uncertainty of the whole data sample moment into consideration, which is overlooked in \cite{fan2022nearly}. The uncertainty brought by the whole data sample moments may cause a loss of efficiency if not accounted for properly \citep{zhang2020generalized}.

Third, the proposed estimator is potentially more efficient. The proposed method is developed under the conditional density model and the integration of whole data sample moments in this paper can make use of  the model information. Specifically, the standard MAS estimator is asymptotically equivalent to the estimator in \cite{fan2022nearly} when the uniform subsampling probability is used and sampling fraction $n/N$ tends to zero.
Thus, the asymptotic variance of our modified MAS estimator is no larger, and possibly smaller, than that of the estimator in \cite{fan2022nearly} when the subsampling ratio is less than $1/2$ (see Theorem \ref{theo: MAS compare} in the main text). 

Finally, \cite{fan2022nearly}'s method focuses on improving the weight of each observation in the subsample and hence can only improve the inverse probability weighted subsampling estimator. In contrast, our method improves efficiency by combining estimating functions and is able to improve any subsampling estimators that can be expressed as the solution of an estimating equation, including the inverse probability weighted subsampling estimator and the maximum sampled conditional likelihood subsampling estimator as special cases.

\section{Additional Numerical Results}\label{app: additional simul res}

\subsection{Approximation Details for the Optimal Moment Function}\label{app: approxi details for glmm}

Here we provide the approximation details for the optimal moment function under mixed effects logistic model.
Note that the log likelihood function $l(\btheta)$ of the mixed effects logistic model has the following form
\[
\sum_{i=1}^{N}\left[\left(\sum_{j=1}^{3}Y_{i}^{(j)}\bbeta^{\T}\bX_{i}^{(j)}\right)+\log\left\{\int\kappa(\bX_{i}, \bY_{i}, w; \bbeta)\phi(w,\sigma)dw\right\}\right],
\]
where $\kappa(\bX_{i}, \bY_{i}, w; \bbeta) = \exp(\sum_{j=1}^{3}[Y_{i}^{(j)}w-\log\{1+\exp(\bbeta^{\T}\bX_{i}^{(j)}+w)\}])$ and $\phi(w,\sigma)$ is the density function of the normal distribution with mean zero and variance $\sigma^{2}$.
The likelihood function and hence the corresponding score function involve integrals.
Therefore, we consider an computationally easy approximation of the optimal moment function.
Specifically, we consider a reduced working model, the logistic regression to construct a moment function which is informative for the estimation of $\bbeta$.
The first moment function we considered is the score function under the logistic regression
\[
\bq_{1}(\bx,\by; \bxi_{1}) = \sum_{j=1}^{3}\left\{y^{(j)}-\frac{\exp(\bxi_{1}^{\T}\bx^{(j)})}{1+\exp(\bxi^{\T}\bx^{(j)})}\right\}\bx^{(j)}.
\]
Note that the reduced model does not contain the information of $\sigma$.
We further consider the other moment function $h_{2}(\bx,\by)$ that contains the information of $\sigma$.
We consider a quadratic approximation of $\kappa(\bx, \by, w; \bzero)\approx \kappa^{(0)}(\bx, \by)+\kappa^{(1)}(\bx, \by)w+\kappa^{(2)}(\bx, \by)w^{2}/2$, where $\kappa^{(0)}(\bx, \by) = \kappa(\bx, \by, 0; \bzero)$, $\kappa^{(1)}(\bx, \by) = \partial \kappa(\bx, \by, w; \bzero)/\partial w\mid_{w=0}$, and $\kappa^{(2)}(\bx, \by) = \partial^{2}\kappa(\bx, \by, w; \bzero)/\partial w^{2}\mid_{w=0}$.
Then, we approximate the score function of $\sigma$ as 
\[
q_{2}(\bx,\by; \xi_{2}) = \frac{\kappa^{(2)}(\bx, \by)\xi_{2}}{\kappa^{(0)}(\bx, \by)+\kappa^{(2)}(\bx, \by)\xi_{2}^{2}/2}.
\]
Let $\bq(\bx, \by; \bxi) = (\bq_{1}(\bx, \by; \bxi_{1})^{\T}, q_{2}(\bx, \by; \xi_{2}))^{2}$ where $\bxi = (\bxi_{1}^{\T}, \xi_{2})^{\T}$.
We take the moment function $\tilde{\bh}^{\rm app}(\bx,\by)= \bq(\bx, \by; \tilde{\bxi})$ with $\tilde{\bxi}$ obtained in the way introduced at the end of Section \ref{subsec: approx hopt}.

\subsection{Numerical Comparisons with Newton's Method}\label{app: compr with Newton}

According to the theoretical results in Theorems 5 and 6  and the discussions in Appendix \ref{app: Newton discuss}, the MAS estimators with the optimal moment functions have several unique advantages compared to the one-step Newton estimator, denoted by NW.
On the one hand, the MAS estimator updates the initial estimator with an adaptive step size  which results in a better  performance compared to the Newton estimator when the subsample size is small.
On the other hand, the MAS estimators distinguish themselves from the one-step Newton estimator by their better flexibility since the MAS estimators can accommodate general whole data sample moments with guaranteed efficiency gain.
In this subsection, we first calculate the optimal moment function-based MAS estimators and compare it to the one-step Newton estimator under the logistic regression.
All estimators are calculated based on a uniform subsample.

Let $\bX = (X_{1},\dots,X_{9})$, where $X_{j}$ for $j=1,\dots,9$ are independently and identically distributed from $U(-1,1)$. 
Conditional on $\bX$, the outcome $Y$ is generated by the logistic regression with $P(Y=1\mid \bX)=\exp(\alpha_{0}+\bX^{\T}\bbeta_{0})/\{1+\exp(\alpha_{0}+\bX^{\T}\bbeta_{0})\}$ where $\alpha_{0}=0$ and $\bbeta_{0}=(0.25,0.5,0.75,1,1,1,1,1,1)^{\T}$. 
The parameter of interest is $\btheta_{0}=(\alpha_{0},\bbeta_{0}^{\T})^{\T}$.
We fix the whole data size $N= 10^{7}$ and take the subsample size $n$ to be $100$, $1000$, and $2000$,  respectively. 

\begin{figure}[!ht]
	\centering
	\includegraphics[scale=0.75]{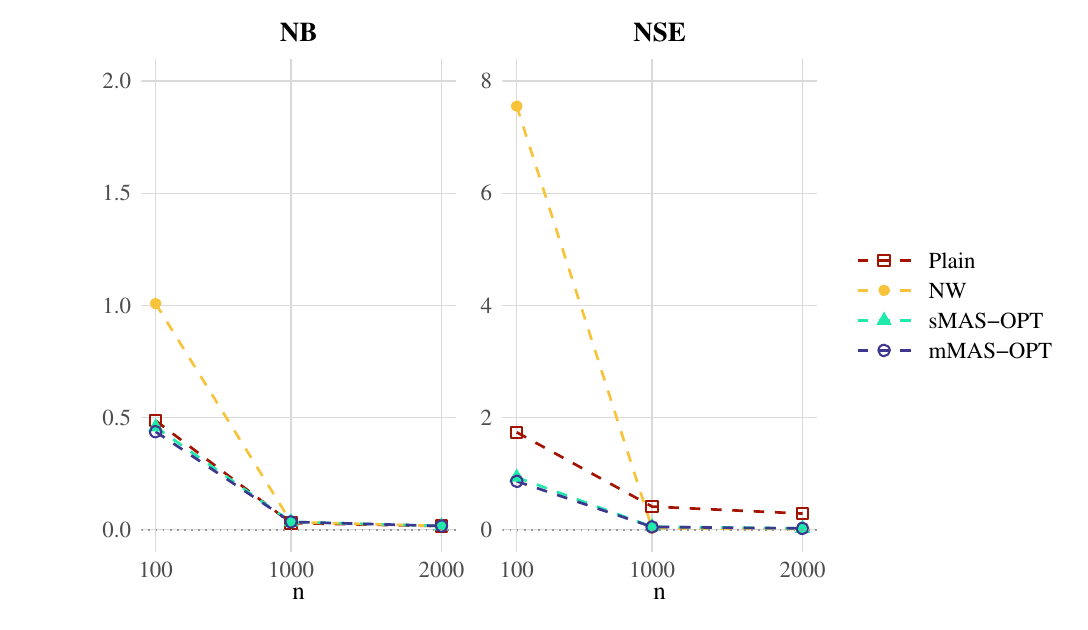}
	\caption{NB and NSE of  different estimators under the logistic regression model.}
	\label{fig: logit bias se nw}
\end{figure}

Figure \ref{fig: logit bias se nw} plots the NB and NSE of different estimators based on $1000$ repetitions. 
From Fig.~\ref{fig: logit bias se nw}, it can be seen that the NB and NSE of the one-step Newton estimator are significantly larger than those of the plain estimator and the MAS estimator for small subsample sizes.
For large subsample sizes, the MAS estimators and the one-step Newton estimator perform similarly and better than the plain estimator.
This is consistent with our discussion about one-step Newton estimator in the main text.
When the subsample size is small, the initial estimator and the estimations of hessian matrix and gradient are not accurate.
The MAS estimators adaptively adjust the updated step to ensure the better performance for small subsample sizes.
The discussion details can be found in Appendix \ref{app: Newton discuss}.
Table \ref{tab: logit nw time n} presents the CPU times of above estimators. The computing time of the MAS estimators is about a third of that of the one-step Newton estimator.

\begin{table}[!ht]
	\centering
	\caption{\label{tab: logit nw time n}CPU times (second) of different estimators under the logistic regression model.}
	\begin{tabular}{ccccccccccccccc}
		\toprule
		$n$&\footnotesize Plain& NW&\footnotesize sMAS-OPT&\footnotesize mMAS-OPT\\
		\midrule
		100 & 0.782 & 6.112 &  2.045 & 2.073\\
		1000 & 0.849 &6.224   & 2.156  &  2.185 \\
		2000   &0.871&6.433    & 2.245  &  2.274\\
		\bottomrule
	\end{tabular}
\end{table}

We next compared the proposed MAS estimators with the one-step Newton estimator under the mixed effects logistic model.
We consider the mixed effects logistic model whose likelihood function involves integrals.
Suppose there are $N$ clusters and $3$ observations in each cluster.
The random effects $W_{i},i=1,\cdots, N$ are independently and identically distributed from $N(0,\sigma_{0}^{2})$ with $\sigma_{0}=1$.
Let $\bX_{i} = (\bX_{i}^{(1)\T},\bX_{i}^{(2)\T},\bX_{i}^{(3)\T})^{\T}$ where $\bX_{i}^{(j)} = (X_{i1}^{(j)},\cdots,X_{i9}^{(j)})^{\T}$ for $j=1,2,3$ and $X_{ik}^{(j)}$ for $k=1,\cdots,9$ are independently and identically distributed from $U(-1,1)$.
Conditional on $W_{i}$ and $\bX_{i}$, the outcome vector $\bY_{i} = (Y_{i}^{(1)},Y_{i}^{(2)},Y_{i}^{(3)})^{\T}$ has binary components generated from the mixed effects logistic model $P(Y_{i}^{(j)}\mid \bX_{i}^{(j)},W_{i}) = \exp(\alpha_{0}+\bX_{i}^{(j)\T}\bbeta_{0}+W_{i})/\{1+\exp(\alpha_{0}+\bX_{i}^{(j)\T}\bbeta_{0}+W_{i})\}$ for $j=1,2,3$ and $i=1,\cdots,N$, where $\alpha_{0}=0$ and $\bbeta_{0}$ is a $9$-dimensional vector with each component equaling $0.2$.
The parameter of interest is $\btheta_{0} = (\sigma_{0},\alpha_{0},\bbeta_{0}^{\T})^{\T}$.
We fix the whole data size $N= 2\times 10^{5}$ and take the subsample size $n$ to be $100$, $1000$, and $2000$,  respectively. 
The computing time of one-step Newton estimator under the subsample size $n=100$ is 1753.56 seconds which took more than half an hour.
In contrast, the computing time of the MAS estimators was less than 30 seconds.
The reason is that the update in one-step Newton estimator requires the calculation of hessian matrix and gradient involving integral operation over the whole data.
The MAS strategy avoids the problem by using an approximation of the optimal moment function.
Therefore, we calculate a variant of the one-step Newton estimator, denoted by NW-APP, which replaces the score function in gradient with the same approximation as the MAS estimators, for fair comparison.

\begin{figure}[!ht]
	\centering
	\includegraphics[scale=0.75]{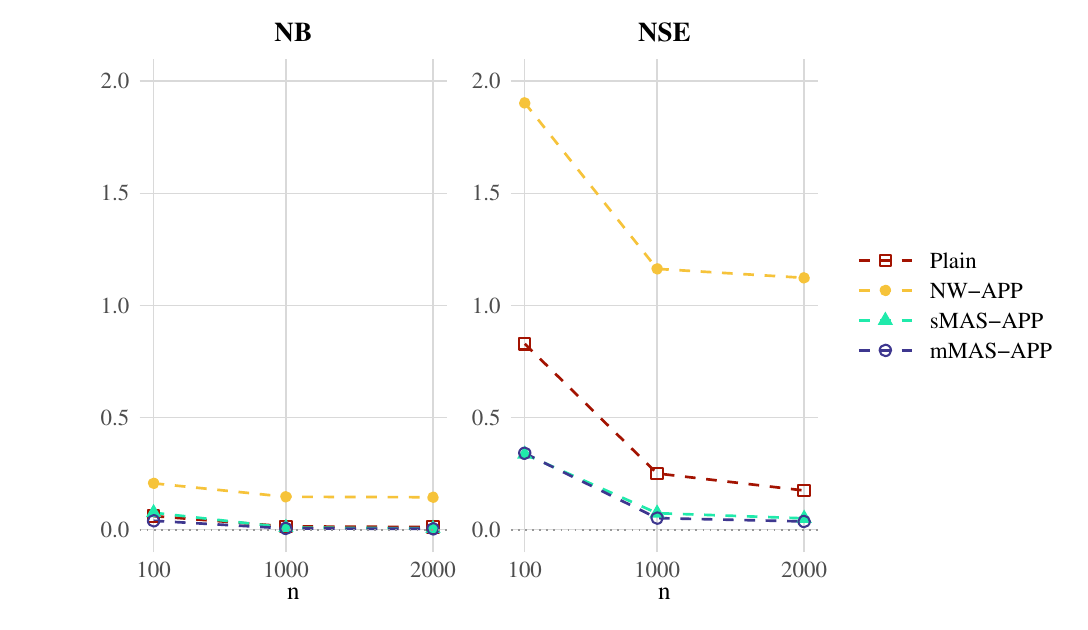}
	\caption{NB and NSE of different estimators under the mixed effects logistic regression.}
	\label{fig: bias se nw}
\end{figure}

Figure~\ref{fig: bias se nw} plots the NB and NSE of the above estimators.
From Fig.~\ref{fig: bias se nw}, it can be seen that the plain estimator and two MAS estimators have similar and small biases for all subsample sizes while NW-APP has quite large biases.
This phenomenon suggests that the consistency of NW-APP cannot be guaranteed.
The NSE of the two MAS estimators is significantly smaller than that of plain subsampling estimator and the estimator NW-APP for all subsample sizes, especially for small subsample size. Table \ref{tab: nw time n} presents the CPU times of the four estimators under different subsample sizes.
The computing time of NW-APP and the MAS estimators is similar to that of the plain estimator.
However, NW-APP performs badly in terms of estimation accuracy.

\begin{table}[!ht]
	\centering
	\caption{\label{tab: nw time n}CPU times (second) of different estimators under the mixed effects logistic regression.}
	\begin{tabular}{ccccccccccccccc}
		\toprule
		$n$&\footnotesize Plain& NW-APP&\footnotesize sMAS-APP&\footnotesize mMAS-APP\\
		\midrule
		100 & 1.480 &1.944  &2.129  & 2.161\\
		1000 & 8.749 &9.497 &10.097 &11.732\\
		2000   &17.356 & 18.076  &19.175&22.392\\
		\bottomrule
	\end{tabular}
\end{table}

\subsection{Optimal Moment Function under Weibull Regression Model}\label{app: sim Weibull}

Here we provide the explicit expression of $\int\tilde{\bh}^{\rm opt}(\bx,\by)f(y\mid \bx; \btheta)dy$ such that the modified MAS estimator with the moment function $\tilde{\bh}^{\rm opt}(\bx,\by)$ can be calculated quickly. Suppose $\tilde{\btheta} = (\tilde{\alpha},\tilde{\bbeta}^{\T})^{\T}$ is the plain subsampling estimator for $\btheta_{0}$.
Under the Weibull regression model, we have 
\begin{equation*}
	\tilde{\bh}^{\rm opt}(\bx,\by)=\begin{pmatrix}
		\frac{1}{\tilde{\alpha}}+\log(y)-y^{\tilde{\alpha}}\log(y)\exp\{(1,\bx^{\T})\tilde{\bbeta}\}\\
		[1-y^{\tilde{\alpha}}\exp\{(1,\bx^{\T})\tilde{\bbeta}\}](1,\bx^{\T})^{\T}
	\end{pmatrix},
\end{equation*}
and
\begin{equation*}
	\begin{split}
		&\int\tilde{\bh}^{\rm opt}(\bx,\by)f(y\mid \bx; \btheta)dy\\
		&=\begin{pmatrix}
			\frac{1}{\tilde{\alpha}}-\frac{\gamma}{\alpha}-\frac{(1,\bx^{\T})\bbeta}{\alpha}-\{\frac{\Gamma(1+\tilde{\alpha}/\alpha){\rm Psi}(1+\tilde{\alpha}/\alpha)}{\alpha}-\frac{(1,\bx^{\T})\bbeta}{\alpha}\}\exp\{-\frac{(1,\bx^{\T})\bbeta}{\alpha}\tilde{\alpha}+(1,\bx^{\T})\tilde{\bbeta}\}\\
			[1-\Gamma(1+\tilde{\alpha}/\alpha)\exp\{-\frac{(1,\bx^{\T})\bbeta}{\alpha}\tilde{\alpha}+(1,\bx^{\T})\tilde{\bbeta}\}](1,\bx^{\T})^{\T}
		\end{pmatrix},
	\end{split}
\end{equation*}
where $\gamma$ is Euler-Mascheroni constant, $\Gamma(\cdot)$ is Gamma function, and ${\rm Psi}(\cdot)$ is Psi function.

\subsection{Model Details of the Two Level GLMM for Car Crash Data}\label{app: detail car crash data}

Suppose there are $N$ crashes. 
For the $i$th crash, there are $n_{i}$ cars.
For the $j$th car in the $i$th crash, there are $n_{ij}$ people.
Denote the outcome and covariate vector of the $k$th people in the $j$th car of the $i$th crash as $Y_{i}^{(jk)}$ and $\bX_{i}^{(jk)}$, respectively.
Let 
\[
\bY_{i} = \left(Y_{i}^{(11)},\cdots,Y_{i}^{(1n_{i1})},\cdots,Y_{i}^{(n_{i}1)},\cdots,Y_{i}^{(n_{i}n_{in_{i}})}\right)^{\T}
\] 
and 
\[
\bX_{i}=\left(\bX_{i}^{(11)\T},\cdots,\bX_{i}^{(1n_{i1})\T},\cdots,\bX_{i}^{(n_{i}1)\T},\cdots,\bX_{i}^{(n_{i}n_{in_{i}})\T},n_{i},n_{i1},\cdots,n_{in_{i}}\right)^{\T}.
\]
We build a two-level mixed effects logistic model as follows
\begin{equation*}
	f(\by_{i}\mid\bx_{i};\btheta)=\int\prod_{j=1}^{n_{i}}\int\prod_{k=1}^{n_{ij}}\exp\left[y_{i}^{(jk)}\eta_{i}^{(jk)}(u,v)-b\{\eta_{i}^{jk}(u,v)\}\right]q(v;\sigma_{1})dv q(u;\sigma_{2})du,
\end{equation*}
where $\btheta = (\bbeta^{\T},\sigma_{1},\sigma_{2})^{\T}$, $\eta_{i}^{(jk)}(u,v) = \bx_{i}^{(jk)}\bbeta+v+u,b(x) = \log(1+e^x)$, $q(\cdot,\sigma)$ is the density function of normal  distribution with mean zero and standard deviations $\sigma$.
The corresponding quasi-log likelihood function $l(\btheta)$ is
\[
\begin{aligned}
	\sum_{i=1}^{N}\log\int\prod_{j=1}^{n_{i}}\int\exp\left\{\sum_{k=1}^{n_{ij}}\left(\eta_{i}^{(jk)}(u,v)-\log[1+\exp\{\eta_{i}^{(jk)}(u,v)\}]\right)\right\}q(v;\sigma_{1})dv q(u;\sigma_{2})du.
\end{aligned}
\]
It can be seen that the likelihood function and hence the score function involves double integral operations.
The whole data-based MLE is quite time-consuming which has been demonstrated in the main text.
Under the two level mixed effects logistic regression, the whole data sample moment based on the optimal moment function also involves the integral operation.
We consider its approximation by using the score function of the logistic regression, a reduced model of the mixed effects logistic regression.
Let
\[
\bq(\bx, \by; \bxi) = \sum_{j=1}^{n_{i}} \sum_{k=1}^{n_{ij}}\left[y_{i}^{(jk)}\bxi^{\T}\bx_{i}^{(jk)}-\log\{1+\exp(\bxi^{\T}\bx_{i}^{(jk)})\}\right].
\]
We take the moment function $\tilde{\bh}^{\rm app}(\bx,\by)= \bq(\bx, \by; \tilde{\bxi})$ with $\tilde{\bxi}$ obtained in the way introduced in Section \ref{subsec: approx hopt}.

\vskip 0.2in

\bibliographystyle{asa}
\bibliography{mas}
\end{document}